\documentclass[12pt]{article}
\pdfoutput=1
\usepackage{putex}
\usepackage{mathtools}
\usepackage{mathrsfs}
\usepackage{graphicx}
\usepackage{caption}
\usepackage{subcaption}
\usepackage{epstopdf}
\usepackage{enumerate}
\usepackage{cite}
\usepackage{youngtab}
\usepackage{tensor}
\usepackage{slashed}
\usepackage[aligntableaux=center]{ytableau}
\usepackage{url}
\usepackage{bbold}

\usepackage{enumitem} 
\setlist[itemize]{leftmargin=*}
\setlist[enumerate]{leftmargin=*}

\numberwithin{equation}{section}
\allowdisplaybreaks 
\newcommand{\abs}[1]{\left\lvert #1 \right\rvert}

\newcommand {\be} {\begin {equation}}
\newcommand {\ee} {\end {equation}}

\newcommand {\bes} {\begin {equation*}}
\newcommand {\ees} {\end {equation*}}

\newcommand{\es}[2] {\begin{equation} \label{#1} \begin{split} #2 \end{split} \end{equation}}

\newcommand{\Z}{\mathbb{Z}}

\newcommand{\R}{\mathbb{R}}
\newcommand{\C}{\mathbb{C}}

\def\Tr{\mop{Tr}}

\newcommand{\beq}{\begin{equation}}
\newcommand{\eeq}{\end{equation}}

\def\<{\langle}
\def\>{\rangle}

\newcommand{\bC}{\ensuremath{\mathbb{C}}}

\newcommand{\cA}{\ensuremath{\mathcal{A}}}
\newcommand{\cB}{\ensuremath{\mathcal{B}}}

\newcommand{\cD}{\ensuremath{\mathcal{D}}}

\newcommand{\cH}{\ensuremath{\mathcal{H}}}
\newcommand{\cI}{\ensuremath{\mathcal{I}}}
\newcommand{\cJ}{\ensuremath{\mathcal{J}}}

\newcommand{\cM}{\ensuremath{\mathcal{M}}}
\newcommand{\cN}{\ensuremath{\mathcal{N}}}
\newcommand{\cO}{\ensuremath{\mathcal{O}}}
\newcommand{\cP}{\ensuremath{\mathcal{P}}}
\newcommand{\cQ}{\ensuremath{\mathcal{Q}}}
\newcommand{\cR}{\ensuremath{\mathcal{R}}}

\newcommand{\cU}{\ensuremath{\mathcal{U}}}
\newcommand{\cV}{\ensuremath{\mathcal{V}}}
\newcommand{\cW}{\ensuremath{\mathcal{W}}}

\newcommand{\dd}{\mathrm{d}}

\newcommand{\tpsi}{\widetilde{\psi}}
\newcommand{\tq}{\widetilde{q}}
\newcommand{\tQ}{\widetilde{Q}}

\newcommand{\vphi}{\varphi}

\usepackage{amsthm}

\theoremstyle{plain}
\newtheorem{theorem}{Theorem}
\newtheorem{proposition}{Proposition}
\newtheorem*{example}{Example}

\theoremstyle{definition}
\newtheorem{definition}{Definition}

\begin{document}

\preprint{CALT-TH 2018-056\\ PUPT-2579}

\institution{Caltech}{Walter Burke Institute for Theoretical Physics, California Institute of Technology, \cr Pasadena, CA 91125, USA}
\institution{PU}{Department of Physics, Princeton University, Princeton, NJ 08544, USA}
\institution{Weizmann}{Department of Particle Physics and Astrophysics, Weizmann Institute of Science, \cr Rehovot 76100, Israel}

\title{Coulomb Branch Quantization and \\ Abelianized Monopole Bubbling}

\authors{Mykola Dedushenko,\worksat{\Caltech} Yale Fan,\worksat{\PU} Silviu S.~Pufu,\worksat{\PU} and Ran Yacoby\worksat{\Weizmann}}

\abstract{
We develop an approach to the study of Coulomb branch operators in 3D $\mathcal{N}=4$ gauge theories and the associated quantization structure of their Coulomb branches. This structure is encoded in a one-dimensional TQFT subsector of the full 3D theory, which we describe by combining several techniques and ideas. The answer takes the form of an associative and noncommutative star product algebra on the Coulomb branch. For ``good'' and ``ugly'' theories (according to the Gaiotto-Witten classification), we also exhibit a trace map on this algebra, which allows for the computation of correlation functions and, in particular, guarantees that the star product satisfies a truncation condition. This work extends previous work on abelian theories to the non-abelian case by quantifying the monopole bubbling that describes screening of GNO boundary conditions. In our approach, monopole bubbling is determined from the algebraic consistency of the OPE. This also yields a physical proof of the Bullimore-Dimofte-Gaiotto abelianization description of the Coulomb branch.
}

\date{}

\maketitle

\tableofcontents
\setlength{\unitlength}{1mm}

\newpage

\section{Introduction}

Gauge theories in three dimensions contain special local defect operators called monopole operators, which are defined by requiring certain singular behavior of the gauge field close to the insertion point \cite{Borokhov:2002ib}.  These operators play important roles in the dynamics of these theories, and in particular in establishing various interesting properties such as infrared (IR) dualities between theories with different ultraviolet (UV) descriptions (see \cite{Son:2015xqa, Aharony:2015mjs, Karch:2016sxi, Murugan:2016zal, Seiberg:2016gmd, Hsin:2016blu, Radicevic:2016wqn, Kachru:2016rui, Kachru:2016aon, Karch:2016aux, Metlitski:2016dht, Aharony:2016jvv, Benini:2017dus, Komargodski:2017keh} for some recent examples).  Because these operators are not polynomial in the Lagrangian fields, they are notoriously difficult to study, and most studies so far have focused on determining only their quantum numbers \cite{Murthy:1989ps, Borokhov:2002ib, Metlitski:2008dw, Pufu:2013vpa, Dyer:2013fja, Dyer:2015zha, Chester:2015wao, 2013PhRvL.111m7202B, 2015arXiv150205128K, Benini:2009qs, Benini:2011cma, Imamura:2011su, Kim:2009wb, Aharony:2015pla, Kapustin:2009kz}.  The goal of this paper is to present the first direct computations of operator product expansion (OPE) coefficients and correlation functions of monopole operators in 3D non-abelian gauge theories.

We focus on a class of 3D gauge theories with $\cN  = 4$ supersymmetry (eight Poincar\'e supercharges) constructed by coupling a vector multiplet with gauge group $G$ to a matter hypermultiplet that transforms in some representation of $G$.\footnote{These theories do not allow the presence of Chern-Simons terms.  While it is possible to construct $\cN = 4$ Chern-Simons-matter theories \cite{Gaiotto:2008ak,Aharony:2008ug, Aharony:2008gk,Imamura:2008dt, Hosomichi:2008jd}, we do not study them here.}  For a matter representation of sufficiently large dimension, these theories flow in the IR to interacting superconformal field theories (SCFTs), whose correlation functions are generally intractable.  However, as shown in \cite{Chester:2014mea,Beem:2016cbd}, these theories also contain one-dimensional protected subsectors whose correlation functions are topological, and one may hope that computations in these protected subsectors become tractable.  This is indeed the case, as was shown in \cite{Dedushenko:2016jxl, Dedushenko:2017avn} and as will be explored in further detail here.  While 3D $\cN = 4$ SCFTs have in general two inequivalent protected topological sectors, one associated with the Higgs branch and one with the Coulomb branch, it is the Coulomb branch sector that contains monopole operators and that will therefore be the focus of our work. (The Higgs branch sector was studied in \cite{Dedushenko:2016jxl}.) From the 3D SCFT point of view, the information contained in either of the two protected sectors is equivalent to that contained in the $(n \leq 3)$-point functions of certain half-BPS local operators in the SCFT \cite{Chester:2014mea, Beem:2016cbd, Dedushenko:2016jxl, Dedushenko:2017avn}. 

The Coulomb branch protected sector consists of operators that belong to the cohomology of a supercharge $\cQ^C$ that is a linear combination of a Poincar\'e and a conformal supercharge.\footnote{Similar statements hold about the Higgs branch protected sector if one replaces $\cQ^C$ with another supercharge $\cQ^H$.}  As such, one may think that the protected sector mentioned above is emergent at the IR fixed point, and therefore inaccessible in the UV description.  This is indeed true for SCFTs defined on $\R^3$.  However, as was shown in \cite{Dedushenko:2016jxl,Dedushenko:2017avn}, if one defines the QFT on a round $S^3$ instead of on $\R^3$, then the protected sector becomes accessible in the UV because on $S^3$, the square of $\cQ^C$ does not contain special conformal generators.  Indeed, Poincar\'e and special conformal generators are mixed together when mapping a CFT from $\R^3$ to $S^3$.  As we will explain, the square of $\cQ^C$ includes an isometry of $S^3$ that fixes a great circle, and this is the circle where the 1D topological quantum field theory (TQFT) lives.

Previous work \cite{Dedushenko:2016jxl} used the idea of defining the QFT on $S^3$ together with supersymmetric localization to solve the 1D Higgs branch theory by describing a method for computing its structure constants.  The Coulomb branch case is much more complicated because it involves monopole operators.  A complete solution of the 1D Coulomb branch theory was obtained for abelian gauge theories in \cite{Dedushenko:2017avn}.  Building on the machinery developed in \cite{Dedushenko:2017avn}, we describe how to compute all observables within the 1D Coulomb branch topological sector of an arbitrary non-abelian 3D $\mathcal{N} = 4$ gauge theory by constructing ``shift operators'' whose algebra is a representation of the OPE of the 1D TQFT operators (also known as twisted(-translated) Coulomb branch operators, for reasons that will become clear). 

The mathematical physics motivation for studying the 1D TQFT is that it provides a ``quantization'' of the ring of holomorphic functions defined on the Coulomb branch $\cM_C$.  This can be explained as follows.  The 3D theories that we study have two distinguished branches of the moduli space of vacua:  the Higgs branch and the Coulomb branch.  These are each parametrized, redundantly, by VEVs of gauge-invariant chiral operators whose chiral ring relations determine the branches as generically singular complex algebraic varieties.  While the Higgs branch chiral ring relations follow from the classical Lagrangian, those for the Coulomb branch receive quantum corrections.  The Coulomb branch is constrained by extended SUSY to be a generically singular hyperk\"ahler manifold of quaternionic dimension equal to the rank of $G$, which, with respect to a choice of complex structure, can be viewed as a complex symplectic manifold.  The half-BPS operators that acquire VEVs on the Coulomb branch, to be referred as Coulomb branch operators (CBOs), consist of monopole operators, their dressings by vector multiplet scalars, and operators built from the vector multiplet scalars themselves. (Monopole operator VEVs encode those of additional scalar moduli, the dual photons.) All of the holomorphic functions on $\cM_C$ are given by VEVs of the subset of CBOs that are chiral with respect to an $\cN=2$ subalgebra. Under the OPE, these operators form a ring, which is well-known to be isomorphic to the ring $\bC[\cM_C]$ of holomorphic functions on $\cM_C$.  It was argued in \cite{Beem:2016cbd} that because the operators in the 1D TQFT are in one-to-one correspondence with chiral ring CBOs, the 1D TQFT is a deformation quantization of $\bC[\cM_C]$.  Indeed, the 1D OPE induces an associative but noncommutative product on $\bC[\cM_C]$ referred to as a star product, which in the limit $r \to \infty$ ($r$ being the radius of $S^3$) reduces to the ordinary product of the corresponding holomorphic functions, and that at order $1/r$ gives the Poisson bracket of the corresponding holomorphic functions.

Note that both the quantization of \cite{Beem:2016cbd} in the ``$Q+S$'' cohomology and our quantization on a sphere are realizations of the older idea of obtaining a lower-dimensional theory by passing to the equivariant cohomology of a supercharge, which originally appeared in the context of the $\Omega$-deformation in 4D theories \cite{Nekrasov:2002qd, Nekrasov:2009rc, Nekrasov:2010ka} and was also applied to 3D theories in \cite{Yagi:2014toa,Bullimore:2016nji,Bullimore:2016hdc}.

\begin{figure}[t!]
\centering
\includegraphics[width=1.0\textwidth]{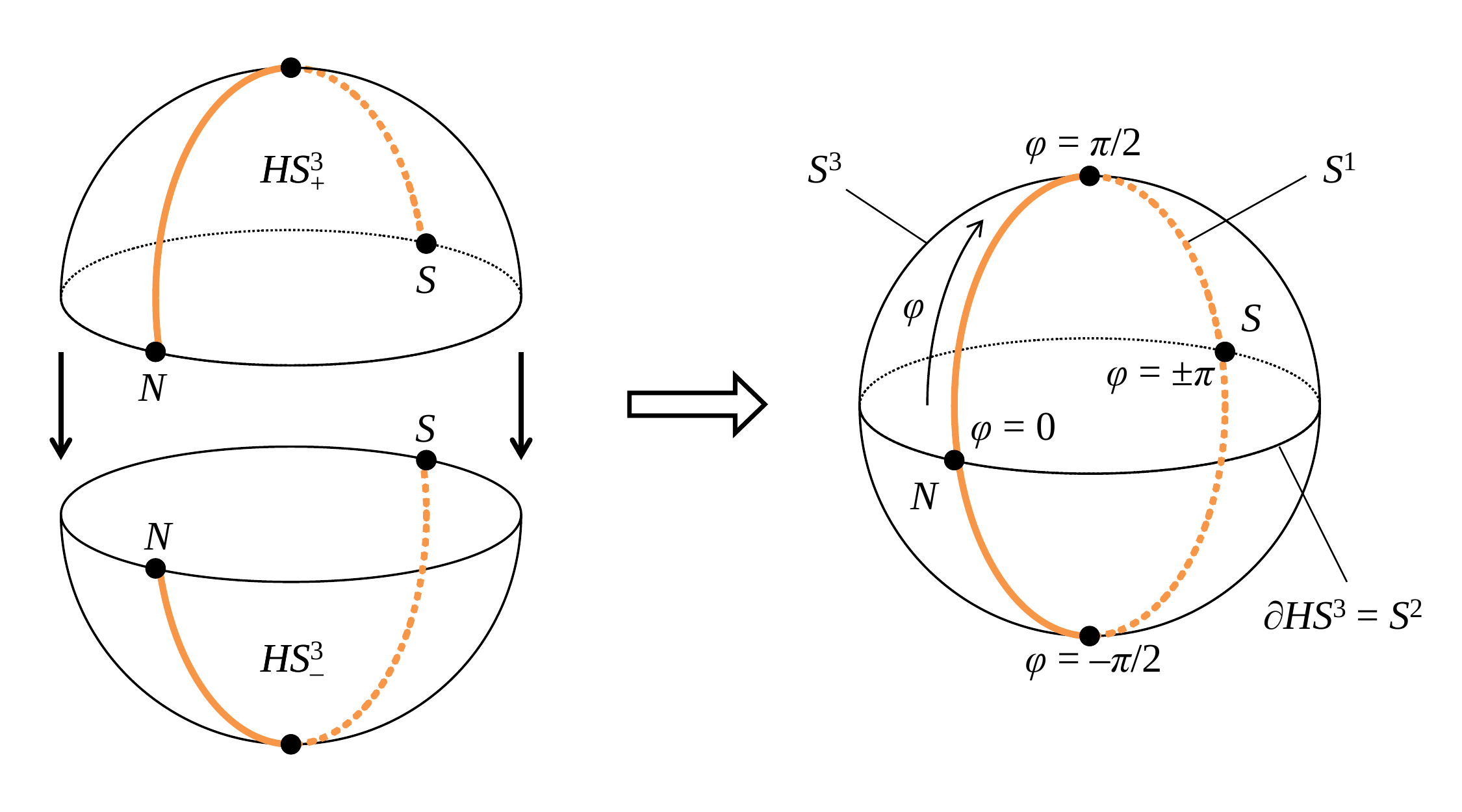}
\caption{\label{fig:sphere} A schematic depiction of $S^3$, obtained by gluing two hemispheres $HS^3_{\pm}\cong B^3$. The 1D TQFT lives on the $S^1$ parametrized by the angle $\varphi$ (thick orange line). The 1D TQFT circle intersects the equatorial $S^2=\partial HS^3_{\pm}$ at two points identified with its North ($N$) and South ($S$) poles.}
\end{figure}

Our procedure for solving the 1D Coulomb branch theory uses a combination of the cutting and gluing axioms, supersymmetric localization, and a consistency requirement that we refer to as polynomiality.  We first cut $S^3$ into two hemispheres $HS^3_{\pm}$ along an equatorial $S^2=\partial HS^3_{\pm}$ orthogonal to the circle along which the 1D operators live (see Figure~\ref{fig:sphere}). Correlators are then represented by an inner product of wavefunctions generated by the path integral on $HS^3_{\pm}$ with insertions of twisted CBOs. In \cite{Dedushenko:2017avn, Dedushenko:2018tgx}, it was shown that it suffices to consider such wavefunctions $\Psi_{\pm}(\cB_{\mathrm{BPS}})$ with operator insertions only at the tip of $HS^3_{\pm}$, and evaluated on a certain class of half-BPS boundary conditions $\cB_{\mathrm{BPS}}$. Insertions of twisted CBOs anywhere on the great semicircles of $HS^3_{\pm}$ can then be realized, up to irrelevant $\cQ^C$-exact terms, as simple shift operators acting on this restricted class of wavefunctions. It was shown in \cite{Dedushenko:2017avn} that these shift operators can be fully reconstructed from general principles and knowledge of $\Psi_{\pm}(\cB_{\mathrm{BPS}})$. Moreover, their algebra provides a faithful representation of the star product.  Finally, one can determine expectation values (which is to say, more abstractly, that one can define an evaluation map on $\bC[\cM_C]$, known as the trace map in deformation quantization) by gluing $\Psi_+(\cB_{\mathrm{BPS}})$ and $\Psi_-(\cB_{\mathrm{BPS}})$ with an appropriate measure, as will be reviewed in Section~\ref{sec:glue_f}.

The fact that the star product algebra can be determined independently of evaluating correlators is very useful. First, calculating correlators using the above procedure involves solving matrix integrals, which can be complicated for gauge groups of high rank. On the other hand, the star product can be inferred from the comparatively simple calculation of the wavefunctions $\Psi_{\pm}(\cB_{\mathrm{BPS}})$. Second, the matrix models representing correlators diverge for ``bad'' theories in the sense of Gaiotto and Witten \cite{Gaiotto:2008ak}.\footnote{In bad theories \cite{Yaakov:2013fza, Assel:2017jgo, Dey:2017fqs, Assel:2018exy}, the IR superconformal R-symmetry is not visible in the UV, which invalidates the usual localization logic \cite{Jafferis:2010un, Hama:2010av}.} Nevertheless, as we will see, the $HS^3$ wavefunctions and the star product extracted from them are well-defined even in those cases. Therefore, we emphasize that our formalism works perfectly well even for bad theories, as far as the Coulomb branch and its deformation quantization are concerned. However, correlation functions cannot be computed for such theories, and the star products might not satisfy the truncation property introduced in \cite{Beem:2016cbd}. 

On a more technical note, we provide a new way of analyzing ``monopole bubbling'' \cite{Kapustin:2006pk}.   Monopole bubbling is a phenomenon whereby the charge of a singular monopole is screened to a lower one by small 't Hooft-Polyakov monopoles.  In our setup, this phenomenon manifests itself through the fact that our shift operators for a monopole of given charge contain contributions proportional to those of monopoles of smaller charge, with coefficients that we refer to as bubbling coefficients.  While we do not know of a localization-based algorithm for obtaining these coefficients in general, we propose that the requirement that the OPE of any two 1D TQFT operators should be a polynomial in the 1D operators uniquely determines the bubbling coefficients, up to operator mixing ambiguities.  In Section~\ref{sec:poly}, we provide many examples of gauge theories of small rank where we explicitly carry out our algorithm to determine the shift operators and bubbling coefficients. These results are also interesting for the purpose of comparison with the literature on direct localization computations of bubbling in 4D, e.g., \cite{Gomis:2011pf, Ito:2011ea, Gang:2012yr}, which were subsequently refined by \cite{Brennan:2018yuj, Brennan:2018moe, Brennan:2018rcn}.
 
The main mathematical content of this work is a construction of deformation quantizations of Coulomb branches of 3D $\cN=4$ theories that also satisfy the truncation condition of \cite{Beem:2016cbd} in the case of good or ugly theories, as a consequence of the existence of the natural trace map (the one-point function).\footnote{Such star products are also called ``short'' in ongoing mathematical work on their classification, as we learned from P.~Etingof.} By taking the commutative limit, we recover the ordinary Coulomb branch of the theory in the form of the ``abelianization map'' proposed by \cite{Bullimore:2015lsa}. Therefore, our approach also provides a way to prove the abelianization proposal of \cite{Bullimore:2015lsa} starting from basic physical principles.  Moreover, the knowledge of bubbling coefficients that our approach provides vastly expands the domain of applicability of abelianization to all Lagrangian 3D $\cN=4$ theories of cotangent type. Finally, we expect that translating our approach into a language that uses the mathematical definition of the Coulomb branch \cite{Nakajima:2015txa,Braverman:2016wma,Braverman:2016pwk,Braverman:2017ofm} might be of independent interest in the study of deformation quantization.
 
The rest of this paper is organized as follows.  Section~\ref{sec:ShiftOps} contains a review of the setup of our problem as well as a derivation of the shift operators without taking bubbling into account.  Section~\ref{sec:dress_and_bubble} discusses the dressing of monopole operators with vector multiplet scalars and sets up the computation of the bubbling coefficients.  In Section~\ref{sec:poly}, we provide explicit examples of shift operators and bubbling coefficients in theories of small rank.  In Section~\ref{sec:Appl}, we discuss several applications of our formalism:  to determining chiral rings, to chiral ring quantization, and to computing correlation functions of monopole operators and performing checks of non-abelian mirror symmetry.  Many technical details, further examples, and comments on connections between our approach and existing ones can be found in the appendices.

\section{Shift Operators}\label{sec:ShiftOps}

\subsection{Setup}

\subsubsection{Theories}

We study 3D $\cN=4$ gauge theories of cotangent type, which are the same theories whose quantized Higgs branches were the subject of \cite{Dedushenko:2016jxl}. Coulomb branches of abelian gauge theories were scrutinized in \cite{Dedushenko:2017avn} using different techniques, and here we extend those techniques to the case of general gauge groups $G\cong \prod_i G_i$, where each $G_i$ is either simple or abelian. As the construction of such theories was detailed in \cite{Dedushenko:2016jxl,Dedushenko:2017avn}, we only briefly describe it here.

These theories are built from a 3D $\cN=4$ vector multiplet $\cV$ taking values in the Lie algebra $\mathfrak{g}={\rm Lie}(G)$ and from a 3D $\cN=4$ hypermultiplet $\cH$ valued in a (generally reducible) representation $\cR$ of $G$.  $\cH$ can be written in terms of half-hypermultiplets taking values in $\cR\oplus \overline{\cR}$, which is the meaning of the term ``cotangent type.''  More general representations of half-hypermultiplets should also be possible to address using our techniques, but we do not consider them in the present work.

Our focus is on such theories supersymmetrically placed on the round $S^3$ of radius $r$. There are several good reasons for choosing this background. One is that compactness makes the application of supersymmetric localization techniques more straightforward. But the most important reason, as should be clear to readers familiar with \cite{Dedushenko:2016jxl, Dedushenko:2017avn}, is that the sphere is a natural setting for deformation quantization of moduli spaces: the Coulomb and Higgs branches in such a background can be viewed as noncommutative, with $1/r$ playing the role of a quantization parameter.  As with the 2D $\Omega$-background in flat space \cite{Bullimore:2015lsa}, the result is an effective compactification of spacetime to a line.

Furthermore, quantized Coulomb and Higgs branch chiral rings are directly related to physical correlation functions, and in particular encode the OPE data of the BPS operators in the IR superconformal theory, whenever it exists. This relation equips the noncommutative star product algebra of observables with a natural choice of ``trace'' operation --- the one-point function of the QFT --- as well as natural choices of basis corresponding to operators that are orthogonal with respect to the two-point function and have well-defined conformal dimensions at the SCFT point. These extra structures are a significant advantage of quantization using the spherical background, and they are responsible for much of the progress that we make in this paper.

The $\cN=4$ supersymmetric background on $S^3$ is based on the supersymmetry algebra $\mathfrak{s}=\mathfrak{su}(2|1)_\ell \oplus \mathfrak{su}(2|1)_r$, which also admits a central extension $\widetilde{\mathfrak{s}}=\widetilde{\mathfrak{su}(2|1)}_\ell \oplus \widetilde{\mathfrak{su}(2|1)}_r$, with central charges corresponding to supersymmetric mass and FI deformations of the theory. In the flat-space limit $r\to\infty$, this algebra becomes the usual $\cN=4$ super-Poincar\'e algebra, implying that all results of this paper should have a good $r\to\infty$ limit. All of the necessary details on the SUSY algebra $\mathfrak{s}$, and how the vector and hypermultiplets transform under it, can be found either in Section 2 of \cite{Dedushenko:2016jxl} or in Section 2.1 and Appendix A.2 of \cite{Dedushenko:2017avn}. Supersymmetric actions for $\cV$, $\cH$, and their deformations by mass and FI terms can also be found in those sections.

The SUSY algebra $\widetilde{\mathfrak s}$ contains two interesting choices of supercharge, $\cQ^H$ and $\cQ^C$. They satisfy the following relations:
\begin{equation}
(\cQ^H)^2 = \frac{4i}{r} (P_\tau + R_C + ir\widehat{\zeta}), \qquad (\cQ^C)^2 = \frac{4i}{r} (P_\tau + R_H + ir\widehat{m}),
\label{Qsquared}
\end{equation}
where $P_\tau$ denotes a $U(1)$ isometry of $S^3$ whose fixed-point locus is a great circle parametrized by $\varphi\in(-\pi,\pi)$: call it $S_\varphi^1\subset S^3$.\footnote{Concretely, $\tau$ is the fiber coordinate in an $S^1$ fibration over the disk $D^2$, i.e., $S_\tau^1\to S^3\to D^2$.  After conformally mapping to flat space, $P_\tau$ would be a rotation that fixes the image of $S_\varphi^1$, which is a line.} Here, $R_C$ and $R_H$ are the Cartan generators of the usual $SU(2)_C\times SU(2)_H$ R-symmetry of $\cN=4$ SUSY, which in terms of the inner $U(1)_\ell\times U(1)_r$ R-symmetry of $\mathfrak{s}$ are identified as:
\begin{align}
R_H = \frac12 (R_\ell + R_r),\qquad R_C=\frac12 (R_\ell - R_r).
\end{align}
The notations $\widehat{\zeta}$ and $\widehat{m}$ stand for the FI and mass deformations, i.e., central charges of $\widetilde{\mathfrak{s}}$.

The most important features of $\cQ^H$ and $\cQ^C$ are that if we consider their actions on the space of local operators and compute their equivariant cohomologies, the answers have very interesting structures. The operators annihilated by $\cQ^H$ are the so-called twisted-translated Higgs branch operators (HBOs), whose OPE encodes a quantization of the Higgs branch; such operators for the theories of interest were fully studied in \cite{Dedushenko:2016jxl}. Correspondingly, the cohomology of $\cQ^C$ contains twisted-translated Coulomb branch operators, whose structure has so far been explored only for abelian theories \cite{Dedushenko:2017avn}. Such operators must be inserted along the great circle $S^1_\varphi$ fixed by $(\cQ^C)^2$, and their OPE encodes a quantization of the Coulomb branch. More details on twisted-translated operators are given in Appendix \ref{sec:TTops}.

\subsubsection{Observables}\label{sec:observables}

The purpose of this work is to study the cohomology of $\cQ^C$ and associated structures for general non-abelian gauge theories of cotangent type. The operators annihilated by $\cQ^C$ are constructed from monopole operators and a certain linear combination of scalars in the vector multiplet.  Recall that the vector multiplet contains an $SU(2)_C$ triplet of scalars $\Phi_{\dot a \dot b}=\Phi_{\dot b\dot a}$.  Using the notation of \cite{Dedushenko:2017avn}, the following linear combination is annihilated by $\cQ^C$:
\begin{equation}
\Phi(\varphi)=\Phi_{\dot a\dot b}(\varphi)v^{\dot a}v^{\dot b}, \qquad v=\frac{1}{\sqrt 2}\left(\begin{matrix}
e^{i\varphi/2}\cr e^{-i\varphi/2}
\end{matrix} \right),
\label{twistedscalar}
\end{equation}
whenever this operator is inserted along $S^1_\varphi\subset S^3$.  On the other hand, (bare) BPS monopole operators are defined as defects imposing special boundary conditions on the gauge field and on $\Phi_{\dot a\dot b}$. They were first defined for 3D $\cN=4$ theories in \cite{Borokhov:2002cg}, while the non-supersymmetric version was introduced earlier in \cite{Borokhov:2002ib}. The twisted-translated monopole operators that we study --- which are essentially those of \cite{Borokhov:2002cg} undergoing an additional $SU(2)_C$ rotation as we move along $S^1_\varphi$ --- were described in detail in \cite{Dedushenko:2017avn}.  Their definition is rather intricate, so it will be helpful to review it, with an eye toward the additional complications that arise in non-abelian gauge theories.

First recall that in a $U(1)$ gauge theory, a (bare) non-supersymmetric monopole operator is a local defect operator that sources magnetic flux at a point in 3D spacetime. In a non-abelian gauge theory, the quantized charge $b$ is promoted to a matrix, or more precisely, a cocharacter of $G$ (referred to as the GNO charge \cite{Goddard:1976qe}).\footnote{This is a more refined notion than the topological charge labeled by $\pi_1(G)$ (when it exists): such charges correspond to global symmetries of the Coulomb branch whose conserved currents in the UV are the abelian field strengths and which may be enhanced in the IR.}  A cocharacter is an element of ${\rm Hom}(U(1) ,G)/G \cong {\rm Hom}(U(1),\mathbb{T})/\cW$. Passing from the element of ${\rm Hom}(U(1),\mathbb{T})/\cW$ to the map of algebras $\R\to\mathfrak{t}$, we see that cocharacters can also be identified with Weyl orbits in the coweight lattice $\Lambda^\vee_w\subset\mathfrak{t}$ of $G$, i.e., in the weight lattice of the Langlands dual group $^LG$. Since every Weyl orbit contains exactly one dominant weight (lying in the fundamental Weyl chamber), it is conventional to label monopole charges by dominant weights of $^LG$ \cite{Bullimore:2015lsa}. Let $b\in\mathfrak{t}$ be such a dominant weight of $^LG$.  Then a bare monopole operator is defined by a sum over $\cW b$, the Weyl orbit of $b$, of path integrals with singular boundary conditions defined by elements of $\cW b$.  Specifically, the insertion of a twisted-translated monopole operator at a point $\varphi\in S^1_\varphi$ is defined by the following singular boundary conditions for $F_{\mu\nu}$ and $\Phi_{\dot a \dot b}$:
\begin{align}
\label{monsing}
\ast F\sim b \frac{y_\mu \dd y^\mu}{|y|^3},\qquad \Phi_{\dot1 \dot1}=-(\Phi_{\dot2\dot2})^\dagger\sim - \frac{b}{2|y|}e^{-i\varphi},\qquad \Phi_{\dot1\dot2}\sim 0,
\end{align}
where it is understood that one must compute not a single path integral, but rather a sum of path integrals over field configurations satisfying \eqref{monsing} with $b$ ranging over the full Weyl orbit of a given dominant weight.  Here, ``$\sim$'' means ``equal up to regular terms'' and $y^\mu$ are Riemann normal coordinates centered at the monopole insertion point.  The origin of \eqref{monsing} is that twisted-translated monopoles are chiral with respect to the $\mathcal{N} = 2$ subalgebra defined by the polarization vector in \eqref{twistedscalar} at any given $\varphi$. This requires that the real scalar in the $\mathcal{N} = 2$ vector multiplet diverge as $\frac{b}{2|y|}$ near the monopole \cite{Borokhov:2002cg} and results in nontrivial profiles for the $\mathcal{N} = 4$ vector multiplet scalars near the insertion point.

We denote such twisted-translated monopole operators by $\cM^b(\varphi)$, or simply $\cM^b$. The $\cQ^C$-cohomology, in addition to $\cM^b(\varphi)$ and gauge-invariant polynomials in $\Phi(\varphi)$, contains monopole operators dressed by polynomials $P(\Phi)$, or dressed monopoles, which we denote by $[P(\Phi)\cM^b]$. Note that because monopoles are really given by sums over Weyl orbits, the notation $[P(\Phi)\cM^b]$ is not merely a product of $P(\Phi)$ and $\cM^b$, but rather:
\begin{equation}
\label{dressedmono}
\left[P(\Phi)\cM^b\right] = \frac1{|\cW_b|}\sum_{{\rm w}\in\cW}P(\Phi^{\rm w})\times (\text{insertion of a charge-(${\rm w}\cdot b$) monopole singularity}),
\end{equation}
where $\Phi^{\rm w}$ means that as we sum over the Weyl orbit, we act on the $P(\Phi)$ insertion as well. Because $\cM^b$ breaks the gauge group at the insertion point down to the subgroup $G_b\subset G$ that preserves $b$, $P(\Phi)$ must be invariant under the $G_b$ action.\footnote{We will see that after localization, $\Phi$ takes values in the Cartan subalgebra $\mathfrak{t}_\C$, in which case the $G_b$-invariance of $P(\Phi)$ boils down to $\cW_b$-invariance, where $\cW_b$ is the Weyl group of $G_b$. But then, because \eqref{dressedmono} includes summation over the Weyl orbit of $\cW$, there is no real need to require $\cW_b$-invariance of $P(\Phi)$, as it will be automatically averaged over the subgroup $\cW_b\subset \cW$ upon this summation. Therefore, later on, when we write formulas in terms of $\Phi\in\mathfrak{t}_\C$, we can insert arbitrary polynomials $P(\Phi)$ in $[P(\Phi)\cM^b]$.\label{Wnoninv}} Also, to avoid overcounting, we must divide by the order of the stabilizer of $b$ in $\cW$.\footnote{The factor $|\cW_b|^{-1}$ only appears when we sum over elements of $\cW$, while equations written directly in terms of a sum over the Weyl orbit do not need such a factor.}

At this point, we pause to discuss a few subtleties inherent to the above definition. They are important for precise understanding, and ultimately for performing computations correctly, but may be skipped on first reading.

First, let us ask ourselves what exactly $\Phi^{\rm w}$ is. After all, the Weyl group acts canonically on the Cartan subalgebra $\mathfrak{t}$, but it does not have a natural action on the full Lie algebra $\mathfrak{g}$ where $\Phi$ is valued. Indeed, from the identification $\cW=N(\mathbb{T})/Z(\mathbb{T})$, a Weyl group element is interpreted as an element of the normalizer $N(\mathbb{T})\subset G$ of the maximal torus $\mathbb{T}$, up to an element of the centralizer $Z(\mathbb{T})\subset G$. On $\mathfrak{t}$, the centralizer $Z(\mathbb{T})$ acts trivially, but it certainly acts nontrivially on the full algebra, making the action of $\cW$ on $\mathfrak{g}$ ambiguous. However, the action of $\cW$ on a $G_b$-invariant polynomial $P(\Phi)$ is nevertheless unambiguous. To understand this, note that the magnetic charge $b\in\mathfrak{t}$ is obviously preserved by $Z(\mathbb{T})$, so the group $G_b$ includes $Z(\mathbb{T})$ as a subgroup. In particular, it means that $P(\Phi)$ is $Z(\mathbb{T})$-invariant, and hence the action of ${\rm w}\in N(\mathbb{T})/Z(\mathbb{T})$ on $P(\Phi)$ is unambiguous --- this is the action that appears in \eqref{dressedmono}.

This is not the only subtlety to take care of. It is also worth noting that the action of the Weyl group on the dressing factor is different from its action on $b$. The fundamental reason for this is that $\Phi$ represents a ``non-defect'' observable (given by an insertion of fundamental fields in the path integral), while $b$ characterizes the defect: namely, it describes the strength of the monopole singularity that plays the role of a boundary condition for the fundamental fields. In Appendix B.1 of \cite{Dedushenko:2017avn}, it was explained how symmetries act on observables of these two types (it was emphasized for global symmetries there, but the argument is exactly the same for gauge symmetries): the actions are inverses of each other. Namely, in our case, both $b$ and $\Phi$ are elements of $\mathfrak{g}$, and if the gauge symmetry acts on $\Phi$ by $U$ ($\Phi \mapsto U\Phi U^{-1}$), then it acts on $b$ by $U^{-1}$ ($b \mapsto U^{-1}b U$). The monopole singularity is labeled by $b\in\mathfrak{t}$, and the Weyl group has a natural action on it coming from the identification $\cW=N(\mathbb{T})/Z(\mathbb{T})$. The dressing factor is a $G_b$-invariant polynomial $P(\Phi)$ that is also acted on by $\cW$, as we just explained. If we act on $b$ by ${\rm w}\in\cW$ (that is, $b\mapsto {\rm w}\cdot b$), then we should act on $P(\Phi)$ by ${\rm w}^{-1}$: $P(\Phi^{\rm w})={\rm w}^{-1}\cdot P(\Phi)$. After restricting $\Phi$ to take values in $\mathfrak{t}_\C$ (which happens after localization and gauge fixing), it is convenient to note that $\cW$ acts on $\mathfrak{t}$ by orthogonal matrices, and hence the left action by ${\rm w}^{-1}$ is the same as the right action by ${\rm w}$. This provides a convenient way to perform actual calculations: the Weyl group acts from the left on $b\in\mathfrak{t}$ and from the right on $\Phi\in\mathfrak{t}$ once we represent them as a column vector $(b^i)$ and a row vector $(\Phi_i)$, respectively, in some orthonormal basis of $\mathfrak{t}$.

Another convention that we choose to follow is that by $b$ in $[P(\Phi)\cM^b]$, we mean some weight of $^LG$ within the given Weyl orbit, though not necessarily the dominant one.  Whenever we label monopoles by dominant weights, we explicitly say so. The polynomial $P(\Phi)$ appearing inside the square brackets is always the one attached to the charge-$b$ singularity (whether or not $b$ is dominant), while the Weyl-transformed singularities ${\rm w}\cdot b$ are multiplied by Weyl-transformed polynomials, as in \eqref{dressedmono}.

In this paper, we develop methods for computing correlation functions of dressed monopole operators of the form \eqref{dressedmono}. There are several techniques that we combine in order to achieve our results: cutting and gluing techniques \cite{Dedushenko:2018aox,Dedushenko:2018tgx}, localization, and algebraic consistency of the resulting OPE. In what follows, we describe each of them and what role they play in the derivation.

\subsection{Gluing Formula}\label{sec:glue_f}

The cutting and gluing property \cite{Dedushenko:2018aox,Dedushenko:2018tgx} holds in any local quantum field theory, and it has already been applied to the abelian version of our problem in \cite{Dedushenko:2017avn}. This is also one of the key ingredients in the non-abelian generalization here.  We can motivate its application as follows.  As explained in \cite{Dedushenko:2017avn}, only a very restricted class of configurations of twisted CBOs on $S^3$ is amenable to a direct localization computation. A less direct approach is to endow the path integral on $S^3$ with extra structure by dividing it into path integrals on two open halves.  These path integrals individually prepare states in the Hilbert space of the theory on $S^2$.  The advantage of this procedure is that it allows for operator insertions within $S^3$ to be implemented by acting on these boundary states with operators on their associated Hilbert spaces.

Specifically, the round $S^3$ is glued from two hemispheres, $HS^3_+$ and $HS^3_-$, and we need to know how this procedure is represented at the level of quantum field theories living on them. Recall that gluing corresponds to taking $\langle\Psi_-|\Psi_+\rangle$, where $|\Psi_+\rangle\in \cH_{S^2}$ and $\langle\Psi_-|\in\cH_{S^2}^\vee$ are states generated at the boundaries of the two hemispheres. Furthermore, in Lagrangian theories with no more than two derivatives, this operation is represented by an integral over the space of polarized boundary conditions \cite{Dedushenko:2018aox,Dedushenko:2018tgx} for a choice of polarization on $\cP(S^2)$, the phase space associated with $S^2=\partial HS^3$. For a special choice of supersymmetry-preserving polarization, this integral can be localized to the finite-dimensional subspace of half-BPS boundary conditions of a certain type, which results in a simple gluing formula \cite{Dedushenko:2017avn,Dedushenko:2018tgx}:
\begin{align}
\label{glueForm}
\langle\Psi_-|\Psi_+\rangle = \frac1{|\cW|}\sum_{B\in \Lambda_w^\vee}\int_{\mathfrak t}\, \dd^r\sigma\, \mu(\sigma,B)\, \langle\Psi_-|\sigma,B\rangle \langle\sigma,B|\Psi_+\rangle.
\end{align}
Here, the integration goes over the Cartan $\mathfrak{t}\subset\mathfrak{g}$, $\Lambda_w^\vee\subset \mathfrak{t}$ is the coweight lattice,  $\mu(\sigma,B)$ is the gluing measure given by the one-loop determinant on $S^2$,
\begin{align}
\mu(\sigma, B)&=Z^{\rm c.m.}_\text{one-loop}(\sigma,B)Z^{\rm v.m.}_\text{one-loop}(\sigma,B)\cJ(\sigma,B), \nonumber \\
Z^{\rm v.m.}_\text{one-loop}(\sigma,B)\cJ(\sigma,B)&=\prod_{\alpha\in\Delta^+}(-1)^{\alpha\cdot B}\left[\left( \frac{\alpha\cdot \sigma}{r} \right)^2 + \left( \frac{\alpha\cdot B}{2r} \right)^2\right], \nonumber \\
Z^{\rm c.m.}_\text{one-loop}(\sigma,B)&=\prod_{w\in\cR} (-1)^{\frac{|w\cdot B|-w\cdot B}{2}} \frac{\Gamma\left(\frac12 + iw\cdot\sigma +\frac{|w\cdot B|}{2} \right)}{\Gamma\left(\frac12 - iw\cdot\sigma +\frac{|w\cdot B|}{2} \right)}, \label{gluingmeasure}
\end{align}
and $\langle\Psi_-|\sigma,B\rangle$, $\langle\sigma,B|\Psi_+\rangle$ are the hemisphere partition functions with prescribed boundary conditions determined by $\sigma\in\mathfrak{t}$ and $B\in\Lambda_w^\vee\subset\mathfrak{t}$.\footnote{In \eqref{gluingmeasure}, $\mathcal{J}$ is a standard Vandermonde determinant and we have omitted an overall power of $r$ from the logarithmic running of the 2D FI parameters.} We think of $\langle\Psi_-|\sigma,B\rangle$, $\langle\sigma,B|\Psi_+\rangle$ as wavefunctions on $\mathfrak{t}\times \Lambda_w^\vee$: they are elements of an appropriate functional space, such as $L^2(\mathfrak{t}\times \Lambda_w^\vee)$, a precise identification of which is not important. The boundary conditions parametrized by $\sigma, B$ are half-BPS boundary conditions on bulk fields preserving 2D $(2,2)$ SUSY on $S^2$, namely an $\mathfrak{su}(2|1)$ subalgebra of $\mathfrak{s}$ containing $\mathcal{Q}^C$. In terms of the on-shell components of the multiplets $\cH=(q_a, \tq^a, \psi_{\alpha\dot a}, \tpsi_{\alpha\dot a})$ and $\cV=(A_\mu, \Phi_{\dot a\dot b}, \lambda_{\alpha a \dot a})$, as well as the variables $q_\pm\equiv q_1\pm iq_2$ and $\tq_\pm\equiv \tq_1\pm i\tq_2$, these boundary conditions are given by:
\begin{gather}
q_+\big|=\tq_-\big| = \left(\cD_\perp q_- +\frac{\Phi_{\dot1\dot1}-\Phi_{\dot2\dot2}}{2}q_- \right)\bigg| =\left(\cD_\perp \tq_+ +\frac{\Phi_{\dot1\dot1}-\Phi_{\dot2\dot2}}{2}\tq_+ \right)\bigg| =0, \nonumber \\
(\psi_{\dot1} -\sigma_3 \psi_{\dot 2})\big| = (\tpsi_{\dot1} +\sigma_3 \tpsi_{\dot 2})\big|=0, \vphantom{\bigg(\bigg)} \nonumber \\
A_\parallel\big| = \pm \frac{B}2 (\sin\theta-1)\dd\tau, \quad \frac{\Phi_{\dot1\dot1}+\Phi_{\dot2\dot2}}{2i}\bigg| =\frac{B}{2r}, \quad \Phi_{\dot1\dot2}\big|=\frac{\sigma}{r}, \vphantom{\bigg(\bigg)} \nonumber \\
(\lambda_{1\dot2}-i\lambda_{2\dot2}+\sigma_3(\lambda_{1\dot1}-i\lambda_{2\dot1}))\big|=(\lambda_{1\dot2}+i\lambda_{2\dot2}-\sigma_3(\lambda_{1\dot1}+i\lambda_{2\dot1}))\big|=0. \label{BPSbc}
\end{gather}
Note that such boundary conditions specify the magnetic flux $B\in\Lambda_w^\vee$ through the boundary $S^2$. Thus we could alternatively think of $B$ as the corresponding cocharacter, i.e., the full Weyl orbit $\cW B$, in which case the sum in \eqref{glueForm} would run over the set of cocharacters (allowed magnetic charges) $\Gamma_m=\Lambda_w^\vee/\cW$. In such a case, the boundary conditions above would have to be understood in the same way as the definition of the monopole operator: one would have to evaluate the hemisphere partition function for every element of the Weyl orbit $\cW B\subset\Lambda_w^\vee$ and sum the results. We find it more convenient to treat $B$ as an element of $\Lambda_w^\vee$, in which case we simply sum over $B\in\Lambda_w^\vee$ in the gluing formula and there is no need for a separate sum over Weyl reflections.

The gluing formula \eqref{glueForm} holds as long as the states $\Psi_\pm$ are supersymmetric, i.e., annihilated by $\cQ^C$ \cite{Dedushenko:2017avn,Dedushenko:2018tgx}. This is true for the state generated at the boundary of the empty hemisphere, and remains valid if we start inserting $\cQ^C$-closed observables inside. Such insertions will modify the hemisphere partition function, and can be represented as certain operators acting on the empty hemisphere partition function. In this paper, we are only concerned with local observables, described above as gauge-invariant polynomials in $\Phi(\varphi)$ and dressed monopole operators. Such local observables form an OPE algebra $\cA_C$, which will turn out to be a quantization of the Coulomb branch. Therefore, all we need to do is find how $\Phi(\varphi)$ and dressed monopoles act on the hemisphere partition function.

\subsection{Input from Localization}

An important step is to compute the hemisphere partition function with insertions of local $\cQ^C$-closed observables. As explained in \cite{Dedushenko:2017avn}, because correlation functions do not depend on the positions of the insertions, we can move them all to the tip of the hemisphere and replace them by an equivalent composite operator located there. In the abelian case, the GNO charge of the twisted CBO at the tip is equal to the sum of the GNO charges of all insertions, while in the non-abelian case, it is determined by taking tensor products of representations of $^LG$. It suffices to consider a bare monopole at the tip, as it is trivial to include insertions of (gauge-invariant monomials in) the scalar $\Phi(\varphi)$ anywhere along $S^1_\varphi$.

The hemisphere partition function can be computed using supersymmetric localization. In fact, half of the computation that we need has already been performed in \cite{Dedushenko:2017avn}, whose conventions we closely follow. Recall that the round sphere is parametrized by $0\leq\theta\leq\pi/2$, $0\leq\varphi \leq2\pi$, and $-\pi\leq\tau\leq\pi$, and $S^1_\varphi$ is located at $\theta=\pi/2$, where the $\tau$-circle shrinks. The sphere is cut into two hemispheres along the $S^2$ located at $\varphi=0$ and $\varphi=\pm\pi$. It is also sometimes convenient to use spherical coordinates $(\eta,\psi,\tau)$, which are related to the ``fibration'' coordinates $(\theta,\varphi,\tau)$ by
\begin{equation}
(\cos\theta, \sin\theta\cos\varphi, \sin\theta\sin\varphi) = (\sin\eta\sin\psi, -\sin\eta\cos\psi, \cos\eta), \quad \tau = \tau,
\end{equation}
where $\eta,\psi\in[0,\pi]$. In terms of such coordinates, the cut is located at $\eta=\pi/2$.

We place the monopole of charge $b$ at $(\theta, \varphi)=(\pi/2, \pi/2)$, which is the tip of the hemisphere, by imposing \eqref{monsing} there. In spherical coordinates, the monopole insertion point is $\eta=0$. We also impose the conditions \eqref{BPSbc} at the boundary of the hemisphere. The BPS equations that follow from $\cQ^C$ can be conveniently written in terms of
\begin{align}
R\equiv\sin\theta,\quad \Phi_r\equiv {\rm Re}(R e^{i\varphi}\Phi_{\dot1\dot1}),\quad \Phi_i\equiv {\rm Im}(R e^{i\varphi}\Phi_{\dot1\dot1}),
\end{align}
and they take the form:
\begin{gather}
[\Phi_{\dot1\dot2},\Phi_i]=[\Phi_{\dot1\dot2},\Phi_r]=0, \nonumber \\
\textstyle D_{12}={\rm Re}(D_{11})=0,\quad {\rm Im}(D_{11})=-\frac1r \Phi_{\dot1\dot2}, \nonumber \\
\cD_\mu\Phi_{\dot1\dot2}=\cD_\tau\Phi_i=0,\quad \cD_\tau\Phi_r = ir[\Phi_r,\Phi_i], \nonumber \\
R\cD_R\Phi_i + \cD_\varphi\Phi_r=0,\quad R(1-R^2)\cD_R\Phi_r - \cD_\varphi\Phi_i=0, \nonumber \\
\textstyle F_{\mu\nu}=\sqrt{g}\epsilon_{\mu\nu\rho}\kappa_\rho\cD^\rho\Phi_r, \text{ where } \kappa_\rho = \left(1, 1, \frac1{\sin^2\theta}\right)_\rho. \label{BPS}
\end{gather}
In the last equation, the index $\rho$ is summed over, and indices are raised and lowered using the metric in \cite{Dedushenko:2017avn}.  These equations have a straightforward (non-bubbling) solution that only exists if the boundary (flux) coweight $B$ matches one of the coweights in the Weyl orbit corresponding to the monopole charge. In other words, if the monopole's dominant coweight is $b$, then the straightforward solution exists iff $B=wb$ for some $w\in\cW$. This solution has vanishing fields in the hypermultiplet as well as vanishing fermions in the vector multiplet, while the bosons in the vector multiplet take the form:
\begin{gather}
\label{AbSol}
D_{12}=0,\quad \Phi_{\dot1\dot2}=irD_{11}=irD_{22}=\frac{\sigma}{r}\in\mathfrak{t},\cr
\Phi_{\dot1\dot1}=\Phi_{\dot2\dot2}=\frac{iB}{2r\sin\eta}=\frac{iB}{2r\sqrt{\cos^2\theta + \sin^2\theta \cos^2\varphi}},\cr 
A^\pm=-\frac{B}{2}(\cos\psi \mp 1)\, \dd\tau=\frac{B}2 \left(\frac{\sin\theta \cos\varphi}{\sqrt{1-\sin^2\theta \sin^2\varphi}}\pm1 \right)\dd\tau,
\end{gather}
where $A^-$ is defined everywhere on the hemisphere except the interval $\pi/2\leq \varphi\leq \pi$ at $\theta=\pi/2$; similarly, $A^+$ is defined everywhere except on $0\leq \varphi\leq \pi/2,\ \theta=\pi/2$. Here, $D_{ab}$ are the auxiliary fields in the vector multiplet.

The above straightforward solution is a direct generalization of the abelian one from \cite{Dedushenko:2017avn}.  Therefore, \eqref{AbSol} can be called the ``abelian solution.'' Indeed, since $B\in\mathfrak{t}$, we see that only components valued in the maximal torus of the gauge group have VEVs. It is known, however, that in the non-abelian case, the equations \eqref{BPS} might have additional loci of solutions. They correspond to screening effects that go by the name of ``monopole bubbling'' \cite{Kapustin:2006pk}. In particular, one notices that close to the special circle $\theta=\pi/2$, the last equation in \eqref{BPS} becomes the Bogomolny equation, and the bubbling loci in the moduli spaces of Bogomolny equations have been an active area of study. We will discuss bubbling in more detail soon, but for now let us focus on \eqref{AbSol}.

The abelian solution \eqref{AbSol} has the feature that all fields with nontrivial VEVs on the localization locus are vector multiplet fields valued in $\mathfrak{t}$. In other words, the VEVs look as though the gauge group were actually $\mathbb{T}$, the maximal torus of $G$. This is essentially how the ``abelianization'' of \cite{Bullimore:2015lsa} makes an appearance in our approach.

Note that since the Yang-Mills action is $\cQ^C$-exact \cite{Dedushenko:2016jxl, Dedushenko:2017avn}, one can use it for localization and to compute the relevant determinants in the weak-coupling limit $g_{\rm YM}\to 0$. The action (with boundary terms included such that the sum of the bulk and boundary pieces is $\cQ^C$-exact \cite{Dedushenko:2017avn}) vanishes on the localization locus, and it remains only to compute the one-loop determinants in the background of \eqref{AbSol}.

The action for hypermultiplets in the background of \eqref{AbSol} becomes quadratic, so there is no need to localize them separately: one can directly integrate them out. Furthermore, this action is simply that of free hypermultiplets coupled to the $\mathbb{T}$-valued gauge background. Each representation $\cR$ of $G$ gives a set of abelian charges under $\mathbb{T}$ given by the weights $w\in\cR$. Therefore, we can borrow the corresponding one-loop determinant from the previous work \cite{Dedushenko:2017avn}, where the abelian case was studied:
\es{Z1loopHyper}{
	Z_\text{1-loop}^{\rm hyper}=\prod_{w\in\cR}\frac1{r^{\frac{|w\cdot B|}{2}}} \frac{\Gamma\left(\frac{1+|w\cdot B|}{2}-iw\cdot\sigma \right)}{\sqrt{2\pi}} \,.
}
The only novelty in the computation of non-abelian one-loop determinants is that vector multiplets contribute: we need to include the contributions of W-bosons and gaugini. An indirect derivation of these determinants will be presented in Section \ref{sec:Schur}. The answer is given by
\es{Z1loopVector}{
	Z_\text{1-loop}^{\rm vec}=\prod_{\alpha\in\Delta}r^{\frac{|\alpha\cdot B|}{2}}\frac{\sqrt{2\pi}}{\Gamma\left(1+\frac{|\alpha\cdot B|}{2}-i\alpha\cdot\sigma \right)} \,.
}
Therefore, the contribution from the abelian solution to the hemisphere partition function with a monopole labeled by a coweight $b \in \Lambda_w^\vee\subset\mathfrak{t}$ inserted at the tip is given by
\begin{align}
\label{Z_unbub}
Z(b; \sigma, B)=\sum_{b'\in \cW b} \delta_{B,b'}\frac{\prod_{w\in\cR}\frac{1}{\sqrt{2\pi}r^{\frac{|w\cdot b'|}{2}}}\Gamma\left(\frac{1+|w\cdot b'|}{2}-iw\cdot\sigma \right)}{\prod_{\alpha\in\Delta}\frac{1}{\sqrt{2\pi}r^{\frac{|\alpha\cdot b'|}{2}}}\Gamma\left(1+\frac{|\alpha\cdot b'|}{2}-i\alpha\cdot\sigma \right)}\equiv \sum_{b'\in\cW b}Z_0(b'; \sigma,B),
\end{align}
where the $\delta_{B,b'}$ enforces flux conservation: the flux sourced by the monopole equals the flux exiting through $S^2$. We have introduced the notation $Z_0$ for an ``incomplete'' partition function that does not include a sum over the Weyl orbit of $b$. Such a quantity does not represent a physical monopole operator, but it will prove to be convenient in the following sections.\footnote{We do not keep careful track of the overall sign of the hemisphere wavefunction, as it cancels in the gluing formula.}

In general, $Z$ as given above is not the full answer, because there are contributions from additional loci in the localization computation. We now discuss them.

\subsection{Monopole Bubbling}

Close to the monopole insertions, our BPS equations behave as Bogomolny equations on $\R^3$ with a monopole singularity at the origin. Such equations are known to have ``screening solutions'' in addition to the simple abelian ``Dirac monopole'' solution described in the previous subsection. The main property of such solutions is that while at the origin of $\R^3$ they have a monopole singularity characterized by $b\in\Lambda_w^\vee$, at infinity they behave as Dirac monopoles of different charges $v\in\Lambda_w^\vee$. It is also known that such solutions only exist when $v$ is a weight in the representation determined by the highest weight $b$ such that $|v| < |b|$ (in which case $v$ is said to be ``associated to'' $b$, sometimes written simply as $v < b$). Let $\cM(b,v)$ denote the moduli space of such screening solutions. For given $b$ and $v$, let $\rho$ be the length scale over which the screening takes place. It is one of the moduli for solutions of the Bogomolny equations, and taking $\rho\to 0$ corresponds to going to the boundary of $\cM(b,v)$. In this limit, the solution approaches a Dirac monopole of charge $v$ everywhere on $\R^3$ except for an infinitesimal neighborhood of the origin where the non-abelian screening takes place.  This solution can be thought of as a singular (Dirac) monopole screened by coincident and infinitesimally small smooth ('t Hooft-Polyakov) monopoles; the latter have GNO charges labeled by coroots.  It is natural to suppose that such solutions also exist on $S^3$: while at finite $\rho$ they are expected to receive $1/r$ corrections compared to the flat-space case, in the $\rho\to 0$ limit, they should be exactly the same bubbling solutions as on $\R^3$.

Notice that our general BPS equations require solutions to be abelian away from the insertion point. However, within the radius $\rho$, the screening solutions to the Bogomolny equations are essentially non-abelian. Therefore, smooth screening solutions corresponding to generic points of $\cM(b,v)$ cannot give new solutions to the BPS equations. However, boundary components of $\cM(b,v)$ where $\rho \to 0$ can give new, singular solutions to the BPS equations, which fail to be abelian only at the insertion point of the monopole operator. They should therefore be taken into account in the localization computation. Since such a solution behaves as an abelian Dirac monopole of charge $v$ everywhere except at the insertion point, it is convenient to factor out $Z(v; \sigma, B)$ computed in the previous subsection, and to say that the full contribution from the ``$b\to v$'' bubbling locus is given roughly by
\begin{equation}
Z_{\rm mono}(b, v; \sigma, B)Z(v; \sigma,B),
\end{equation}
where $Z_{\rm mono}$ characterizes the effect of monopole bubbling. We call it the bubbling factor. In fact, such a simple presentation is not quite correct, and we need to be more precise here. Recall that the monopole insertion is not just defined by a single singular boundary condition \eqref{monsing}: rather, one sums over the Weyl orbit of such singular boundary conditions. Therefore, the above expression is expected to have sums over such orbits for both $b$ and $v$. A more general expectation, which turns out to be correct, is that the contribution of the bubbling locus takes the form
\begin{equation}
\sum_{\substack{b'\in \cW b\cr v'\in\cW v}} Z^{\rm ab}_{\rm mono}(b', v'; \sigma, B)Z_0(v'; \sigma, B)
\end{equation}
where, as before, $b$ and $v$ are understood to be (dominant) coweights representing magnetic charges, and we sum over their Weyl orbits. The new quantity appearing in this equation,
\begin{equation}
Z^{\rm ab}_{\rm mono}(b', v'; \sigma, B),
\end{equation}
is called the ``abelianized bubbling factor.'' It depends on coweights $b', v'\in \Lambda_w^\vee\subset \mathfrak{t}$ rather than on cocharacters, while physical answers in the full non-abelian theory depend on cocharacters and thus always include sums over Weyl orbits. The abelianized bubbling factors introduced here prove to be of great importance for the formalism of this paper.  Later on, we will provide more rigorous evidence for their relevance based purely on group theory arguments that are independent of the heuristic path integral--inspired explanation of this section. Note also that the abelianized bubbling factors are expected to behave under Weyl reflections in the following way:\footnote{Here, all variables take values in $\mathfrak{t}$, so the action of $\cW$ is unambiguous.}
\begin{equation}
\label{heu_inv}
Z^{\rm ab}_{\rm mono}({\rm w}\cdot b, {\rm w}\cdot v; {\rm w}\cdot \sigma, {\rm w}\cdot B)=Z^{\rm ab}_{\rm mono}(b, v; \sigma, B),\quad {\rm w}\in\cW.
\end{equation}
Now we can write the complete answer for the hemisphere partition function:
\begin{equation}
\label{Zbubbled}
\langle \sigma, B|\Psi_b\rangle = Z(b;\sigma, B) + \sum_{|v|<|b|}\sum_{\substack{b'\in \cW b\cr v'\in\cW v}} Z^{\rm ab}_{\rm mono}(b', v'; \sigma, B)Z_0(v'; \sigma,B),
\end{equation}
where $\Psi_b$ represents the state generated at the boundary of the hemisphere with a physical monopole of charge $b$ inserted at the tip. Here, the first sum goes over dominant coweights $v$ satisfying $|v|<|b|$, while the second sum goes over the corresponding Weyl orbits.

The localization approach to the computation of $Z_{\rm mono}(b, v; \sigma, B)$ is quite technical, having been a subject of several works in the past \cite{Gomis:2011pf, Ito:2011ea, Gang:2012yr}, and more recently \cite{Brennan:2018yuj, Brennan:2018rcn}. In the current paper, we do not attempt a direct computation of $Z_{\rm mono}(b, v; \sigma, B)$ or $Z^{\rm ab}_{\rm mono}(b, v; \sigma, B)$. Instead, we describe a roundabout way to find them from the algebraic consistency of our formalism.   We find that the $Z^{\rm ab}_{\rm mono}(b, v; \sigma, B)$ are always given by certain rational functions, but we do not need to assume anything about their form.

\subsection{Shift Operators}

In this section, we derive how insertions of local $\cQ^C$-closed observables are represented by operators acting on the hemisphere wavefunction (up to the so-far unknown bubbling factors). The easiest ones are polynomials in $\Phi(\varphi)$. Just like in \cite{Dedushenko:2017avn}, we can think of them as entering the hemisphere either through the North pole $(\varphi, \theta)=(0, \pi/2)$ or through the South pole $(\varphi, \theta)=(\pi, \pi/2)$. Then we simply substitute the solution \eqref{AbSol} into the definition of $\Phi(\varphi)$ either for $0<\varphi<\pi/2$ or for $\pi/2 < \varphi < \pi$. We find that for the North pole,
\begin{align}
\Phi(\varphi=0)&=\frac12 (\Phi_{\dot1\dot1}+2\Phi_{\dot1\dot2}+\Phi_{\dot2\dot2})=\frac1r\left( \sigma + \frac{i}{2}B \right).
\end{align}
Similarly, for the South pole,
\begin{align}
\Phi(\varphi=\pi)&=\frac12 (-\Phi_{\dot1\dot1}+2\Phi_{\dot1\dot2}-\Phi_{\dot2\dot2})=\frac1r\left( \sigma - \frac{i}{2}B \right).
\end{align}
This operator simply measures the values of $\sigma$ and $B$ away from the monopole insertion, and bubbling is accounted for trivially. In particular, on the unbubbled locus, $B$ evaluates to $b$, while for the bubbling loci it evaluates to the corresponding $B=v$. Thus we conclude that $\Phi(\varphi)$ is represented by the following North and South pole operators:
\begin{equation}
\Phi_N=\frac1r\left( \sigma + \frac{i}{2}B \right)\in\mathfrak{t}_\C,\qquad \Phi_S=\frac1r\left( \sigma - \frac{i}{2}B \right)\in\mathfrak{t}_\C,
\end{equation}
where $B$ should be thought of as measuring $B\in\Lambda_w^\vee$ at the boundary $S^2$, i.e., it multiplies the wavefunction $\Psi(\sigma,B)$ by $B$, and thus is simply a diagonal multiplication operator.

It is also not too hard to obtain the generalizations of the abelian shift operators from \cite{Dedushenko:2017avn} that represent insertions of non-abelian monopoles. From the structure of the partition functions above, it is clear that they take the following form:
\begin{align}
\label{shiftGeneral}
\cM^b=\sum_{b'\in\cW b} M^{b'} + \sum_{|v|<|b|}\sum_{\substack{b'\in\cW b\cr v'\in\cW v}}Z_{\rm mono}^{\rm ab}(b',v'; \sigma,B)M^{v'}.
\end{align}
Here, $M^b$ is an abelianized (non-Weyl-averaged) shift operator which represents the insertion of a bare monopole singularity characterized by the coweight (not cocharacter!) $b$, and whose definition ignores bubbling phenomena. The inclusion of abelianized bubbling coefficients $Z_{\rm mono}^{\rm ab}$ takes care of screening effects, and summing over Weyl orbits corresponds to passing to cocharacters, i.e., true physical magnetic charges.

The expression \eqref{shiftGeneral} is evident from the structure of the hemisphere partition function with a monopole inserted, as described in the previous subsections. Indeed, away from the monopole insertion, its effect must be represented by a sum over bubbling sectors, and within each bubbling sector, the contribution must take the form of a sum over the Weyl reflections of the basic contribution. The expression \eqref{shiftGeneral}, in fact, represents nothing else but the abelianization map proposed in \cite{Bullimore:2015lsa}: the full non-abelian operator $\cM^b$ is written in terms of the abelianized monopoles $M^b$ acting on $\Psi(\sigma, B)$, wavefunctions on $\mathfrak{t}\times \Lambda_w^\vee$.

It remains to determine the expressions for $M^b$ acting on wavefunctions $\Psi(\sigma,B)$. Just like in \cite{Dedushenko:2017avn}, there are separate sets of operators that implement insertions through the North and South poles.  These generate isomorphic algebras, and they are uniquely determined by the following set of consistency conditions:
\begin{enumerate}[label=\arabic*)]
	\item They should shift the magnetic flux at which $\Psi(\sigma,B)$ is supported by $b\in\Lambda_w^\vee$.
	\item They should commute with $\Phi$ at the opposite pole, i.e., $[M_N^b,\Phi_S]=[M_S^b,\Phi_N]=0$.
	\item They should commute with another monopole at the opposite pole, i.e., $[M_N^b,M_S^{b'}]=0$.
	\item When acting on the vacuum (empty hemisphere) wavefunction, the result should agree with \eqref{Z_unbub}.
\end{enumerate}
This set of conditions determines the North shift operator to be
\begin{equation}
\label{general_shift_N}
M^b_N = \frac{\prod_{w\in\cR}\left[\frac{(-1)^{(w\cdot b)_+}}{r^{|w\cdot b|/2}} \left(\frac12 +i rw\cdot \Phi_N \right)_{(w\cdot b)_+} \right]}{\prod_{\alpha\in\Delta}\left[\frac{(-1)^{(\alpha\cdot b)_+}}{r^{|\alpha\cdot b|/2}} \left(i r \alpha\cdot \Phi_N\right)_{(\alpha\cdot b)_+} \right]} e^{-b\cdot(\frac{i}2 \partial_\sigma +\partial_B)},
\end{equation}
where $(a)_+\equiv a$ if $a\geq0$ and $(a)_+\equiv 0$ otherwise, $(x)_n$ stands for the Pochhammer symbol $\Gamma(x+n)/\Gamma(x)$, and $x\cdot y$ represents the canonical pairing $\mathfrak{t}^\ast \times \mathfrak{t}\to\R$. The analogous South pole operator is
\begin{equation}
M^b_S = \frac{\prod_{w\in\cR}\left[\frac{(-1)^{(-w\cdot b)_+}}{r^{|w\cdot b|/2}} \left(\frac12 +i rw\cdot \Phi_S \right)_{(-w\cdot b)_+} \right]}{\prod_{\alpha\in\Delta}\left[\frac{(-1)^{(-\alpha\cdot b)_+}}{r^{|\alpha\cdot b|/2}} \left(i r \alpha\cdot \Phi_S\right)_{(-\alpha\cdot b)_+} \right]} e^{b\cdot(\frac{i}2 \partial_\sigma -\partial_B)}.
\end{equation}
By counting powers of $r^{-1}$ in the general expression \eqref{general_shift_N}, we find that the dimension of a charge-$b$ monopole is
\begin{equation}
\Delta_b = \frac{1}{2}\left(\sum_{w\in\cR} |w\cdot b| - \sum_{\alpha\in\Delta} |\alpha\cdot b|\right).
\label{dimensionformula}
\end{equation}
This dimension formula will come in handy later.

The shift operators satisfy an important multiplication property, which later on will allow us to generate monopoles of arbitrary charge starting from a few low-charge monopoles:
\begin{equation}
\label{product_rule}
M^{b_1}_N \star M^{b_2}_N = P_{b_1,b_2}(\Phi) M^{b_1 + b_2}_N \text{ for dominant $b_1$ and $b_2$,}
\end{equation}
and similarly for the South pole operators, where $P_{b_1,b_2}(\Phi)$ is some polynomial in $\Phi$. We use $\star$ to denote products \emph{as operators} (in particular, shift operators act on the $\Phi$-dependent prefactors in $M_{N,S}$), emphasizing that they form an associative noncommutative algebra. In fact, \eqref{product_rule} holds slightly more generally than for dominant weights: if $\Delta_+$ is some choice of positive roots (determined by a hyperplane in $\mathfrak{t}^\ast$), then \eqref{product_rule} holds whenever the condition $(b_1\cdot \alpha)(b_2\cdot \alpha)\geq 0$ is satisfied for all $\alpha\in\Delta_+$. The property \eqref{product_rule} ensures that in the product of two physical bare monopoles, the highest-charge monopole appears without denominators. If in addition, $b_1$ and $b_2$ satisfy the property that $(b_1\cdot w)(b_2\cdot w)\geq 0$ for all matter weights $w\in\cR$, then a stronger equality holds: 
\begin{equation}
M^{b_1}_N \star M^{b_2}_N =  M^{b_1 + b_2}_N.
\end{equation}
Finally, for general $b_1$ and $b_2$, we have:
\begin{equation}
M^{b_1}_N \star M^{b_2}_N =\frac{\prod_{w\in\cR}(-iw\cdot\Phi_N)^{(w\cdot b_1)_+ + (w\cdot b_2)_+ - (w\cdot (b_1+b_2))_+}}{\prod_{w\in\Delta}(-i\alpha\cdot\Phi_N)^{(\alpha\cdot b_1)_+ + (\alpha\cdot b_2)_+ - (\alpha\cdot (b_1+b_2))_+}}  M^{b_1 + b_2}_N + O(1/r).
\label{generalproduct}
\end{equation}
These are precisely the abelian chiral ring relations of \cite{Bullimore:2015lsa}.

In addition to defining $M_{N,S}$, we also need to properly define dressed monopoles.  This is an important and somewhat subtle consideration, especially due to the interplay with bubbling. We discuss it in Section \ref{sec:dress_and_bubble}. Before doing so, let us first fill a gap in the above discussion by comparing our results to supersymmetric indices in four dimensions, which provide a way to derive the vector multiplet one-loop determinant.

\subsection{Reduction of Schur Index}\label{sec:Schur}

Our setup has a natural uplift to a supersymmetric index of 4D $\cN=2$ theories on $S^3\times S^1$. The operators constructed from $\Phi(\varphi)$ lift to supersymmetric Wilson loops wrapping the $S^1$, while monopole operators correspond to supersymmetric 't Hooft loops on $S^1$. Certain questions relevant to this 4D setup have been studied in the literature in great detail, and we can use the answers to determine the unbubbled partition functions. By shrinking the $S^1$ factor, the 4D results allow us to infer the unbubbled one-loop determinants mentioned in the previous subsections. Doing this for the bubbling contributions is more subtle and is discussed in Appendix \ref{appen:Zmono}, where we find agreement with our method of deriving bubbling terms in cases where the 4D results are known.  For simplicity, let us first set the radius $r$ of $S^3$ to $1$, and let us denote the circumference of $S^1$ by $\beta$.  To restore $r$, we simply send $\beta \to \beta / r$ in all formulas.

Since the one-loop determinant for hypermultiplets is already known, we concern ourselves only with determining the vector multiplet contribution.  This can be done in a theory with any conveniently chosen matter content. We can always choose the matter content in such a way that both the 4D $\cN=2$ and the 3D $\cN=4$ theories are conformal. The corresponding 4D index is known as the Schur index.  The Schur index is defined as 
\es{SchurDef}{
	{\cal I}(p) = {\rm Tr}_{\cH_{S^3}}(-1)^F p^{E-R}
}
where the trace is taken over the Hilbert space of the 4D theory on $S^3$ and $R$ is the Cartan generator of the $\mathfrak{su}(2)$ R-symmetry, normalized so that the allowed charges are quantized in half-integer units. In the path integral description, ${\cal I}(p)$ evaluated when $p = e^{-\beta}$ is given by an $S^3\times S^1$ partition function, with $S^1$ of circumference $\beta$ and with an R-symmetry twist by $e^{\beta R}$ as we go once around the $S^1$.  This $S^3\times S^1$ partition function is invariant under all 4D superconformal generators that commute with $E-R$, or in other words, that have $E = R$.  One can easily list these generators and check that they form an $\mathfrak{su}(2|1) \oplus \mathfrak{su}(2|1)$ superalgebra.  Thus the superconformal index \eqref{SchurDef} is invariant under $\mathfrak{su}(2|1) \oplus \mathfrak{su}(2|1)$.  It is also invariant under all continuous deformations of the superconformal theory: in particular, it is independent of $g_\text{YM}$ and can be computed at weak coupling.

One can additionally insert an 't Hooft loop of GNO charge $b$ (taken to be a dominant coweight) wrapping $S^1$ at one pole of $S^3$ and the oppositely charged loop at the opposite pole of $S^3$. The answer for this modified index in a general 4D $\cN=2$ gauge theory, up to a sign and ignoring the bubbling effect, can be found in \cite{Gang:2012yr}:\footnote{We set $\eta_a = 1$ and $x = \sqrt{p}$ in Equation (3.44) of \cite{Gang:2012yr}.}
\begin{equation}
	{\cal I}_b(p) = \frac{1}{\abs{{\cal W}_b}}  \int \left( \prod_{i=1}^{\operatorname{rank}(G)} \frac{d\lambda_i}{2 \pi} \right)\, \left[ \prod_{\alpha \in \Delta} 
	\left(1 - e^{i \alpha \cdot \lambda} p^{\frac{\abs{\alpha \cdot b}}{2}}  \right) \right] \text{P.E.}[ I_v(e^{i \lambda_i}, p) ]  \, \text{P.E.}[ I_h(e^{i \lambda_i}, p) ], \label{IN2}
\end{equation}
where $\Delta$ is the set of all roots, $\text{P.E.}$ is the plethystic exponential defined as $\text{P.E.}[f(x)]\equiv \exp \left[ \sum_{n=1}^\infty  \frac{f(x^n)}{n}  \right]$, $I_v$ is the contribution from the ${\cal N} = 2$ vector, and $I_h$ is the contribution from the ${\cal N} = 2$ hyper in the representation $\cR$:
\begin{align}
	I_v(e^{i \lambda_i}, p) &= -2 \sum_{\alpha \in \text{adj}} \frac{p^{1 + \frac{\abs{\alpha \cdot b}}{2}}}{1 - p} e^{i \alpha \cdot \lambda} \,, \nonumber \\
	I_h (e^{i \lambda_i}, p) &= \sum_{w \in \cR} \frac{p^{\frac 12 + \frac{\abs{w \cdot b}}{2}} }{1 - p} (e^{i w\cdot \lambda} + e^{-i w \cdot \lambda}) \,. \label{Iv}
\end{align}
Using the identity $\exp\left[-\sum_{n=1}^\infty  \frac{a^n}{n(1 - q^n)} \right] = (a; q)$ where $(a; q) \equiv \prod_{n=0}^\infty (1 - a q^n) $ is the $q$-Pochhammer symbol, we can rewrite ${\cal I}_b(p)$ as 
\es{Index}{
	\cI_b(p)&= \frac{ (p;p)^{2 \operatorname{rank}(G)} }{\abs{\cW_b}}  \int_{-\pi}^\pi \left(\prod_{i=1}^{\operatorname{rank}(G)} \frac{d\lambda_i}{2\pi} \right)
	\frac { \prod_{\alpha\in\Delta}  \left[  \left(1-e^{i\alpha\cdot \lambda}p^{\frac{|\alpha\cdot b|}{2}}\right) (e^{i\alpha\cdot \lambda}p^{1+\frac{|\alpha \cdot b|}{2}}; p)^2 \right]}{\prod_{w\in\cR} (e^{iw \cdot \lambda}p^{\frac 12+\frac{|w\cdot b|}{2}}; p)(e^{-iw\cdot \lambda}p^{\frac 12 +\frac{|w \cdot b|}{2}}; p)} \,.
}
We would like to determine the 3D hemisphere partition function.  One way to do so is to use the results of \cite{Dedushenko:2018tgx} to first extract the 4D half-index from \eqref{Index}, and then dimensionally reduce it.  Alternatively (and this is how we proceed), we first reduce the index \eqref{Index} to 3D to find the $S^3$ partition function, and then use the gluing formula from Section~\ref{sec:glue_f} to recover the hemisphere partition function as the square root of the absolute value of the integrand.  One can then fix signs by consistency with gluing.

To reduce \eqref{Index} down to three dimensions, we take the $\beta\to 0$ limit, where in addition to setting $p=e^{-\beta}$, we scale the integration variable accordingly:
\begin{equation}
\lambda = \beta\sigma \in\mathfrak{t}.
\end{equation}
The angular variable $\lambda$ (parametrizing the maximal torus $\mathbb{T}\subset G$) then ``opens up'' into an affine variable $\sigma\in\mathfrak{t}$.  To take the limit, one needs the following identities (see \cite{Dedushenko:2017avn}):
\es{identities}{
	\frac{1}{(p^x;p)} &= e^{\frac{\pi^2}{6\beta}}\beta^{x-\frac{1}{2}}\frac{1}{\sqrt{2\pi}}\Gamma(x)(1+O(\beta)) \,, \qquad
	(p;p) = \sqrt{\frac{2\pi}{\beta}}e^{-\frac{\pi^2}{6\beta}}(1+O(\beta))\,,
}
which give
\begin{gather}
	\cI_b \approx 
	\frac{e^{-\frac{\pi^2 r}{3\beta}\left(\dim G -\sum_{I=1}^{N_f}\dim R_I\right)}}{|\cW_b|} \int_{-\infty}^{\infty} \left( \prod_{i=1}^{\text{rank}(G)}  d\sigma_i \right) \prod_{\alpha\in \Delta^+}\left( (\alpha\cdot \sigma)^2+\frac{|\alpha\cdot b|^2}{4}\right) \nonumber \\
	\times\frac{\prod_{w\in\cR}\left|\frac{\beta^{\frac{|w \cdot b|}{2}}}{\sqrt{2\pi} r^{\frac{|w \cdot b|}{2}}}\Gamma\left(\frac{1+|w \cdot b|}{2}-iw \cdot \sigma \right)\right|^2}{\prod_{\alpha\in \Delta} \left|\frac{\beta^{\frac{|\alpha \cdot b|}{2}}}{\sqrt{2\pi} r^{\frac{|\alpha \cdot b|}{2}}}\Gamma\left(1+\frac{|\alpha \cdot b|}{2}-i\alpha \cdot \sigma\right)\right|^2} \label{IbredTmp}
\end{gather}
as $\beta \to 0$. In \eqref{IbredTmp}, we restored the radius $r$ of $S^3$ by dimensional analysis.

The exponential prefactor in \eqref{IbredTmp} is precisely the Cardy behavior of \cite{DiPietro:2014bca}.  In the integrand, we recognize the one-loop contribution of the hypermultiplet to the $S^3$ partition function,
\es{Z1loopHyperS3}{
	Z_\text{1-loop, $S^3$}^\text{hyper} (\sigma) = \prod_{w\in\cR}\left|\frac{1}{\sqrt{2\pi}r^{\frac{|w \cdot b|}{2}}}\Gamma\left(\frac{1+|w \cdot b|}{2}-iw \cdot \sigma \right)\right|^2 \,,
}
multiplied by $\beta^{\frac{|w \cdot b|}{2}}$.   The remaining factor in the integrand must be proportional to the one-loop contribution of the vector multiplet to the $S^3$ partition function.  Assuming that the one-loop vector multiplet contribution comes multiplied by $\beta^{-\frac{|\alpha \cdot b|}{2}}$ (by analogy with the hypermultiplet factor), we conclude that it is equal to
\es{Z1loopVectorS3}{
	Z_\text{1-loop, $S^3$}^\text{vec} (\sigma) =
	\frac{\prod_{\alpha\in \Delta^+}\left( (\alpha\cdot \sigma)^2+\frac{|\alpha\cdot b|^2}{4}\right)}
	{\prod_{\alpha\in \Delta} \left|\frac{1}{\sqrt{2\pi}r^{\frac{|\alpha \cdot b|}{2}}}\Gamma\left(1+\frac{|\alpha \cdot b|}{2}-i\alpha \cdot \sigma\right)\right|^2} \,.
}
The $S^3$ partition function is then given by the expression
\es{ZbS3}{
	Z_b =  \frac{1}{|\cW_b|} \int_{-\infty}^{\infty} \left( \prod_{i=1}^{\text{rank}(G)} d\sigma_i \right) 
	Z_\text{1-loop, $S^3$}^\text{vec} (\sigma)  Z_\text{1-loop, $S^3$}^\text{hyper} (\sigma) \,.
}
Note that using this method, the overall normalization of $Z_b$ is ambiguous, but we propose that the correct expression is given by \eqref{ZbS3}.  This expression passes the check that when $b=0$, it reduces to the $S^3$ partition function derived in \cite{Kapustin:2009kz}, namely
\es{Z0}{
	Z=Z_0 = \frac{1}{|\cW|}\int_{-\infty}^{\infty} \left( \prod_{i=1}^{\text{rank}(G)} d\sigma_i \right)  \frac{\prod_{\alpha\in \Delta^+}4 \sinh^2 (\pi\alpha \cdot \sigma)}{\prod_{w\in\cR }2\cosh(\pi w \cdot \sigma)} \,.
}
What remains to be done is to use \eqref{Z1loopHyperS3} and \eqref{Z1loopVectorS3} to verify the hemisphere one-loop determinants given in \eqref{Z1loopHyper} and \eqref{Z1loopVector}.  To do so, we use the gluing formula \eqref{glueForm} as well as the explicit expression for the gluing measure in \eqref{gluingmeasure}.  It immediately follows that the hypermultiplet and vector multiplet contribute \eqref{Z1loopHyper} and \eqref{Z1loopVector} to the hemisphere partition function, respectively.  The hypermultiplet contribution \eqref{Z1loopHyper} was previously determined by an explicit computation of the one-loop determinant on the hemisphere \cite{Dedushenko:2017avn}.  It would be interesting to carry out the analogous computation for the non-abelian vector multiplet, which we have bypassed by means of the above argument.\footnote{Note that the hemisphere and the $\mathcal{Q}^C$-invariant background \eqref{AbSol} with a monopole at the tip $\eta = 0$ preserve an $\mathcal{N} = 2$ subalgebra $\mathfrak{su}(2|1)$, generated by what are called $Q_1^\pm$ and $Q_2^\pm$ in \cite{Dedushenko:2017avn}.  The suggestive form of \eqref{Z1loopVector} then leads one to wonder whether it can be explained by a Higgsing argument familiar in the study of theories with four supercharges (see, e.g., \cite{Benini:2016qnm}).  Namely, with respect to the aforementioned $\mathcal{N} = 2$ subalgebra, the hypermultiplet decomposes into $\mathcal{N} = 2$ chiral multiplets of R-charge $1/2$ and the vector multiplet decomposes into an $\mathcal{N} = 2$ vector multiplet and an adjoint chiral multiplet of R-charge 1.  Suppose that one could deform the action in such a way as to accommodate arbitrary R-charge $q$ for the chiral multiplets transforming in representations $\mathcal{R}, \overline{\mathcal{R}}$ of $G$ (as in, e.g., \cite{Jafferis:2010un, Hama:2010av}) while preserving the $\mathcal{N} = 2$ superpotential coupling that descends from the $\smash{\Phi^{\dot{a}\dot{b}}}\Phi_{\dot{a}\dot{b}}$ term in the $\mathcal{N} = 4$ Lagrangian.  Then one might expect the corresponding chiral multiplet one-loop determinant on the hemisphere to take the form of a product over weights $w\in \mathcal{R}$ of
	\begin{equation}
	Z_\text{chiral}^q(w\cdot \sigma)\sim \Gamma\left(1 - q + \frac{|w\cdot B|}{2} - iw\cdot \sigma\right),
	\end{equation}
	so that the numerator of $Z_0(b'; \sigma, B)$ in \eqref{Z_unbub} comes from $Z_\text{chiral}^{1/2}(w\cdot \sigma)$ and the denominator from
	\begin{equation}
	Z_\text{vector}(\alpha\cdot \sigma) = \frac{1}{Z_\text{chiral}^0(-\alpha\cdot \sigma)},
	\label{Higgsing}
	\end{equation}
	by reflection symmetry of the roots $\alpha$.  Here, \eqref{Higgsing} follows from the Higgs mechanism and $Z_\text{vector}(\alpha\cdot \sigma)$ denotes the contribution to the vector multiplet one-loop determinant from a mode in the $\alpha$-direction.
	
	It would be interesting to make this intuition precise.  However, due to our choice of $\mathcal{N} = 2$ superalgebra on $S^3$, ours is not the standard $\mathcal{N} = 2$ Coulomb branch localization, where chiral multiplet fields vanish on the localization locus.  Indeed, \eqref{AbSol} implies a nontrivial background for the scalar in the adjoint chiral multiplet (i.e., $\sigma/r$) as well as for the scalar in the $\mathcal{N} = 2$ vector multiplet (i.e., $-B/r\sin\eta$).  In particular, $\sigma\in \mathfrak{t}$ is not the standard Coulomb branch parameter: it labels the scalar zero mode of the adjoint chiral and \emph{not} of the vector.}

\section{Dressing and Abelianized Bubbling}\label{sec:dress_and_bubble}

We have now derived the structure of bare monopoles, up to the bubbling factors. In this section, we extend this construction to more general dressed monopoles. Recall that the magnetic charge $b$ breaks the gauge group at the insertion point down to $G_b$, the centralizer of $b$. As is well-known in the literature \cite{Bullimore:2015lsa} and as reviewed in Section \ref{sec:observables}, one may dress the monopole operator by some $G_b$-invariant polynomial $P(\Phi)$ in the variable $\Phi(\varphi)$.

If $P(\Phi)$ is invariant under the full gauge group $G$, then it is a valid $\cQ^C$-closed observable on its own. This makes the definition of the corresponding dressed monopole essentially trivial: we simply ``collide'' two separate observables $P(\Phi)$ and $\cM^b$, which within our formalism means multiplying them \emph{as operators} acting on the hemisphere wavefunction. Using the star product notation for such multiplication, we thus have:
\begin{equation}
\left[P(\Phi)\cM^b\right] := P(\Phi)\star \cM^b.
\end{equation}
When the polynomial $P(\Phi)$ is only invariant under a subgroup $G_b$ rather than the full gauge group $G$, we must proceed differently because $P(\Phi)$ only makes sense as part of $[ P(\Phi)\cM^b]$, not as a separate observable. Had bubbling not been an issue, the solution would again be straightforward: we would simply define $[P(\Phi)\cM^b] = |\cW_b|^{-1}\sum_{{\rm w}\in\cW} P(\Phi^{\rm w})M^{{\rm w}\cdot b}$. In general, however, the presence of bubbling makes such a simple definition incomplete.

For the remainder of this section, we focus on the case of a simple gauge group $G$. The generalization to the situation where $G$ is a simple factor of a larger gauge group is straightforward: different simple factors couple to each other only through the matter multiplets, and representation-theoretic issues can be addressed for each simple factor separately. The final conclusion of this section, Theorem \ref{thetheorem}, holds for general $G$ with the understanding that for non-simple gauge groups, bubbling terms for a monopole operator magnetically charged under one simple factor might also depend on scalars $\Phi$ from other simple factors. From the point of view of a given simple factor $G$, $\Phi$'s valued in other simple factors $G'$ act as twisted masses for $G'$.

\subsection{Dressed Monopoles and Invariant Theory}

A general dressed monopole operator takes the form
\begin{equation}
\label{gen_dress}
\left[P(\Phi)\cM^b \right]=\frac1{|\cW_b|}\sum_{{\rm w}\in\cW} P(\Phi^{\rm w})M^{{\rm w}\cdot b} + \cdots
\end{equation}
where the ellipses stand for bubbling contributions. It is intuitively clear that such observables, constructed for all possible choices of $P(\Phi)$, cannot all be independent: there should exist a minimal set of dressed monopoles, a basis in some sense, from which all other dressed monopoles follow. In this subsection, we make this intuition precise by rigorously proving that for a given magnetic charge $b$, there exists a set of \emph{primitive} dressed monopoles that accomplish this.

\begin{definition}

Dressed monopoles $[P_1(\Phi)\cM^b], [P_2(\Phi)\cM^b], \ldots, [P_p(\Phi)\cM^b]$ are called primitive (of magnetic charge $b$) if they form a basis for the (free) module of dressed charge-$b$ monopoles over the ring of $G$-invariant polynomials. This means that by taking linear combinations
\begin{equation}
\label{lin_prim}
\sum_{i=1}^p Q_i(\Phi) \star \left[P_i(\Phi)\cM^b \right]
\end{equation}
where the $Q_i(\Phi)$ are $G$-invariant polynomials, we obtain dressed monopoles with all possible leading terms of the form \eqref{gen_dress}, and furthermore, that $p$ is the minimum number that makes this possible. We will always assume $P_1(\Phi)=1$, so that the first primitive monopole is the bare monopole itself.

\end{definition}

\begin{example}

In $SU(2)$ gauge theory, the Weyl group is $\Z_2$, which simply flips $b\to -b$ and $\Phi\to -\Phi \in\mathfrak{t}_\C$. A dressed monopole of charge $b$ takes the form $P(\Phi)M^b + P(-\Phi)M^{-b} + \text{bubbling}$. In this case, there are only two primitive dressed monopoles for each $b$: 
\begin{align}
\cM^b &= M^b + M^{-b} + \text{bubbling}, \nonumber \\
\left[\Phi\cM^b\right] &= \Phi(M^b - M^{-b}) + \text{bubbling}.
\end{align}
By writing
\begin{equation}
P(\Phi)=\frac{P(\Phi)+P(-\Phi)}{2} + \frac{P(\Phi)-P(-\Phi)}{2\Phi} \Phi,
\label{arbitraryP}
\end{equation}
it becomes obvious that any other dressed monopole can be defined as:
\begin{equation}
\left[ P(\Phi)\cM^b \right] := \frac{P(\Phi)+P(-\Phi)}{2} \star \cM^b + \frac{P(\Phi)-P(-\Phi)}{2\Phi} \star \left[\Phi\cM^b\right].
\end{equation}

\end{example}

To describe primitive monopoles for general gauge groups, it is enough to focus on the leading term of \eqref{gen_dress}, as we do in this subsection. Bubbling contributions will be analyzed from this point of view in the next subsection.

The leading term in \eqref{gen_dress} is constructed to be invariant under the Weyl group action.  Therefore, we can classify such leading terms by identifying invariants of the Weyl group in the corresponding (reducible) representations of $\cW$. Alternatively, we could achieve this by focusing on the dressing factors and classifying polynomials $P(\Phi)$ invariant under $G_b$. Since $\Phi\in\mathfrak{t}_\C$ after localization and gauge fixing, it is enough to impose invariance under $\cW_b$ (the Weyl group of $G_b$).  Thus dressed monopoles can be classified by $\cW_b$-invariant polynomials in $\Phi$.\footnote{$\cW_b$-invariant polynomials on $\mathfrak{t}$ can be uniquely extended to $G_b$-invariant polynomials on $\mathfrak{g}$.} Nevertheless, we find it more convenient to study the invariants of $\cW$ directly.

\begin{proposition}

Let $G$ be a simple Lie group, $\cW$ its Weyl group, and $b$ a dominant coweight (a magnetic charge). Then there exists a set of primitive monopoles (of magnetic charge $b$) $[P_1(\Phi)\cM^b], [P_2(\Phi)\cM^b], \ldots, [P_p(\Phi)\cM^b]$, where $p=|\cW b|$ is the size of the Weyl orbit of $b$.

\end{proposition}

The remainder of this subsection is devoted to proving this proposition using classical facts from invariant theory. Less mathematically inclined readers are free to skip it.

\begin{proof} Consider $\rho^b$, a representation of $\cW$ spanned as a $\C$-linear space by the Weyl orbit of the coweight $b$.  We write it in terms of shift operators $M^{{\rm w}\cdot b}$, ${\rm w}\in\cW$, as
\begin{equation}
\rho^b := \operatorname{Span}_\C\{M^{{\rm w}\cdot b} \,|\, {\rm w}\in\cW\}.
\end{equation}
This representation is reducible: in particular, it contains a trivial subrepresentation spanned by $\sum_{b'\in\cW b} M^{b'}$, which is the simplest invariant corresponding to the bare monopole operator.

The Cartan subalgebra $\mathfrak{t}$ itself is also a $\cW$-module: $\cW$ acts on it in an irreducible $r$-dimensional representation, where $r={\rm rank}(G)$. We will denote such a representation simply by $\mathfrak{t}$.  The variable $\Phi=\sum_{a=1}^r\Phi_a h^a$ clearly takes values in this representation.

However, recall from the discussion after \eqref{dressedmono} that when ${\rm w}\in\cW$ acts on a dressed monopole operator by transforming the weight according to $b \mapsto {\rm w}\cdot b$, physics tells us that the dressing factor should be acted on by ${\rm w}^{-1}$: $\Phi \mapsto \Phi^{\rm w}= {\rm w}^{-1} \cdot \Phi$. Since ${\rm w}$ is represented by an orthogonal matrix on $\mathfrak{t}$, this is the same as acting with ${\rm w}^{\rm T}$ from the left or with ${\rm w}$ from the right. This is how one acts in a dual representation.  Thus in a dressed monopole operator, we think of $\Phi$ as transforming in the dual representation $\mathfrak{t}^\ast$. The dressing factor $P(\Phi)$ entering \eqref{gen_dress}, being a polynomial in $\Phi$, transforms in $S\mathfrak{t}^\ast$, the symmetric algebra of $\mathfrak{t}^\ast$, or equivalently, the algebra $\C[\mathfrak{t}]$ of polynomial functions on $\mathfrak{t}$. This implies that any dressed monopole operator is determined by an invariant vector inside the following $\cW$-module:
\begin{equation}
\mathfrak{R}_b:=S\mathfrak{t}^\ast \otimes \rho^b\cong \C[\mathfrak{t}]\otimes \rho^b.
\end{equation}
Thus the leading terms in dressed monopoles of charge $b$ are classified by invariants $\mathfrak{R}_b^\cW$.

Questions of this sort have been studied extensively in the ancient subject of invariant theory (see, for example, \cite{invariants}). To begin, let us understand the structure of $S\mathfrak{t}^\ast\cong \C[\mathfrak{t}]$ as a representation of $\cW$, in particular its isotypic decomposition.  It is well-known that the ring of invariants for a reflection group (such as the Weyl group) has the structure of another polynomial ring (see \cite[Part V]{invariants}, in particular \cite[Section 18-1]{invariants}):
\begin{align}
\C[\mathfrak{t}]^\cW \cong \C[f_1, \dots, f_r], \text{ where } r=\dim \mathfrak{t}={\rm rank}(G).
\end{align}
Here, the $f_i$ are invariant homogeneous polynomials whose degrees $d_i$ satisfy $\prod_{i=1}^r d_i = |\cW|$.  Another well-known object is the ring of covariants \cite[Part VII]{invariants}, which is defined as follows. Consider an ideal in $\C[\mathfrak{t}]$ generated by non-constant invariant polynomials:
\begin{equation}
I=\left( \C[\mathfrak{t}]^\cW_{\deg>0} \right)=(f_1, \ldots, f_r).
\end{equation}
The ring of covariants is defined as
\begin{equation}
\C[\mathfrak{t}]_\cW = \C[\mathfrak{t}]/I.
\end{equation}
It is again well-known \cite[Section 24-1]{invariants} that $\C[\mathfrak{t}]_\cW\cong \C[\cW]$ as a $\cW$-module, where $\C[\cW]$ is the regular representation. Since $\cW$ maps $I$ to itself, Maschke's theorem implies that one can find a $\cW$-invariant subspace $M_\cW\subset \C[\mathfrak{t}]$ such that $\C[\mathfrak{t}] \cong I \oplus M_\cW$, and this $M_\cW\cong \C[\mathfrak{t}]_\cW\cong\C[\cW]$. Finally, \cite[Section 18-3]{invariants} implies that $\C[\mathfrak{t}]$ is a free $\C[\mathfrak{t}]^\cW$-module generated by the basis of $M_\cW$: $\C[\mathfrak{t}]\cong \C[\mathfrak{t}]^\cW \otimes_\C M_\cW$. To summarize, the structure of $S\mathfrak{t}^*\cong \C[\mathfrak{t}]$ as a $\cW$-module is
\begin{equation}
\C[\mathfrak{t}] \cong \C[\mathfrak{t}]^\cW \otimes_\C \C[\cW],
\label{isotypic}
\end{equation}
where $\C[\cW]$ is realized on polynomials from $M_\cW\subset \C[\mathfrak{t}]$. Equation \eqref{isotypic} also encodes the isotypic decomposition since every $m$-dimensional irrep of $\cW$ appears in $\C[\cW]$ precisely $m$ times.

With this knowledge, our representation of interest becomes
\begin{equation}
\mathfrak{R}_b \cong \C[\mathfrak{t}]^\cW \otimes_\C \C[\cW] \otimes_\C \rho^b.
\end{equation}
Now the problem of identifying $\mathfrak{R}_b^\cW$ simplifies substantially:
\begin{equation}
\mathfrak{R}_b^\cW \cong \C[\mathfrak{t}]^\cW \otimes_\C \left( \C[\cW] \otimes_\C \rho^b\right)^\cW.
\end{equation}
Namely, we must find an invariant subspace in $\C[\cW] \otimes_\C \rho^b$, which is a product of two finite-dimensional representations of $\cW$. Any other element of $\mathfrak{R}_b^\cW$ is obtained by multiplication with invariant polynomials from $\C[\mathfrak{t}]^\cW = \C[f_1,\dots ,f_r]$.

What we have proven so far is the following: $\mathfrak{R}_b^\cW$ is a free $\C[\mathfrak{t}]^\cW$-module, and any $\C$-basis of $\left(\C[\cW] \otimes_\C \rho^b\right)^\cW$ gives a $\C[\mathfrak{t}]^\cW$-basis of $\mathfrak{R}_b^\cW$, i.e., a set of primitive dressed monopoles of magnetic charge $b$.

To compute $\left( \C[\cW] \otimes_\C \rho^b\right)^\cW$, we simply decompose each of the two representations into irreducible components and pair up dual representations. Indeed, by Schur's lemma, only tensor products like $V\otimes V^*$, where $V$ is some irrep and $V^*$ is its dual, can contain invariant subspaces. We can also easily find the dimension of $\left( \C[\cW] \otimes_\C \rho^b\right)^\cW$. Since $\C[\cW]$ contains each irreducible representation $\rho_i$ of $\cW$ exactly $\dim(\rho_i)$ times,
\begin{equation}
\left( \C[\cW] \otimes_\C \rho_i\right)^\cW \cong \C^{\dim(\rho_i)}.
\end{equation}
Decomposing $\rho^b$ into irreducible components as $\rho^b \cong \oplus_{i\in I(b)}\rho_i$, this obviously gives:
\begin{equation}
\left( \C[\cW] \otimes_\C \rho^b\right)^\cW \cong \oplus_{i\in I(b)} \left( \C[\cW] \otimes_\C \rho_i\right)^\cW\cong \C^{\dim(\rho^b)}.
\end{equation}
Hence there are exactly $\dim(\rho^b)=|\cW b|$ primitive dressed monopoles of charge $b$. \end{proof}

We have now classified the leading terms in dressed monopoles. Any such leading term must be extended by the appropriate bubbling contributions to give a complete physical dressed monopole, and primitive monopoles are no exception:
\begin{align}
\left[P_i(\Phi)\cM^b\right] = \frac1{|\cW_b|}\sum_{{\rm w}\in\cW} P_i(\Phi^{\rm w})M^{{\rm w}\cdot b} + \text{bubbling}, \quad i=1, \ldots, |\cW b|.
\end{align}
We now turn to the study of these bubbling contributions.

\subsection{Abelianized Monopole Bubbling}

Suppose we have found a set of polynomials $P_1, \ldots, P_{|\cW b|}$ such that the dressed monopoles $[P_i(\Phi)\cM^b]$ form the primitive set for a given magnetic charge $b$, in the sense explained in the previous subsection.  That is, $|\cW_b|^{-1}\sum_{{\rm w}\in\cW} P_i(\Phi^{\rm w})M^{{\rm w}\cdot b}$ for $i=1, \ldots, |\cW b|$ form a basis for $\mathfrak{R}_b^\cW$ (the space of dressed charge-$b$ monopoles) over $\C[\mathfrak{t}]^\cW$ (the algebra of gauge-invariant polynomials in $\Phi$). In this subsection, we will show that there exists a special \emph{bubbled} and abelianized monopole shift operator $\widetilde{M}^b = M^b +\cdots$ such that
\begin{equation}
\label{eqn_on_Z}
\left[P_i(\Phi)\cM^b\right] = \frac{1}{|\cW_b|}\sum_{{\rm w}\in\cW} P_i(\Phi^{\rm w})\widetilde{M}^{{\rm w}\cdot b}.
\end{equation}
The left-hand side has the following structure: for each $i$,
\begin{equation}
\label{prim_struct}
\left[P_i(\Phi)\cM^b\right] = \frac{1}{|\cW_b|}\sum_{{\rm w}\in\cW} P_i(\Phi^{\rm w})M^{{\rm w}\cdot b} + \frac1{|\cW_b|}\sum_{|v|<|b|}\sum_{{\rm w}\in\cW} V^{b\to v}_i(\Phi^{\rm w})M^{{\rm w}\cdot v}.
\end{equation}
Here, the first sum corresponds to the sector with no screening effects, and the remaining terms describe monopole bubbling, with $V_i^{b\to v}$ given by some rational functions of $\Phi\in\mathfrak{t}_\C$ that encode the bubbling data (because the $V_i$ are not yet known, the $|\cW_b|^{-1}$ in the second term is optional). By equating the right-hand sides of \eqref{eqn_on_Z}  and \eqref{prim_struct}, we obtain a system of linear equations for $\widetilde{M}^{{\rm w}\cdot b}$, ${\rm w}\in\cW$:
\begin{equation}
\label{lin_eqn_Z}
\sum_{{\rm w}\in\cW} P_i(\Phi^{\rm w})\widetilde{M}^{{\rm w}\cdot b} = \sum_{{\rm w}\in\cW} P_i(\Phi^{\rm w})M^{{\rm w}\cdot b} + \sum_{|v|<|b|}\sum_{{\rm w}\in\cW} V^{b\to v}_i(\Phi^{\rm w})M^{{\rm w}\cdot v}.
\end{equation}
Its solution provides the definition of $\widetilde{M}^b$, but first we need to show that such a solution exists, i.e., that the matrix of coefficients $P_i(\Phi^{\rm w})$ is nondegenerate. This essentially follows from the primitivity of $[P_i(\Phi)\cM^b]$, and the proof is given in Appendix \ref{sec:nondeg}.

The solution to \eqref{lin_eqn_Z} takes the form
\begin{equation}
\label{Shift}
\widetilde{M}^b = M^b + \sum_{|v|< |b|}Z^{\rm ab}_{b\to v}(\Phi)M^v,
\end{equation}
where the first term has an obvious origin: it is the shift operator that describes the sector without bubbling. Here $b$ is a fixed coweight, whereas the sum in the second term is taken over \emph{all coweights} whose length is less than that of $|b|$.

The functions $Z^{\rm ab}_{b\to v}(\Phi)$ are some rational functions of $\Phi\in\mathfrak{t}_\C$ that encode the bubbling phenomena. They do not have any invariance property under the action of $\cW$. We may extend them to non-dominant $b$ by postulating the following transformation property:
\begin{equation}
Z^{\rm ab}_{{\rm w}\cdot b\to {\rm w}\cdot v}(\Phi) = Z^{\rm ab}_{b\to v}(\Phi^{\rm w}),
\end{equation}
consistent with \eqref{heu_inv}. These functions are what we refer to as \emph{abelianized bubbling factors}. Recall that we previously argued for their existence using heuristic path integral reasoning. We have now rigorously proven their existence by relying solely on group theory.

As also mentioned in Appendix \ref{sec:nondeg}, the expression for $\widetilde{M}^{{\rm w}\cdot b}$ can be obtained from the expression for $\widetilde{M}^b$ by a Weyl reflection:
\begin{equation}
\widetilde{M}^{{\rm w}\cdot b} = M^{{\rm w}\cdot b} + \sum_{|v|< |b|}Z^{\rm ab}_{b\to v}(\Phi^{\rm w})M^{{\rm w}\cdot v}=M^{{\rm w}\cdot b} + \sum_{|v|< |b|}Z^{\rm ab}_{{\rm w}\cdot b\to {\rm w}\cdot v}(\Phi)M^{{\rm w}\cdot v}.
\end{equation}
Having established the existence of abelianized and bubbled monopoles $\widetilde{M}^b$, one can very easily construct arbitrary dressed monopoles. In fact, this proves the following theorem.

\begin{theorem} \label{thetheorem}

A shift operator describing an arbitrary physical dressed monopole of magnetic charge $b$ can be constructed as
\begin{equation}
\frac1{|\cW_b|}\sum_{{\rm w}\in\cW} F(\Phi^{\rm w})\widetilde{M}^{{\rm w}\cdot b},
\end{equation}
where $F(\Phi)$ is a polynomial in $\Phi\in\mathfrak{t}_\C$.

\end{theorem}

Such an expression will automatically produce, in the leading term, $F$ averaged over $\cW_b$, the stabilizer of $b$ in $\cW$, as well as generating the appropriate subleading terms describing bubbling. 

The abelianized bubbling coefficients prove to be very useful below.

\subsection{Relation to the Abelianization Map}

Before putting the notion of abelianized bubbling to work, let us comment on how it fits into the context of previous studies.

One approach to understanding the geometry of the Coulomb branch of a 3D $\mathcal{N} = 4$ theory was proposed in \cite{Bullimore:2015lsa}. Let $\cM_C^\text{abel}\subset \cM_C$ denote the generic points on the Coulomb branch where the gauge group $G$ is broken to its maximal torus $\mathbb{T}$. Using the fact that the chiral ring is independent of gauge couplings, it was argued in \cite{Bullimore:2015lsa} that the abelianized chiral ring $\bC[\cM_C^\text{abel}]$ can be determined by integrating out the massive W-bosons at one loop, ignoring nonperturbative effects. This ring is generated by (VEVs of) dressed chiral monopole operators of $\mathbb{T}$, the complex scalars $\Phi_a$ ($a=1,\ldots,r$), and the inverses of the W-boson complex masses $\alpha(\Phi)$ for all roots $\alpha\in \Delta$. At points on $\cM_C$ where a non-abelian subgroup of $G$ is restored, some $\alpha(\Phi)\to 0$ and hence $\bC[\cM_C^\text{abel}]$ becomes ill-defined. Nonperturbative effects cannot be ignored at such points.

These nonperturbative effects are encoded in the so-called abelianization map, which expresses a chiral monopole operator $\mathcal{M}$ in the non-abelian theory as a linear combination of monopole operators $M$ in the low-energy abelian gauge theory, with coefficients being meromorphic functions of the complex abelian vector multiplet scalars.  In our notation, this map takes precisely the form \eqref{shiftGeneral} or, before Weyl-averaging, \eqref{Shift}.  The abelianization map realizes $\bC[\cM_C]$ as the subring of $\bC[\cM_C^\text{abel}]$ generated by the operators on the RHS of \eqref{shiftGeneral}.  To obtain $\bC[\cM_C]$, all we need are the abelianized bubbling coefficients $Z^\text{ab}$.  These bubbling coefficients ensure that $\bC[\cM_C]$ closes without needing to include $\alpha(\Phi)^{-1}$, so that it is well-defined everywhere on $\cM_C$.

The shift operators that we construct allow us to directly compute the OPE of chiral monopole operators within a cohomological truncation of a given 3D $\mathcal{N} = 4$ gauge theory.  As explained in the introduction, this OPE encodes information about the geometry of the Coulomb branch beyond the chiral ring data.  In particular, our shift operators give a concrete realization of the abelianization map of \cite{Bullimore:2015lsa}, allowing us to determine the bubbling coefficients from the bottom up.  In our approach, the bubbling coefficients so obtained can further be used as input to calculate SCFT correlators via the gluing formula of Section \ref{sec:glue_f}.

In fact, as explained in Appendix \ref{versus}, previous formulations of the abelianization map do not distinguish between the abelianized bubbling coefficients $Z^\text{ab}$ and certain coarser, Weyl-averaged counterparts thereof (denoted by $Z_\text{mono}$), as written in \eqref{MwithweylaveragedZmono}.  To our knowledge, even the $Z_\text{mono}$ remain inaccessible to direct localization computations except in a few classes of examples, namely $G = U(N)$ with fundamental and adjoint hypermultiplets \cite{Ito:2011ea}.  The fact that the previously considered Weyl-averaged bubbling coefficients $Z_\text{mono}$ can be written in terms of the more basic $Z^\text{ab}$ is one of the key observations of our work, and the computability of $Z^\text{ab}$ is one of our main results.\footnote{Some hints as to the necessity of the refined quantities $Z^\text{ab}$ were obtained for $USp(2N)$ theories with fundamental matter in \cite{Assel:2018exy}.  There, a proposal was made for extending the abelianized chiral ring relations to the full Coulomb branch, relying on a subtle change of variables for abelian monopole operators ((2.3) in \cite{Assel:2018exy}).  While not phrased in the language of monopole bubbling, this proposal should really be understood as a conjecture for the abelianized bubbling coefficients of this class of theories (in the commutative limit).  Indeed, the prediction of \cite{Assel:2018exy} agrees with what we explicitly derive in Section \ref{ranktwo} for the case $N = 2$: compare to the $r\to\infty$ limit of our \eqref{compareusp4}.  To match conventions, note that our abelianized monopoles whose charges are the four Weyl images of $\omega_2^\vee$ in $USp(4)$ gauge theory correspond to what are called $u_a^\pm$ ($a = 1, 2$) in \cite{Assel:2018exy}.  Our perspective puts the proposal of \cite{Assel:2018exy} into the more general context of abelianized monopole bubbling.}

For a bare monopole, decomposing $Z_\text{mono}$ into $Z^\text{ab}$ is merely a rewriting of the Weyl sum.  However, the refinement of bubbling by abelianized bubbling turns out to be crucial for constructing dressed monopoles.  Given a bare monopole, its abelianized bubbling coefficients allow us to construct all of its dressings in a way that guarantees closure of the star algebra.  As we discuss next, our claim is that the closure of this algebra, or ``polynomiality,'' determines $Z^\text{ab}$ uniquely up to operator mixing, in a sense to be made precise.  By taking star products of (dressed) monopoles whose bubbling coefficients are known, one can inductively extract $Z_\text{mono}$ for all pairs of monopole charges $(b, v)$ with $v < b$.

\section{Bubbling from Polynomiality}\label{sec:poly}

The algebra of quantum Coulomb branch operators, $\cA_C$, is believed to consist of gauge-invariant polynomials $P(\Phi)$ in the $\cQ^C$-closed variable $\Phi(\varphi)$ and dressed monopole operators $[F(\Phi)\cM^b]$, where the dressing factor $F(\Phi)$ is a $G_b$-invariant polynomial in $\Phi(\varphi)$. Note that the subleading (bubbling) terms in $[F(\Phi)\cM^b]$ can involve rational functions of $\Phi$, but the leading term must be built solely from the polynomial $F(\Phi)$. Such an assumption has also been made in the recent literature on 3D $\cN=4$ Coulomb branches \cite{Bullimore:2015lsa, Dimofte:2018abu}. One of the reasons that we expect this to be true is that VEVs of such operators should be algebraic functions on the Coulomb branch.  Thus it would be unnecessary (and problematic) to introduce poles by choosing $P(\Phi)$ or $F(\Phi)$ rational. The appearance of rational functions in the OPE can be ruled out using similar reasoning.

In good or ugly theories, we can make this argument slightly more explicit. The Coulomb branch in such theories is expected to be a hyperk\"ahler cone. Furthermore, because conformal dimensions are bounded from below, there are only finitely many operators below any fixed conformal dimension, and because $1/r$ has dimension one, only finitely many operators can appear on the right in any OPE.  In particular, this should hold for star products in $\cA_C$, which is simply a sector of the OPE algebra in the IR CFT. This excludes denominators of the form $\left(1/r + P(\Phi)\right)^{-1}$ where $P(0)=0$, as such denominators, when expanded in $1/r$, give infinitely many terms. The remaining possibility is to have denominators of the form $1/P(\Phi)$ where $P(0)=0$. But such operators blow up at the origin of the cone: they are not part of the coordinate ring and thus should not appear in the algebra.

In general, it is hard to give a more rigorous argument for polynomiality of observables due to the absence of a mathematical definition of the QFTs that concern us here. Nevertheless, we proceed under the assumption that polynomiality holds, using the above heuristic reasoning and support from the existing literature as good evidence for it. Furthermore, the results that we describe are in complete agreement with this assumption, implying that the algebra $\cA_C$ constructed to satisfy polynomiality is self-consistent.

One important observation is that if we neglect to include bubbling terms in the definition of $[P(\Phi)\cM^b]$, then polynomiality in general fails: operator products of such observables produce denominators that do not cancel. Therefore, one role of the bubbling terms is to guarantee polynomiality. In this section, we argue that polynomiality actually fully determines the algebra $\cA_C$, up to the natural ambiguity of operator mixing.

\subsection{Mixing Ambiguity and Deformation Quantization}

In quantum field theories, an arbitrarily chosen basis of observables need not be diagonal with respect to the two-point function, nor does it need to diagonalize the dilatation operator in the case of a CFT. Observables can mix with others of the same dimension, and on curved spaces, they can also mix with those of lower dimension, the difference being compensated for by powers of background (super)gravity fields. The mixing patterns often depend on short-distance effects, in particular how we define composite observables, creating ambiguities that must be resolved in the end by diagonalizing the two-point function.

For our theories on $S^3$, the Riemann curvature is proportional to $1/r^2$. Mixing with odd powers of $1/r$ might not necessarily be generated by coupling to background SUGRA, but we include it in the formalism because it helps with the polynomiality argument in the following sections. It could be that imposing some other requirement along with polynomiality would allow us to determine bubbling coefficients uniquely up to mixing with only even powers of $1/r$. However, we will not need to do so in this paper: mixing ambiguities can still be resolved in the end. The presence of operator mixing implies that in our problem, it is natural to make $r$-dependent basis changes of the form
\begin{equation}
\label{mixing_amb}
\cO \mapsto \cO + \sum_{n\geq0} \frac1{r^n} \cO_n
\end{equation}
where if $\cO$ has dimension $\Delta$, then $\cO_n$ has dimension $\Delta-n$. Other quantum numbers, if present, should also be preserved by such transformations.

One might recognize redefinitions of the form \eqref{mixing_amb} as typical ``gauge'' transformations considered in (equivariant) deformation quantization. In the present context, they were discussed in \cite{Beem:2016cbd}, where the problem was posed for SCFTs in flat space and transformations of the form \eqref{mixing_amb} were less relevant due to the absence of a natural ``mixing'' parameter like $1/r$. In the $S^3$ setup, however, \eqref{mixing_amb} does naturally arise due to mixing. Such transformations first appeared in the deformation quantization literature \cite{Lichnerowicz1979, fedosov1994, Nest1995, NEST1995151, SB_1993-1994__36__389_0, Kontsevich:1997vb, formality}, where the classification of quantizations often drastically simplifies once the problem is studied modulo \eqref{mixing_amb}. It is therefore reasonable to first solve our problem of constructing $\cA_C$ modulo transformations of the form \eqref{mixing_amb} (or rather, similar ones defined in the next paragraph). After that, the mixing ambiguities can be resolved.  In an SCFT, this can be achieved by diagonalizing the two-point function: the diagonalization determines a preferred basis of ``SCFT operators'' in the algebra $\cA_C$. Alternatively, it might sometimes be enough to have an answer given in \emph{some} basis, not necessarily the diagonal one (especially in bad theories, where one cannot straightforwardly compute correlators).

For the study of Coulomb branch operators, transformations of the general form \eqref{mixing_amb} might not be the most adequate choice. We wish to think of the leading (i.e., no-bubbling) term of a dressed monopole $[P(\Phi)\cM^b]$ as canonically defined, while the subleading bubbling terms might be ambiguous. If $P(\Phi)$ has large enough degree, one can find other monopole operators in the theory that have higher magnetic charge but lower dimension. According to \eqref{mixing_amb}, they can mix with $[P(\Phi)\cM^b]$. This can indeed happen in physical operator mixing. However, for studying the structure of monopole operators, such a mixing is too crude, as it would alter the leading term of $[P(\Phi)\cM^b]$.  We therefore define a class of less general transformations that respect the GNO charge. Namely, if $\cO$ is a monopole operator of GNO charge $b$, we only consider mixing with operators corresponding to GNO charges $v$ (including zero) such that $b$ can bubble into $v$. Recall that this means $|v|<|b|$ and that $v$ belongs to the $^LG$-representation of highest weight $b$. Such a relation determines a partial order on the set of monopole operators, and we denote by $|\cO_n|<|\cO|$ the situation where the GNO charge of $\cO$ ``can bubble'' into the GNO charge of $\cO_n$. Then we may consider more restrictive transformations of the form
\begin{equation}
\label{mixing_GNO}
\cO \mapsto \cO + \sum_{\substack{n\geq 0\\ |\cO_n|<|\cO|}} \frac1{r^n}\cO_n,
\end{equation}
where as before, the dimension of $\cO_n$ is $n$ units smaller than that of $\cO$.

We wish to first study monopole operators modulo such transformations. This means that for a given monopole $[P(\Phi)\cM^b]$, the bubbling terms are not uniquely determined. We can shift $[P(\Phi)\cM^b]$ by a linear combination of dressed monopoles of lower magnetic charge and lower dimension (differences in dimension being compensated for by powers of $1/r$), in this way obtaining a valid, though different, definition of a dressed monopole operator. We refer to \eqref{mixing_GNO} as the mixing ambiguity later on in this paper. The more general mixing \eqref{mixing_amb} would only be relevant in an SCFT if we were to look for an orthonormal basis of observables in the end.

Such shifts significantly alter the bubbling coefficients $V_i^{b\to v}(\Phi)$ appearing in the definition of $[P(\Phi)\cM^b]$: they can be shifted by polynomials or even by multiples of other bubbling terms, which translates into complicated rational ambiguities of \emph{abelianized} bubbling coefficients $Z_{b\to v}^{\rm ab}(\Phi)$. Any concrete expressions for bubbling coefficients available in the literature always implicitly refer to some choice of basis, thus resolving the mixing ambiguity in the algebra of observables. The presence of such ambiguities inherent to $\cA_C$ means that there is no chance of determining $\cA_C$ simply from polynomiality. In particular, this gives a negative answer to a question raised, e.g., in \cite{Dimofte:2018abu} on whether structural properties of $\cA_C$ (polynomiality and gauge invariance) determine it uniquely. We argue, however, that the next simplest possibility holds: $\cA_C$ is uniquely determined by polynomiality precisely up to mixing ambiguities of the form \eqref{mixing_GNO}. We start by proving this claim in the simplest cases.

\subsection{Baby Case: Theories with Minuscule Monopoles}

The simplest case is that in which the algebra $\cA_C$ is fully generated by monopole operators in minuscule representations of $^LG$. Such monopoles cannot bubble because for minuscule coweights $b$, there are no $v$ such that both $|v|<|b|$ holds and $b-v$ is a coroot. For such monopole operators, we have the following simple expressions:
\begin{equation}
\left[P(\Phi)\cM^b\right] = \frac1{|\cW_b|}\sum_{{\rm w}\in\cW} P(\Phi^{\rm w})M^{{\rm w}\cdot b}.
\end{equation}
Higher-charge monopole operators might contain bubbling terms, but they are easily determined by taking products of lower-charge monopoles. Such cases were previously addressed in the literature using different methods, and essentially comprise the main examples in \cite{Bullimore:2015lsa} because abelianization has a simpler structure in these cases.

Theories with minuscule generators include those with the gauge group $PSU(N) = SU(N)/\allowbreak\Z_N$, whose Langlands dual is $SU(N)$: the fundamental weights of $SU(N)$ are minuscule and thus cannot bubble. Another example is $U(N)$ gauge theory, since $U(N)$ is self-dual and its fundamental weights are also minuscule.\footnote{One can also use the $U(N)$ results to solve the $SU(N)$ theory, even though the latter has no minuscule monopoles. This point will be discussed later.} We discuss further aspects of these theories in Section \ref{sec:Appl} and Appendix \ref{CHIRALAPPENDIX}. We now move on to the more interesting (and novel) case of theories with no minuscule generators, starting from the lower-rank gauge groups.

\subsection{Rank-One Theories}\label{sec:rankone}

The only rank-one gauge theory with no minuscule generators is $SU(2)$ gauge theory. The dual group is $SO(3)$, so the lowest monopole operator corresponds to a root, i.e., the vector representation of $SO(3)$. In a normalization where the weights of $SU(2)$ are half-integers and products of weights with monopole charges (cocharacters, or dominant coweights) are integers, the minimal monopole has $b=2$. It can bubble to the zero-charge sector, because $0<|b|$ and $b-0$ is a root. The abelianized monopole operator takes the form
\begin{equation}
\widetilde{M}^2 = M^2 + Z(\Phi)
\end{equation}
with a single abelianized bubbling term, a function $Z(\Phi)$. Knowledge of $Z(\Phi)$ allows one to construct arbitrary dressed monopole operators of charge $2$, and ultimately, by taking star products of the latter, monopoles of arbitrary charge.

For the sake of generality, we may suppose that $SU(2)$ is a simple factor in a larger gauge group $G=SU(2)\times\cdots$. Therefore, we implicitly assume that $Z(\Phi)$ might also depend on scalars $\Phi$ valued in other simple factors of $G$, which from the point of view of a given $SU(2)$ factor play the role of masses.

The basic shift operator of charge $b\in\Z$ is (we work in the North picture from now on, so we drop the ``$N$'' subscripts):
 \es{MNorth}{
M^b = \frac{\prod_{w\in\cR}\left[\frac{(-1)^{(w\cdot b)_+}}{r^{|w\cdot b|/2}} \left(\frac12 +i rw\cdot \Phi \right)_{(w\cdot b)_+} \right]}{\prod_{\alpha\in\{+1, -1\}}\left[\frac{(-1)^{(\alpha\cdot b)_+}}{r^{|\alpha\cdot b|/2}} \left(i r \alpha\cdot \Phi\right)_{(\alpha\cdot b)_+} \right]} e^{-b\cdot(\frac{i}2 \partial_\sigma +\partial_B)}.
 }
By counting powers of $r^{-1}$, we read off the dimension of a bare monopole of charge $b$:
\begin{equation}
\Delta_b = \sum_{w\in\cR} \frac{|w\cdot b|}{2} - |b|.
\end{equation}
The dressed monopole is constructed as\footnote{If $Z(\Phi)$ and/or $P(\Phi)$ depend on $\Phi$'s valued in other simple factors, we only reverse the sign of $\Phi$ valued in $SU(2)$, as we are only concerned with the action of the Weyl group of $SU(2)$ here.}
\begin{align}
\left[P(\Phi)\cM^2\right] &= P(\Phi)\widetilde{M}^2 + P(-\Phi)\widetilde{M}^{-2}\cr
&=P(\Phi)M^2 + P(-\Phi)M^{-2} + P(\Phi)Z(\Phi) + P(-\Phi)Z(-\Phi).
\end{align}
Since an arbitrary $P(\Phi)$ can be written as \eqref{arbitraryP}, we clearly see that the primitive dressed monopoles in this case are:
\begin{align}
\cM^2 &= M^2 + M^{-2} + Z(\Phi) + Z(-\Phi),\cr
\left[\Phi\cM^2\right] &= \Phi(M^2 - M^{-2}) + \Phi(Z(\Phi) - Z(-\Phi)).
\end{align}
We then compute the following star products of these primitive monopoles with the Weyl-invariant polynomial $\Phi^2$:
\begin{align}
\cM^2\star\Phi^2 &=\left[\left(\Phi-\frac{2i}{r}\right)^2\cM^2\right]  +\frac4{r^2} \left( Z(\Phi) + Z(-\Phi)\right) + \frac{4i}{r}\Phi\left(Z(\Phi)-Z(-\Phi) \right),\nonumber \\
\left[\Phi\cM^2\right]\star \Phi^2 &=\left[\left(\Phi-\frac{2i}{r}\right)^2\Phi\cM^2\right] +\frac{4}{r^2}\Phi \left(Z(\Phi)-Z(-\Phi)\right) + \frac{4i}{r}\Phi^2 \left(Z(\Phi)+Z(-\Phi)\right),
\end{align}
with the first terms on the right being dressed monopoles with dressing factors $\left(\Phi-\frac{2i}{r}\right)^2$ and $\left(\Phi-\frac{2i}{r}\right)^2\Phi$, respectively. The polynomiality condition implies that the remaining terms must be Weyl-invariant polynomials in $\Phi\in \mathfrak{su}(2)$ (and possibly other simple factors):
\begin{align}
\label{poly_eqn_rank1}
\frac4{r^2} \left( Z(\Phi) + Z(-\Phi)\right) + \frac{4i}{r}\Phi\left(Z(\Phi)-Z(-\Phi) \right)&\equiv\frac1{r} A_0(\Phi^2) \in \C[\Phi^2],\cr
\frac{4}{r^2}\Phi \left(Z(\Phi)-Z(-\Phi)\right) + \frac{4i}{r}\Phi^2 \left(Z(\Phi)+Z(-\Phi)\right)&\equiv\frac1{r} A_1(\Phi^2)\in\C[\Phi^2].
\end{align}
Recall that the operator mixing ambiguity allows one to shift the bubbling factors $Z(\Phi) + Z(-\Phi)$ and $\Phi\left(Z(\Phi)-Z(-\Phi) \right)$ by arbitrary Weyl-invariant polynomials whose degrees are fixed by dimensional analysis. Using the freedom to shift $\Phi\left(Z(\Phi)-Z(-\Phi) \right)$, we can make $A_0(\Phi^2)$ vanish. After doing so, we solve Equation \eqref{poly_eqn_rank1} for $Z(\Phi)$:
\begin{equation}
\label{Zsol_almost_rank1}
Z(\Phi)=-\frac{i A_1(\Phi^2)}{8\Phi(\Phi-\frac{i}{r})}.
\end{equation}
We have not yet used the ambiguity to shift $Z(\Phi) + Z(-\Phi)$ by a Weyl-invariant polynomial. Such shifts that leave $\frac1r(Z(\Phi) + Z(-\Phi))+i\Phi\left(Z(\Phi)-Z(-\Phi) \right)$ invariant (because we have fixed the latter expression by demanding $A_0(\Phi)=0$) give one the freedom to shift $Z(\Phi)$ by
\begin{equation}
\Delta Z(\Phi)=\frac{\Phi+\frac{i}{r}}{2\Phi}V(\Phi^2),
\end{equation}
with $V(\Phi^2)$ an arbitrary Weyl-invariant polynomial. Adding this ambiguity to \eqref{Zsol_almost_rank1} gives:
\begin{equation}
Z(\Phi)=-i\frac{A_1(\Phi^2) + 4i(\Phi^2 + \frac1{r^2})V(\Phi^2)}{8\Phi(\Phi-\frac{i}{r})}.
\end{equation}
For any $A_1(\Phi^2)$, there exists a unique polynomial $V(\Phi^2)$ such that the numerator $A_1(\Phi^2) + 4i(\Phi^2 + \frac1{r^2})V(\Phi^2)\equiv 8ic$ does not depend on $\Phi\in\mathfrak{su}(2)$, where $c$ is a dimensionful constant.\footnote{However, it can still depend on $\Phi$ valued in other simple factors.} Therefore, by completely fixing the mixing ambiguity, we find that:
\begin{equation}
Z(\Phi)=\frac{c}{\Phi(\Phi-\frac{i}{r})}.
\end{equation}
It remains to determine $c$. To this end, we compute the following expression:
\begin{equation}
\label{poly_last_step}
\cM^2 \star \left[\Phi\cM^2\right] - \left[(\Phi-2i/r)\cM^2\right] \star \cM^2,
\end{equation}
which must satisfy the polynomiality constraint. This is where the answer starts to depend on the precise matter content of the theory (all previous steps apply equally well to all matter representations $\cR$). Assume that the gauge group is precisely $SU(2)$ (with no other simple factors), and that the theory has $N_f$ fundamental and $N_a$ adjoint hypermultiplets. The dimension of a charge-$b$ monopole is hence
\begin{equation}
\label{dim_rank1}
\Delta_b=\frac{|b|}{2}N_f + |b|(N_a-1).
\end{equation}
A straightforward computation with shift operators gives
\begin{align}
\eqref{poly_last_step}&=\frac{8ic^2 r^3}{(1+r^2\Phi^2)^2} + \left[\frac1{2\Phi}\left(\frac{i}{2r} +\frac{\Phi}{2}\right)^{2(N_f-1)} \left(\frac{3i}{2r}+\Phi\right)^{2N_a}\left(\frac{i}{2r}+\Phi\right)^{2N_a} + (\Phi\leftrightarrow -\Phi)\right].
\end{align}
At this point, we see that the precise answer for $c$ depends on whether $N_f\geq1$ or $N_f=0$. If $N_f\geq1$, then the second term on the right is a Weyl-invariant polynomial and the only non-polynomial piece is $\frac{8ic^2 r^3}{(1+r^2\Phi^2)^2}$, implying that only $c=0$ is consistent with polynomiality. However, if $N_f=0$, then one finds that the poles at $\Phi=\pm i/r$ (whose presence would violate polynomiality) vanish when
\begin{equation}
c^2=(2r)^{-4N_a} \implies c= \pm (2r)^{-2N_a}.
\end{equation}
The sign of $c$ remains undetermined, and indeed, the algebra is consistent for both signs of $c$. In fact, it is not hard to see that flipping the sign of $c$ has the same effect on the algebra $\cA_C$ as flipping the overall sign of $\cM^2$, which is simply a change of basis. This, in particular, shows that after performing Gram-Schmidt orthogonalization, the algebra is unaffected, and the physical correlation functions do not depend on the sign of $c$.

We will soon see that, quite curiously, such a sign ambiguity is not present in higher-rank cases. In the present case, there exists a convenient way to fix the sign. Notice that a theory with only adjoint matter admits two possible global forms of the gauge group: either $SU(2)$ or $SO(3)$. They differ by the spectrum of allowed monopole operators. While $\cM^2$ is the lowest monopole in the $SU(2)$ case, the $SO(3)$ gauge theory also admits $\cM^1$. Indeed, the Langlands dual of $SO(3)$ is $SU(2)$, and $\cM^1$ is in the fundamental representation. Because $\cM^1$ is minuscule, it contains no bubbling term:
\begin{align}
\cM^1 &= M^1 + M^{-1},\cr
\left[\Phi\cM^1\right] &= \Phi(M^1 - M^{-1}).
\end{align}
We can then define $\cM^2 = \cM^1\star \cM^1$ and $\left[\Phi\cM^2\right]=\left[\Phi\cM^1\right]\star \cM^1$, and calculate the bubbling term generated in this way. This gives the following value of $c$ for the $SO(3)$ gauge theory:
\begin{equation}
\label{c_ans}
c=(-4r^2)^{-N_a}.
\end{equation}
One could wonder whether the $SU(2)$ global form corresponds to a different sign, but this is not the case. There exists another trick to access bubbling terms in $SU(2)$ (and more generally, in $SU(N)$) gauge theory. It consists of studying the $U(2)$ theory first, and then gauging the $U(1)_{\rm top}$ symmetry that rotates the dual photon in the diagonal $U(1)$ gauge group (this approach was also used in \cite{Dey:2017fqs, Assel:2018exy}). In a $U(2)$ gauge theory, monopole charges are labeled by two integers $(n,m)\in \Z^2$, and some of them are minuscule. In particular, $\cM^{(1,0)}$ and $\cM^{(-1,0)}$ are minuscule, and their product can be used to determine the non-minuscule $\cM^{(1,-1)}$. After gauging $U(1)_{\rm top}$, the latter becomes $\cM^2$ of the $SU(2)$ gauge theory. Proceeding along these lines gives the same value for $c$ as in \eqref{c_ans}.

So in the end, we find that the bubbling coefficient in $SU(2)$ (or $SO(3)$, when possible) gauge theory with $N_f$ fundamentals and $N_a$ adjoints, up to the operator mixing ambiguity, takes the form
\begin{align}
Z(\Phi)=\begin{cases}
0 & \text{if $N_f>0$}, \\
\frac{(-4r^2)^{-N_a}}{\Phi(\Phi-\frac{i}{r})} & \text{if $N_f=0$},
\end{cases}
\end{align}
which then determines $\widetilde{M^2} = M^2 + Z(\Phi)$.

Let us now generalize to the case where $SU(2)$ is a simple factor in a larger gauge group, that is, $G=SU(2)\times G'$. Then the $N_f$ fundamentals of $SU(2)$ form some generally reducible representation $\cR'_f$ of $G'$, while the $N_a$ adjoints of $SU(2)$ form another representation $\cR'_a$ of $G'$. This modifies the computation of \eqref{poly_last_step} as follows:
\begin{align}
\eqref{poly_last_step}&=\frac{8ic^2 r^3}{(1+r^2\Phi^2)^2} + \Bigg[\frac{1}{2\Phi\left(\frac{i}{2r}+\frac{\Phi}{2}\right)^2}\prod_{w\in\cR'_f}\left(\left(\frac{i}{2r}+\frac{\Phi}{2}\right)^2 - (w\cdot \Phi')^2\right) \nonumber \\
&\times \prod_{w\in\cR'_a} \left(\left(\frac{3i}{2r}+\Phi\right)^2 - (w\cdot\Phi')^2\right)\left(\left(\frac{i}{2r}+\Phi\right)^2-(w\cdot\Phi')^2\right) + (\Phi\leftrightarrow -\Phi)\Bigg],
\end{align}
where $\Phi\in\mathfrak{t}\subset\mathfrak{su}(2)$ and $\Phi'\in\mathfrak{t}'\subset{\rm Lie}(G')$. The cancellation of poles determines $c$:
\begin{align}
\label{massive_rank1}
c=\prod_{w\in\cR'_a} \left(-\frac1{4r^2} - (w\cdot\Phi')^2 \right)\prod_{w\in\cR'_f} (-iw\cdot\Phi'),
\end{align}
where the sign was fixed by passing to the $U(2)$ theory and applying the ``gauging $U(1)_{\rm top}$'' trick. This shows that in a general gauge theory with gauge group $G=SU(2)\times G'$, the abelianized bubbling term for monopoles magnetically charged under the $SU(2)$ factor takes the same form $\frac{c}{\Phi(\Phi-\frac{i}{r})}$ where $\Phi\in\mathfrak{t}\subset\mathfrak{su}(2)$, while $c$ is no longer a constant, but rather a nontrivial function of $\Phi'$ from the $G'$ vector multiplets. This last result is enough to study the algebra $\cA_C$ and corresponding correlators for arbitrary quivers of $SU(2)$ gauge groups.

\subsection{Rank-Two Theories} \label{ranktwo}

In this subsection, we repeat the above analysis for rank-two gauge groups, namely $SU(3)$, $PSU(3)$, $USp(4)\cong Spin(5)$, $SO(5)$, and $G_2$, demonstrating how polynomiality determines bubbling coefficients. This will further clarify the general procedure, which was applied to rank-one theories in the previous subsection.

\subsubsection{$A_2$ Theories}

Consider the $A_2$ gauge theories, i.e., those based on either $SU(3)$ or $PSU(3)=SU(3)/\Z_3$ gauge group. The $PSU(3)$ case is trivial, as the theory admits monopoles in fundamental representations of the dual group $SU(3)$.  Such monopoles are minuscule, so they do not bubble, and being the generators, they fully determine the algebra. In the $SU(3)$ gauge theory, however, the monopole charges take values in the weight lattice of $PSU(3)$, which coincides with its root lattice. Letting $\alpha_1$ and $\alpha_2$ denote simple roots of $SU(3)$, the coroots $\alpha_1^\vee = 2\alpha_1/(\alpha_1,\alpha_1)$ and $\alpha_2^\vee = 2\alpha_2/(\alpha_2,\alpha_2)$ generate the root lattice of $PSU(3)$, and physical monopole charges (cocharacters) correspond to Weyl orbits in this lattice. 

The minimal monopole operator corresponds to the Weyl orbit of $\alpha_1^\vee$, which coincides with the root system of $PSU(3)$. In standard conventions, $\alpha_1^\vee + \alpha_2^\vee$ is the dominant coroot, so we could use it to label the minimal-charge monopole operator $[P(\Phi)\cM^{\alpha_1^\vee+\alpha_2^\vee}]$. In practice, we find it slightly more convenient to label it by $\alpha_1^\vee$. Such a monopole can only bubble to a trivial representation, since the only weight shorter than $|\alpha_1^\vee|$ is a zero weight, and it belongs to the highest-weight representation generated by $\alpha_1^\vee + \alpha_2^\vee$. Therefore, there exists only one bubbling coefficient in this case, $Z(\Phi)$, which determines the abelianized version of the minimal monopole and its dressings:
\begin{align}
\widetilde{M}^{\alpha_1^\vee} &= M^{\alpha_1^\vee} + Z(\Phi),\cr
\left[P(\Phi)\cM^{\alpha_1^\vee}\right] &= \sum_{{\rm w}\in\cW} P(\Phi^{\rm w})\widetilde{M}^{{\rm w}\cdot \alpha_1^\vee}.
\end{align}
In the $A_2$ case, $\Phi=(\Phi_1, \Phi_2)$ and $\cW=S_3$; the ring of invariants can be described as
\begin{equation}
\C[\Phi_1, \Phi_2]^\cW = \C\left[f_1,f_2\right] \text{ where } f_1=\Phi_1^2+\Phi_2^2,\ f_2=\Phi_2(\Phi_2^2 - 3\Phi_1^2).
\end{equation}
There are six primitive dressed monopoles of minimal charge that generate the space of dressed monopoles (of minimal charge) as a $\C[\Phi_1, \Phi_2]^\cW$-module. They can be chosen as:
\begin{align}
\cM^{\alpha_1^\vee},\quad \left[\Phi_1\cM^{\alpha_1^\vee}\right],\quad \left[\Phi_2\cM^{\alpha_1^\vee}\right], \nonumber \\
\left[\Phi_1^2\cM^{\alpha_1^\vee}\right],\quad \left[\Phi_1\Phi_2\cM^{\alpha_1^\vee}\right],\quad \left[\Phi_1^3\cM^{\alpha_1^\vee}\right].
\label{primitivesforsu3}
\end{align}
The next step, just like in the rank-one case, is to compute star products of these with the lowest invariant polynomial $\Phi_1^2 + \Phi_2^2$ (often referred to as the quadratic Casimir in physics literature). A straightforward computation for general dressed $[P(\Phi)\cM^{\alpha_1^\vee}]$ gives:\footnote{Our conventions are that $\mathcal{W} = \{1, {\rm w}_a, {\rm w}_a^2, {\rm w}_b, {\rm w}_b {\rm w}_a, {\rm w}_b {\rm w}_a^2\}$ where
\begin{equation}
{\rm w}_a = \left(\begin{array}{cc} -1/2 & -\sqrt{3}/2 \\ \sqrt{3}/2 & -1/2 \end{array}\right), \quad {\rm w}_b = \left(\begin{array}{cc} -1 & 0 \\ 0 & 1 \end{array}\right),
\end{equation}
and ${\rm w} : \left(\begin{array}{cc} \Phi_1 & \Phi_2 \end{array}\right)\mapsto \left(\begin{array}{cc} \Phi_1 & \Phi_2 \end{array}\right){\rm w}$ for ${\rm w}\in \mathcal{W}$.}
\begin{align}
\left[P(\Phi)\cM^{\alpha_1^\vee}\right] \star (\Phi_1^2+\Phi_2^2) &= \left[((\Phi_1-2i/r)^2+\Phi_2^2)P(\Phi)\cM^{\alpha_1^\vee}\right] \nonumber \\
&\phantom{==} + \sum_{{\rm w}\in\cW} \left(\frac4{r^2} + \frac{4i}{r}\Phi_1^{\rm w} \right)P(\Phi^{\rm w})Z(\Phi^{\rm w}). \label{product_with_inv}
\end{align}
The last term above must be a Weyl-invariant polynomial for all possible polynomials $P$. It is enough to impose this requirement for $P = 1, \Phi_1, \Phi_2, \Phi_1^2, \Phi_1\Phi_2, \Phi_1^3$.  Recall that
\begin{equation}
\left[\Phi_1^j\Phi_2^k \cM^{\alpha_1^\vee}\right] = \sum_{{\rm w}\in\cW} (\Phi_1^{\rm w})^j(\Phi_2^{\rm w})^k M^{{\rm w}\cdot \alpha_1^\vee} + V_{jk}(\Phi), \quad V_{jk}(\Phi)\equiv\sum_{{\rm w}\in\cW} (\Phi_1^{\rm w})^j(\Phi_2^{\rm w})^k Z(\Phi^{\rm w}).
\end{equation}
We see that the last term in \eqref{product_with_inv} for $P = 1, \Phi_1, \Phi_2, \Phi_1^2, \Phi_1\Phi_2, \Phi_1^3$ is simply:
\begin{align}
\sum_{{\rm w}\in\cW} \left(\frac4{r^2} + \frac{4i}{r}\Phi_1^{\rm w} \right)Z(\Phi^{\rm w}) &= \frac{4}{r^2}V_{00}(\Phi) + \frac{4i}{r}V_{10}(\Phi), \nonumber \\
\sum_{{\rm w}\in\cW} \left(\frac4{r^2} + \frac{4i}{r}\Phi_1^{\rm w} \right)\Phi_1^{\rm w} Z(\Phi^{\rm w}) &= \frac{4}{r^2}V_{10}(\Phi) + \frac{4i}{r}V_{20}(\Phi), \nonumber \\
\sum_{{\rm w}\in\cW} \left(\frac4{r^2} + \frac{4i}{r}\Phi_1^{\rm w} \right)\Phi_2^{\rm w} Z(\Phi^{\rm w}) &= \frac{4}{r^2}V_{01}(\Phi) + \frac{4i}{r}V_{11}(\Phi), \nonumber \\
\sum_{{\rm w}\in\cW} \left(\frac4{r^2} + \frac{4i}{r}\Phi_1^{\rm w} \right)(\Phi_1^{\rm w})^2 Z(\Phi^{\rm w}) &= \frac{4}{r^2}V_{20}(\Phi) + \frac{4i}{r}V_{30}(\Phi), \label{linear_sys_A2} \\
\sum_{{\rm w}\in\cW} \left(\frac4{r^2} + \frac{4i}{r}\Phi_1^{\rm w} \right)\Phi_1^{\rm w}\Phi_2^{\rm w} Z(\Phi^{\rm w}) &= \frac{4}{r^2}V_{11}(\Phi) + \frac{4i}{r}V_{21}(\Phi), \nonumber \\
\sum_{{\rm w}\in\cW} \left(\frac4{r^2} + \frac{4i}{r}\Phi_1^{\rm w} \right)(\Phi_1^{\rm w})^3 Z(\Phi^{\rm w}) &= \frac{4}{r^2}V_{30}(\Phi) + \frac{4i}{r}V_{40}(\Phi). \nonumber
\end{align}
The right-hand side of each of these equations should be a Weyl-invariant polynomial.  This linear system can be solved for $Z(\Phi)$, but we will do better if we first use the operator mixing freedom.  Recall that $V_{00}$, $V_{10}$, $V_{01}$, $V_{20}$, $V_{11}$, $V_{30}$, being the bubbling terms in \eqref{primitivesforsu3}, can be shifted by Weyl-invariant polynomials in $\Phi_1, \Phi_2$ (and $r^{-1}$).  Using such shifts of $V_{10}$, $V_{20}$, $V_{11}$, $V_{30}$, we can make the right-hand sides of the first four equations in \eqref{linear_sys_A2} vanish, while those of the fifth and sixth ones should be Weyl-invariant polynomials.  In other words, we obtain:
\begin{align}
\frac{4}{r^2}V_{00}(\Phi) + \frac{4i}{r}V_{10}(\Phi) &= 0, & \frac{4}{r^2}V_{10}(\Phi) + \frac{4i}{r}V_{20}(\Phi) &= 0, \nonumber \\
\frac{4}{r^2}V_{01}(\Phi) + \frac{4i}{r}V_{11}(\Phi) &= 0, & \frac{4}{r^2}V_{20}(\Phi) + \frac{4i}{r}V_{30}(\Phi) &= 0, \label{vanishing} \\
\frac{4}{r^2}V_{11}(\Phi) + \frac{4i}{r}V_{21}(\Phi) &= \frac{1}{r}A(\Phi), & \frac{4}{r^2}V_{30}(\Phi) + \frac{4i}{r}V_{40}(\Phi) &= \frac{1}{r}B(\Phi), \nonumber
\end{align}
where $A$ and $B$ are Weyl-invariant polynomials (hence polynomials in $f_1$ and $f_2$).  Inserting this into \eqref{linear_sys_A2}, we solve the resulting linear system for $Z(\Phi)$ to find that
\begin{equation}
Z(\Phi) = \frac{i(\Phi_2 A(\Phi) - B(\Phi))}{6\Phi_1(\Phi_1 - i/r)(\Phi_1^2 - 3\Phi_2^2)}.
\end{equation}
So far, we have not used the freedom to shift $V_{00}$ and $V_{01}$ by Weyl-invariant polynomials $F_{00}(\Phi) = F_{00}(f_1,f_2)$ and $F_{01}(\Phi) = F_{01}(f_1,f_2)$.  To preserve the first four equations in \eqref{vanishing}, such shifts should be accompanied by
\begin{equation}
V_{k0}\to V_{k0} + \left(\frac{i}{r}\right)^k F_{00} \quad (k = 1, 2, 3), \quad V_{11}\to V_{11} + \frac{i}{r}F_{01}.
\end{equation}
Solving another linear system, namely
\begin{equation}
\begin{gathered}
\sum_{{\rm w}\in \mathcal{W}} \Delta Z(\Phi^{\rm w}) = F_{00}(\Phi), \quad \sum_{{\rm w}\in \mathcal{W}} \Phi_2^{\rm w}\Delta Z(\Phi^{\rm w}) = F_{01}(\Phi), \\
\sum_{{\rm w}\in \mathcal{W}} (\Phi_1^{\rm w})^k\Delta Z(\Phi^{\rm w}) = \left(\frac{i}{r}\right)^k F_{00}(\Phi) \quad (k = 1, 2, 3), \quad \sum_{{\rm w}\in \mathcal{W}} \Phi_1^{\rm w}\Phi_2^{\rm w}\Delta Z(\Phi^{\rm w}) = \frac{i}{r}F_{01}(\Phi),
\end{gathered}
\end{equation}
we find that such shifts trace back to the following shift in $Z(\Phi)$:
\begin{equation}
\Delta Z(\Phi) = -\frac{\Phi_2[(f_1 + 4/r^2)F_{01}(\Phi) - f_2 F_{00}(\Phi)] + (f_1 + 1/r^2)(f_1 + 4/r^2)F_{00}(\Phi) - f_2 F_{01}(\Phi)}{6\Phi_1(\Phi_1 - i/r)(\Phi_1^2 - 3\Phi_2^2)}.
\end{equation}
Shifting $Z$ by such an expression is equivalent to shifting $A(\Phi) = A(f_1, f_2)$ and $B(\Phi) = B(f_1, f_2)$ by
\begin{align}
\Delta A(f_1, f_2) &= i(f_1 + 4/r^2)F_{01}(f_1, f_2) - if_2 F_{00}(f_1, f_2), \nonumber \\
\Delta B(f_1, f_2) &= if_2 F_{01}(f_1, f_2) - i(f_1 + 1/r^2)(f_1 + 4/r^2)F_{00}(f_1, f_2).
\end{align}
We can use such shifts to eliminate the $f_2$-dependence of $A$ and $B$.  Indeed, we first choose $F_{00}$ to eliminate the $f_2$-dependence of $A$.  We then choose $F_{01}(f_1, f_2) = F(f_1)$ to preserve the condition that $A(f_1, f_2) = A(f_1)$.  Noting that shifts of the form
\begin{align}
\Delta F_{00}(f_1, f_2) &= (f_1 + 4/r^2)P(f_1, f_2), \nonumber \\
\Delta F_{01}(f_1, f_2) &= f_2 P(f_1, f_2)
\end{align}
leave $\Delta A(f_1, f_2)$ invariant, we still have the freedom to shift $B$ by
\begin{equation}
\Delta B(f_1, f_2) = if_2 F(f_1) + i(f_2^2 - (f_1 + 1/r^2)(f_1 + 4/r^2)^2)P(f_1, f_2).
\end{equation}
The $P$-dependent part of this shift can be used to make $B$ at most linear in $f_2$ (by polynomial long division), whereupon $F$ can be chosen to eliminate the remaining $f_2$-dependence of $B$.  Having completely used the mixing freedom in this way, we find that the abelianized bubbling term takes the form:
\begin{equation}
Z(\Phi) = \frac{i(\Phi_2 A(f_1) - B(f_1))}{\Phi_1(\Phi_1 - i/r)(\Phi_1^2 - 3\Phi_2^2)},\quad f_1=\Phi_1^2 + \Phi_2^2,\quad f_2=\Phi_2(\Phi_2^2 - 3\Phi_1^2).
\label{su3result}
\end{equation}
We have now reached the limits of what can be done based on the gauge group only. The concrete expressions for the polynomials $A(f_1)$ and $B(f_1)$ depend on the matter content, as in the rank-one case. For simplicity, let us consider only the case of an $SU(3)$ vector multiplet coupled to $N_f$ fundamental flavors. We then compute the following star product:
\begin{equation}
\label{su3_nopole}
\cM^{\alpha_1^\vee}\star \left[\Phi_1\cM^{\alpha_1^\vee}\right] - \left[(\Phi_1- 2i/r)\cM^{\alpha_1^\vee}\right]\star \cM^{\alpha_1^\vee} = \left[P(\Phi)\cM^{\alpha_1^\vee}\right] + R(\Phi).
\end{equation}
The combination above is devised in such a way that $\cM^{2\alpha_1^\vee}$ does not show up on the right. The monopole $\cM^{\alpha_2^\vee + 2\alpha_1^\vee}$ would be present for more general matter representations (e.g., if we included adjoint matter), but it does not show up in our case either, which is why the theory with only fundamental matter is somewhat simpler. Here, $P(\Phi)$ is some polynomial dressing factor, while $R(\Phi)$ is a Weyl-invariant polynomial.

The expressions for $P$ and $R$ are lengthy, so we do not provide them here for brevity. Polynomiality of $P(\Phi)$ --- that is, cancellation of poles --- determines the unknown terms $A(f_1)$ and $B(f_1)$. We find that
\begin{equation}
\label{su3_answer1}
\begin{aligned}
&\text{for even $N_f$:} & A(f_1) &= 0, & B(f_1) &= \textstyle -4i\left(\frac{-i}{2\sqrt{3}}\right)^{N_f}(f_1 + 1/r^2)^{N_f/2}, \\[5 pt]
&\text{for odd $N_f$:} & B(f_1) &= 0, & A(f_1) &= \textstyle 4i\left(\frac{-i}{2\sqrt{3}}\right)^{N_f}(f_1 + 1/r^2)^{(N_f-1)/2}.
\end{aligned}
\end{equation}
We have thus determined \eqref{su3result}.

As in the rank-one case, this result can be generalized to a gauge group $G=SU(3)\times G'$ and $SU(3)$-valued monopoles. If the $N_f$ fundamentals of $SU(3)$ form a representation $\cR'$ of $G'$, then the no-pole condition encodes the polynomials $A(f_1)$ and $B(f_1)$ as follows:
\begin{equation}
xA(x^2 - r^{-2}) - B(x^2 - r^{-2}) = 4i\prod_{w\in\cR'}\left(-\frac{ix}{2\sqrt{3}} - iw\cdot \Phi'\right).
\end{equation}
Here, $\Phi'$ corresponds to scalars from the $G'$ vector multiplet. This formula reproduces \eqref{su3_answer1} if we take $\cR'=\C^{N_f}$ to be a trivial representation of $G'$, that is, all weights $w$ to be zero.

This final result allows one to study quivers of $SU(3)$ groups in which every gauge node only couples to fundamental matter; it also allows for the inclusion of masses by treating $\Phi'$ as a background.

\paragraph{Higher magnetic charges.}

Finally, we would like to explain how to construct monopoles of other magnetic charges. This is straightforward in the theory with $PSU(3)$ gauge group: the dual group is $SU(3)$, so both fundamental representations of $SU(3)$ give allowed monopole charges. Their tensor products generate arbitrary representations of $SU(3)$. In the case of $SU(3)$ gauge theory, things are slightly more involved, but still tractable. 

We have derived an expression for $Z(\Phi)$, which is enough to build a dressed monopole $[P(\Phi_1,\Phi_2)\cM^{\alpha_1^\vee}]$ corresponding to the Weyl orbit of $\alpha_1^\vee$, with arbitrary polynomial $P$. Is it enough to construct all allowed monopoles in the theory? After all, the coroots take values in a two-dimensional lattice spanned by $\alpha_1^\vee, \alpha_2^\vee$, and merely on representation-theoretic grounds, one cannot construct all representations labeled by dominant weights in this lattice just from tensor products of the adjoint representation. However, by taking star products of dressed monopoles $[P(\Phi_1,\Phi_2)\cM^{\alpha_1^\vee}]$, one can actually generate everything else.

From \eqref{dimensionformula}, we see that in an $SU(3)$ theory with $N_f$ fundamentals, the dimensions of three lowest bare monopoles are
\begin{equation}
\Delta_{\alpha_1^\vee}=N_f -4,\quad
\Delta_{2\alpha_1^\vee}=2N_f -8,\quad
\Delta_{\alpha_2^\vee+2\alpha_1^\vee}=2N_f-6.
\end{equation}
Since $2\Delta_{\alpha_1^\vee}<\Delta_{\alpha_2^\vee+2\alpha_1^\vee}$, the product of two bare monopoles $\cM^{\alpha_1^\vee}$ cannot generate $\cM^{\alpha_2^\vee+2\alpha_1^\vee}$. However, the latter can appear if we compensate for the mismatch in dimensions by dressing the monopoles with an appropriate number of $\Phi$'s. For example, the star product
\begin{equation}
\left[\Phi_1(\Phi_1-2i/r)\cM^{\alpha_1^\vee}\right]\star \cM^{\alpha_1^\vee}- \left[\Phi_1\cM^{\alpha_1^\vee}\right]\star \left[\Phi_1\cM^{\alpha_1^\vee}\right]
\end{equation}
has $\cM^{\alpha_2^\vee+2\alpha_1^\vee}$ as a leading term, and can therefore serve as a definition of $\cM^{\alpha_2^\vee+2\alpha_1^\vee}$. Similarly taking products of monopoles dressed by higher-degree polynomials, we can obtain dressed versions of $\cM^{\alpha_2^\vee+2\alpha_1^\vee}$. Having constructed in this way both $\cM^{\alpha_1^\vee}$, $\cM^{\alpha_2^\vee+2\alpha_1^\vee}$ and their dressed versions, we can generate all other allowed monopoles.

\subsubsection{$B_2\cong C_2$ Theories}

There are two compact rank-two gauge groups that correspond to the $B_2\cong C_2$ Lie algebra: $USp(4)\cong Spin(5)$ and $SO(5)\cong USp(4)/\Z_2$, which are Langlands dual to each other. The group $USp(4)$ is often called $Sp(2)$, but we will use the former notation. The root lattice is generated by the short simple root $\alpha$ and the long simple root $\beta$. In our conventions, we write them in Cartesian coordinates as $\alpha=(1,0)$ and $\beta=(-1,1)$. The coroots are $\alpha^\vee=2\alpha=(2,0)$ and $\beta^\vee=\beta=(-1,1)$, so that $\alpha^\vee$ is a long coroot.

\paragraph{$SO(5)$ gauge theory.} First consider the $SO(5)$ gauge theory. The monopoles are labeled by (Weyl orbits of) the dominant weights of the dual group $USp(4)$, whose root lattice is generated by $\alpha^\vee$ and $\beta^\vee$. The group $USp(4)$ has two fundamental representations: the four-dimensional defining representation and the five-dimensional vector representation of $SO(5)=USp(4)/\Z_2$. The four-dimensional representation has weights
\begin{gather}
\omega_1^\vee=\frac{1}{2}\alpha^\vee=(1,0), \quad \omega_1^\vee+\beta^\vee=(0,1), \nonumber \\
\omega_1^\vee-\alpha^\vee=(-1,0), \quad \omega_1^\vee-\alpha^\vee-\beta^\vee=(0,-1).
\end{gather}
This representation is minuscule, so the smallest monopole of the model does not bubble:
\begin{equation}
\left[P(\Phi)\cM^{\omega_1^\vee}\right]=\sum_{{\rm w}\in\cW} P(\Phi^{\rm w}) M^{{\rm w}\cdot \omega_1^\vee}.
\end{equation}
The five-dimensional representation of $USp(4)$ is not minuscule.  Its weights are
\begin{gather}
\omega_2^\vee=\alpha^\vee + \beta^\vee=(1,1), \quad \omega_2^\vee-\alpha^\vee=\beta^\vee=(-1,1), \nonumber \\
-\beta^\vee=(1,-1), \quad -\beta^\vee-\alpha^\vee=(-1,-1), \quad (0,0).
\end{gather}
We see that the charge-$\omega_2^\vee$ monopole can bubble to zero magnetic charge. Therefore, the abelianized monopole takes the form
\begin{equation}
\label{w2_abel_B2}
\widetilde{M}^{\omega_2^\vee} = M^{\omega_2^\vee} + Z(\Phi).
\end{equation}
This $Z(\Phi)$ can be deduced by computing star products involving only the minimal monopole $[P(\Phi)\cM^{\omega_1^\vee}]$. On the other hand, it can also be found using our algorithmic polynomiality approach (which is really a different application of the same idea, namely consistency of the OPE algebra). Let us determine it using such an approach --- both for practice, and because it will soon be useful for the study of the $USp(4)$ gauge theory.

The charge-$\omega_2^\vee$ dressed monopoles are constructed as
\begin{equation}
\label{abel_mono_reminder}
\left[P(\Phi)\cM^{\omega_2^\vee}\right]=\frac12\sum_{{\rm w}\in\cW} P(\Phi^{\rm w})\widetilde{M}^{{\rm w}\cdot\omega_2^\vee}.
\end{equation}
As before, $\Phi=(\Phi_1,\Phi_2)\in\mathfrak{t}_\C$. The Weyl group is $D_4=\Z_4 \rtimes \Z_2$, and the ring of invariants is
\begin{equation}
\C[\Phi_1,\Phi_2]^\cW=\C[f_1, f_2] \text{ where } f_1=\Phi_1^2 + \Phi_2^2,\ f_2=\Phi_1^2\Phi_2^2.
\end{equation}
Notice that $\omega_2^\vee$ is preserved by the subgroup of $\cW$ that switches $\Phi_1\leftrightarrow \Phi_2$, which explains the $\frac12$ in \eqref{abel_mono_reminder}. Therefore, such monopoles can only be dressed by polynomials symmetric under $\Phi_1\leftrightarrow \Phi_2$ (this happens automatically once we apply \eqref{abel_mono_reminder}). Dressing by a symmetric polynomial only depends on the symmetric part of $Z(\Phi_1,\Phi_2)$. Therefore, we may assume that $Z(\Phi_1, \Phi_2)=Z(\Phi_2,\Phi_1)$.

Since the Weyl orbit of $\omega_2^\vee$ has four elements, there are four primitive dressed monopoles that generate all dressed charge-$\omega_2^\vee$ monopoles as a $\C[\Phi_1,\Phi_2]^\cW$-module. Choose them to be:
\begin{align}
\cM^{\omega_2^\vee},\quad \left[(\Phi_1+\Phi_2)\cM^{\omega_2^\vee}\right],\quad \left[(\Phi_1+\Phi_2)^2\cM^{\omega_2^\vee}\right],\quad \left[(\Phi_1+\Phi_2)^3\cM^{\omega_2^\vee}\right].
\end{align}
The next step is to compute their star products with $f_1$. For arbitrary $P(\Phi)$, we find:
\begin{align}
\label{B2_poly_step1}
\left[P(\Phi)\cM^{\omega_2^\vee}\right]\star (\Phi_1^2 + \Phi_2^2)&-\left[\left(\left(\Phi_1 -\frac{i}{r}\right)^2 + \left(\Phi_2 -\frac{i}{r}\right)^2 \right)P(\Phi)\cM^{\omega_2^\vee}\right]\cr
&=\sum_{{\rm w}\in\cW}\left(\frac{1}{r^2} + \frac{i}{r}(\Phi_1^{\rm w} + \Phi_2^{\rm w}) \right)P(\Phi^{\rm w})Z(\Phi^{\rm w}).
\end{align}
We require that the second line be a polynomial, in particular for $P=1$, $\Phi_1+\Phi_2$, $(\Phi_1+\Phi_2)^2$, and $(\Phi_1+\Phi_2)^3$. Using notation similar to that in the $SU(3)$ case,
\begin{equation}
\left[(\Phi_1+\Phi_2)^k\cM^{\omega_2^\vee}\right]=\frac12\sum_{{\rm w}\in\cW} (\Phi_1^{\rm w}+\Phi_2^{\rm w})^k\widetilde{M}^{{\rm w}\cdot\omega_2^\vee} + V_k(\Phi), \quad V_k(\Phi)\equiv\frac12\sum_{{\rm w}\in\cW}(\Phi_1^{\rm w}+\Phi_2^{\rm w})^k Z(\Phi^{\rm w}),
\end{equation}
we identify the last term in \eqref{B2_poly_step1} for $P(\Phi)=(\Phi_1+\Phi_2)^k$ as $\frac{2}{r^2}V_k(\Phi) + \frac{2i}{r}V_{k+1}(\Phi)$, which we demand to be a Weyl-invariant polynomial. Using the operator mixing freedom to shift $V_1$, $V_2$, and $V_3$, we can make the first three of these polynomials vanish, but not the last:
\begin{align}
\label{linsys_B2}
\frac{2}{r^2}V_k(\Phi) + \frac{2i}{r}V_{k+1}(\Phi)&=0 \quad (k = 0, 1, 2),\cr
\frac{2}{r^2}V_3(\Phi) + \frac{2i}{r}V_{4}(\Phi)&=\frac1{r}A(\Phi_1^2+\Phi_2^2, \Phi_1^2\Phi_2^2)\in\C[\Phi_1^2+\Phi_2^2, \Phi_1^2\Phi_2^2].
\end{align}
Using the expressions $V_k(\Phi)=\sum_{{\rm w}\in\cW}(\Phi_1^{\rm w}+\Phi_2^{\rm w})^k Z(\Phi^{\rm w})$, we solve this system of four equations under the assumption that $Z(\Phi_1,\Phi_2)=Z(\Phi_2,\Phi_1)$ to find:
\begin{equation}
\label{Z_first_sol_B2}
Z(\Phi)=-\frac{iA(\Phi_1^2+\Phi_2^2, \Phi_1^2\Phi_2^2)}{32\Phi_1 \Phi_2(\Phi_1+\Phi_2)(\Phi_1+\Phi_2-i/r)}.
\end{equation}
The next step is to fix the remaining mixing freedom, which allows for shifts of $V_0$ by Weyl-invariant polynomials $F(\Phi_1^2+\Phi_2^2, \Phi_1^2\Phi_2^2)$, namely $V_0 \to V_0 + F$. To preserve the form of the equations \eqref{linsys_B2}, we also shift $V_k \to V_k + \left(\frac{i}{r}\right)^k F$ for $k = 1, 2, 3$. This can be solved for the corresponding shift $\Delta Z(\Phi)$ of $Z(\Phi)$:
\begin{equation}
\Delta Z(\Phi)=-\frac{(\Phi_1+\Phi_2+i/r)(\Phi_1-\Phi_2+i/r)(\Phi_1-\Phi_2-i/r)F\left(\Phi_1^2+\Phi_2^2, \Phi_1^2\Phi_2^2\right)}{16\Phi_1\Phi_2(\Phi_1+\Phi_2)}.
\end{equation}
Comparing with \eqref{Z_first_sol_B2}, we see that such shifts are equivalent to shifting $A$ by
\begin{align}
\Delta A =\left(\frac{2}{r^4}+\frac{4f_1}{r^2}+f_1^2 - 4f_2 \right)F(f_1, f_2),
\end{align}
where $f_1=\Phi_1^2+\Phi_2^2$ and $f_2=\Phi_1^2\Phi_2^2$. Because the expression in parentheses is no more than linear in $f_2$, such shifts can completely eliminate the $f_2$-dependence from $A$. Indeed, for an arbitrary polynomial $A(f_1,f_2)$, there is a unique $F(f_1,f_2)$ such that $A(f_1,f_2)+ \Delta A(f_1,f_2)$ depends only on $f_1$. This fully fixes the mixing freedom, so that in the end, we have
\begin{equation}
\label{Z_sol_B2}
Z(\Phi)=\frac{a(\Phi_1^2+\Phi_2^2)}{\Phi_1 \Phi_2(\Phi_1+\Phi_2)(\Phi_1+\Phi_2-i/r)}.
\end{equation}
Determining $a$ requires computing an appropriate star product. The answer depends on the matter content, and in this case it is not too hard to include both $N_f$ five-dimensional flavors of $SO(5)$ and $N_a$ adjoint flavors. We consider the following star product of minimal monopoles, which is enough to generate the next-to-minimal monopole of charge $\omega_2^\vee$:
\begin{align}
\label{so5_step2}
\cM^{\omega_1^\vee}\star \left[\Phi_1^3\cM^{\omega_1^\vee}\right] - \left[(\Phi_1-i/r)^3\cM^{\omega_1^\vee}\right]\star \cM^{\omega_1^\vee}.
\end{align}
The reason for including $\Phi_1^3$ can be seen from the dimensions of the monopoles:
\begin{align}
\Delta_{\omega_1^\vee}=N_f + 3(N_a-1),\quad
\Delta_{\omega_2^\vee}=2N_f + 4(N_a-1).
\end{align}
Only with the insertion of (at least) $\Phi_1^3$ do we find that the dimension of \eqref{so5_step2}, given by $2\Delta_{\omega_1^\vee}+3=2N_f + 6N_a - 3$, exceeds $\Delta_{\omega_2^\vee}$ for all values of $N_a$, thus allowing the monopole of charge $\omega_2^\vee$ to appear on the right. It indeed appears, in bare form for $N_a=0$ and in dressed form for $N_a\neq 0$. We subtract it from the above star product and look at the free (charge-zero) term, demanding its polynomiality. This determines $a(\Phi_1^2+\Phi_2^2)$. For brevity, we do not present the cumbersome intermediate formulas and only give the final answer:
\begin{equation}
\label{a_SO5}
a(x)=-\left(\frac{x}{2}+\frac1{4r^2} \right)^{N_f}\left(-\frac{x}{8r^2}-\frac1{16r^4}\right)^{N_a}.
\end{equation}
This expression determines $Z(\Phi)$, from which we can construct arbitrary dressed monopoles of charge $\omega_2^\vee$. With the two monopoles corresponding to fundamental weights of $USp(4)$ in hand, we can construct arbitrary monopoles in the $SO(5)$ gauge theory. Notice that it was clear from the beginning that the charge-$\omega_1^\vee$ monopole suffices to generate the algebra.

Like in all cases so far, it is not hard to generalize to a non-simple gauge group $G=SO(5)\times G'$, assuming that the $N_f$ fundamentals of $SO(5)$ form a representation $\cR_f'$ of $G'$ while the $N_a$ adjoints transform in $\cR_a'$ of $G'$. Modifying the above calculation appropriately gives
\begin{equation}
a(x)=-\prod_{w\in\cR_f'}\left(\frac{x}{2} + \frac{1}{4r^2} - (w\cdot\Phi')^2\right)\prod_{w\in\cR_a'}\left[\left(-\frac1{4r^2}-(w\cdot \Phi')^2\right)\left(\frac{x}{2}+\frac1{4r^2}-(w\cdot \Phi')^2\right)\right].
\end{equation}
Here, as before, $\Phi'$ is valued in the $G'$ vector multiplets. This answer for $a(x)$ of course reduces to \eqref{a_SO5} when all weights in $\cR_f'$ and $\cR_a'$ vanish. As usual, $\Phi'$ plays the role of a mass matrix if we treat $G'$ as a global symmetry.

\paragraph{$USp(4)$ gauge theory.} Consider the $USp(4)$ gauge theory. It has the same simple roots $\alpha,\beta$ and simple coroots $\alpha^\vee$, $\beta^\vee$ as in the $SO(5)$ case. Only the lattice of allowed weights of matter representations is different. 

The dual group is $SO(5)$, which has no minuscule representations. The minimal monopole has charge $\omega_2^\vee$, like the next-to-minimal monopole of the $SO(5)$ gauge theory. It is defined by the same equations \eqref{w2_abel_B2} and \eqref{abel_mono_reminder}. Further steps involving the ring of invariants, the primitive dressed monopoles, and ultimately the answer \eqref{Z_sol_B2} are applicable to the $USp(4)$ case as well --- they do not depend on the global form of the gauge group. To proceed, we need to find the polynomial $a$ entering the abelianized bubbling coefficient \eqref{Z_sol_B2}. This step depends on the matter content, and hence on the global form of the gauge group.

Let us consider for simplicity a theory which only has matter in $N_4$ copies of the four-dimensional representation of $USp(4)$. We compute the star product
\begin{equation}
\cM^{\omega_2^\vee}\star \left[(\Phi_1+\Phi_2)\cM^{\omega_2^\vee}\right]-\left[(\Phi_1+\Phi_2-2i/r)\cM^{\omega_2^\vee}\right]\star \cM^{\omega_2^\vee}.
\end{equation}
The charges $2\omega_2^\vee$ and $2\omega_1^\vee$ cancel from the result. Polynomiality of the remainder requires $a(x)$ to be a constant, and determines its square. We describe the answer in the more general case of $G=USp(4)\times G'$, assuming that the $N_4$ fundamentals of $USp(4)$ transform in $\cR_4'$ of $G'$:
\begin{equation}
a=-\prod_{w\in\cR_4'} (-i w\cdot\Phi'),
\end{equation}
which determines the bubbling coefficient
\begin{equation}
Z(\Phi)=\frac{a}{\Phi_1 \Phi_2(\Phi_1+\Phi_2)(\Phi_1+\Phi_2-i/r)}.
\label{compareusp4}
\end{equation}
The situation here is reminiscent of the $SU(2)$ case: without an extra group $G'$, all weights $w$ vanish, and we find that $a=0$. In other words, in the $USp(4)$ gauge theory with $N_4>0$, in the absence of extra gaugings and masses, the bubbling coefficient of the minimal monopole can be removed using operator mixing. If $N_4=0$, then $a=-1$, where the sign was chosen to agree with the $SO(5)$ answer (in the $N_4=0$ case, the absence of matter allows for both $USp(4)$ and $SO(5)$ gauge groups, and we can determine the $USp(4)$ answer from the $SO(5)$ answer: the theories differ only by a $\Z_2$ gauging). For other values of $N_4$, we fixed the sign of $a$ arbitrarily, since it only affects the algebra $\cA_C$ up to a change of basis.

Finally, let us add that at this point, using some physics intuition, we can easily guess the answer for $a(x)$ in the more general case where we have matter in a representation $[\mathbf{4}\otimes\cR_4'] \oplus [\mathbf{5}\otimes\cR_5'] \oplus [\mathbf{adj}\otimes \cR_a']$ of the gauge group $G=USp(4)\times G'$. Here, $\mathbf{4}$, $\mathbf{5}$, and $\mathbf{adj}$ are the four-dimensional, five-dimensional, and adjoint representations of $USp(4)$, and $\cR_4'$, $\cR_5'$, $\cR_a'$ are some representations of $G'$. We have seen before that contributions of different matter multiplets enter the answer for $a(x)$ multiplicatively. This makes sense from the localization point of view: bubbling terms are given by one-loop determinants around fixed points in the bubbling loci, and one-loop determinants of various matter multiplets contribute multiplicatively. So it is natural to expect that the answer in this general case should be given by
\begin{align}
a(x) &= -\prod_{w\in\cR_4'} (-i w\cdot\Phi')\prod_{w\in\cR_5'}\left(\frac{x}{2} + \frac{1}{4r^2} - (w\cdot\Phi')^2\right) \nonumber \\
&\times \prod_{w\in\cR_a'}\left[\left(-\frac1{4r^2}-(w\cdot \Phi')^2\right)\left(\frac{x}{2}+\frac1{4r^2}-(w\cdot \Phi')^2\right)\right].
\end{align}
Above, we borrowed the contributions of $\mathbf{5}$ and $\mathbf{adj}$ from the subsection on the $SO(5)$ case, as the theory with only these types of matter allows for either $SO(5)$ or $USp(4)$ gauge group.

\subsubsection{$G_2$ Theories}\label{sec:G2_general}

The remaining rank-two simple gauge group is $G_2$. It has only one compact form, which is of course centerless and Langlands dual to itself, meaning that we do not have to study various cases as before. We describe the root system $\Delta$ of $G_2$ in Cartesian coordinates such that the short simple root is $\alpha=(1,0)$ and the long simple root is $\beta=(-\frac32, \frac{\sqrt 3}{2})$. The corresponding coroots are $\alpha^\vee=2\alpha=(2,0)$ and $\beta^\vee=\frac23 \beta=(-1,\frac{\sqrt3}{3})$, which are now long and short, respectively, and generate the root system $\Delta^\vee$ of the dual $G_2$. It is convenient to describe $\Delta^\vee$ in terms of another pair of simple coroots, which we define as $\alpha_{\rm mon}=\alpha^\vee+2\beta^\vee=(0,\frac2{\sqrt 3})$ and $\beta_{\rm mon}=-2\alpha^\vee-3\beta^\vee=(-1,-\sqrt{3})$, where now $\alpha_{\rm mon}$ is short and $\beta_{\rm mon}$ is long.

The smallest irreducible representation is 7-dimensional: its weights are given by a zero weight $(0,0)$ and the six short roots in $\Delta$, namely $\alpha$ and its Weyl images. Because of the zero weight, the representation is not minuscule. The next-smallest irreducible representation is the 14-dimensional adjoint representation. The $\mathbf{7}$ and $\mathbf{14}$ are fundamental representations, but we will refer only to $\mathbf{7}$ as the fundamental, and to $\mathbf{14}$ as the adjoint.

The minimal monopole charge is described by the nonzero weights in the $\mathbf{7}$ of the dual $G_2$, i.e., by the Weyl orbit of the short coroot $\alpha_{\rm mon}$ (or equivalently, by $\beta^\vee$, which belongs to the same Weyl orbit). Because $\mathbf{7}$ is not minuscule, it can bubble into the identity:
\begin{equation}
\widetilde{M}^{\alpha_{\rm mon}} = M^{\alpha_{\rm mon}} + Z(\Phi),
\end{equation}
and the physical dressed monopole of minimal charge is defined by
\begin{equation}
\left[P(\Phi)\cM^{\alpha_{\rm mon}}\right]=\frac12\sum_{{\rm w}\in\cW} P(\Phi^{\rm w})\widetilde{M}^{{\rm w}\cdot \alpha_{\rm mon}}.
\end{equation}
The Weyl group is $\cW=D_6 = \Z_6\rtimes \Z_2$, the group of symmetries of a hexagon. As usual, $\Phi=(\Phi_1,\Phi_2)\in\mathfrak{t}_\C$, and the ring of invariants is
\begin{equation}
\C[\Phi_1,\Phi_2]^\cW = \C[f_1, f_2] \text{ where } f_1=\Phi_1^2 + \Phi_2^2,\ f_2=\Phi_2^2(\Phi_2^2-3\Phi_1^2)^2.
\end{equation}
Because the Weyl orbit of $\alpha_{\rm mon}$ has order $6$, there are six primitive dressed monopoles:
\begin{align}
\cM^{\alpha_{\rm mon}},\quad \left[\Phi_2\cM^{\alpha_{\rm mon}}\right],\quad \left[\Phi_2^2\cM^{\alpha_{\rm mon}}\right],\cr
\left[\Phi_2^3\cM^{\alpha_{\rm mon}}\right],\quad \left[\Phi_2^4\cM^{\alpha_{\rm mon}}\right],\quad \left[\Phi_2^5\cM^{\alpha_{\rm mon}}\right].
\end{align}
The next few steps are exactly the same as before.  Namely, we compute the star product
\begin{align}
\label{G2_step1}
\left[P(\Phi)\cM^{\alpha_{\rm mon}}\right] \star (\Phi_1^2 + \Phi_2^2) &- \left[\left(\Phi_1^2 + \left(\Phi_2 - \frac{2i}{r\sqrt{3}} \right)^2 \right)P(\Phi)\cM^{\alpha_{\rm mon}}\right]\cr
&=\frac12\sum_{{\rm w}\in\cW} \left(\frac{4}{3r^2} + \frac{4i}{r\sqrt{3}}\Phi_2^{\rm w} \right)P(\Phi^{\rm w})Z(\Phi^{\rm w})
\end{align}
and demand polynomiality for $P=1,\Phi_2,\ldots, \Phi_2^5$. Because $\alpha_{\rm mon}$ is preserved by the Weyl reflection $(\Phi_1, \Phi_2) \to (-\Phi_1, \Phi_2)$, it is sufficient to consider only dressings by polynomials invariant under such a reflection (which also explains the factor of $\frac12$ in the definition of the monopole). Therefore, one can assume from the beginning that
\begin{equation}
Z(\Phi_1, \Phi_2)=Z(-\Phi_1, \Phi_2).
\end{equation}
Polynomiality of the last term in \eqref{G2_step1} and the operator mixing freedom almost completely determine $Z(\Phi)$. To avoid repetition, we simply state the final answer:
\begin{equation}
Z(\Phi)=\frac{A(\Phi_1^2 + \Phi_2^2)}{\Phi_2 (\sqrt{3}\Phi_2 -\frac{i}{r})(\Phi_1^2 - 3\Phi_2^2)(3\Phi_1^2 - \Phi_2^2)}.
\end{equation}
Finally, to determine the polynomial $A	$, we compute another star product:
\begin{align}
\label{G2_step2}
\cM^{\alpha_{\rm mon}}\star \left[\Phi_2\cM^{\alpha_{\rm mon}}\right] - \left[\left(\Phi_2-\frac{2i}{r\sqrt{3}}\right)\cM^{\alpha_{\rm mon}}\right] \star \cM^{\alpha_{\rm mon}}.
\end{align}
At this point, we limit ourselves to the theory with $N_f$ seven-dimensional flavors of $G_2$. In this case, higher magnetic charges $2\alpha_{\rm mon}$ and $\beta_{\rm mon}$ cancel from the above expression. The monopole of charge $2\alpha_{\rm mon}$ cancels because \eqref{G2_step2} is specifically constructed to ensure its cancellation, while the charge $\beta_{\rm mon}$ cannot appear for dimensional reasons. Indeed, the dimensions of the lowest monopoles are
\begin{align}
\Delta_{\alpha_{\rm mon}}=2N_f-6,\quad
\Delta_{\beta_{\rm mon}}=4N_f-10. \label{dimg2}
\end{align}
The dimension of \eqref{G2_step2} is $2\Delta_{\alpha_{\rm mon}}+1<\Delta_{\beta_{\rm mon}}$, so the monopole of dimension $\Delta_{\beta_{\rm mon}}$ cannot appear on the right. However, the dressed monopole of charge $\alpha_{\rm mon}$ appears, and demanding polynomiality of its dressing factor determines $A$ to be
\begin{equation}
A(x)=\frac{16}{3\sqrt{3}}\left(\frac{x}{4} + \frac1{12 r^2} \right)^{N_f}.
\end{equation}

It is also not difficult to generalize to the case of a gauge group $G = G_2 \times G'$, assuming that the $N_f$ fundamentals of $G_2$ transform in a representation $\cR'$ of $G'$:
\begin{equation}
A(x)=\frac{16}{3\sqrt{3}}\prod_{w\in \cR'}\left(\frac{x}{4} + \frac1{12 r^2} - (w\cdot \Phi')^2 \right).
\end{equation}

\subsection{General Case}

The detailed exploration of the lower-rank theories in the above subsections should give the reader a sense of what polynomiality-based computations look like. Further, it shows a clear pattern and allows us to formulate a strategy that should work for general gauge groups.

To begin, one identifies the set of minimal monopoles that are expected to generate the algebra. They can either be minuscule or bubble into the charge-zero sector (which we refer to as ``bubbling into the identity''). They cannot bubble into smaller nonzero charges, as that would contradict their minimality. If all of them are minuscule, we are done: it only remains to make sure that they indeed generate everything, and to determine the relations. 

If there exists a minimal monopole of charge $\omega$ that is not minuscule, then it can bubble into the identity, and we should determine the corresponding abelianized bubbling factor $Z(\Phi)$. First, we use invariant theory to identify the set of primitive dressed monopoles of charge $\omega$. Then we compute their star products with the quadratic Casimir $f_1 = \sum_{i=1}^{r}\Phi_i^2$. By demanding polynomiality of the answer and using the operator mixing freedom, we almost completely determine the bubbling factor $Z(\Phi)$, up to an unknown Weyl-invariant polynomial $A$. These steps clearly work in an arbitrary gauge theory. The next step is the most challenging one: we need to construct a star product that determines the unknown polynomial $A$. We have seen that at this step, sometimes $A$ is uniquely determined, and sometimes it is only determined up to a sign, which is a harmless ambiguity that can be related to a change of basis in $\cA_C$.

We consider the above procedure as strong evidence that polynomiality fully determines the algebra $\cA_C$ (if not a proof, at a physical level of rigor). It would still be desirable to find a more elegant and mathematically illuminating way to reach this conclusion.

\section{Applications and Examples}\label{sec:Appl}

We now demonstrate the applications of our shift operator formalism in a number of simple examples.  More elaborate examples can be found in the appendices.

\subsection{Chiral Rings and Coulomb Branches}

In the commutative limit ($r\to\infty$), the quantum algebra $\cA_C$ reduces to the Coulomb branch chiral ring. Because finite-$r$ computations, as shown above, allow one to determine bubbling coefficients and thus $\cA_C$ in any theory, this provides a simple way to construct Coulomb branches even when other approaches face difficulties. However, finite-$r$ computations can be very hard, so it is convenient to first develop the commutative version of shift operators. This is the subject of this section, and the answer takes the form of abelianization as in \cite{Bullimore:2015lsa}.

We begin by noting that the shift operator $M^b_N$ from \eqref{general_shift_N} has a well-defined $r\to\infty$ limit. First, because the operator $e^{-b\cdot (\frac{i}{2}\partial_\sigma + \partial_B)}$ acts on $\Phi$ by a shift
\begin{equation}
\Phi \mapsto \Phi - \frac{i}{r}b,
\end{equation}
this shift vanishes in the $r\to\infty$ limit, so that $e^{-b\cdot (\frac{i}{2}\partial_\sigma + \partial_B)}$ no longer acts on $\Phi$-dependent terms. Instead, it effectively turns into a generator of the group ring $\C[\Lambda_w^\vee]$ associated to the lattice of coweights (considered as an abelian group). Such generators, denoted by $e[b]$, are subject to the relations
\begin{equation}
e[b_1]e[b_2]=e[b_1+b_2].
\end{equation}
Next, we observe that the $\Phi$-dependent rational prefactor in the definition \eqref{general_shift_N} of $M^b_N$ also has a well-defined $r\to\infty$ limit. Denoting the commutative limit of $M^b_N$ by $v^b$, we find that it looks as follows:\footnote{The $(w\cdot b)_+$ in the exponent is not a typo. It was previously the lower index of a Pochhammer symbol, but in the commutative limit, it turns into a power.  Note also that \eqref{abel_mon}, as written, holds for semisimple gauge groups, for which the sum of the weights $w$ vanishes.  Otherwise, it should include an additional factor of $r^{b\Sigma/2}$ where $\Sigma$ is the sum of all $U(1)$ charges of hypermultiplets in the theory, as can be seen by writing $(w\cdot b)_+ = (|w\cdot b| + w\cdot b)/2$ in \eqref{general_shift_N}.}
\begin{equation}
\label{abel_mon}
v^b=\frac{\prod_{w\in\cR} \left(-i w\cdot \Phi \right)^{(w\cdot b)_+}}{\prod_{\alpha\in\Delta} \left(-i \alpha\cdot \Phi \right)^{(\alpha\cdot b)_+}} e[b].
\end{equation}
This expression includes the case where some matter multiplets have masses, in which case $\cR$ is considered to be a representation of both gauge and flavor groups, and some $\Phi$'s are VEVs of the background vector multiplets (that is, masses).  Note that \eqref{abel_mon} immediately implies \eqref{generalproduct}.

This \eqref{abel_mon} is precisely as in \cite{Bullimore:2015lsa}, showing that we indeed recover their abelianization map in the $r\to\infty$ limit. A bonus of our formalism is that the abelianized bubbling coefficients of Sections \ref{sec:dress_and_bubble} and \ref{sec:poly} are known, and also have a well-defined $r\to\infty$ limit. Introducing the notation
\begin{equation}
z_{b\to v}(\Phi)\equiv \lim_{r\to\infty} Z_{b\to v}^{\rm ab}(\Phi)
\end{equation}
and the corresponding notation for the commuting abelianized monopole shift operator,
\begin{equation}
\label{tilde_commut}
\widetilde{v}^b \equiv \lim_{r\to\infty} \widetilde{M}^b = v^b + \sum_{|u|<|b|} z_{b\to u}(\Phi)v^u,
\end{equation}
we conclude that the commuting versions of general physical dressed monopoles are given by
\begin{equation}
\label{dressed_commut}
\left[P(\Phi)V^b\right]=\frac1{|\cW_b|}\sum_{{\rm w}\in\cW} P_i(\Phi^{\rm w})\widetilde{v}^{{\rm w}\cdot b}.
\end{equation}
Let us consider a few examples of Coulomb branches determined using this technique.

\subsubsection{$SU(2)$ with $N_f$ Fundamentals and $N_a$ Adjoints} \label{SU2CHIRAL}

In Section \ref{sec:rankone}, we showed that in the $SU(2)$ gauge theory with $N_f>1$ fundamentals and any number $N_a$ of adjoints, the abelianized bubbling coefficient $Z_{2\to 0}^{\rm ab}(\Phi)$ is a polynomial.  Hence, up to operator mixing, we can take $Z_{2\to 0}^{\rm ab}(\Phi)=0$. The same is then true for its $r\to\infty$ limit, $z_{2\to 0}(\Phi)=0$. 

When $N_f=0$, the bubbling term is a nontrivial rational function,
\begin{equation}
Z_{2\to 0}(\Phi)=\frac{(-4r^2)^{-N_a}}{\Phi(\Phi-\frac{i}{r})}.
\end{equation}
However, we see that its $r\to\infty$ limit is zero unless $N_a=0$. Hence we can again take $z_{2\to 0}(\Phi)=0$, except in a pure gauge theory, which will be treated separately.

Since the Cartan is one-dimensional, we write the Cartan-valued $\Phi$ simply as a complex number. The two primitive monopoles of minimal charge $b=2$ in the commuting limit take the form:
\begin{align}
v^2 + v^{-2}&=\left(-i\frac{\Phi}{2}\right)^{N_f}(i\Phi)^{2(N_a-1)}(e[2]+(-1)^{N_f}e[-2]),\cr
\Phi(v^2 - v^{-2})&=\Phi\left(-i\frac{\Phi}{2}\right)^{N_f}(i\Phi)^{2(N_a-1)}(e[2]-(-1)^{N_f}e[-2]).
\end{align}
In addition, we have the variable $\Phi^2$. Define:
\begin{equation}
\cU=2^{N_f-1}(v^2 + v^{-2}), \qquad \cV=-i2^{N_f-1}\Phi(v^2 - v^{-2}), \qquad \cW=\Phi^2.
\end{equation}
The only relation between these variables follows from $e[2]e[-2]=1$ and takes the form
\begin{equation}
\label{D_singular}
\cV^2 + \cU^2\cW = \cW^{N_f+2N_a-1},
\end{equation}
which is the defining equation of a $D_{N_f + 2N_a}$ singularity. According to Equation \eqref{dim_rank1}, the dimension of the lowest monopole operator is $\Delta_2=N_f + 2N_a-2$. We see that the theory is good whenever $N_f+2N_a>2$. Precisely for such values, \eqref{D_singular} determines a cone. For $N_f+2N_a=2$, $\cU$ has dimension (or rather R-charge) zero, while for $N_f+2N_a=1$, that is, $N_a=0$ and $N_f=1$, the monopole has negative R-charge --- in both of these cases, the theory is bad and \eqref{D_singular} is not a cone.

It is also straightforward to include masses by turning on background VEVs for flavor symmetries. In such a case, the bubbling term remains nontrivial in the $r\to\infty$ limit, as we know from \eqref{massive_rank1}, and is given by
\begin{equation}
z_{2\to 0}(\Phi)=\frac{\prod_{a=1}^{N_a} \left(- M_a^2 \right)\prod_{i=1}^{N_f} (-iM_i)}{\Phi^2}
\end{equation}
where $M_a$ and $M_i$ are the masses of the adjoint and fundamental hypers, respectively. The expressions for the commuting shift operators are also modified (as follows from coupling to the background multiplet):
\begin{align}
v^2&=\frac{\prod_{i=1}^{N_f}\left(-i\frac{\Phi}{2} - iM_i\right)\prod_{a=1}^{N_a}(i\Phi + iM_a)^2}{(i\Phi)^2}e[2],\cr
v^{-2}&=\frac{\prod_{i=1}^{N_f}\left(i\frac{\Phi}{2} - iM_i\right)\prod_{a=1}^{N_a}(-i\Phi + iM_a)^2}{(i\Phi)^2}e[-2].
\end{align}
Using the variables
\begin{equation}
\cU=2^{N_f-1}(v^2 + v^{-2}+2z_{2\to0}(\Phi)), \qquad \cV=-i2^{N_f-1}\Phi(v^2 - v^{-2}), \qquad \cW=\Phi^2
\label{UVWFlat}
\end{equation}
and the relation $e[2]e[-2]=1$, we find:
\begin{equation}
\cV^2\cW + \left(\cU \cW - \prod_{a=1}^{N_a} \left(- M_a^2 \right)\prod_{i=1}^{N_f} (-2iM_i) \right)^2=\prod_{i=1}^{N_f}\left(\cW-4M_i^2\right)\prod_{a=1}^{N_a}\left(\cW - M_a^2 \right)^2,
\end{equation}
which at $N_a=0$ agrees with the result in \cite{Assel:2018exy} found by gauging $U(1)_{\rm top}$ of the $U(2)$ theory.

\subsubsection{Pure $SU(2)$}
For a pure $SU(2)$ gauge theory, the bubbling term in the commutative limit is
\begin{equation}
z_{2\to0}(\Phi)=\frac1{\Phi^2},
\end{equation}
so the abelianized shift operators are
\begin{align}
\widetilde{v}^{\pm2} = v^{\pm2} + \frac1{\Phi^2}.
\end{align}
The primitive monopoles take the form
\begin{align}
\widetilde{v}^{2}+\widetilde{v}^{-2}&=-\frac1{\Phi^2}(e[2]+e[-2])+\frac2{\Phi^2},\nonumber \\[5 pt]
\Phi(\widetilde{v}^{2}-\widetilde{v}^{-2})&=-\frac1{\Phi}(e[2]-e[-2]).
\end{align}
If we define
\begin{equation}
\cU=\frac12(\widetilde{v}^{2}+\widetilde{v}^{-2}), \qquad \cV=\frac12\Phi(\widetilde{v}^{2}-\widetilde{v}^{-2}), \qquad \cW=\Phi^2,
\end{equation}
then we find that $e[2]e[-2]=1$ implies the relation
\begin{equation}
\cV^2=\cU^2\cW-2\cU,
\end{equation}
which does not belong to the series \eqref{D_singular} and agrees with (A.9) in \cite{Assel:2018exy}.

\subsubsection{$G_2$ with $N_f$ Fundamentals}

To demonstrate the effectiveness of our formalism, we now discuss the theory with gauge group $G_2$ and $N_f$ hypermultiplets in the seven-dimensional fundamental representation of $G_2$. Recall from Section \ref{sec:G2_general} that the lattice of coweights is generated by a short coroot $\alpha_{\rm mon}$ and a long coroot $\beta_{\rm mon}$. At zero magnetic charge, there are two Casimir invariants
\begin{equation}
f_1=\Phi_1^2+\Phi_2^2,\qquad f_2=\Phi_2^2(\Phi_2^2-3\Phi_1^2)^2,
\end{equation}
and at magnetic charge $\alpha_{\rm mon}$, there are six primitive dressed monopoles, which in the commutative limit give six primitive commutative monopoles:
\begin{alignat}{3}
m_0&=V^{\alpha_{\rm mon}}, \quad & m_1&=[\Phi_2 V^{\alpha_{\rm mon}}], \quad & m_2&=[\Phi_2^2 V^{\alpha_{\rm mon}}],\nonumber \\
m_3&=[\Phi_2^3 V^{\alpha_{\rm mon}}], \quad & m_4&=[\Phi_2^4 V^{\alpha_{\rm mon}}], \quad & m_5&=[\Phi_2^5 V^{\alpha_{\rm mon}}].
\end{alignat}
In Section \ref{sec:G2_general}, we found the abelianized bubbling factor $Z(\Phi)$ for ``$\alpha_{\rm mon}\to 0$.'' Its $r\to\infty$ limit is
\begin{equation}
\label{commut_G2_Z}
z(\Phi)=\frac{4^{2-N_f}(\Phi_1^2 + \Phi_2^2)^{N_f}}{9\Phi_2^2(\Phi_1^2 - 3\Phi_2^2)(3\Phi_1^2 - \Phi_2^2)}.
\end{equation}
We now have the ingredients in place to determine the chiral ring.

We first observe from \eqref{dimg2} that by taking the products
\begin{align}
\label{generate_beta}
m_1^2 - m_2 m_0,\quad m_2m_1 - m_3 m_0,\quad m_3 m_1 - m_4m_0,\cr
m_4m_1 - m_5 m_0,\quad m_4m_2 - m_5m_1,\quad m_4m_3-m_5m_2,
\end{align}
we can obtain all six primitive dressed monopoles of magnetic charge $\beta_{\rm mon}$. Because $\alpha_{\rm mon}$ and $\beta_{\rm mon}$ are fundamental coweights, they obviously generate the rest of the charges. Furthermore, \eqref{generate_beta} implies that monopoles of charge $\beta_{\rm mon}$ are generated from those of charge $\alpha_{\rm mon}$. Therefore, the six monopoles $m_0, \ldots, m_5$ and two Casimirs $f_1$ and $f_2$ generate the full chiral ring.

It remains to determine their relations. They follow from the relations in $\C[\Lambda_w^\vee]$:
\begin{align}
\label{e_rels}
e[\alpha_{\rm mon}]e[-\alpha_{\rm mon}]=1,\cr
e[\beta_{\rm mon}+\alpha_{\rm mon}]e[-\beta_{\rm mon}-\alpha_{\rm mon}]=1,\cr
e[\beta_{\rm mon}+2\alpha_{\rm mon}]e[-\beta_{\rm mon}-2\alpha_{\rm mon}]=1,\cr
e[\beta_{\rm mon}+\alpha_{\rm mon}]e[\alpha_{\rm mon}]-e[\beta_{\rm mon}+2\alpha_{\rm mon}]=0.
\end{align}
These relations can easily be seen to follow, in turn, from linear dependences between the short (co)roots of $G_2$. Moreover, they generate a complete set of relations in $\C[\Lambda_w^\vee]$: short (co)roots generate the full (co)weight lattice, and relations between the short (co)roots determine everything.

Using the definition \eqref{abel_mon} of commuting shift operators, incorporating the abelianized bubbling factor \eqref{commut_G2_Z} according to \eqref{tilde_commut} and \eqref{dressed_commut}, and using the relations \eqref{e_rels}, one can derive the relations
\begin{equation}
\label{G2_c1}
L_i = [P_i(\Phi)V^{\alpha_{\rm mon}}] + F_i(\Phi) \quad (i = 1, 2, 3, 4)
\end{equation}
between the chiral ring generators, where
\begin{align}
\label{G2_c2}
L_1 &\equiv m_2^2 +\frac12 m_4 m_0 -\frac32 m_3m_1 + \frac38 (m_1^2 -m_2m_0)f_1, \nonumber \\
L_2 &\equiv m_3m_2+m_5 m_0-2m_4m_1+\frac34 (m_2m_1-m_3m_0)f_1, \\
L_3 &\equiv m_3^2 +\frac12 m_5 m_1 -\frac32 m_4m_2 + \frac3{16} (m_3m_1-m_4m_0)f_1 - \frac9{64}(m_1^2-m_2m_0)f_1^2, \nonumber \\
L_4 &\equiv 2m_5m_2-m_4m_3-\frac1{16}m_1m_0f_2 - \frac34 (3m_4m_1 - m_5m_0)f_1 + \frac9{16}(2m_2m_1-m_3m_0)f_1^2, \nonumber
\end{align}
and $P_i$ and $F_i$ are $N_f$-dependent polynomials in $\Phi$ that can be expressed in terms of known generators. The simplest case is $N_f=0$, where most of the right-hand sides vanish:
\begin{equation}
\left([P_i(\Phi)V^{\alpha_{\rm mon}}] + F_i(\Phi)\right)_{i = 1, 2, 3, 4} = \left(0, 0, -\frac13 m_0, 0\right).
\end{equation}
For $N_f=1$, the answer is:
\begin{equation}
\left([P_i(\Phi)V^{\alpha_{\rm mon}}] + F_i(\Phi)\right)_{i = 1, 2, 3, 4} = \left(\frac13 m_0, \, \frac23 m_1, \, -\frac{11}{24}f_1 m_0 +\frac13 m_2, \, f_1m_1-\frac23 m_3\right).
\end{equation}
Note that we have not checked whether this is a complete set of equations, i.e., whether the Coulomb branch is a complete intersection, though it should be possible to do so from a more careful analysis of the relations. However, these equations are locally independent, so the Coulomb branch is at least a local complete intersection.

\subsection{Quantized Chiral Rings}

Having explained how our formalism can be used to derive the Coulomb branch chiral rings of the gauge theories under study, we now turn to a more refined observable: the OPE of the Coulomb branch 1D sector.  As explained in \cite{Beem:2016cbd,Dedushenko:2016jxl}, the OPE
\begin{equation}
\mathcal{O}_i(\varphi)\mathcal{O}_j(0) \xrightarrow{\varphi\to 0^-} \sum_k \frac{c_{ij}{}^k\mathcal{O}_k(0)}{r^{\Delta_i+\Delta_j-\Delta_k}}
\end{equation}
can be interpreted as a noncommutative star product
\begin{equation}
\mathcal{O}_i\star \mathcal{O}_j = \sum_k \frac{c_{ij}{}^k\mathcal{O}_k}{r^{\Delta_i+\Delta_j-\Delta_k}} \label{star}
\end{equation}
on the chiral ring that reduces to ordinary commutative multiplication of the corresponding holomorphic functions as we take $r\to \infty$:
\begin{equation}
\mathcal{O}_i\star \mathcal{O}_j|_{O(r^0)} = \mathcal{O}_i\mathcal{O}_j = \mathcal{O}_j\mathcal{O}_i.
\end{equation}
Here, $\mathcal{O}_i(\varphi)$ denotes a twisted CBO on $S^1_\varphi$, and the $S^3$ radius $r$ keeps track of differences in conformal dimension.  The topological property of cohomology classes of $\cQ^C$ ensures that \eqref{star} is position-independent.  This star product has the interpretation as a quantization of the ring of holomorphic functions on the Coulomb branch, with $1/r$ serving as the quantization parameter.  In particular, the terms of order $1/r$ in the OPE are interpreted as the Poisson bracket of the holomorphic functions from the $r \to \infty$ limit, induced by the holomorphic symplectic form on $\cM_C$:
\begin{equation}
[\mathcal{O}_i, \mathcal{O}_j]_\ast|_{O(r^{-1})} = \{\mathcal{O}_i, \mathcal{O}_j\}.
\end{equation} 

\subsubsection{$SU(2)$ with $N_f$ Fundamentals and $N_a$ Adjoints}

To illustrate that the OPE indeed gives more information than the chiral ring, let us present an example where distinct 3D theories have the same Coulomb branch chiral ring but different star products.  Such an example was in fact already encountered in Section~\ref{SU2CHIRAL}:  it is  the $SU(2)$ gauge theory with $N_f$ fundamental and $N_a$ adjoint hypermultiplets.  In the previous subsection, we showed that the Coulomb branch is a $D_{N_f + 2 N_a}$ singularity, so it depends only on the combination $n = N_f + 2N_a$.  We now show that the OPE does not depend only on this combination, so that for any fixed $n$, we obtain $\lceil n/2 \rceil$ distinct quantizations of the ring of holomorphic functions on the cone over the $D_n$ singularity.

We restrict to the case $N_f + 2 N_a > 2$, where the theory is good, and to $N_f > 0$, where all bubbling coefficients can be set to zero.  The operators of dimension $\Delta_\cU = N_f + 2 N_a -2$, $\Delta_\cV = N_f + 2N_a - 1$, and $\Delta_\cW = 2$ whose flat-space limits are given in \eqref{UVWFlat} are
\begin{equation}
\cU= 2^{N_f - 1} (M^2 + M^{-2}), \qquad \cV=- i 2^{N_f - 1} \Phi(M^2 - M^{-2}), \qquad \cW=\Phi^2.
\label{UVW}
\end{equation}
Using the corresponding shift operators obtained from \eqref{MNorth}, we then find
 \es{Relation}{ 
  \mathcal{V}^2 + \mathcal{U}\star \mathcal{W}\star \mathcal{U} = P(\mathcal{W}) + \frac 2r \mathcal{U}\star \mathcal{V} \,,
 }
where all products are understood to be star products and $P(\mathcal{W}) = \mathcal{W}^{N_f + 2N_a - 1} + O(1/r)$ is the following polynomial in $\mathcal{W}$:
\begin{equation}
P(\mathcal{W})\equiv \frac{ \left(\sqrt{\mathcal{W}} +  \frac{2i}{r}\right)\left(\sqrt{\mathcal{W}} +  \frac{i}{r}\right)^{2(N_f - 1)}\left[\left(\sqrt{\mathcal{W}} +  \frac{i}{2 r}\right)\left(\sqrt{\mathcal{W}} +  \frac{3i}{2 r}\right)\right]^{2N_a}}{2\sqrt{\mathcal{W}}}
 + (i\leftrightarrow -i)
\end{equation}
(despite appearances, this expression is indeed a polynomial).  To leading order in $1/r$, we reproduce \eqref{D_singular}.  We can also compute various OPEs such as the antisymmetrized OPEs of the Coulomb branch chiral ring generators \eqref{UVW}:
 \es{OPEsSU2}{
  [\mathcal{U}, \mathcal{W}]_\star &= \frac{4}{r} \mathcal{V} - \frac{4}{r^2} \cU \,, \\
  [\mathcal{V}, \mathcal{W}]_\star &= - \frac{4}{r} \mathcal{W}\star \mathcal{U} - \frac{4}{r^2} \cV\,,  \\
  [\mathcal{U}, \mathcal{V}]_\star &= - \frac 2r \cU^2 + Q(\cW) \,,
 }
where $Q(\cW)$ is a polynomial in $\cW$ given by
 \es{QDef}{
  Q(\cW) \equiv \frac{i  \left( \sqrt{\cW} + \frac{i}{r}  \right)^{2 (N_f-1)} 
    \left[  \left( \sqrt{\cW} + \frac{i}{2r}  \right) \left( \sqrt{\cW} + \frac{3i}{2r}  \right) \right]^{2 N_a} }{2 \sqrt{\cW} }
     + (i\leftrightarrow -i)
 }
(this expression is again a polynomial in $\cW$, despite its appearance).  

We see that \eqref{Relation} and \eqref{OPEsSU2} do not depend only on the combination $N_f + 2N_a$ that determines the Coulomb branch, thus providing an example of different quantizations of the same chiral ring.\footnote{One may ask, however, whether a change of basis for the generators $\cU$, $\cV$, and $\cW$ could render \eqref{Relation} and \eqref{OPEsSU2} dependent only on $N_f + 2 N_a$.  For changes of basis where we only allow ourselves to redefine each operator by adding operators of strictly lower dimension multiplied by appropriate factors of $1/r$, it is impossible to make \eqref{Relation} and \eqref{OPEsSU2} depend only on $N_f + 2 N_a$.}  Note, however, that the $1/r$ terms in \eqref{OPEsSU2}, like the chiral ring relation \eqref{D_singular}, do depend only on $N_f + 2 N_a$: thus the Poisson structure on $D_{N_f + 2 N_a}$ is the same for all of the distinct quantizations.

For other examples where our formalism can be used to determine the quantization of the Coulomb branch chiral ring, see Appendix~\ref{CHIRALAPPENDIX}. 

\subsubsection{$G_2$ with $N_f$ Fundamentals}

Let us make a few comments on the theory with gauge group $G_2$, which appeared as one of our earlier examples. At the very least, the same two Casimirs and six primitive monopoles of minimal charge $\alpha_{\rm mon}$ are expected to generate the noncommutative algebra $\cA_C$:
\begin{alignat}{3}
f_1&=\Phi^2 + \Phi_2^2, & f_2&=\Phi_2^2(\Phi_2^2-3\Phi_1^2)^2, && \nonumber \\
m_0&=\cM^{\alpha_{\rm mon}}, \qquad & m_1&=[\Phi_2 \cM^{\alpha_{\rm mon}}], \qquad & m_2&=[\Phi_2^2 \cM^{\alpha_{\rm mon}}],\nonumber \\
m_3&=[\Phi_2^3 \cM^{\alpha_{\rm mon}}], \qquad & m_4&=[\Phi_2^4 \cM^{\alpha_{\rm mon}}], \qquad & m_5&=[\Phi_2^5 \cM^{\alpha_{\rm mon}}].
\end{alignat}
They satisfy the same relations as in \eqref{G2_c1}, \eqref{G2_c2}, with the left-hand side written in terms of the star product and the right-hand side receiving $1/r$ corrections.

Interestingly, however, one can find simple relations that identify a much smaller set of generators of $\cA_C$ as a noncommutative algebra, or alternatively (but not equivalently in general), as a commutative Poisson algebra. Namely, we find that
\begin{align}
m_i \star f_1 - f_1\star m_i=-\frac{4i}{r\sqrt{3}}m_{i+1} - \frac{4}{3r^2}m_i, \quad i=0,\ldots,5,
\end{align}
implying that it is enough to have $f_1$, $f_2$, and $m_0$ to generate the rest of the algebra through star products. The above equation also implies the Poisson bracket $\{m_i, f_1\}=-\frac{4i}{\sqrt{3}}m_{i+1}$.  In order to compute star products, we must use the bubbling factor derived in Section \ref{sec:G2_general}.

\subsection{Correlation Functions and Mirror Symmetry} \label{correlators}

We now demonstrate the utility of the shift operator formalism for computing correlation functions of twisted CBOs, with applications to non-abelian 3D mirror symmetry \cite{Intriligator:1996ex,deBoer:1996mp,deBoer:1996ck}.  We first review the general setup for the computation of correlation functions before giving an example. 

\subsubsection{Matrix Model}

The three ingredients for computing correlation functions are the vacuum hemisphere wavefunction, the gluing measure, and the shift operators.  The vacuum hemisphere wavefunction $\Psi_0(\sigma, B)$ (where $\sigma$ is valued in the Cartan of $\mathfrak{g}$ and $B$ in the coweight lattice) can be read off from \eqref{Z_unbub} by setting $b=0$:
\begin{equation}
\Psi_0(\sigma, B)\equiv Z_0(\vec{0}; \sigma, B) = \delta_{B, \vec{0}}\frac{\prod_{w\in \mathcal{R}} \frac{1}{\sqrt{2\pi}}\Gamma(\frac{1}{2} - iw\cdot \sigma)}{\prod_{\alpha\in \Delta} \frac{1}{\sqrt{2\pi}}\Gamma(1 - i\alpha\cdot \sigma)} \,,
\label{vacuum}
\end{equation}
where $\mathcal{R}$ denotes the weights of the hypermultiplet representation $\mathcal{R}$ of $G$ and $\Delta$ denotes the roots of $G$.  The gluing measure $\mu(\sigma, B)$ is as in \eqref{gluingmeasure}, namely
 \es{GluingAgain}{
\mu(\sigma, B)&= \prod_{\alpha\in\Delta^+}(-1)^{\alpha\cdot B}\left[\left( \frac{\alpha\cdot \sigma}{r} \right)^2 + \left( \frac{\alpha\cdot B}{2r} \right)^2\right]
 \prod_{w\in\cR} (-1)^{\frac{|w\cdot B|-w\cdot B}{2}} \frac{\Gamma\left(\frac12 + iw\cdot\sigma +\frac{|w\cdot B|}{2} \right)}{\Gamma\left(\frac12 - iw\cdot\sigma +\frac{|w\cdot B|}{2} \right)} 
 }
(note that $\prod_{\alpha\in \Delta^+} (-1)^{\alpha\cdot B} = e^{2\pi i\rho\cdot B}$ where $\rho$ is the Weyl vector).  The shift operators are given by \eqref{eqn_on_Z} combined with \eqref{Shift} (see also Appendix~\ref{appen:conv}).  Without loss of generality, we work in the North picture, where
 \es{shiftAgain}{
M^b_N &= \frac{\prod_{w\in\cR}\left[\frac{(-1)^{(w\cdot b)_+}}{r^{|w\cdot b|/2}} \left(\frac12 +i rw\cdot \Phi_N \right)_{(w\cdot b)_+} \right]}{\prod_{\alpha\in\Delta}\left[\frac{(-1)^{(\alpha\cdot b)_+}}{r^{|\alpha\cdot b|/2}} \left(i r \alpha\cdot \Phi_N\right)_{(\alpha\cdot b)_+} \right]} e^{-b\cdot(\frac{i}2 \partial_\sigma +\partial_B)} \,, \quad
 \Phi_N = \frac{1}{r} \left( \sigma + i \frac{B}{2} \right)  \,,
 }
and therefore drop the $N$ subscripts.

With these ingredients, the matrix model expression for the correlator of twisted CBOs $\cO_i(\vphi_i)$, $i = 1, \ldots, n$, inserted at points $\vphi_i$ obeying $0 < \varphi_1 < \cdots < \varphi_n < \pi$, takes the form of an inner product (see also \cite{Dedushenko:2017avn})
\begin{equation}
\langle\mathcal{O}_1(\varphi_1)\cdots \mathcal{O}_n(\varphi_n)\rangle_{S^3} = \frac{1}{|\mathcal{W}|Z_{S^3}}\sum_{\vec{B}}\int d\vec{\sigma}\, \mu(\vec{\sigma}, \vec{B})\Psi_0(\vec{\sigma}, \vec{B})\widehat{\mathcal{O}}_1\cdots \widehat{\mathcal{O}}_n\Psi_0(\vec{\sigma}, \vec{B}) \,,
\label{CBmatrixmodel}
\end{equation}
where $\widehat{\cO}_i$ are the shift operators corresponding to $\cO_i$ and $Z_{S^3}$ is the vacuum $S^3$ partition function by which we divide to obtain a normalized correlator:
 \es{Gluing}{
Z_{S^3} = \frac{1}{|\mathcal{W}|}\sum_{\vec{B}} \int d\vec{\sigma}\, \mu(\vec{\sigma}, \vec{B})\Psi_0(\vec{\sigma}, \vec{B})^2.
 }
This is a special case of the gluing formula of Section \ref{sec:glue_f}.  From here on, we drop the hats on the shift operators and therefore do not make a notational distinction between a shift operator and the twisted CBO that it represents.

\subsubsection{An $\mathcal{N} = 8$ Example} \label{n8example}

As a concrete example, let us consider the $U(N_c)$ gauge theory with one adjoint hypermultiplet and one fundamental hypermultiplet.  This theory has $\cN = 8$ SUSY enhancement (being IR dual to $\mathcal{N} = 8$ $U(N_c)$ SYM) and is therefore self-mirror \cite{Bashkirov:2010kz}.  This theory is ugly in the sense of Gaiotto and Witten \cite{Gaiotto:2008ak}, so the monopoles of lowest dimension saturate the unitarity bound $\Delta = 1/2$ and generate a free subsector.  The $S^3$ partition function is
\begin{align}
Z_{S^3} &= \frac{1}{N_c!}\int \prod_{I=1}^{N_c} d\sigma_I\, \frac{\prod_{I<J} 4\sinh^2(\pi\sigma_{IJ})}{\prod_{I, J} 2\cosh(\pi\sigma_{IJ})\prod_I 2\cosh(\pi\sigma_I)} \,, \label{ZN8} 
\end{align}
where $I, J = 1, \ldots, N_c$ and $\sigma_{IJ} \equiv \sigma_I - \sigma_J$. 

The weight lattice $\mathbb{Z}^{N_c}$ is generated by the $N_c$ fundamental weights $(1, \smash{\vec{0}}), \ldots, (\smash{\vec{0}}, 1)$, and the $N_c^2 - N_c$ roots are the pairwise differences of these fundamental weights.  Since $U(N_c)$ is its own Langlands dual, we can think of $\vec{\sigma}$ and $\vec{B}$ as taking values in $\Z^{N_c}$.   The vacuum wavefunction \eqref{vacuum} in this theory simplifies to
\begin{equation}
\Psi_0(\vec{\sigma}, \vec{B}) = \delta_{\vec{B}, \vec{0}} \prod_{I=1}^{N_c} \frac{1}{\sqrt{2\pi}}\Gamma\left(\frac{1}{2} - i\sigma_I \right) 
 \prod_{I, J} \frac{1}{\sqrt{2\pi}}\Gamma\left(\frac{1}{2} - i\sigma_{IJ} \right)
  \prod_{I<J} \frac{2 \sinh(\pi\sigma_{IJ})}{\sigma_{IJ}},
\end{equation}
and omitting factors of $r$ for convenience, the gluing measure is\footnote{The adjoint hyper contributes a sign $(-1)^{|B_{IJ}|} = (-1)^{(B_{IJ})_+ + (-B_{IJ})_+}$ to the $\prod_{I<J}$, which cancels with the $(-1)^{B_{IJ}}$ in \eqref{gluingmeasure}.}
\begin{equation}
\mu(\vec{\sigma}, \vec{B}) = \prod_{I<J} \left(\sigma_{IJ}^2 + \frac{1}{4}B_{IJ}^2\right)\prod_{I=1}^{N_c} (-1)^{(-B_I)_+}\frac{\Gamma(\frac{1 + |B_I|}{2} + i\sigma_I)}{\Gamma(\frac{1 + |B_I|}{2} - i\sigma_I)} \,.
\end{equation}
The partition function obtained by the gluing formula \eqref{Gluing} then reproduces \eqref{ZN8}.

The operators in the 1D Higgs branch sector of the same theory can be written as $U(N_c)$-invariant products of antiperiodic adjoint scalars $Q(\varphi)$ and $\tQ(\varphi)$ on $S^1_\varphi$.  The correlation functions of these operators can be computed via the prescription of \cite{Dedushenko:2016jxl} (in particular, see Section 7.3 of \cite{Dedushenko:2016jxl}).  They reduce to calculations in a free theory  with 1D propagator
 \es{Wick}{
\langle   Q_i{}^j(\varphi_1) \widetilde{Q}_{i'}{}^{j'}(\varphi_2)\rangle_\sigma = -\delta_{ii'}\delta^{jj'}\frac{\operatorname{sign}(\varphi_{12}) + \tanh(\pi \sigma_{ij})}{8\pi r}e^{-\sigma_{ij}\varphi_{12}} \,,
 }
where $\sigma_{ij} \equiv \sigma_i - \sigma_j$.  Here, $\langle\rangle_\sigma$ denotes an auxiliary correlator from which the full correlator $\langle\rangle$ is obtained by an appropriate integral over $\sigma$.  In particular, for operators $\cO^i$ constructed from $Q$ and $\tQ$, the correlation function is
\begin{align}
  \langle \cO^1(\vphi_1) \cdots \cO^n(\vphi_n) \rangle = \frac{1}{Z_{S^3} N_c!}\int &\prod_{i=1}^{N_c} d\sigma_i \, \frac{\prod_{i<j} 4\sinh^2(\pi\sigma_{ij})}{\prod_{i, j} 2\cosh(\pi\sigma_{ij})\prod_i 2\cosh(\pi\sigma_i)} \nonumber \\
   &\times \langle \cO^1(\vphi_1) \cdots \cO^n(\vphi_n) \rangle_\sigma \,, \label{CorrFnHiggs}
\end{align}
where $\langle \cO^1(\vphi_1) \cdots \cO^n(\vphi_n) \rangle_\sigma$ is computed via Wick contractions with the propagator \eqref{Wick}.

\subsubsection{$U(2)$ with $N_a = N_f = 1$} \label{u2n8example}

Consider the case of smallest rank, $U(2)$ with $N_a = N_f = 1$.  The Coulomb branch chiral ring operators of lowest dimension are $\mathcal{M}^{(\pm 1, 0)}$ (with $\Delta = 1/2$) and $\tr\Phi$, $\mathcal{M}^{\pm(1, 1)}$, $\mathcal{M}^{(1, -1)}$, $\mathcal{M}^{(\pm 2, 0)}$ (with $\Delta = 1$).  Particular linear combinations of these operators comprise the chiral ring generators, namely $\mathcal{M}^{\pm(1, 0)}$, $\mathcal{M}^{\pm(1, 1)}$, and $-\mathcal{M}^{(1, -1)} - 2i\tr\Phi$.  They satisfy the single relation (see Appendix \ref{sym2c2relationderivation})
\begin{equation}
[(\mathcal{M}^{(-1, 0)})^2 - 4\mathcal{M}^{(-1, -1)}][(\mathcal{M}^{(1, 0)})^2 - 4\mathcal{M}^{(1, 1)}] = (-\mathcal{M}^{(1, -1)} - 2i\tr\Phi)^2.
\label{sym2c2relation}
\end{equation}
The products in this equation are commutative chiral ring products, not star products.  Thus the Coulomb branch factorizes into free and interacting sectors as $\operatorname{Sym}^2(\mathbb{C}^2)\cong \mathbb{C}^2\times (\mathbb{C}^2/\mathbb{Z}_2)$.

By matching all two- and three-point functions of these lowest-dimension twisted CBOs and HBOs across mirror symmetry, computed within their respective 1D topological sectors, we can derive the mirror map (see Appendix~\ref{mirrorappendix})
  \begin{align}
\smash{\frac{1}{(4\pi)^{1/2}}}\mathcal{M}^{(\mp 1, 0)} \vphantom{\frac{1}{4\pi}} &\leftrightarrow \tr Q, \tr\tilde Q, \label{mirrormap1} \\
\frac{1}{4\pi}\mathcal{M}^{(\mp 2, 0)} &\leftrightarrow (\tr Q)^2, (\tr\tilde Q)^2, \label{mirrormap2} \\
\frac{1}{2\pi}\mathcal{M}^{\mp(1, 1)} &\leftrightarrow \tr Q^2, \tr\tilde Q^2, \label{mirrormap3}\\
\frac{1}{4\pi}\left(\mathcal{M}^{(1, -1)} - \frac{1}{r}\right) &\leftrightarrow \tr Q\tr\tilde Q, \label{mirrormap4} \\
-\frac{i}{4\pi}\tr\Phi &\leftrightarrow \tr Q\tilde Q.  \label{mirrormap5}
  \end{align}
The operators on the LHS in the Coulomb branch TQFT have precisely the same correlation functions as the operators on the RHS in the Higgs branch TQFT\@.

A complementary way of deriving the mirror map, which does not require computing all correlation functions to a given order, is as follows.  First, we match certain ``basic'' operators by computing their correlation functions.  Next, we generate composite operators from these basic operators via the star product and use the fact that the structure of the star product is the same on both sides to deduce the map between these composite operators (whose one-point functions can then be matched, as a further consistency check; in our basis, mixing with the identity renders one-point functions nonzero).  This point of view emphasizes that the shift operators themselves, which generate the star product via composition, are more fundamental than the correlators that they compute in that one can write all correlators as expectation values of composite operators obtained via the OPE\@.  For an illustration of this procedure, see Appendix \ref{mirrorappendix}.

\subsubsection{$U(N_c)$ with $N_a = N_f = 1$}

We do not study the case $N_c > 2$ in detail, but let us point out that the mirror map in this case takes
 \es{mirrormapUN}{
\frac{1}{(4\pi)^{1/2}}\mathcal{M}^{(-1, \vec{0})}\leftrightarrow \tr Q, \quad \frac{1}{(4\pi)^{1/2}}\mathcal{M}^{(1, \vec{0})}\leftrightarrow \tr\tilde Q,
 }
with normalizations being fixed by the two-point functions
\begin{equation}
\frac{1}{4\pi}\langle\mathcal{M}^{(-1, \vec{0})}(\varphi_1)\mathcal{M}^{(1, \vec{0})}(\varphi_2)\rangle = \langle\tr Q(\varphi_1)\tr\tilde Q(\varphi_2)\rangle = -\frac{N_c\operatorname{sign}\varphi_{12}}{8\pi r}.
\end{equation}
By taking star products, it also follows that the suitably normalized monopoles $\smash{\mathcal{M}^{(\mp p, \vec{0})}}$ (which can bubble) map to $(\tr Q)^p$ and $(\tr\tilde Q)^p$.\footnote{More generally, we expect the Higgs branch operators $\tr(Q_1\cdots Q_{N_f})$ and $\tr(\tilde{Q}_{N_f}\cdots \tilde{Q}_1)$ in the $U(N_c)^{N_f}$ necklace quiver with one fundamental node to map to monopole operators of GNO charge $(\mp 1, \vec{0})$ in $U(N_c)$ SQCD with $N_a = 1$ and $N_f\geq 1$, but for $N_f > 1$, the insertion on the Coulomb branch side does not simplify so easily, and correlators on the Higgs branch side also become difficult to compute.}

Going beyond the free sector, it is natural to conjecture that the monopoles $\mathcal{M}^{(\mp\vec{1}_p, \vec{0}_{N_c - p})}$ (which do not bubble) map to $\tr Q^p$ and $\tr\tilde Q^p$ for $p = 1, \ldots, N_c$, although we have been unable to demonstrate this analytically.  These monopoles are special for several reasons.  First, assuming the correctness of the stated map, they correspond to all of the independent traces of powers of $ Q$ and $\tilde Q$ individually.  Second, it seems that they comprise the minimal set of bare monopoles needed to generate all other bare monopoles via star products.

\section{Discussion}

\subsection{Summary}

This work ties various loose ends together.  First, it extends the formalism of shift operators for Coulomb branch operators \cite{Dedushenko:2017avn} to arbitrary non-abelian 3D $\cN=4$ gauge theories with hypermultiplet matter.  In the process, it provides an alternative approach to the abelianization description of the Coulomb branch and clarifies the meaning of the abelianization map \cite{Bullimore:2015lsa}.  In particular, it can be seen as a derivation of the latter from first principles.

Our approach additionally allows for the computation of correlation functions of Coulomb branch operators in good and ugly theories, thus providing natural choices of basis that relate the noncommutative star product algebra $\cA_C$ of Coulomb branch operators to these correlation functions\@.  The relation between $\cA_C$ and correlation functions seems to become transparent only when quantizing the Coulomb branch by placing the $\cN = 4$ theory on a sphere rather than by studying it in an $\Omega$-background: the latter route to quantization has a less straightforward connection to SCFT operators.

Finally, on our way to achieving these goals, we gained an improved understanding of monopole bubbling phenomena, which are crucial nonperturbative effects in the description of magnetic defects. Our approach to bubbling is purely algebraic in nature, based on symmetries and algebraic consistency of the OPE\@. It avoids the technicalities of previous analytic bubbling computations \cite{Gomis:2011pf, Ito:2011ea, Brennan:2018rcn, Brennan:2018yuj}, which involve equivariant integration over the moduli space of bubbling solutions to the Bogomolny equation, therefore serving as a good check and testing ground for them.  Our approach further allows for the determination of previously unknown bubbling coefficients, such as in theories without minuscule monopoles.

While the focus of this paper is mostly on developing the general formalism, we also provide some explicit applications and examples in theories of small rank. In Section~\ref{sec:poly}, we derive the ``abelianized bubbling coefficients'' for a large family of rank-one and rank-two gauge theories, which can be used to extract data on the Coulomb branch operators of these theories (including the algebra $\cA_C$ and its correlators) in a completely straightforward and algorithmic fashion. We then illustrate these results in Section \ref{sec:Appl}. While the abelian examples in \cite{Dedushenko:2017avn} provide quantizations of $A_N$ singularities, we present in Section~\ref{sec:Appl} the example of $SU(2)$ gauge theory with fundamental and adjoint matter, resulting in many inequivalent quantizations of the $D_N$ singularity. For the purpose of illustration, we also apply our formalism to the $G_2$ gauge theory, as no other techniques are available in this case.  Looking ahead, shift operators provide a method to potentially determine previously unknown Coulomb branch chiral rings and their quantizations, such as those of bad theories.  Finally, we use our matrix model for correlation functions of twisted CBOs to derive, in some cases, how Higgs and Coulomb branch chiral ring operators map across non-abelian 3D mirror symmetry.  Using our formalism, we are able to derive the precise normalization factors in the mirror map and distinguish operators that could mix on the basis of symmetries.  Further applications are gathered in the appendices.

\subsection{Future Directions and Open Problems}

At a computational level, there exist numerous directions in which the discussion of Section \ref{correlators} could be generalized.  The most well-known families of mirror theories are those of $ADE$ type \cite{Intriligator:1996ex}, as well as the higher-rank counterparts of those of $A$- and $D$-type \cite{deBoer:1996mp, deBoer:1996ck}.  In \cite{Dedushenko:2017avn}, the abelian $A$ series (i.e., the mirror duality between SQED$_{N_f}$ and the affine $A_{N_f - 1}$ quiver) was analyzed in our formalism, resulting in a derivation of the precise mirror map as a refinement of the known mapping of charge matrices \cite{deBoer:1996ck}.  The self-mirror duality considered in Section \ref{n8example} is only a special case of the non-abelian generalizations of the $A$-type mirror symmetries derived in \cite{deBoer:1996mp}.\footnote{Namely, the $U(N_c)^n$ necklace quiver with $v_i\geq 0$ fundamentals charged under each gauge group is dual to the $U(N_c)^v$ necklace quiver where $v = \smash{\sum_i v_i}$ and for every $i$, there is a fundamental charged under the $j^\textrm{th}$ gauge group where $j(i > 1) = \sum_{\ell=1}^{i-1} v_\ell$ and $j(1) = \sum_{\ell=1}^n v_\ell$.}  Aside from the $ADE$ examples that we have not investigated, more examples can be found in \cite{Cremonesi:2013lqa}, and further examples can be generated via the procedure of \cite{Dey:2014tka}, in the same spirit as the constructive approach of \cite{Kapustin:1999ha} to abelian mirror symmetry.  It remains to be seen what further lessons for 3D mirror symmetry can be extracted from our formalism.

Besides further applications of our formalism to gather more data on various 3D $\cN=4$ gauge theories (in particular SCFTs), or to check or discover new dualities, there are a number of conceptual questions that present interesting avenues for future work:
\begin{itemize}
\item It would be interesting to extend our construction to more general gauge theories, namely gauge theories that also have charged matter in half-hypermultiplets, those that involve both ordinary and twisted multiplets at once, and/or theories with Chern-Simons couplings. Understanding the moduli spaces of vacua, their quantization, and the corresponding correlation functions in such theories, if possible, are among the outstanding questions to address.

\item It would be interesting to compare the bubbling terms obtained using our method to those coming from the dimensional reduction of the 4D bubbling terms computed in \cite{Ito:2011ea}.  We performed a few preliminary comparisons (summarized in Appendix \ref{appen:Zmono}) and found that the two agree up to operator mixing and various normalization factors, but a more systematic study is needed.  Furthermore, it has been observed (already in \cite{Ito:2011ea}, and later in \cite{Brennan:2018yuj}) that the results of \cite{Ito:2011ea} sometimes involve discrepancies with those obtained using the AGT correspondence, particularly in 4D $\mathcal{N} = 2$ superconformal QCD ($SU(N_c)$ with $N_f = 2N_c$).  A fix was recently proposed in \cite{Brennan:2018rcn}. Based on our preliminary checks, it appears that all of the subtleties in 4D involving integration over monopole moduli space disappear upon reduction to 3D, and it would be nice to understand why.  Likewise, the relation of our construction to the Moyal product of \cite{Ito:2011ea} and its implications for the line operator OPE in 4D remain to be understood.

\item It would be interesting to recast our construction of shift operators and bubbling coefficients (or equivalently, abelianization) in a way that uses the mathematical definition of the Coulomb branch \cite{Nakajima:2015txa,Braverman:2016wma,Braverman:2016pwk,Braverman:2017ofm}.  It could also be of interest to understand whether the abelianized bubbling terms introduced in this work, which provide an algebraic decomposition of the Weyl-averaged bubbling terms considered heretofore in the literature, have a corresponding geometric interpretation in terms of a decomposition of monopole moduli space.\footnote{We thank T.\ Dimofte for this last remark.}

\item It would be interesting to understand more conceptually whether there exists a relation between quantization on $S^3$ and quantization via the $\Omega$-background \cite{Yagi:2014toa,Bullimore:2016nji,Bullimore:2016hdc} (see also \cite{Beem:2018fng}). Similar relations are abundant in various dimensions for problems involving a supercharge (equivariant differential) $Q$ such that $Q^2$ is a vector field with fixed points.  See, for instance, the recent work \cite{Festuccia:2018rew} for the case of isolated fixed points.

\item More broadly, our work fits into the larger program of constructing and classifying deformation quantizations arising from 3D $\cN=4$ quantum field theories.\footnote{We thank P.~Etingof for a discussion about this topic.} While our construction is certainly derived starting from a Lagrangian description, one may wonder whether it can be generalized to non-Lagrangian theories (such as various classes of SCFTs from \cite{Balasubramanian:2018pbp}), and/or whether Lagrangian theories play a special role in the broader classification program of deformation quantizations.
\end{itemize}

\section*{Acknowledgements}

We thank T.\ Daniel Brennan, Anindya Dey, Tudor Dimofte, Pavel Etingof, Amihay Hanany, Petr Kravchuk, Dominik Miketa, Hiraku Nakajima, and Takuya Okuda for various discussions and correspondence. YF thanks the members of QMAP, UC Davis for their hospitality. The work of MD was supported by the Walter Burke Institute for Theoretical Physics and the U.S. Department of Energy, Office of Science, Office of High Energy Physics, under Award No.\ DE-SC0011632, as well as the Sherman Fairchild Foundation. The work of YF was supported in part by the NSF GRFP under Grant No.\ DGE-1656466 and by the Graduate School at Princeton University. The work of SSP was supported in part by the US NSF under Grant No.~PHY-1820651, by the Simons Foundation Grant No.~488651, and by an Alfred P.~Sloan Research Fellowship. The work of RY was supported in part by a grant from the Israel Science Foundation Center for Excellence, by the Minerva Foundation with funding from the Federal German Ministry for Education and Research, and by the ISF within the ISF-UGC joint research program framework (grant no.~1200/14).

\appendix
\section{Conventions}\label{appen:conv}

Here, we summarize our conventions and notation. Unless otherwise stated, $G$ is assumed to be a simple gauge group, $\mathfrak{g}$ its Lie algebra, $\mathfrak{t}$ its Cartan subalgebra, and $\mathfrak{t}_\C=\mathfrak{t}\otimes\C$ its complexification. The root system is denoted by $\Delta$, the weight lattice by $\Lambda_w$, and the coweight lattice (the weight lattice of $^LG$) by $\Lambda^\vee_w$. The matter representation is $\cR\oplus \overline\cR$.

The abelian North pole shift operator \eqref{general_shift_N} is denoted by $M_N^b$, while its South pole analog is denoted by $M_S^b$. These operators do not incorporate monopole bubbling effects.  Sometimes, we simply write $M^b$, in which case it is assumed to be the North pole operator. Here, $b\in\Lambda^\vee_w$ is a coweight of $G$. The abelianized shift operator (including bubbling) is denoted by $\widetilde{M}^b_N$, with the same remark concerning N/S:
\begin{equation}
\widetilde{M}^b= M^b + \sum_{|v|<|b|}Z^{\rm ab}_{b\to v}(\Phi) M^v,
\end{equation}
where the sum is over coweights shorter than $b$ and $Z^{\rm ab}_{b\to v}(\Phi)$ are abelianized bubbling coefficients. The commutative ($r\to\infty$) limit of the shift operator $M^b_N$ is denoted by $v^b$, and the same for $M^b_S$, as the N/S distinction disappears in the commutative limit. Similarly, the abelianized bubbling factor in this limit is denoted by $z_{b\to v}(\Phi)$, and the abelianized commutative shift operator is:
\begin{equation}
\widetilde{v}^b = v^b + \sum_{|u|<|b|}z_{b\to u}(\Phi)v^u.
\end{equation}
We deal with a number of objects that involve sums over Weyl orbits. If a quantity $F(b)$ depends on the coweight $b$, then we employ the following convention in summing over its Weyl orbit:
\begin{equation}
\sum_{b'\in \cW b}F(b') \equiv \frac1{|\cW_b|}\sum_{{\rm w}\in\cW}F({\rm w}\cdot b),
\end{equation}
where $\cW_b\subset\cW$ is the stabilizer of $b$.  In particular, we use it to define the Weyl-averaged shift operator, the bare monopole operator, and the dressed monopole operator:
\begin{align}
\overline{M^b}&=\sum_{b'\in\cW b} M^{b'},\cr
\cM^b&=\sum_{b'\in\cW b} \widetilde{M}^{b'} =\frac1{|\cW_b|} \sum_{{\rm w}\in\cW}\left( M^{{\rm w}\cdot b} + \sum_{|v|<|b|}Z^{\rm ab}_{b\to v}(\Phi^{\rm w}) M^{{\rm w}\cdot v}  \right),\cr
[P(\Phi)\cM^b]&=\frac1{|\cW_b|} \sum_{{\rm w}\in\cW}\left( P(\Phi^{\rm w})M^{{\rm w}\cdot b} + P(\Phi^{\rm w})\sum_{|v|<|b|}Z^{\rm ab}_{b\to v}(\Phi^{\rm w}) M^{{\rm w}\cdot v}  \right).
\end{align}
As explained in the main text, $\Phi^{\rm w}={\rm w}^{-1}\cdot \Phi$, and $\Phi$ takes values in $\mathfrak{t}_\C=\mathfrak{t}\otimes\C$. If a monopole of GNO charge $b$ cannot bubble, then $\mathcal{M}^b = \overline{M^b}$.  The Weyl-averaged shift operator $\overline{M^b}$ defined here does not appear in the main text, but it plays a certain role in the appendices. The dressed commuting monopole operator is defined as:
\begin{equation}
[P(\Phi)V^b]=\frac1{|\cW_b|}\sum_{{\rm w}\in\cW}P(\Phi^{\rm w})\widetilde{v}^{{\rm w}\cdot b}.
\end{equation}
Using the transformation property
\begin{equation}
Z^{\rm ab}_{{\rm w}\cdot b\to{\rm w}\cdot v}(\Phi) = Z^{\rm ab}_{b\to v}(\Phi^{\rm w}),
\end{equation}
we might also introduce 
\begin{equation}
Z_\text{mono}^{b\to v}(\Phi) = \sum_{b'\in\cW b}Z^{\rm ab}_{b'\to v}(\Phi),
\label{weylaveragedZmono}
\end{equation}
so that the bare monopole becomes
\begin{equation}
\cM^b = \overline{M^b} + \sum_{|v|<|b|}Z_\text{mono}^{b\to v}(\Phi) M^v.
\label{MwithweylaveragedZmono}
\end{equation}
In the appendices, we sometimes omit brackets [] around dressed monopoles when no risk of confusion is present.

\section{Twisted-Translated Operators}\label{sec:TTops}

Here is a brief, qualitative review of twisted operators and their corresponding topological sectors.  Let us first recall the setup in $\mathbb{R}^3$.

In 3D $\mathcal{N} = 4$ SCFTs, half-BPS operators are labeled by their charges $(\Delta, j, j_H, j_C)$ under the bosonic subalgebra $\mathfrak{so}(3, 2)\oplus \mathfrak{su}(2)_H\oplus \mathfrak{su}(2)_C$ of the 3D $\mathcal{N} = 4$ superconformal algebra $\mathfrak{osp}(4|4)$.  They are Lorentz scalars ($j = 0$) and can be classified as either HBOs ($\Delta = j_H$, $j_C = 0$) or CBOs ($\Delta = j_C$, $j_H = 0$), which we write abstractly with $\mathfrak{su}(2)_{H/C}$ spinor indices as $\mathcal{O}_{(a_1\cdots a_{2j_H})}$ and $\mathcal{O}_{(\dot{a}_1\cdots \dot{a}_{2j_C})}$.  Hence $\mathfrak{su}(2)_H$ and $\mathfrak{su}(2)_C$ are spontaneously broken on the Higgs and Coulomb branches, respectively.  In a Lagrangian theory, the vector multiplet contains adjoint scalars $\Phi_{\dot{a}\dot{b}}$ in the triplet of $\mathfrak{su}(2)_C$ and the hypermultiplet contains scalars $q_a, \tilde{q}_a$ in the doublet of $\mathfrak{su}(2)_H$ and in $\mathcal{R}, \overline{\mathcal{R}}$ of $G$.  Then HBOs are precisely gauge-invariant polynomials in $q_a, \tilde{q}_a$ while CBOs consist of $\Phi_{\dot{a}\dot{b}}$ and (dressed) monopole operators $\mathcal{M}_{\dot{a}_1\cdots \dot{a}_{2j_C}}^b$.

The key fact is that \emph{twisted} HBOs/CBOs, defined as
\begin{equation}
\mathcal{O}(x) = u^{a_1}(x)\cdots u^{a_{2j_H}}(x)\mathcal{O}_{a_1\cdots a_{2j_H}}(x), \quad \mathcal{O}(x) = v^{\dot{a}_1}(x)\cdots v^{\dot{a}_{2j_C}}(x)\mathcal{O}_{\dot{a}_1\cdots \dot{a}_{2j_C}}(x)
\label{twistedops}
\end{equation}
with appropriate position-dependent R-symmetry polarization vectors $u$ and $v$, have topological correlation functions when the coordinate $x$ is restricted to a line in $\mathbb{R}^3$.  This is because they represent equivariant cohomology classes of certain supercharges $\mathcal{Q}^{H/C}\in \mathfrak{osp}(4|4)$.  In particular, they are annihilated by $(\mathcal{Q}^{H/C})^2$, and operators $\mathcal{O}(x)$ at different $x$ are related by $\mathcal{Q}^{H/C}$-exact operations called \emph{twisted translations}.  Hence the $\mathcal{Q}^{H/C}$-cohomology class of a twisted-translated operator $\mathcal{O}(x)$ is independent of its position $x$ along the line.  It follows that each supercharge $\mathcal{Q}^{H/C}$ has an associated 1D topological sector of cohomology classes: the OPE of these twisted HBOs/CBOs is an associative but noncommutative product, since there exists an ordering along the line.

The setup on $S^3$, where we localize with respect to $\mathcal{Q}^C$, is essentially the same (up to subtleties involving the ``branch point'' at infinity, discussed at length in \cite{Dedushenko:2017avn}): the distinguished line is stereographically mapped to a great circle $S_\varphi^1$, so that twisted operators are parametrized by $\varphi$ rather than $x$ in \eqref{twistedops}, and the deformation parameter $r$ (implicit in the definitions of $\mathcal{Q}^{H/C}$) becomes the $S^3$ radius.  Taking $g_\text{YM}\to\infty$ at fixed $r$ gives an SCFT on $S^3$ whose correlators are equivalent to those of the IR SCFT in flat space by stereographic projection.  The non-conformal 3D $\mathcal{N} = 4$ superalgebra $\mathfrak{s}$ contains the $\mathfrak{su}(2)_\ell\oplus \mathfrak{su}(2)_r$ isometries of $S^3$ as well as $\mathfrak{u}(1)_\ell\oplus \mathfrak{u}(1)_r$ R-symmetries.  The supercharges $\mathcal{Q}^{H/C}$ each contain terms from both $\mathfrak{su}(2|1)$ factors of $\mathfrak{s}$, as required by the fact that they square to isometries with nontrivial fixed points.  The corresponding twisted translations take the form $P_\varphi + R_{H/C} = \{\mathcal{Q}^{H/C}, \ldots\}$.  Finally, the embedding of $\mathfrak{s}$ into $\mathfrak{osp}(4|4)$, as well as the polarization vectors in \eqref{twistedops}, are specified by Cartan embeddings of the $\mathfrak{u}(1)$ R-symmetries into $\mathfrak{su}(2)_H$ and $\mathfrak{su}(2)_C$.

These twisted operators are interesting for at least two reasons:
\begin{itemize}
	\item Their two- and three-point functions fix those of HBOs and CBOs in the full 3D theory, by conformal symmetry and R-symmetry (roughly, conformal symmetry suffices to put any two or three operators on a great circle).
	\item At any fixed $\varphi$, twisted operators in the cohomology of $\mathcal{Q}^{H/C}$ are in one-to-one correspondence with elements of the Higgs/Coulomb branch chiral ring.  The R-symmetry polarization vector $u$ or $v$ fixes a complex structure on the corresponding branch, so that the operators are chiral with respect to an $\mathcal{N} = 2$ superconformal subalgebra of $\mathfrak{osp}(4|4)$ whose embedding depends on the vector.
\end{itemize}
We focus on twisted CBOs representing nontrivial $\mathcal{Q}^C$-cohomology classes, namely: the twisted scalar $\Phi(\varphi) = v^{\dot{a}}(\varphi)v^{\dot{b}}(\varphi)\Phi_{\dot{a}\dot{b}}(\varphi)$, twisted bare monopoles $\mathcal{M}^b(\varphi)$, and twisted dressed monopoles $[P(\Phi)\mathcal{M}^b(\varphi)]$ (composite operators formed by monopoles and scalars).

\section{Matrix Nondegeneracy and Abelianized Bubbling}\label{sec:nondeg}

Here, we prove that the matrix determining the linear system \eqref{lin_eqn_Z} is nondegenerate, thus implying that \eqref{lin_eqn_Z} has a unique solution.

The Weyl group might not act freely on the orbit of a general (dominant) coweight $b$, meaning that $|\cW b| = \dim_\C(\rho^b) < |\cW|$. Each ${\rm w}\cdot b\in\cW b$ has a possibly nontrivial stabilizer ${\rm Stab}_{{\rm w}\cdot b}\equiv\cW_{{\rm w}\cdot b}\subset \cW$, and as a result, $\widetilde{M}^{{\rm w}\cdot b}$ in Equation \eqref{lin_eqn_Z} is multiplied by $\sum_{{\rm w'}\in {\rm Stab}_{{\rm w}\cdot b} } \allowbreak P_i(\Phi^{{\rm w'w}})$. For brevity, let us denote $P_i(\Phi^{\rm w})$ averaged over ${\rm Stab}_{{\rm w}\cdot b}$ by $\overline{P_i(\Phi^{\rm w})}$. Let us also pick representatives ${\rm w}_1, \ldots, {\rm w}_{\dim(\rho^b)}$ of classes in $\cW/{\rm Stab}_{b}$, so that the basis of $\rho^b$ is given by $M^{b}=M^{{\rm w}_1\cdot b}, M^{{\rm w}_2\cdot b}, \ldots, M^{{\rm w}_{\dim(\rho^b)}\cdot b}$ (we assume that ${\rm w}_1={\rm id}$ represents the trivial class). Then Equation \eqref{eqn_on_Z} can be written in matrix form as
\begin{align}
\left( \begin{matrix}
\cM^b\cr
\left[P_2\cM^b\right]\cr
\vdots\cr
\left[P_{\dim(\rho^b)}\cM^b\right]
\end{matrix} \right)=|\cW_b|^{-1}\mathbf{P} \left( \begin{matrix}
\widetilde{M}^{{\rm w}_1\cdot b}\cr
\widetilde{M}^{{\rm w}_2\cdot b}\cr
\vdots\cr
\widetilde{M}^{{\rm w}_{\dim(\rho^b)}\cdot b}
\end{matrix} \right),
\end{align}
where
\begin{equation}
\mathbf{P}=\left( \begin{matrix}
\overline{P_1(\Phi^{\rm w_1})} & \overline{P_1(\Phi^{\rm w_2})} & \overline{P_1(\Phi^{\rm w_3})} & \cdots & \overline{P_1(\Phi^{\rm w_{\dim(\rho^b)}})}\cr
\overline{P_2(\Phi^{\rm w_1})} & \overline{P_2(\Phi^{\rm w_2})} & \overline{P_2(\Phi^{\rm w_3})} & \cdots & \overline{P_2(\Phi^{\rm w_{\dim(\rho^b)}})}\cr
\vdots & \vdots & \vdots & \ddots & \vdots\cr
\overline{P_{\dim(\rho^b)}(\Phi^{\rm w_1})} & \overline{P_{\dim(\rho^b)}(\Phi^{\rm w_2})} & \overline{P_{\dim(\rho^b)}(\Phi^{\rm w_3})} & \cdots & \overline{P_{\dim(\rho^b)}(\Phi^{\rm w_{\dim(\rho^b)}})}\cr
\end{matrix} \right).
\end{equation}
In fact, this matrix is nondegenerate, meaning that its determinant is given by a polynomial in $\Phi$ that is not identically zero, as we now show. By construction, $|\cW_b|^{-1}\sum_{{\rm w}\in\cW} P_i(\Phi^{\rm w})M^{{\rm w}\cdot b}$ for $i=1, \ldots, \dim(\rho^b)$ form a basis over $\C[\mathfrak{t}]^\cW$. This implies that the rows of the matrix $\mathbf{P}$ are linearly independent over $\C[\mathfrak{t}]^\cW$, i.e., over Weyl-invariant polynomials.

Let us assume that the matrix is nonetheless degenerate: this means that one of the rows, say the $j^\text{th}$ row, is a linear combination of the other rows with coefficients being rational and generally non-Weyl-invariant functions $Q_i$:
\begin{equation}
\label{linear_dep}
\overline{P_j(\Phi^{\rm w_a})}=\sum_{\substack{i=1, \ldots, \dim(\rho^b)\cr i\neq j}} Q_i(\Phi)\overline{P_i(\Phi^{\rm w_a})} \text{ for all } a=1,\ldots, \dim(\rho^b). 
\end{equation}
This should hold as an identity for all $a=1,\ldots, \dim(\rho^b)$, with $Q_i(\Phi)$ independent of $a$. Acting with an element of the Weyl group ${\rm w}\in\cW$ on \eqref{linear_dep} should give another valid identity for all $a=1,\ldots, \dim(\rho^b)$. On the other hand, doing so simply permutes the columns of $\mathbf{P}$.  Thus it permutes the equations in \eqref{linear_dep}, at the same time replacing $Q_i(\Phi)$ by $Q_i(\Phi^{\rm w})$. Doing this for every element of $\cW$ and averaging implies that we can replace $Q_i(\Phi)$ in \eqref{linear_dep} by its Weyl-averaged version. So we may assume that $Q_i$ are Weyl-invariant rational functions. Every such $Q_i$ is a ratio
\begin{equation}
Q_i(\Phi) = \frac{A_i(\Phi)}{B_i(\Phi)}
\end{equation}
of polynomials $A_i$ and $B_i$. Let us consider
\begin{equation}
D(\Phi)=\prod_{\substack{i=1, \ldots, \dim(\rho^b)\cr i\neq j}}B_i(\Phi),
\end{equation}
which is the common denominator (not necessarily the minimal one) of all the $Q_i$. Even if this $D_i$ is not Weyl-invariant, one can define another polynomial that is:
\begin{equation}
D^\cW(\Phi)=\prod_{{\rm w}\in\cW} D(\Phi^{\rm w}).
\end{equation}
If we now multiply relation \eqref{linear_dep} by this $D^\cW(\Phi)$, we obtain: 
\begin{equation}
\label{lin_dep_2}
D^\cW(\Phi)\overline{P_j(\Phi^{\rm w_a})}=\sum_{\substack{i=1, \ldots, \dim(\rho^b)\cr i\neq j}} D^\cW(\Phi)Q_i(\Phi)\overline{P_i(\Phi^{\rm w_a})} \text{ for all } a=1,\ldots, \dim(\rho^b). 
\end{equation}
This cancels all denominators of $Q_i$. Furthermore, since both $Q_i(\Phi)$ and $D^\cW(\Phi)$ are Weyl-invariant, their product $D^\cW(\Phi)Q_i(\Phi)$ is a Weyl-invariant polynomial. So \eqref{lin_dep_2} says that the rows of $\mathbf{P}$ are linearly dependent over the ring of Weyl-invariant polynomials $\C[\mathfrak{t}]^\cW$. This is a contradiction, which proves that the matrix $\mathbf{P}$ is nondegenerate.

Having proven that $\mathbf{P}$ is nondegenerate, we can solve \eqref{eqn_on_Z}:
\begin{align}
\left( \begin{matrix}
\widetilde{M}^{{\rm w}_1\cdot b}\cr
\widetilde{M}^{{\rm w}_2\cdot b}\cr
\vdots\cr
\widetilde{M}^{{\rm w}_{\dim(\rho^b)}\cdot b}
\end{matrix} \right)= |\cW_b|\mathbf{P}^{-1} \left( \begin{matrix}
\cM^b\cr
\left[P_2\cM^b\right]\cr
\vdots\cr
\left[P_{\dim(\rho^b)}\cM^b\right]
\end{matrix} \right).
\end{align}
Since the Weyl group simply permutes the columns of $\mathbf{P}$, it is enough to have an expression for $\widetilde{M}^b$, with all other $\widetilde{M}^{{\rm w}\cdot b}$ obtained as Weyl images thereof. Remembering that the leading term in $\left[P_i(\Phi)\cM^b\right]$ takes the form $|\cW_b|^{-1}\sum_{{\rm w}\in\cW}P_i(\Phi^{\rm w})M^{{\rm w}\cdot b}$, we write the solution as:
\begin{align}
\widetilde{M}^b = M^b + \sum_{i=1}^{\dim(\rho^b)}(\textbf{P}^{-1})_1{}^i\sum_{|v|<|b|}\sum_{{\rm w}\in\cW} V_i^{b\to v}(\Phi^{\rm w})M^{{\rm w}\cdot v}.
\end{align}
Introducing the notation $Z^{\rm ab}_{b\to v}(\Phi)$ for the second term, the solution takes the form \eqref{Shift}.

\section{Bubbling Coefficients from 4D}\label{appen:Zmono}

In $U(N)$ gauge theories, there exist known methods for computing monopole bubbling coefficients in the 4D $\mathcal{N} = 2$ context \cite{Ito:2011ea}, the results of which can be used to infer bubbling coefficients in the corresponding 3D $\mathcal{N} = 4$ theories in a specific basis.  However, the na\"ive dimensional reduction prescription sometimes requires supplementing these known results with nontrivial normalization factors to ensure polynomiality, at least for monopoles of sufficiently high charge.  Here, we comment on these subtleties, leaving a more complete understanding for future work.

\subsection{The IOT Algorithm} \label{versus}

A systematic procedure for computing monopole bubbling coefficients relevant to the line operator index of 4D $\mathcal{N} = 2$ $U(N)$ gauge theories with fundamental or adjoint hypermultiplets was developed in \cite{Ito:2011ea} and adapted to $S^3\times S^1$ in \cite{Gang:2012yr}.  We refer to it as the ``IOT algorithm.''  It produces a function that we call $Z_\text{mono, IOT}^{b\to v}$ as follows.  Consider the quantity $Z_\text{mono}^{\mathbb{R}^3}(b, v)$ defined in \cite{Gang:2012yr}:\footnote{Alternatively denoted by $Z_\text{mono}^N(b, v)$ or $Z_\text{mono}^S(b, v)$, the square of which is $Z_\text{mono}^{S^3}(b, v)$ in \cite{Gang:2012yr}.} with all flavor symmetry fugacities $\eta_i$ set to 1, it is a function of the thermal fugacity $x$ (related to the size of the thermal circle by $x^2 = p = e^{-\beta}$) and the Cartan variables $\lambda_i$ (related to ours by $\lambda_i = \beta\sigma_i$).  We set
\begin{equation}
Z_\text{mono, IOT}^{b\to v}(\beta, \sigma)\equiv Z_\text{mono}^{\mathbb{R}^3}(b, v; x, \lambda, \eta = 1).
\end{equation}
The bubbling coefficients $Z_\text{mono, IOT}^{b\to v}$, and those considered in previous literature on 4D $\mathcal{N} = 2$ theories, share the property that they are Weyl-invariant in $b$ but not in $v$:
\begin{equation}
Z_\text{mono, IOT}^{b\to v}(\beta, \sigma) = Z_\text{mono, IOT}^{{\rm w}\cdot b\to v}(\beta, \sigma) = Z_\text{mono, IOT}^{b\to {\rm w}\cdot v}(\beta, {\rm w}\cdot \sigma).
\label{symmetryproperty}
\end{equation}
Hence they should be identified with our abelianized bubbling coefficients, which obey \eqref{heu_inv} and are Weyl-invariant with respect to neither $b$ nor $v$, only after Weyl-averaging over $b$ as in \eqref{weylaveragedZmono}.\footnote{To illustrate the difference between Weyl-averaged and abelianized bubbling, consider a monopole whose charge is a simple coroot and that can bubble only into the identity:
\begin{equation}
\mathcal{M}^b = \overline{M^b} + Z_\text{mono}^{b\to\text{id}}(\Phi).
\end{equation}
The bar denotes Weyl averaging, and the bubbling term is a Weyl-symmetric rational function: $\overline{Z_\text{mono}^{b\to\text{id}}(\Phi)} = Z_\text{mono}^{b\to\text{id}}(\Phi)$.  If $Z_\text{mono}^{b\to\text{id}}(\Phi)$ were the whole story, then it would be natural to guess that
\begin{equation}
[P(\Phi)\mathcal{M}^b] \stackrel{!}{=} \overline{P(\Phi)M^b} + \overline{P(\Phi)}Z_\text{mono}^{b\to\text{id}}(\Phi)
\end{equation}
where $P(\Phi)$ is a (not necessarily Weyl-invariant) polynomial.  However, such a guess turns out to be inconsistent with polynomiality of the Coulomb branch algebra.  Properly defining a dressed monopole instead requires a more elementary bubbling term $Z^\text{ab}_{b\to\text{id}}(\Phi)$ satisfying $\overline{Z^\text{ab}_{b\to\text{id}}(\Phi)} = Z_\text{mono}^{b\to\text{id}}(\Phi)$:
\begin{equation}
[P(\Phi)\mathcal{M}^b] = \overline{P(\Phi)M^b} + \overline{P(\Phi)Z^\text{ab}_{b\to\text{id}}(\Phi)},
\end{equation}
where in general, $Z^\text{ab}_{b\to\text{id}}(\Phi)\neq Z_\text{mono}^{b\to\text{id}}(\Phi)$.}  As described in Section \ref{sec:Schur}, this identification involves a dimensional reduction in which we keep only the leading term in $Z_\text{mono, IOT}^{b\to v}$ as a power series expansion in $\beta$, which occurs at order $\beta^{\Delta_b - \Delta_v}$.  We then make an appropriate substitution $\beta\to -2i/r$ (with the constant of proportionality determined empirically) to restore dimensions.  However, our $Z_\text{mono}^{b\to v}$ is a function of $\Phi$ that is defined to multiply monopole shift operators on the left, while $Z_\text{mono, IOT}^{b\to v}$ as given above is a function of $\sigma$.  To adjust for this discrepancy, we substitute $\Phi$ for $\sigma$ after subtracting the $B$-dependent term evaluated in the appropriate monopole background, i.e., the flux created by the charge-$v$ monopole that the bubbling coefficient multiplies.  In the end, we arrive at the prescription
\begin{equation}
Z_\text{mono}^{b\to v}(\Phi) = Z_\text{mono, IOT}^{b\to v}(\beta, \sigma)|_{O(\beta^{\Delta_b - \Delta_v}), \beta\to -2i/r, \sigma\to r\Phi - iv/2}
\label{betaprescription}
\end{equation}
for going from the dimensionally reduced $Z_\text{mono, IOT}^{b\to v}$ to our Weyl-averaged bubbling coefficients $Z_\text{mono}^{b\to v}$.  It should be kept in mind that the Weyl-averaged bubbling coefficients \eqref{betaprescription} computed via the IOT algorithm single out a basis of Coulomb branch operators, whose star algebra must then be consistent with polynomiality.  By contrast, our polynomiality-based approach, which applies to a far more general class of theories, determines bubbling coefficients up to a choice of basis.

\subsection{Bubbling Patterns}

It turns out that \eqref{betaprescription} is not always sufficient to guarantee that the resulting 3D monopoles satisfy polynomiality.  We conjecture that in general, this prescription should be supplemented by signs and combinatorial factors unrelated to Weyl symmetrization, the latter of which involve dividing by an integer that depends on the number of simple roots subtracted to get from the bare monopole $b$ to the bubbled monopole $v$.  The pattern of combinatorial factors depends on the ``depth'' to which the $b$ monopole can bubble, i.e., the maximum of the number of simple roots that must be subtracted from $b$ to obtain any monopole charge to which $b$ can bubble.  On the other hand, the possible dressing signs depend on the rank and matter content of the theory, as well as on the monopole charges.  Schematically, we see from a number of examples that one possibility for the general bubbling pattern is
\begin{align}
\mathcal{M}^b &= \overline{M^b} + \frac{\epsilon_{1, 1}}{c_{1, 1}}\sum_{v\in \mathcal{W}v_1} Z_\text{mono}^{b\to v}(\Phi)M^v, \label{depth1} \\
\mathcal{M}^b &= \overline{M^b} + \left(\frac{\epsilon_{2, 1}}{c_{2, 1}}\sum_{v\in \mathcal{W}v_1} {} + \frac{\epsilon_{2, 2}}{c_{2, 2}}\sum_{v\in \mathcal{W}v_2} {}\right) Z_\text{mono}^{b\to v}(\Phi)M^v, \label{depth2} \\
\mathcal{M}^b &= \overline{M^b} + \left(\frac{\epsilon_{3, 1}}{c_{3, 1}}\sum_{v\in \mathcal{W}v_1} {} + \frac{\epsilon_{3, 2}}{c_{3, 2}}\sum_{v\in \mathcal{W}v_2} {} + \frac{\epsilon_{3, 3}}{c_{3, 3}}\sum_{v\in \mathcal{W}v_3} {}\right) Z_\text{mono}^{b\to v}(\Phi)M^v \label{depth3}
\end{align}
at depth $1, 2, 3$, respectively.  Here, the subscript $i$ on $v_i$ indicates the number of simple roots by which it differs from $b$, the $c_{i, j}$ are positive integers with
\begin{equation}
c_{1, 1} = c_{2, 1} = c_{3, 1} = 1, \quad 2|c_{2, 2}, \quad 4|c_{3, 2}, \quad 24|c_{3, 3},
\end{equation}
and $\epsilon_{i, j}\in \{\pm 1\}$.  It would be interesting to determine the pattern for arbitrary depth, but we are limited to relatively small charges $b$ by our implementation of the IOT algorithm.\footnote{The IOT algorithm for $U(N)$ constructs $Z_\text{mono, IOT}^{b\to v}$ as a sum of functions labeled by $N$-tuples of Young diagrams, and the bottleneck lies in enumerating these diagrams.  Namely, given $b, v$ in the Cartan of $U(N)$ where $b$ bubbles into $v$, we define the matrix $K = \operatorname{diag}(K_1, \ldots, K_k)$ by
\begin{equation}
\Tr e^{2\pi ib\nu} = \Tr e^{2\pi iv\nu} + (e^{2\pi i\nu} + e^{-2\pi i\nu} - 2)\Tr e^{2\pi iK\nu},
\end{equation}
where $\nu$ is a dummy variable.  This condition can be understood by fixing a Weyl ordering of $v$ and ordering the entries of $K$ from greatest to least.  Then the algorithm requires finding all $N$-tuples of Young diagrams with $k$ boxes in all, colored with the numbers $s = 1, \ldots, k$, such that $K_s = v_{\alpha(s)} + j_s - i_s$ where $\alpha(s) = 1, \ldots, N$ labels the diagram to which $s$ belongs and $i_s, j_s$ are the row and column positions.}  Let us present some evidence for this conjecture in a few simple examples where the IOT algorithm applies.

\subsection{Examples}

\subsubsection{$U(2)$ with $N_a = N_f = 1$}

We start with $U(2)$ SQCD with one adjoint and one fundamental hypermultiplet, which appears as a prototypical example throughout these appendices.  Our conventions for $U(N)$ are as in Section \ref{n8example}.  For convenience, we first record some useful star products of CBOs (to be derived below).  The free sector is generated by the $\Delta = 1/2$ operators $\mathcal{M}^{(\pm 1, 0)}$, whose quadratic star products generate the $\Delta = 1$ operators $\mathcal{M}^{(\pm 2, 0)}$ and $\mathcal{M}^{(1, -1)}$:
 \begin{align}
(\mathcal{M}^{(\pm 1, 0)})^2 &= \mathcal{M}^{(\pm 2, 0)}, \label{pm10squared} \\
\mathcal{M}^{(-1, 0)}\star \mathcal{M}^{(1, 0)} &= \mathcal{M}^{(1, 0)}\star \mathcal{M}^{(-1, 0)} + \frac{2}{r} = \mathcal{M}^{(1, -1)}. \label{10withm10}
 \end{align}
On the other hand, the $\Delta = 1$ operators $\tr\Phi$ and $\mathcal{M}^{\pm(1, 1)}$ satisfy the quadratic relations
\begin{align}
\mathcal{M}^{\pm(1, 1)}\star \mathcal{M}^{\pm(1, 1)} &= \mathcal{M}^{\pm(2, 2)}, \\
\mathcal{M}^{\pm(1, 1)}\star \mathcal{M}^{\mp(1, 1)} &= \left(\frac{1}{2r}\pm i\Phi_1\right)\left(\frac{1}{2r}\pm i\Phi_2\right) = \frac{1}{4r^2}\pm \frac{i}{2r}\tr\Phi + \frac{1}{2}[\tr\Phi^2 - (\tr\Phi)^2], \label{pm11withmp11} \\[2 pt]
[\tr\Phi, \mathcal{M}^{\pm(1, 1)}]_\star &= \pm\frac{2i}{r}\mathcal{M}^{\pm(1, 1)},
\end{align}
where we have defined the commutator $[\cdot, \cdot]_\star$ with respect to the star product.  In the mixed sector, we have the relations
\begin{align}
\mathcal{M}^{\pm(1, 1)}\star \mathcal{M}^{(\pm 1, 0)} = \mathcal{M}^{(\pm 1, 0)}\star \mathcal{M}^{\pm(1, 1)} &= \mathcal{M}^{\pm(2, 1)}, \\
\mathcal{M}^{\pm(1, 1)}\star \mathcal{M}^{(\mp 1, 0)}\pm \frac{1}{2r}\mathcal{M}^{(\pm 1, 0)} &= -i[\Phi_2\mathcal{M}^{(\pm 1, 0)}], \\
\mathcal{M}^{(\mp 1, 0)}\star \mathcal{M}^{\pm(1, 1)}\mp \frac{1}{2r}\mathcal{M}^{(\pm 1, 0)} &= -i[\Phi_2\mathcal{M}^{(\pm 1, 0)}],
\end{align}
as well as the miscellaneous relations
\begin{gather}
\mathcal{M}^{(1, -1)}\star \mathcal{M}^{(1, 0)} = \mathcal{M}^{(1, 0)}\star \mathcal{M}^{(1, -1)} + \frac{2}{r}\mathcal{M}^{(1, 0)} = \mathcal{M}^{(2, -1)}, \label{misc1} \\
\mathcal{M}^{(2, 0)}\star \mathcal{M}^{(1, 0)} = (\mathcal{M}^{(1, 0)})^3 = \mathcal{M}^{(3, 0)}, \label{misc2} \\
\mathcal{M}^{(1, -1)}\star \mathcal{M}^{(1, -1)} = \mathcal{M}^{(2, -2)} - \frac{2}{r}\mathcal{M}^{(1, -1)}. \label{complicatedrelation}
\end{gather}
Implicit in the above relations is a choice of basis (monopole bubbling coefficients), which we now specify.

In $U(2)$ gauge theory, a monopole of charge $(b_1, b_2)$ can bubble if and only if $|b_1 - b_2|\geq 2$.  In particular, the following monopoles cannot bubble:
\begin{equation}
\mathcal{M}^{(1, 0)}, \quad \mathcal{M}^{(1, 1)}, \quad \mathcal{M}^{(2, 1)}.
\end{equation}
To examine some monopoles of small charge that can bubble, we compute that the IOT algorithm yields the following bubbling coefficients in $U(2)$ SQCD with $N_a = N_f = 1$:
\begin{align}
Z_\text{mono, IOT}^{(2, 0)\to (1, 1)} &= 2 - \frac{1}{2(1 + \sigma_{12}^2)} + O(\beta), \\
Z_\text{mono, IOT}^{(1, -1)\to (0, 0)} &= \frac{\beta}{2}\left[1 - \frac{1}{8(1 - i\sigma_{12})} - \frac{1}{8(1 + i\sigma_{12})} - \frac{1}{i(\sigma_1 + \sigma_2)}\right](\sigma_1 + \sigma_2) + O(\beta^2), \label{zmono1m1to00} \\
Z_\text{mono, IOT}^{(3, 0)\to (2, 1)} &= 3 - \frac{3}{9 + 4\sigma_{12}^2} + O(\beta), \\
Z_\text{mono, IOT}^{(2, -1)\to (1, 0)} &= i\beta\left[1 - i(\sigma_1 + \sigma_2) + \frac{i\sigma_1}{2} - \frac{1 - i\sigma_1}{4(3 - 2i\sigma_{12})} + \frac{1 + i\sigma_1}{4(3 + 2i\sigma_{12})}\right] + O(\beta^2). \label{zmono2m1to10}
\end{align}
In this theory, the leading power of $\beta$ in $Z_\text{mono, IOT}^{(b_1, b_2)\to (v_1, v_2)}$ is $\Delta_{(b_1, b_2)} - \Delta_{(v_1, v_2)} = \frac{1}{2}(|b_1| + |b_2| - |v_1| - |v_2|)$.  Applying the prescription \eqref{betaprescription}, we obtain
\begin{align}
Z_\text{mono}^{(2, 0)\to (1, 1)}(\Phi_1, \Phi_2) &= 2 - \frac{1}{2(1 + r^2\Phi_{12}^2)}, \\
Z_\text{mono}^{(1, -1)\to (0, 0)}(\Phi_1, \Phi_2) &= -i\left[1 - \frac{1}{8(1 - ir\Phi_{12})} - \frac{1}{8(1 + ir\Phi_{12})} - \frac{1}{ir\tr\Phi}\right]\tr\Phi, \label{c1m1to00} \\
Z_\text{mono}^{(3, 0)\to (2, 1)}(\Phi_1, \Phi_2) &= 3 - \frac{1}{4(1 - ir\Phi_{12})} - \frac{1}{4(2 + ir\Phi_{12})}, \\
Z_\text{mono}^{(2, -1)\to (1, 0)}(\Phi_1, \Phi_2) &= \frac{1}{2r}\left[3 - 4ir\tr\Phi + 2ir\Phi_1 - \frac{1 - 2ir\Phi_1}{4(1 - ir\Phi_{12})} + \frac{3 + 2ir\Phi_1}{4(2 + ir\Phi_{12})}\right],
\end{align}
which enter into the shift operators
\begin{align}
\mathcal{M}^{(2, 0)} &= \overline{M^{(2, 0)}} + Z_\text{mono}^{(2, 0)\to (1, 1)}(\Phi_1, \Phi_2)M^{(1, 1)}, \\
\mathcal{M}^{(1, -1)} &= \overline{M^{(1, -1)}} + Z_\text{mono}^{(1, -1)\to (0, 0)}(\Phi_1, \Phi_2), \\
\mathcal{M}^{(3, 0)} &= \overline{M^{(3, 0)}} + Z_\text{mono}^{(3, 0)\to (2, 1)}(\Phi_1, \Phi_2)M^{(2, 1)} + Z_\text{mono}^{(3, 0)\to (2, 1)}(\Phi_2, \Phi_1)M^{(1, 2)}, \\
\mathcal{M}^{(2, -1)} &= \overline{M^{(2, -1)}} + Z_\text{mono}^{(2, -1)\to (1, 0)}(\Phi_1, \Phi_2)M^{(1, 0)} + Z_\text{mono}^{(2, -1)\to (1, 0)}(\Phi_2, \Phi_1)M^{(0, 1)}.
\end{align}
Using these shift operators, we can then reproduce the star products for this theory given in \eqref{pm10squared}, \eqref{10withm10}, \eqref{misc1}, \eqref{misc2}.  Hence these results of the IOT algorithm are consistent with polynomiality.

All of the bubbling monopoles considered in the previous paragraph bubble to depth 1, and clearly satisfy \eqref{depth1} with $\epsilon_{1, 1} = 1$.  To see some more complicated examples, let us first consider monopoles whose charges take the form $(a, -a)$.  We have already seen how $\mathcal{M}^{(1, -1)}$ bubbles.  For the star product relation \eqref{complicatedrelation} to hold, it turns out that we must define
\begin{align}
\mathcal{M}^{(2, -2)} = \overline{M^{(2, -2)}} &+ Z_\text{mono}^{(2, -2)\to (1, -1)}(\Phi_1, \Phi_2)M^{(1, -1)} + Z_\text{mono}^{(2, -2)\to (1, -1)}(\Phi_2, \Phi_1)M^{(-1, 1)} \nonumber \\
&+ \frac{1}{2}Z_\text{mono}^{(2, -2)\to (0, 0)}(\Phi_1, \Phi_2) \label{m2m2bubbled}
\end{align}
where $Z_\text{mono}^{(2, -2)\to (1, -1)}(\Phi_1, \Phi_2)$ and $Z_\text{mono}^{(2, -2)\to (0, 0)}(\Phi_1, \Phi_2)$ are obtained via \eqref{betaprescription} from
\begin{align}
Z_\text{mono, IOT}^{(2, -2)\to (1, -1)} &= \beta\left[2i + \sigma_1 + \sigma_2 - \frac{\sigma_1 + \sigma_2}{4(4 + \sigma_{12}^2)}\right] + O(\beta^2), \\
Z_\text{mono, IOT}^{(2, -2)\to (0, 0)} &= \frac{\beta^2}{2}\bigg[\sigma_1^2 + \sigma_2^2 + 4\sigma_1\sigma_2 + 4i(\sigma_1 + \sigma_2) - 3 + \frac{9(3 - 4\sigma_1\sigma_2)}{32(4 + \sigma_{12}^2)} \\
&\hspace{1.5 cm} + \frac{15 - 32i(\sigma_1 + \sigma_2) - 60\sigma_1\sigma_2}{32(1 + \sigma_{12}^2)}\bigg] + O(\beta^3).
\end{align}
The factor of $1/2$ in \eqref{m2m2bubbled} is not accounted for by Weyl symmetrization.  As for $\mathcal{M}^{(3, -3)}$, we must define
\begin{align}
\mathcal{M}^{(3, -3)} = \overline{M^{(3, -3)}} &+ Z_\text{mono}^{(3, -3)\to (2, -2)}(\Phi_1, \Phi_2)M^{(2, -2)} + Z_\text{mono}^{(3, -3)\to (2, -2)}(\Phi_2, \Phi_1)M^{(-2, 2)} \nonumber \\
&+ \frac{1}{8}[Z_\text{mono}^{(3, -3)\to (1, -1)}(\Phi_1, \Phi_2)M^{(1, -1)} + Z_\text{mono}^{(3, -3)\to (1, -1)}(\Phi_2, \Phi_1)M^{(-1, 1)}] \nonumber \\
&+ \frac{1}{24}Z_\text{mono}^{(3, -3)\to (0, 0)}(\Phi_1, \Phi_2)
\end{align}
to ensure closure of the star algebra, namely for the relation
\begin{equation}
(\mathcal{M}^{(1, -1)})^3 = \mathcal{M}^{(3, -3)} - \frac{6}{r}\mathcal{M}^{(2, -2)} + \frac{4}{r^2}\mathcal{M}^{(1, -1)}
\end{equation}
to hold.  The bubbling coefficients are obtained from
\begin{align}
Z_\text{mono, IOT}^{(3, -3)\to (2, -2)} &= \frac{3\beta}{2}\left[\sigma_1 + \sigma_2 + 3i - \frac{\sigma_1 + \sigma_2}{4(9 + \sigma_{12}^2)}\right] + O(\beta^2), \\
Z_\text{mono, IOT}^{(3, -3)\to (1, -1)} &= 3\beta^2\bigg[2\sigma_1^2 + 2\sigma_2^2 + 6\sigma_1\sigma_2 + 12i(\sigma_1 + \sigma_2) - 15 + \frac{9(2 - \sigma_1\sigma_2)}{8(9 + \sigma_{12}^2)} \nonumber \\
&\hspace{1.5 cm} - \frac{31(\sigma_1\sigma_2 - 1) + 24i(\sigma_1 + \sigma_2)}{8(4 + \sigma_{12}^2)}\bigg] + O(\beta^3), \\
Z_\text{mono, IOT}^{(3, -3)\to (0, 0)} &= 3\beta^3\bigg[\sigma_1^3 + \sigma_2^3 + 9\sigma_1\sigma_2(\sigma_1 + \sigma_2) + 9i(\sigma_1^2 + \sigma_2^2 + 4\sigma_1\sigma_2) - \frac{89}{4}(\sigma_1 + \sigma_2) - 15i \nonumber \\
&\hspace{1.5 cm} + \frac{225(\sigma_1 + \sigma_2)(5 - 4\sigma_1\sigma_2)}{1024(9 + \sigma_{12}^2)} + \frac{3(108i + 35(\sigma_1 + \sigma_2))(3 - 4\sigma_1\sigma_2)}{128(4 + \sigma_{12}^2)} \nonumber \\
&\hspace{1.5 cm} + \frac{9(480i + 683(\sigma_1 + \sigma_2) - 1920i\sigma_1\sigma_2 - 380\sigma_1\sigma_2(\sigma_1 + \sigma_2))}{1024(1 + \sigma_{12}^2)}\bigg] \nonumber \\
&\hspace{0.5 cm} + O(\beta^4).
\end{align}
These examples exhibit the combinatorial factors in \eqref{depth2} and \eqref{depth3}, with all signs $\epsilon_{i, j} = 1$.

To substantiate this pattern, we now consider monopoles with charges of the form $(a, 0)$.  For such monopoles, bubbling always occurs to operators of the same dimension, and the polynomiality criterion $(\mathcal{M}^{(1, 0)})^a = \mathcal{M}^{(a, 0)}$ fixes the combinatorial factors.  For $\mathcal{M}^{(4, 0)}$, we must define
\begin{align}
\mathcal{M}^{(4, 0)} = \overline{M^{(4, 0)}} &+ Z_\text{mono}^{(4, 0)\to (3, 1)}(\Phi_1, \Phi_2)M^{(3, 1)} + Z_\text{mono}^{(4, 0)\to (3, 1)}(\Phi_2, \Phi_1)M^{(1, 3)} \nonumber \\
&+ \frac{1}{2}Z_\text{mono}^{(4, 0)\to (2, 2)}(\Phi_1, \Phi_2)M^{(2, 2)}
\end{align}
where the bubbling coefficients are obtained from
\begin{align}
Z_\text{mono, IOT}^{(4, 0)\to (3, 1)} &= 4 - \frac{1}{4 + \sigma_{12}^2} + O(\beta), \\
Z_\text{mono, IOT}^{(4, 0)\to (2, 2)} &= 12 - \frac{15i}{8(i + \sigma_{12})} - \frac{15i}{8(i - \sigma_{12})} - \frac{9i}{16(2i + \sigma_{12})} - \frac{9i}{16(2i - \sigma_{12})} + O(\beta).
\end{align}
For $\mathcal{M}^{(5, 0)}$, we must define
\begin{align}
\mathcal{M}^{(5, 0)} = \overline{M^{(5, 0)}} &+ Z_\text{mono}^{(5, 0)\to (4, 1)}(\Phi_1, \Phi_2)M^{(4, 1)} + Z_\text{mono}^{(5, 0)\to (4, 1)}(\Phi_2, \Phi_1)M^{(1, 4)} \nonumber \\
&+ \frac{1}{4}[Z_\text{mono}^{(5, 0)\to (3, 2)}(\Phi_1, \Phi_2)M^{(3, 2)} + Z_\text{mono}^{(5, 0)\to (3, 2)}(\Phi_2, \Phi_1)M^{(2, 3)}]
\end{align}
where the bubbling coefficients are obtained from
\begin{align}
Z_\text{mono, IOT}^{(5, 0)\to (4, 1)} &= 5 - \frac{5}{25 + 4\sigma_{12}^2} + O(\beta), \\
Z_\text{mono, IOT}^{(5, 0)\to (3, 2)} &= 40 - \frac{115i}{12(3i + 2\sigma_{12})} - \frac{115i}{12(3i - 2\sigma_{12})} - \frac{9i}{4(5i + 2\sigma_{12})} - \frac{9i}{4(5i - 2\sigma_{12})} \nonumber \\
&\hspace{0.5 cm} + O(\beta).
\end{align}
For $\mathcal{M}^{(6, 0)}$, we must define
\begin{align}
\mathcal{M}^{(6, 0)} = \overline{M^{(6, 0)}} &+ Z_\text{mono}^{(6, 0)\to (5, 1)}(\Phi_1, \Phi_2)M^{(5, 1)} + Z_\text{mono}^{(6, 0)\to (5, 1)}(\Phi_2, \Phi_1)M^{(1, 5)} \nonumber \\
&+ \frac{1}{8}[Z_\text{mono}^{(6, 0)\to (4, 2)}(\Phi_1, \Phi_2)M^{(4, 2)} + Z_\text{mono}^{(6, 0)\to (4, 2)}(\Phi_2, \Phi_1)M^{(2, 4)}] \nonumber \\
&+ \frac{1}{24}Z_\text{mono}^{(6, 0)\to (3, 3)}(\Phi_1, \Phi_2)M^{(3, 3)}
\end{align}
where the bubbling coefficients are obtained from
\begin{align}
Z_\text{mono, IOT}^{(6, 0)\to (5, 1)} &= 6 - \frac{3}{2(9 + \sigma_{12}^2)} + O(\beta), \\
Z_\text{mono, IOT}^{(6, 0)\to (4, 2)} &= 120 - \frac{93i}{8(2i + \sigma_{12})} - \frac{93i}{8(2i - \sigma_{12})} - \frac{9i}{4(3i + \sigma_{12})} - \frac{9i}{4(3i - \sigma_{12})} + O(\beta), \\
Z_\text{mono, IOT}^{(6, 0)\to (3, 3)} &= 480 - \frac{2565i}{32(i + \sigma_{12})} - \frac{2565i}{32(i - \sigma_{12})} - \frac{315i}{8(2i + \sigma_{12})} - \frac{315i}{8(2i - \sigma_{12})} \nonumber \\
&\hspace{1.5 cm} - \frac{225i}{32(3i + \sigma_{12})} - \frac{225i}{32(3i - \sigma_{12})} + O(\beta).
\end{align}
All of these examples are consistent with \eqref{depth2} and \eqref{depth3} for bubbling to depths 2 and 3, again with all signs $\epsilon_{i, j} = 1$.  To explore higher monopole charges, it would be useful to have a more efficient implementation of the IOT algorithm.

Contrary to the general expectation from \eqref{symmetryproperty}, nearly all of the expressions for $Z_\text{mono, IOT}^{b\to v}$ that we have encountered so far in this section are manifestly symmetric in $\sigma_1\leftrightarrow \sigma_2$: an exception is \eqref{zmono2m1to10}.

\subsubsection{$U(2)$ with $N_a = 0$ and $N_f\geq 0$}

In examining these theories, we arrive at examples of two interesting phenomena: the fact that the additional combinatorial factors in \eqref{depth1}--\eqref{depth3} must be supplemented by nontrivial signs, and bubbling into higher-dimension operators.\footnote{Heuristically, monopole bubbling is similar to operator mixing in that both effects are related to renormalization (in the case of bubbling, a renormalization of the GNO charge), and both relate operators with the same global symmetry charges (e.g., topological charge).  However, bubbling differs from mixing in that it does not necessarily occur to monopoles of equal or lower dimension, and the bubbling coefficients that account for differences in dimension are generally rational functions of $\Phi$.}

This class of theories is ugly when $N_f = 3$ and good when $N_f > 3$.  The dimension of the $(a, b)$ monopole is
\begin{equation}
\Delta_{(a, b)} = \frac{N_f}{2}(|a| + |b|) - |a - b|.
\end{equation}
Bubbling occurs to monopoles of smaller $|a - b|$, and we see that a monopole bubbles into monopoles of equal or lower dimension only when $a$ and $b$ have opposite signs.  In particular, the $(n, 0)$ monopoles have minimal dimension $\Delta_{(n, 0)} = |n|(N_f/2 - 1)$ for their topological class, yet nonetheless bubble.

For example, we find that the relation $(\mathcal{M}^{(1, 0)})^n = \mathcal{M}^{(n, 0)}$ holds only after accounting for bubbling into higher-dimension operators.  For $n > 0$, the monopoles $\mathcal{M}^{(n, 0)}$ that bubble to depth at most 2 are given by
\begin{align}
(\mathcal{M}^{(1, 0)})^2 = \overline{M^{(2, 0)}} &- \frac{2r^2}{(ir\Phi_{12} + 1)(ir\Phi_{12} - 1)}M^{(1, 1)}, \\
(\mathcal{M}^{(1, 0)})^3 = \overline{M^{(3, 0)}} &- \frac{3r^2}{(ir\Phi_{12} + 2)(ir\Phi_{12} - 1)}M^{(2, 1)} - \frac{3r^2}{(ir\Phi_{12} + 1)(ir\Phi_{12} - 2)}M^{(1, 2)}, \\
(\mathcal{M}^{(1, 0)})^4 = \overline{M^{(4, 0)}} &- \frac{4r^2}{(ir\Phi_{12} + 3)(ir\Phi_{12} - 1)}M^{(3, 1)} - \frac{4r^2}{(ir\Phi_{12} + 1)(ir\Phi_{12} - 3)}M^{(1, 3)} \nonumber \\
&+ \frac{6r^4}{(ir\Phi_{12} + 2)(ir\Phi_{12} + 1)(ir\Phi_{12} - 1)(ir\Phi_{12} - 2)}M^{(2, 2)}, \\
(\mathcal{M}^{(1, 0)})^5 = \overline{M^{(5, 0)}} &- \frac{5r^2}{(ir\Phi_{12} + 4)(ir\Phi_{12} - 1)}M^{(4, 1)} - \frac{5r^2}{(ir\Phi_{12} + 1)(ir\Phi_{12} - 4)}M^{(1, 4)} \nonumber \\
&+ \frac{10r^4}{(ir\Phi_{12} + 3)(ir\Phi_{12} + 2)(ir\Phi_{12} - 1)(ir\Phi_{12} - 2)}M^{(3, 2)} \nonumber \\
&+ \frac{10r^4}{(ir\Phi_{12} + 2)(ir\Phi_{12} + 1)(ir\Phi_{12} - 2)(ir\Phi_{12} - 3)}M^{(2, 3)}.
\end{align}
Using \eqref{betaprescription}, all of these rational functions are accounted for by bubbling coefficients for $\mathcal{M}^{(n, 0)}$ derived from the IOT algorithm if we set $\epsilon_{1, 1} = \epsilon_{2, 1} = -1$, $\epsilon_{2, 2} = +1$, and $c_{2, 2} = 2$ in \eqref{depth1} and \eqref{depth2}.

In this theory, we also have
\begin{align}
\mathcal{M}^{(-1, 0)}\star \mathcal{M}^{(1, 0)} &= \overline{M^{(1, -1)}} + \frac{(-1)^{N_f + 1}}{r^{N_f - 2}}\left[\frac{(ir\Phi_1 - 1/2)^{N_f}}{ir\Phi_{12}(ir\Phi_{12} - 1)} + \frac{(ir\Phi_2 - 1/2)^{N_f}}{ir\Phi_{21}(ir\Phi_{21} - 1)}\right], \label{firstline} \\
\mathcal{M}^{(1, 0)}\star \mathcal{M}^{(-1, 0)} &= \overline{M^{(1, -1)}} + \frac{(-1)^{N_f + 1}}{r^{N_f - 2}}\left[\frac{(ir\Phi_1 + 1/2)^{N_f}}{ir\Phi_{12}(ir\Phi_{12} + 1)} + \frac{(ir\Phi_2 + 1/2)^{N_f}}{ir\Phi_{21}(ir\Phi_{21} + 1)}\right]. \label{secondline}
\end{align}
The rational function appearing in the first line \eqref{firstline} is precisely the bubbling coefficient $Z_\text{mono}^{(1, -1)\to (0, 0)}$ computed via the IOT algorithm and \eqref{betaprescription}, after adjusting for a minus sign:
\begin{equation}
\mathcal{M}^{(-1, 0)}\star \mathcal{M}^{(1, 0)} = \mathcal{M}^{(1, -1)} = \overline{M^{(1, -1)}} - Z_\text{mono}^{(1, -1)\to (0, 0)}(\Phi_1, \Phi_2).
\label{minussign}
\end{equation}
The second line \eqref{secondline} simply differs from the first by a polynomial.\footnote{Setting $\Phi_1 = \frac{x + y}{2ir}$ and $\Phi_2 = \frac{y - x}{2ir}$, we check that $[\mathcal{M}^{(1, 0)}, \mathcal{M}^{(-1, 0)}]_\star$ is proportional to the polynomial
\begin{equation}
\frac{(x - 1)[(x + y + 1)^{N_f} - (y - x - 1)^{N_f}] + (x + 1)[(y - x + 1)^{N_f} - (x + y - 1)^{N_f}]}{x(x - 1)(x + 1)}
\end{equation}
in $x$ and $y$ (moreover, this polynomial is even in $x$, which guarantees that it is Weyl-invariant with respect to $\Phi_1\leftrightarrow \Phi_2$).}  Similarly, we find it necessary to define
\begin{align}
\mathcal{M}^{(2, -2)} = \overline{M^{(2, -2)}} &- Z_\text{mono}^{(2, -2)\to (1, -1)}(\Phi_1, \Phi_2)M^{(1, -1)} - Z_\text{mono}^{(2, -2)\to (1, -1)}(\Phi_2, \Phi_1)M^{(-1, 1)} \nonumber \\
&+ \frac{1}{2}Z_\text{mono}^{(2, -2)\to (0, 0)}(\Phi_1, \Phi_2)
\end{align}
for star products involving $\mathcal{M}^{(2, -2)}$ to close, where the relevant bubbling coefficients follow from applying \eqref{betaprescription} to
\begin{align}
Z_\text{mono, IOT}^{(2, -2)\to (1, -1)} &= \frac{\beta^{N_f - 2}}{2^{N_f - 2}}\left[\frac{2(\sigma_1 + i)^{N_f}}{\sigma_{12}(\sigma_{12} + 2i)} + (\sigma_1\leftrightarrow \sigma_2)\right] + O(\beta^{N_f - 1}), \\
Z_\text{mono, IOT}^{(2, -2)\to (0, 0)} &= \frac{\beta^{2N_f - 4}}{2^{4N_f - 5}}\left[\frac{(2\sigma_1 + i)^{N_f}}{\sigma_{12}(\sigma_{12} + i)^2}\left(\frac{(2\sigma_1 + 3i)^{N_f}}{\sigma_{12} + 2i} + \frac{2(2\sigma_2 + i)^{N_f}}{\sigma_{12} - i}\right) + (\sigma_1\leftrightarrow \sigma_2)\right] \nonumber \\
&+ O(\beta^{2N_f - 3}).
\end{align}
Hence the shift operators for $\mathcal{M}^{(1, -1)}$ and $\mathcal{M}^{(2, -2)}$ fall in line with the general patterns \eqref{depth1} and \eqref{depth2}.

As an aside, examining some of the simplest relations involving $\mathcal{M}^{(2, -2)}$ and dressings of $\mathcal{M}^{(1, -1)}$ allows us to determine the abelianized bubbling coefficients for $\mathcal{M}^{(1, -1)}$ in the ``IOT basis,'' which are not immediate consequences of the IOT algorithm.  First, we obtain the relation
\begin{equation}
(\mathcal{M}^{(1, -1)})^2 = \mathcal{M}^{(2, -2)} + [P_1(\Phi_1, \Phi_2)\mathcal{M}^{(1, -1)}],
\end{equation}
where we have defined the dressed and bubbled monopole
\begin{equation}
[P_1(\Phi_1, \Phi_2)\mathcal{M}^{(1, -1)}]\equiv P_1(\Phi_1, \Phi_2)M^{(1, -1)} + P_1(\Phi_2, \Phi_1)M^{(-1, 1)} - R_1(\Phi_1, \Phi_2)
\end{equation}
(note the minus sign in front of $R_1$, reflecting \eqref{minussign}) with $P_1$ being the polynomial
\begin{align}
&P_1(\Phi_1, \Phi_2) = \frac{(-1)^{N_f}}{r^{N_f - 2}(ir\Phi_{12} + 1)} \nonumber \\
&\times \left[\frac{(ir\Phi_1 - 1/2)^{N_f} - (ir\Phi_2 - 1/2)^{N_f}}{ir\Phi_{12}} - \frac{(ir\Phi_1 + 1/2)^{N_f} - (ir\Phi_2 - 3/2)^{N_f}}{ir\Phi_{12} + 2}\right]
\end{align}
and $R_1$ being a rational function whose explicit form we omit for brevity.  We also have
\begin{equation}
\mathcal{M}^{(-2, 0)}\star \mathcal{M}^{(2, 0)} = \mathcal{M}^{(2, -2)}.
\end{equation}
On the other hand, reversing the order of the star product gives
\begin{equation}
\mathcal{M}^{(2, 0)}\star \mathcal{M}^{(-2, 0)} = \mathcal{M}^{(2, -2)} + [P_2(\Phi_1, \Phi_2)\mathcal{M}^{(1, -1)}] + Q(\Phi_1, \Phi_2)
\end{equation}
where we have defined the dressed and bubbled monopole
\begin{equation}
[P_2(\Phi_1, \Phi_2)\mathcal{M}^{(1, -1)}]\equiv P_2(\Phi_1, \Phi_2)M^{(1, -1)} + P_2(\Phi_2, \Phi_1)M^{(-1, 1)} - R_2(\Phi_1, \Phi_2),
\end{equation}
again accounting for the minus sign for bubbling at depth 1 in this theory.  Here, the dressing function $P_2$ is a non-Weyl-invariant polynomial, $Q$ (which appears in the star product) is a Weyl-invariant polynomial, and $R_2$ is a rational function, none of which we write explicitly.  We have seen that the correct dressing prescription is to associate a bubbling coefficient to each abelian (non-Weyl-averaged) monopole shift operator.  So we have
\begin{equation}
\widetilde{M}^{(1, -1)} = M^{(1, -1)} + R(\Phi_1, \Phi_2) \implies \mathcal{M}^{(1, -1)} = \overline{M^{(1, -1)}} + R(\Phi_1, \Phi_2) + R(\Phi_2, \Phi_1)
\end{equation}
for some (\emph{a priori}, non-symmetric) rational function $R(\Phi_1, \Phi_2)$.  Given the relations stated above, the bubbling function $R(\Phi_1, \Phi_2)$ must satisfy the overconstrained system of equations
\begin{align}
R(\Phi_1, \Phi_2) + R(\Phi_2, \Phi_1) &= -Z_\text{mono}^{(1, -1)\to (0, 0)}(\Phi_1, \Phi_2), \\
P_i(\Phi_1, \Phi_2)R(\Phi_1, \Phi_2) + P_i(\Phi_2, \Phi_1)R(\Phi_2, \Phi_1) &= -R_i(\Phi_1, \Phi_2) \quad (i = 1, 2),
\end{align}
which require that
\begin{equation}
R(\Phi_1, \Phi_2) = \frac{P_i(\Phi_2, \Phi_1)Z_\text{mono}^{(1, -1)\to (0, 0)}(\Phi_1, \Phi_2) - R_i(\Phi_1, \Phi_2)}{P_i(\Phi_1, \Phi_2) - P_i(\Phi_2, \Phi_1)} \quad (i = 1, 2)
\end{equation}
where $Z_\text{mono}^{(1, -1)\to (0, 0)}$ and $R_i$ are symmetric but $P_i$ is (in general) not.\footnote{If $P_i$ is symmetric, then the above expression is indeterminate and we have simply
\begin{equation}
P_i(\Phi_1, \Phi_2)Z_\text{mono}^{(1, -1)\to (0, 0)}(\Phi_1, \Phi_2) = R_i(\Phi_1, \Phi_2).
\end{equation}}  These equations indeed have a solution: using the known expressions for $P_i, R_i$ ($i = 1, 2$), we deduce that
\begin{equation}
R(\Phi_1, \Phi_2) = \begin{cases} \displaystyle -\frac{(1 - 2ir\Phi_2)^{N_f}}{2^{N_f}r^{N_f - 2}(ir\Phi_{12})(1 + ir\Phi_{12})} & \text{for $N_f\geq 5$}, \\[10 pt] \displaystyle -\frac{1}{2}Z_\text{mono}^{(1, -1)\to (0, 0)}(\Phi_1, \Phi_2) & \text{for $N_f = 0, \ldots, 4$}. \end{cases}
\end{equation}
In particular, $R(\Phi_1, \Phi_2)$ is symmetric for $N_f = 0, \ldots, 4$, in which case
\begin{equation}
R_i(\Phi_1, \Phi_2) = \left[\frac{P_i(\Phi_1, \Phi_2) + P_i(\Phi_2, \Phi_1)}{2}\right]Z_\text{mono}^{(1, -1)\to (0, 0)}(\Phi_1, \Phi_2)
\end{equation}
for $i = 1, 2$ (in words, the dressing factor for the identity is simply the Weyl-averaged dressing polynomial).

\subsubsection{$U(3)$ with $N_a = 0$ and $N_f\geq 0$}

A few new lessons can be learned by going to higher rank.  For $U(3)$ SQCD with fundamental flavors, we will be brief and discuss only the monopoles $\mathcal{M}^{(1, 0, -1)}$, $\mathcal{M}^{(2, -1, -1)}$, $\mathcal{M}^{(1, 1, -2)}$, which bubble to depth at most 2.  The IOT results for the Weyl-averaged bubbling coefficients of these monopoles are as follows (these will be useful references for our analysis of the $SU(3)$ theory with the same matter content).  For $\mathcal{M}^{(1, 0, -1)}$, we have
\begin{equation}
Z_\text{mono, IOT}^{(1, 0, -1)\to (0, 0, 0)} = \left(-\frac{i\beta}{2}\right)^{N_f - 4}\sum_{i=1}^3 \frac{(i\sigma_i - 1/2)^{N_f}}{\prod_{j\neq i} i\sigma_{ij}(i\sigma_{ij} - 1)} + O(\beta^{N_f - 3}).
\end{equation}
For $\mathcal{M}^{(2, -1, -1)}$, we have
\begin{align}
Z_\text{mono, IOT}^{(2, -1, -1)\to (1, 0, -1)} &= \frac{\beta^{N_f - 2}}{2^{2N_f - 5}}\frac{1}{i + 2\sigma_{12}}\left(\frac{2^{N_f - 1}(i + \sigma_1)^{N_f}}{3i + 2\sigma_{12}} + \frac{(i + 2\sigma_2)^{N_f}}{-3i + 2\sigma_{12}}\right) + O(\beta^{N_f - 1}), \\
Z_\text{mono, IOT}^{(2, -1, -1)\to (0, 0, 0)} &= \frac{\beta^{2N_f - 6}}{2^{4N_f - 7}}\sum_{\{i, j, k\}} \frac{(i + 2\sigma_i)^{N_f}(i + 2\sigma_j)^{N_f}}{\sigma_{ij}(i + \sigma_{ij})\sigma_{ik}(i + \sigma_{ik})\sigma_{jk}(i + \sigma_{jk})} + O(\beta^{2N_f - 5}),
\end{align}
where $\{i, j, k\}$ runs over permutations of $\{1, 2, 3\}$ and in the first line, the corresponding expressions for other Weyl orderings of the charges $(1, 0, -1)$ are obtained by taking permutations.  For $\mathcal{M}^{(1, 1, -2)}$, we have
\begin{align}
Z_\text{mono, IOT}^{(1, 1, -2)\to (1, 0, -1)} &= Z_\text{mono, IOT}^{(2, -1, -1)\to (1, 0, -1)}|_{\sigma_1\to \sigma_3}, \\
Z_\text{mono, IOT}^{(1, 1, -2)\to (0, 0, 0)} &= Z_\text{mono, IOT}^{(2, -1, -1)\to (0, 0, 0)}.
\end{align}
Polynomiality shows that in addition to using \eqref{betaprescription} to reproduce the expected bubbling coefficients $Z_\text{mono}^{b\to v}$, we must set $c_{2, 2} = 2$ in \eqref{depth2} for bubbling to depth 2.  Moreover, one can show that the shift operators for $\mathcal{M}^{(1, 0, -1)}$ and $\mathcal{M}^{(2, 0, -2)}$ must be defined by setting $\epsilon_{1, 1} = 1$ and $\epsilon_{2, 1} = \epsilon_{2, 2} = 1$ in \eqref{depth1} and \eqref{depth2}, respectively, whereas the shift operators for $\mathcal{M}^{(2, -1, -1)}$ and $\mathcal{M}^{(1, 1, -2)}$ must be defined by setting $\epsilon_{2, 1} = \epsilon_{2, 2} = -1$ in \eqref{depth2} (note that the monopole of charge $(2, 0, -2)$ bubbles not only into $(1, 0, -1)$ and $(0, 0, 0)$, but also into the higher-dimension monopoles $(2, -1, -1)$ and $(1, 1, -2)$).  These examples illustrate that the signs in \eqref{depth1}--\eqref{depth3} are not only theory-dependent, but also monopole-dependent.

\section{More (Quantized) Chiral Rings} \label{CHIRALAPPENDIX}

In this section, we use our formalism to compute the quantized chiral rings of some simple theories.

Existing approaches to deriving the Coulomb branch chiral rings of 3D $\mathcal{N} = 4$ quiver gauge theories or their quantizations include the Hilbert series \cite{Cremonesi:2013lqa, Hanany:2016ezz}, abelianization \cite{Bullimore:2015lsa, Dimofte:2018abu}, combinations of the aforementioned techniques \cite{Hanany:2018xth}, and incorporating half-BPS local operators into the type IIB brane/S-duality realization \cite{Hanany:1996ie} of 3D mirror symmetry \cite{Assel:2017hck}.  In particular, the Hilbert series can be used to infer the quantum numbers of the generators and their relations (such as for $U$, $USp$, and $SO$ gauge theories, for which the Coulomb branch is a complete intersection \cite{Cremonesi:2013lqa}), but it does not specify numerical coefficients.

When the moduli space is a hyperk\"ahler cone, these known techniques for extracting generators and ring relations work well.  In other situations, such as in bad theories \cite{Yaakov:2013fza, Assel:2017jgo, Dey:2017fqs, Assel:2018exy}, the chiral ring has not been as thoroughly studied.  In particular, the Coulomb branch of $SU(N_c)$ gauge theory with $N_f$ fundamental flavors has no global symmetry, and the case $N_c > 2$ is not yet well-understood; we make some comments on the case $N_c = 3$ below.  We expect such theories to present good opportunities for applications of our formalism.

All of the non-abelian examples in this section take the form of $U(N_c)$ or $SU(N_c)$ gauge theories with fundamental and adjoint matter ($N_f$ and $N_a$).  In these theories, the dimension of a monopole with GNO charge $(b_1, \ldots, b_{N_c})$, computed in the UV, is
\begin{equation}
\Delta = \frac{1}{2}N_f(|b_1| + \cdots + |b_{N_c}|) + (N_a - 1)\sum_{i<j} |b_i - b_j|,
\label{dimensionformulaUandSU}
\end{equation}
where we have used the same conventions for roots and weights of $SU(N_c)$ as for $U(N_c)$ (see Section \ref{n8example}), with the understanding that $b_1 + \cdots + b_{N_c} = 0$ in the former case.  According to the Gaiotto-Witten classification, such a theory is good, bad, or ugly if its minimum $\Delta$ is $\geq 1$, $\leq 0$, or $= 1/2$, respectively (in bad theories, unitarity-violating monopole operators are realized by free scalar fields in an IR dual description \cite{Yaakov:2013fza, Assel:2017jgo}).  For $SU(N_c)$, the monopole of smallest dimension has charge $(1, -1, \vec{0})$ and $\Delta = N_f + 2(N_c - 1)(N_a - 1)$, so the theory is good when
\begin{equation}
N_f + 2(N_c - 1)N_a\geq 2N_c - 1.
\label{goodboundSU}
\end{equation}
It is never ugly because $\Delta$ is an integer.  On the other hand, for $U(N_c)$, the monopoles of smallest dimension have charge $(\pm 1, \vec{0})$ and $\Delta = N_f/2 + (N_a - 1)(N_c - 1)$.  We see that when $N_a = 1$ and $N_f > 0$, both the $SU$ and $U$ theories are never bad.

\subsection{SQED$_N$ versus $U(1)$ with One Hyper of Charge $N$}

Before presenting the non-abelian examples, we start by providing another example of two theories that have the same Coulomb branch but different quantizations.  These theories are SQED$_N$ and $U(1)$ gauge theory with a single hyper of charge $N$ (the $\mathbb{Z}_N$ gauge theory of a free hypermultiplet), which we denote by $U(1) + N$.  The CBOs $\mathcal{M}^{\pm 1}, \Phi$ in either theory are represented by the shift operators
\begin{equation}
\mathcal{M}_N^1 = \begin{cases} \displaystyle \frac{(-1)^N}{r^{N/2}}\left(\frac{1 - B}{2} + i\sigma\right)^N e^{-\frac{i}{2}\partial_\sigma - \partial_B} & \text{in SQED$_N$}, \\[10 pt] \displaystyle \frac{(-1)^N}{r^{N/2}}\left(\frac{1 - NB}{2} + iN\sigma\right)_N e^{-\frac{i}{2}\partial_\sigma - \partial_B} & \text{in $U(1) + N$}, \end{cases}
\end{equation}
as well as by
\begin{equation}
\mathcal{M}_N^{-1} = \frac{e^{\frac{i}{2}\partial_\sigma + \partial_B}}{r^{N/2}}, \quad \Phi_N = \frac{1}{r}\left(\sigma + \frac{i}{2}B\right)\times \begin{cases} 1 & \text{in SQED$_N$}, \\ N & \text{in $U(1) + N$}. \end{cases}
\end{equation}
In both theories, let
\begin{equation}
\mathcal{X} = \frac{1}{(4\pi)^{N/2}}\mathcal{M}^{-1}, \quad \mathcal{Y} = \frac{1}{(4\pi)^{N/2}}\mathcal{M}^1, \quad \mathcal{Z} = -\frac{i}{4\pi}\Phi.
\end{equation}
Then we compute that
 \es{CRRelSQED}{
\mathcal{X}\star \mathcal{Y} = \begin{cases} \displaystyle \left(\mathcal{Z} + \frac{1}{8\pi r}\right)^N & \text{in SQED$_N$}, \\[10 pt] \displaystyle \prod_{k=0}^{N-1} \left(\mathcal{Z} + \frac{2k + 1}{8\pi r}\right) & \text{in $U(1) + N$} \,,  \end{cases}
 }
where multiplication on the RHS is understood to be $\star$.  The two theories have identical Coulomb branches ($\mathbb{C}^2/\mathbb{Z}_N$) and chiral ring relations, as can be seen in the $r \to \infty$ limit of \eqref{CRRelSQED}, but different star products.

\subsection{Theories on D2-Branes}

The first few non-abelian examples that we study can be realized as worldvolume theories on $N_c$ D2-branes \cite{Cremonesi:2013lqa}.  These theories, which all have at least one adjoint hypermultiplet, are particularly straightforward to analyze because the matter contribution cancels the denominator in the abelianized chiral ring relations \eqref{generalproduct}.  We will see explicitly how to reproduce the formalism of \cite{Bullimore:2015lsa} for such theories, working our way up in complexity.

\subsubsection{$U(2)$ with $N_a = N_f = 1$} \label{sym2c2relationderivation}

The Coulomb branch of this theory is known to be $\operatorname{Sym}^2(\mathbb{C}^2)$, which has complex dimension four.  To exhibit the Coulomb branch chiral ring, let us denote the generators of each $\mathbb{C}^2$ by $x_i, y_i$ for $i = 1, 2$.  The coordinate ring of the symmetric product is generated by the five symmetric polynomials
\begin{equation}
x_1 + x_2, \quad y_1 + y_2, \quad x_1 x_2, \quad y_1 y_2, \quad (x_1 - x_2)(y_1 - y_2)
\label{coordcombos}
\end{equation}
(the $S_2 = \mathbb{Z}_2$ by which we quotient acts as $1\leftrightarrow 2$), subject to the single relation
\begin{equation}
[(x_1 + x_2)^2 - 4x_1 x_2][(y_1 + y_2)^2 - 4y_1 y_2] = [(x_1 - x_2)(y_1 - y_2)]^2.
\label{sym2c2coordrelation}
\end{equation}
This relation is precisely that of $\mathbb{C}^2/\mathbb{Z}_2$, whence $\operatorname{Sym}^2(\mathbb{C}^2)\cong \mathbb{C}^2\times (\mathbb{C}^2/\mathbb{Z}_2)$.

We would like to identify the combinations \eqref{coordcombos} with Coulomb branch operators.  Recall from Section \ref{u2n8example} that the candidate such operators are
\begin{equation}
\tr\Phi, \quad \mathcal{M}^{(\pm 1, 0)}, \quad \mathcal{M}^{\pm(1, 1)}, \quad \mathcal{M}^{(1, -1)}, \quad \mathcal{M}^{(\pm 2, 0)}.
\end{equation}
While the monopoles $\mathcal{M}^{(\pm 1, 0)}$ and $\mathcal{M}^{\pm(1, 1)}$ do not bubble and are therefore represented by the na\"ive (unbubbled) shift operators \eqref{shiftAgain}, the monopoles $\mathcal{M}^{(1, -1)}$ and $\mathcal{M}^{(\pm 2, 0)}$ do bubble and are most conveniently constructed as star products of $\mathcal{M}^{(\pm 1, 0)}$: see \eqref{pm10squared} and \eqref{10withm10}.  Thinking of $x_i, y_i$ as having topological charge $\mp 1$, respectively (these conventions are natural from the point of view of correlation functions, as in \cite{Dedushenko:2017avn} and Appendix \ref{mirrorappendix}), we would expect to identify $x_1 + x_2$ and $y_1 + y_2$ with $\mathcal{M}^{\mp(1, 0)}$, $x_1 x_2$ and $y_1 y_2$ with $\mathcal{M}^{\mp(1, 1)}$, and $(x_1 - x_2)(y_1 - y_2)$ with a linear combination of $\mathcal{M}^{(1, -1)}$ and $\tr\Phi$.  Thus from \eqref{sym2c2coordrelation}, we would expect a chiral ring relation of the form
\begin{equation}
(\mathcal{M}^{(-2, 0)} - 4\mathcal{M}^{(-1, -1)})(\mathcal{M}^{(2, 0)} - 4\mathcal{M}^{(1, 1)}) = (\mathcal{M}^{(1, -1)} + c\tr\Phi)^2
\label{sym2c2operatorrelation}
\end{equation}
for some constant $c$, where the products are commutative (not $\star$).  Since the star product gives a quantization of the chiral ring product, we would expect \eqref{sym2c2operatorrelation} to hold with respect to the star product after appropriate ordering, and up to subleading $O(1/r)$ terms.\footnote{There are many different ways of writing the star products that reduce to the same chiral ring relations.  We find the ``democratic'' symmetrization convenient.}  Indeed, letting $()_\star$ denote symmetrization with respect to the star product (e.g., $(\mathcal{O}_1\mathcal{O}_2)_\star = \frac{1}{2}(\mathcal{O}_1\star \mathcal{O}_2 + \mathcal{O}_2\star \mathcal{O}_1)$), we find that with $c = 2i$,
\begin{equation}
(\text{LHS of \eqref{sym2c2operatorrelation}} - \text{RHS of \eqref{sym2c2operatorrelation}})_\star = -\frac{2}{r}\mathcal{M}^{(1, -1)} - \frac{4i}{r}\tr\Phi + \frac{5}{r^2} \xrightarrow{r\to\infty} 0,
\end{equation}
which is equivalent to \eqref{sym2c2operatorrelation} in the chiral ring ($r\to\infty$).  To summarize, we identify
\begin{gather}
x_1 + x_2\leftrightarrow \mathcal{M}^{(-1, 0)}, \quad y_1 + y_2\leftrightarrow \mathcal{M}^{(1, 0)}, \quad x_1 x_2\leftrightarrow \mathcal{M}^{(-1, -1)}, \quad y_1 y_2\leftrightarrow \mathcal{M}^{(1, 1)}, \nonumber \\
(x_1 - x_2)(y_1 - y_2)\leftrightarrow -\mathcal{M}^{(1, -1)} - 2i\tr\Phi
\end{gather}
in the chiral ring, as claimed in \eqref{sym2c2relation}.  Given what we have said so far, the last identification would be consistent with either sign in $\pm(\mathcal{M}^{(1, -1)} + 2i\tr\Phi)$: we explain the above choice of sign in our discussion of the case $N_f > 1$ below.

\subsubsection{$U(2)$ with $N_a = 1$ and $N_f\geq 1$}

The Coulomb branch of this theory is known to be $\operatorname{Sym}^2(\mathbb{C}^2/\mathbb{Z}_{N_f})$.  We denote the generators of each copy of $\mathbb{C}^2/\mathbb{Z}_{N_f}$ by $x_i, y_i, z_i$, which satisfy the relations
\begin{equation}
x_i y_i = z_i^{N_f}
\end{equation}
for $i = 1, 2$.  As in our discussion of the case $N_f = 1$, the coordinate ring is generated by the nine symmetric polynomials
\begin{gather}
x_1 + x_2, \quad y_1 + y_2, \quad z_1 + z_2, \quad x_1 x_2, \quad y_1 y_2, \quad z_1 z_2, \nonumber \\
(x_1 - x_2)(y_1 - y_2), \quad (x_1 - x_2)(z_1 - z_2), \quad (y_1 - y_2)(z_1 - z_2),
\label{sym2quotient}
\end{gather}
which satisfy the five relations
\begin{align}
(x_1 x_2)(y_1 y_2) &= \smash{(z_1 z_2)^{N_f}}, \label{rel1} \\
\textstyle \frac{1}{2}[(x_1 + x_2)(y_1 + y_2) + (x_1 - x_2)(y_1 - y_2)] &= \smash{z_1^{N_f} + z_2^{N_f}}, \label{rel2} \\
[(x_1 + x_2)^2 - 4x_1 x_2][(y_1 + y_2)^2 - 4y_1 y_2] &= [(x_1 - x_2)(y_1 - y_2)]^2, \label{rel3} \\
[(x_1 + x_2)^2 - 4x_1 x_2][(z_1 + z_2)^2 - 4z_1 z_2] &= [(x_1 - x_2)(z_1 - z_2)]^2, \label{rel4} \\
[(y_1 + y_2)^2 - 4y_1 y_2][(z_1 + z_2)^2 - 4z_1 z_2] &= [(y_1 - y_2)(z_1 - z_2)]^2. \label{rel5} 
\end{align}
The RHS of \eqref{rel2} can be written in terms of the generators $z_1 + z_2$ and $z_1 z_2$ using Newton's identities.

To physically interpret the Coulomb branch chiral ring, note that the lowest-dimension ($\Delta = N_f/2$) monopoles are $\mathcal{M}^{(\pm 1, 0)}$.  The relation \eqref{pm10squared} holds equally well when $N_f\geq 1$ as when $N_f = 1$, while the relation \eqref{10withm10} generalizes to
\begin{align}
&\mathcal{M}^{(\pm 1, 0)}\star \mathcal{M}^{(\mp 1, 0)} - \overline{M^{(1, -1)}} \nonumber \\
&= \frac{(-1)^{N_f}}{2^{N_f + 2}r^{N_f}}\left[\frac{(2ir\Phi_1\pm 1)^{N_f}(2ir\Phi_{12}\pm 1)^2}{ir\Phi_{12}(ir\Phi_{12}\pm 1)} + \frac{(2ir\Phi_2\pm 1)^{N_f}(2ir\Phi_{12}\mp 1)^2}{ir\Phi_{12}(ir\Phi_{12}\mp 1)}\right].
\end{align}
One can check that the expressions $\mathcal{M}^{(-1, 0)}\star \mathcal{M}^{(1, 0)}$ and $\mathcal{M}^{(1, 0)}\star \mathcal{M}^{(-1, 0)}$ differ by a Weyl-invariant polynomial in $\tr\Phi = \Phi_1 + \Phi_2$ that vanishes in the chiral ring limit ($r\to\infty$).  We work in a basis where $\mathcal{M}^{(-1, 0)}\star \mathcal{M}^{(1, 0)} = \mathcal{M}^{(1, -1)}$.\footnote{This happens to coincide with the ``IOT basis,'' in which
\begin{equation}
\mathcal{M}^{(-1, 0)}\star \mathcal{M}^{(1, 0)} = \mathcal{M}^{(1, -1)} = \overline{M^{(1, -1)}} + Z_\text{mono}^{(1, -1)\to (0, 0)}(\Phi_1, \Phi_2),
\end{equation}
generalizing \eqref{c1m1to00} for $N_f = 1$.  However, we do not need any of the explicit results of the IOT algorithm in this section (or, indeed, in this paper).}  The coordinates $x_i, y_i$ correspond to topological charge $\mp 1$ and $\Delta = N_f/2$, while the $z_i$ correspond to zero topological charge and $\Delta = 1$.  Hence we still expect to have
\begin{equation}
x_1 + x_2\leftrightarrow \mathcal{M}^{(-1, 0)}, \quad y_1 + y_2\leftrightarrow \mathcal{M}^{(1, 0)}
\end{equation}
with $\Delta = N_f/2$, and from the relation
\begin{equation}
\mathcal{M}^{\pm(1, 1)}\star \mathcal{M}^{\mp(1, 1)} = \left[\left(\mp\frac{1}{2r} - i\Phi_1\right)\left(\mp\frac{1}{2r} - i\Phi_2\right)\right]^{N_f}
\end{equation}
(which is derived in our discussion of arbitrary $N_c$ below), we also expect to have
\begin{equation}
x_1 x_2\leftrightarrow \mathcal{M}^{(-1, -1)}, \quad y_1 y_2\leftrightarrow \mathcal{M}^{(1, 1)}, \quad z_1 z_2\leftrightarrow -\Phi_1\Phi_2 = \frac{1}{2}[\tr\Phi^2 - (\tr\Phi)^2]
\end{equation}
in the chiral ring, where the first two generators have $\Delta = N_f$ and the last has $\Delta = 2$.  The remaining generators can be deduced from charge and dimensional considerations, as well as the relations \eqref{rel1}--\eqref{rel5}.  However, we will take a simpler approach.  Rather than directly identifying which Coulomb branch operators are chiral ring generators, we postulate the following more elementary correspondences between coordinate ring generators of $(\mathbb{C}^2/\mathbb{Z}_{N_f})^2$ (before the quotient by $S_2 = \mathbb{Z}_2$) and non-Weyl-averaged Coulomb branch operators:
\begin{equation}
x_1, x_2\leftrightarrow M^{(-1, 0)}, M^{(0, -1)}, \quad y_1, y_2\leftrightarrow M^{(1, 0)}, M^{(0, 1)}, \quad z_1, z_2\leftrightarrow -i\Phi_1, -i\Phi_2.
\label{basicids}
\end{equation}
The basic identifications \eqref{basicids} are consistent with all of those stated earlier, and allow us to determine the missing ones.  For instance, we have
\begin{equation}
\frac{1}{2}(M^{(\pm 1, 0)}M^{(0, \pm 1)} + M^{(0, \pm 1)}M^{(\pm 1, 0)}) = C(\Phi_1, \Phi_2)\mathcal{M}^{\pm(1, 1)}
\end{equation}
where
\begin{equation}
C(\Phi_1, \Phi_2)\equiv \frac{3 + 4r^2\Phi_{12}^2}{4 + 4r^2\Phi_{12}^2} \xrightarrow{r\to\infty} 1,
\end{equation}
which is consistent with $x_1 x_2\leftrightarrow \mathcal{M}^{(-1, -1)}$ and $y_1 y_2\leftrightarrow \mathcal{M}^{(1, 1)}$ in the chiral ring (since the $M$ are not gauge-invariant chiral ring operators, the above product is not a star product and does not obey polynomiality).  We now easily deduce $(x_1 - x_2)(y_1 - y_2)$ by computing
\begin{equation}
\frac{1}{2}((M^{(-1, 0)} - M^{(0, -1)})(M^{(1, 0)} - M^{(0, 1)}) + ((-1, 0)\leftrightarrow (1, 0))) = -\mathcal{M}^{(1, -1)} + R(\Phi_1, \Phi_2)
\end{equation}
where
\begin{align}
R(\Phi_1, \Phi_2) &\equiv \frac{-i(p_-(\Phi_1) - p_-(\Phi_2)) + r\Phi_{12}(3 + 4r^2\Phi_{12}^2)(p_+(\Phi_1) + p_+(\Phi_2))}{2^{N_f + 3}r^{N_f + 1}(1 + r^2\Phi_{12}^2)\Phi_{12}}, \\
p_\pm(x) &\equiv (-1 - 2irx)^{N_f}\pm 3(1 - 2irx)^{N_f}.
\end{align}
We then have that
\begin{gather}
R(\Phi_1, \Phi_2) \xrightarrow{r\to\infty} 2[(-i\Phi_1)^{N_f} + (-i\Phi_2)^{N_f}] \nonumber \\
\implies (x_1 - x_2)(y_1 - y_2)\leftrightarrow -\mathcal{M}^{(1, -1)} + 2[(-i\Phi_1)^{N_f} + (-i\Phi_2)^{N_f}],
\end{gather}
which is consistent with \eqref{rel2} and reduces to precisely the expected result for $N_f = 1$.  We further compute that
\begin{gather}
-\frac{i}{2}((M^{(-1, 0)} - M^{(0, -1)})\Phi_{12} + \Phi_{12}(M^{(-1, 0)} - M^{(0, -1)})) \nonumber \\
= \left(\frac{1}{2r} - i\Phi_{12}\right)M^{(-1, 0)} + \left(\frac{1}{2r} + i\Phi_{12}\right)M^{(0, -1)};
\end{gather}
in the limit $r\to\infty$, this becomes the dressed monopole $-i\Phi_{12}\mathcal{M}^{(-1, 0)}$, which we identify with $(x_1 - x_2)(z_1 - z_2)$ in the chiral ring.  Similarly, we compute that
\begin{gather}
-\frac{i}{2}((M^{(1, 0)} - M^{(0, 1)})\Phi_{12} + \Phi_{12}(M^{(1, 0)} - M^{(0, 1)})) \nonumber \\
= -\left(\frac{1}{2r} + i\Phi_{12}\right)M^{(1, 0)} - \left(\frac{1}{2r} - i\Phi_{12}\right)M^{(0, 1)};
\end{gather}
in the limit $r\to\infty$, this becomes the dressed monopole $-i\Phi_{12}\mathcal{M}^{(1, 0)}$, which we identify with $(y_1 - y_2)(z_1 - z_2)$ in the chiral ring.  We have now used \eqref{basicids} to identify all of the generators \eqref{sym2quotient} with physical Coulomb branch operators; by construction, the appropriately symmetrized star products of the latter reproduce the chiral ring relations \eqref{rel1}--\eqref{rel5}.

\subsubsection{$U(N_c)$ with $N_a = 1$ and $N_f\geq 1$}

Our approach to the Coulomb branch chiral ring of $U(2)$ with $N_a = 1$ and $N_f\geq 1$ generalizes straightforwardly to higher rank.  To see that the Coulomb branch in this case is $\operatorname{Sym}^{N_c}(\mathbb{C}^2/\mathbb{Z}_{N_f})$, we first identify certain elementary non-Weyl-invariant operators with the coordinates $x_i, y_i, z_i$ of $(\mathbb{C}^2/\mathbb{Z}_{N_f})^{N_c}$ ($i = 1, \ldots, N_c$), along the lines of \eqref{basicids}.  The $\smash{\operatorname{Sym}^{N_c}}$ operation corresponds to Weyl averaging.  One then easily checks that these operators satisfy the ``abelianized'' relations $x_i y_i = \smash{z_i^{N_f}}$, upon which it automatically follows that the full (Weyl-invariant) chiral ring generators satisfy the relations of $\operatorname{Sym}^{N_c}(\mathbb{C}^2/\mathbb{Z}_{N_f})$.  This is essentially the approach advocated in \cite{Bullimore:2015lsa}.\footnote{The same reasoning implies that the Coulomb branch chiral ring of $U(N_c)$ with $N_a\geq 1$ and arbitrary $N_f$ is simply $\operatorname{Sym}^{N_c}(\mathbb{C}^2/\mathbb{Z}_{N_f + 2(N_a - 1)})$.  One can also use our polynomiality results for bubbling coefficients in low-rank theories where the IOT prescription is unavailable to construct the quantized Coulomb branch chiral rings for other theories on $N$ D2-branes \cite{Cremonesi:2013lqa}.  These include $USp(2N)$ with one antisymmetric and $N_f$ fundamentals (whose Coulomb branch is $\smash{\operatorname{Sym}^N(\mathbb{C}^2/D_{N_f})}$), and $SO(2N + 1)$ with one symmetric and $N_f$ fundamentals (whose Coulomb branch is the same as that of $USp(2N)$ with one antisymmetric and $N_f + 3$ fundamentals).}

Of particular interest are the non-bubbling monopoles $\smash{\mathcal{M}^{\pm\vec{k}}}$ with charges
\begin{equation}
\vec{k}\equiv (\underbrace{1, \ldots, 1}_k, 0, \ldots, 0), \quad k = 1, \ldots, N_c,
\end{equation}
which are natural candidates for chiral ring generators ($k = 1$ corresponds to the free sector for $N_f = 1$).  The case $k = N_c$ is special because the sum over Weyl reflections is trivial, so in this case, the star products $\smash{\mathcal{M}^{\pm\vec{k}}}\star \smash{\mathcal{M}^{\mp\vec{k}}}$ contain no monopoles and are expressible purely in terms of $\Phi$.  Namely, in this theory, we have
\begin{align}
\mathcal{M}^{(-1, \ldots, -1)}_N &= \frac{e^{\sum_I (\frac{i}{2}\partial_{\sigma_I} + \partial_{B_I})}}{r^{N_c N_f/2}}, \\
\mathcal{M}^{(1, \ldots, 1)}_N &= \frac{(-1)^{N_c N_f}}{r^{N_c N_f/2}}\left[\prod_I \left(\frac{1 - B_I}{2} + i\sigma_I\right)\right]^{N_f}e^{\sum_I (-\frac{i}{2}\partial_{\sigma_I} - \partial_{B_I})},
\end{align}
from which we compute that
\begin{equation}
\mathcal{M}^{\pm(1, \ldots, 1)}\star \mathcal{M}^{\mp(1, \ldots, 1)} = \left[\prod_I \left(\mp\frac{1}{2r} - i\Phi_I\right)\right]^{N_f} \xrightarrow{r\to\infty} \left[\prod_I \left(-i\Phi_I\right)\right]^{N_f},
\end{equation}
generalizing \eqref{pm11withmp11}.  One can write the symmetric polynomial in $\Phi_I$ on the RHS in terms of traces of powers of $\Phi$ using Newton's identities.

\subsection{Theories with No Adjoints}

We now turn to theories with no adjoints, for which the chiral ring relations do not simplify so easily.  We restrict our attention to theories with $SU$ gauge group (which lack minuscule monopoles) and fundamental matter.

As mentioned in Sections \ref{sec:rankone} and \ref{SU2CHIRAL}, we can obtain $SU(N_c)$ gauge theory by gauging the topological $U(1)$ symmetry of a $U(N_c)$ theory with the same matter content.  At the level of local operators in the $U(N_c)$ theory, this is equivalent to restricting to the sector of zero topological charge and setting $\tr\Phi = 0$,\footnote{This prescription for gauging the $U(1)_\text{top}$ of $U(N)$ results in $SU(N)$ gauge theory as opposed to any of the other global forms of $\mathfrak{su}(N)$, and hence can be performed regardless of matter content.  This is simply because the gauging restricts the lattice of GNO charges of monopoles in the resulting theory, as a sublattice of the coweight lattice of $\mathfrak{su}(N)$, to be the coroot lattice.  Hence the magnetic gauge group is $PSU(N)$, which has trivial center.} and indeed, the Coulomb branch of $SU(N_c)$ gauge theory can be obtained as a hyperk\"ahler quotient of the $U(N_c)$ case by $U(1)_\text{top}$ \cite{Assel:2017jgo, Assel:2018exy}.

Correspondingly, our strategy to obtain the dressed monopoles with the proper abelianized bubbling coefficients in $SU(N_c)$ gauge theory is to start with a $U(N_c)$ gauge theory, where the minuscule monopoles and their dressed versions do not bubble, and then take star products of these minuscule monopoles and descend to $SU(N_c)$.  For example, in $U(N_c)$ with $N_f$ fundamentals, the basic monopole shift operators are given by
\begin{align}
\mathcal{M}^{(-1, \vec{0})} &= \frac{1}{r^{N_f/2 - N_c + 1}}\sum_{i=1}^{N_c} \frac{1}{\prod_{j\neq i} ir\Phi_{ij}}e^{\frac{i}{2}\partial_{\sigma_i} + \partial_{B_i}}, \\
\mathcal{M}^{(1, \vec{0})} &= \frac{(-1)^{N_f - N_c - 1}}{r^{N_f/2 - N_c + 1}}\sum_{i=1}^{N_c} \frac{(1/2 + ir\Phi_i)^{N_f}}{\prod_{j\neq i} ir\Phi_{ij}}e^{-\frac{i}{2}\partial_{\sigma_i} - \partial_{B_i}}.
\end{align}
Their star products (generalizing \eqref{firstline} and \eqref{secondline}) are
\begin{equation}
\mathcal{M}^{(\pm 1, \vec{0})}\star \mathcal{M}^{(\mp 1, \vec{0})} = \overline{M^{(1, -1, \vec{0})}} + \frac{(-1)^{N_f + N_c - 1}}{r^{N_f - 2N_c + 2}}\sum_{i=1}^{N_c} \frac{(ir\Phi_i\pm 1/2)^{N_f}}{\prod_{j\neq i} ir\Phi_{ij}(ir\Phi_{ij}\pm 1)},
\label{uncwithnfstarproducts}
\end{equation}
and we work in a basis where $\mathcal{M}^{(1, -1, \vec{0})} = \mathcal{M}^{(-1, \vec{0})}\star \mathcal{M}^{(1, \vec{0})}$.  In the $SU(N_c)$ theory with $N_f$ fundamentals, the operator $\mathcal{M}^{(1, -1, \vec{0})}$ is then given by the RHS of \eqref{uncwithnfstarproducts} with the bottom sign, after imposing $\sum_{i=1}^{N_c} \Phi_i = 0$.\footnote{When possible, we construct the generators for the $SU(N_c)$ chiral ring as products of dressed $U(N_c)$ minuscule monopoles in such a way as to agree with the bubbling coefficients produced by the IOT algorithm described in Appendix \ref{appen:Zmono}.  This choice of operator basis simply facilitates comparison with the results of \cite{Ito:2011ea}; it is no more privileged than those used in the main text.

For $SU$ theories of low rank, the bubbling coefficients constructed in this way are all consistent with those derived in the main text using polynomiality.  For example, setting $N_c = 2$ and $\Phi_1 = -\Phi_2 = \Phi$ gives
\begin{equation}
\mathcal{M}^{(1, -1)} = \overline{M^{(1, -1)}} + \frac{1}{2^{N_f}r^{N_f - 2}}\left[\frac{(1 - 2ir\Phi)^{N_f - 1} - (1 + 2ir\Phi)^{N_f - 1}}{2ir\Phi}\right].
\end{equation}
The bubbling coefficient in square brackets is a polynomial in $\Phi$ when $N_f\geq 1$, as it must be according to Section \ref{sec:rankone}.}

\subsubsection{$SU(2)$ with $N_f\geq 0$ Revisited}

We first revisit $SU(2)$ with $N_f$ fundamental flavors to demonstrate how to reproduce the results of Section \ref{sec:Appl} for the chiral ring and its quantization using the trick of gauging $U(1)_\text{top}$ (compare to the analysis in Appendix A of \cite{Assel:2018exy}).  The chiral ring relation can be written as
\begin{align}
x^2 + zy^2 + z^{N_f - 1} &= 0 \quad (N_f > 0), \nonumber \\
x^2 + zy^2 + y &= 0 \quad (N_f = 0).
\end{align}
Accounting for monopole bubbling is crucial to obtaining the modified relation for $N_f = 0$.  In general, we must also account for operator mixing relative to the na\"ive identifications
\begin{equation}
x\sim \Phi\mathcal{M}^{(1, -1)}, \quad y\sim \mathcal{M}^{(1, -1)}, \quad z\sim \tr\Phi^2
\end{equation}
of the generators with the dressed monopole, bare monopole, and Casimir invariant.

We start with the operator $\tr\Phi^2$ and the minuscule monopoles $\mathcal{M}^{(1, 0)}$ and $\mathcal{M}^{(0, -1)}$ in $U(2)$ with $N_f$ flavors.  From them, we construct
\begin{equation}
\mathcal{M}^{(1, -1)} = \mathcal{M}^{(0, -1)}\star \mathcal{M}^{(1, 0)}, \quad \Phi_1\mathcal{M}^{(1, -1)} = \Phi_1\mathcal{M}^{(0, -1)}\star \mathcal{M}^{(1, 0)},
\end{equation}
where $\Delta_{(1, -1)} = N_f - 2$.  We reduce to $SU(2)$ by setting $\Phi_1 = -\Phi_2 = \Phi$.  We then compute using the corresponding shift operators that for all $N_f$,
\begin{equation}
(\Phi\mathcal{M}^{(1, -1)})^2 - \frac{1}{2}\mathcal{M}^{(1, -1)}\star \tr\Phi^2\star \mathcal{M}^{(1, -1)} - \frac{i}{r}\mathcal{M}^{(1, -1)}\star \Phi\mathcal{M}^{(1, -1)} = P(\Phi)\mathcal{M}^{(1, -1)}
\label{quantizedrelation}
\end{equation}
where $P(\Phi)\mathcal{M}^{(1, -1)} = P(\Phi_1)\mathcal{M}^{(0, -1)}\star \mathcal{M}^{(1, 0)}|_{\Phi_1 = \Phi}$ and $P(\Phi)$ is a polynomial of degree $N_f$:
\begin{equation}
P(\Phi) = \left(-\frac{1}{2}\right)^{N_f + 1}\frac{(1 - (-1)^{N_f})(2ir\Phi + 1)^{N_f - 1} + (2ir\Phi + 1)^{N_f} + (2ir\Phi - 1)^{N_f}}{r^{N_f}}.
\end{equation}
To deduce the quantized chiral ring relation from \eqref{quantizedrelation}, we must write $P(\Phi)\mathcal{M}^{(1, -1)}$ in terms of generators.  Let us see how to do so explicitly for $N_f = 0, 1, 2, 3$ (all values for which the theory is bad and the minimum for which it is good):
\begin{itemize}
\item When $N_f = 0$, we find that
\begin{equation}
P(\Phi)\mathcal{M}^{(1, -1)} = -\mathcal{M}^{(1, -1)}.
\end{equation}
In the chiral ring, we have $x^2 + zy^2 + y = 0$ with
\begin{equation}
x = \Phi\mathcal{M}^{(1, -1)}, \quad y = \mathcal{M}^{(1, -1)}, \quad z = -\frac{1}{2}(\tr\Phi^2).
\end{equation}
\item When $N_f = 1$, we find that
\begin{equation}
P(\Phi)\mathcal{M}^{(1, -1)} = i\Phi\mathcal{M}^{(1, -1)} + \frac{1}{2r}\mathcal{M}^{(1, -1)}.
\end{equation}
In the chiral ring, we have $x^2 + zy^2 + 1 = 0$ with
\begin{equation}
x = 2\Phi\mathcal{M}^{(1, -1)} - i, \quad y = \mathcal{M}^{(1, -1)}, \quad z = -2\tr\Phi^2.
\end{equation}
\item When $N_f = 2$, we find that
\begin{equation}
P(\Phi)\mathcal{M}^{(1, -1)} = \frac{1}{2}\tr\Phi^2\star \mathcal{M}^{(1, -1)} - \frac{1}{4r^2}\mathcal{M}^{(1, -1)}.
\end{equation}
In the chiral ring, we have $x^2 + zy^2 + z = 0$ with
\begin{equation}
x = \Phi\mathcal{M}^{(1, -1)}, \quad y = i(2\mathcal{M}^{(1, -1)} + 1), \quad z = \frac{1}{8}\tr\Phi^2.
\end{equation}
\item When $N_f = 3$, we find that
\begin{align}
&P(\Phi)\mathcal{M}^{(1, -1)} + \frac{i}{2}\tr\Phi^2\star \Phi\mathcal{M}^{(1, -1)} \nonumber \\
&\hspace{1 cm} + \frac{1}{4r}\tr\Phi^2\star \mathcal{M}^{(1, -1)} - \frac{5i}{4r^2}\Phi\mathcal{M}^{(1, -1)} - \frac{1}{8r^3}\mathcal{M}^{(1, -1)} = 0.
\end{align}
In the chiral ring, we have $x^2 + zy^2 + z^2 = 0$ with
\begin{equation}
x = \Phi\mathcal{M}^{(1, -1)} + \frac{i}{4}\tr\Phi^2, \quad y = \sqrt{2}i\mathcal{M}^{(1, -1)}, \quad z = \frac{1}{4}\tr\Phi^2.
\end{equation}
\end{itemize}

\subsubsection{$SU(3)$ with $N_f\geq 0$}

Let us describe how to determine the chiral ring of the $SU(3)$ gauge theory with $N_f \geq 0$ fundamental flavors.  From \eqref{goodboundSU}, we see that this theory is good for $N_f\geq 5$.  For comparison, the $U(3)$ theory with the same matter content is ugly for $N_f = 5$ and good for $N_f > 5$.

A set of generators for the Coulomb branch chiral ring of the $SU(3)$ theory has been identified in \cite{Hanany:2016ezz}.  Choosing convenient Weyl orderings of the GNO charges, it consists of the (dressed) monopoles
\begin{align}
\Delta &= N_f - 4: & \mathcal{M}^{(1, 0, -1)} &= \mathcal{M}^{(0, 0, -1)}\star \mathcal{M}^{(1, 0, 0)}, \label{straightforwardfirst} \\
\Delta &= 2N_f - 6: & \mathcal{M}^{(2, -1, -1)} &= \mathcal{M}^{(0, -1, -1)}\star (\mathcal{M}^{(1, 0, 0)})^2, \label{straightforwardsecond} \\
\Delta &= 2N_f - 6: & \mathcal{M}^{(1, 1, -2)} &= (\mathcal{M}^{(0, 0, -1)})^2\star \mathcal{M}^{(1, 1, 0)}, \\
\Delta &= N_f - 1: & (\Phi_1^3 + \Phi_2^3)\mathcal{M}^{(1, 0, -1)} &= (\Phi_1^3 + \Phi_2^3)\mathcal{M}^{(0, 0, -1)}\star \mathcal{M}^{(1, 0, 0)}, \\
\Delta &= 2N_f - 5: & (\Phi_1 + \Phi_2)\mathcal{M}^{(-1, -1, 2)} &= (\Phi_1 + \Phi_2)\mathcal{M}^{(-1, -1, 0)}\star (\mathcal{M}^{(1, 0, 0)})^2, \\
\Delta &= 2N_f - 4: & (\Phi_1^2 + \Phi_2^2)\mathcal{M}^{(-1, -1, 2)} &= (\Phi_1^2 + \Phi_2^2)\mathcal{M}^{(-1, -1, 0)}\star (\mathcal{M}^{(1, 0, 0)})^2, \\
\Delta &= 2N_f - 5: & (\Phi_1 + \Phi_2)\mathcal{M}^{(1, 1, -2)} &= (\Phi_1 + \Phi_2)\mathcal{M}^{(0, 0, -1)}\star \mathcal{M}^{(0, 0, -1)}\star \mathcal{M}^{(1, 1, 0)}, \\
\Delta &= 2N_f - 4: & (\Phi_1^2 + \Phi_2^2)\mathcal{M}^{(1, 1, -2)} &= (\Phi_1^2 + \Phi_2^2)\mathcal{M}^{(0, 0, -1)}\star \mathcal{M}^{(0, 0, -1)}\star \mathcal{M}^{(1, 1, 0)}, \label{straightforwardlast}
\end{align}
which are straightforward to write in terms of non-bubbling monopoles of $U(3)$ (as we have done above), the dressed monopoles
\begin{equation}
\Phi_1\mathcal{M}^{(1, 0, -1)}, \quad \Phi_2\mathcal{M}^{(1, 0, -1)}, \quad \Phi_1^2\mathcal{M}^{(1, 0, -1)}, \quad \Phi_2^2\mathcal{M}^{(1, 0, -1)},
\label{lessstraightforward}
\end{equation}
which are less straightforward to construct from elementary dressed monopoles, and the scalars $\tr\Phi^2$ and $\tr\Phi^3$, where we write $\Phi = \operatorname{diag}(\Phi_1, \Phi_2, -(\Phi_1 + \Phi_2))$ in the reduction to $SU(3)$.\footnote{To obtain this generating set, we consider the $N_R = 0$ case of the results in \cite{Hanany:2016ezz}: (8.17) yields the three bare monopole operators, while (8.25), (8.26), (8.27), (8.28), (8.29) yield their dressed versions; $\tr\Phi^2$ and $\tr\Phi^3$ are the relevant Casimir invariants after passing from $U(3)$ to $SU(3)$ (8.20).  See Section 8.4.2 for the Hilbert series.}  Constructing the explicit dressings in \eqref{lessstraightforward} is a delicate procedure, so we elect to use the generators
\begin{align}
\Delta &= N_f - 3: & g_1^{(1, 0, -1)}\equiv \mathcal{M}^{(0, 0, -1)}\star \Phi_1\mathcal{M}^{(1, 0, 0)} &= \Phi_1\mathcal{M}^{(1, 0, -1)} + P_{N_f - 3}(\Phi_1, \Phi_2), \label{ratherfirst} \\
\Delta &= N_f - 2: & g_2^{(1, 0, -1)}\equiv \mathcal{M}^{(0, 0, -1)}\star \Phi_1^2\mathcal{M}^{(1, 0, 0)} &= \Phi_1^2\mathcal{M}^{(1, 0, -1)} + P_{N_f - 2}(\Phi_1, \Phi_2), \\
\Delta &= N_f - 3: & (\Phi_1 + \Phi_2)\mathcal{M}^{(1, 0, -1)} &= (\Phi_1 + \Phi_2)\mathcal{M}^{(0, 0, -1)}\star \mathcal{M}^{(1, 0, 0)}, \\
\Delta &= N_f - 2: & (\Phi_1^2 + \Phi_2^2)\mathcal{M}^{(1, 0, -1)} &= (\Phi_1^2 + \Phi_2^2)\mathcal{M}^{(0, 0, -1)}\star \mathcal{M}^{(1, 0, 0)} \label{ratherlast}
\end{align}
in their stead: along with the Casimir invariants, they clearly generate \eqref{lessstraightforward}.  The degrees $d$ of the unspecified polynomials $P_d$ in \eqref{ratherfirst}--\eqref{ratherlast} follow from the dimension formula \eqref{dimensionformulaUandSU}.  This basis differs slightly from that of \cite{Hanany:2016ezz} and has the benefit that the corresponding shift operators are easier to construct.

To make sense of the formulas above, first recall that in defining dressed monopoles, the Weyl group actions on the GNO charges of the abelian summands and on the dressing factors are opposite to each other.  For example, writing
\begin{equation}
\widetilde{M}^{(1, 0, -1)} = M^{(1, 0, -1)} + Z_{(1, 0, -1)\to (0, 0, 0)}^\text{ab}(\Phi_1, \Phi_2, \Phi_3),
\label{10m1example}
\end{equation}
we have
\begin{align}
P(\Phi_1, \Phi_2, \Phi_3)\mathcal{M}^{(1, 0, -1)} &= P(\Phi_1, \Phi_2, \Phi_3)\widetilde{M}^{(1, 0, -1)} + P(\Phi_1, \Phi_3, \Phi_2)\widetilde{M}^{(1, -1, 0)} \nonumber \\
&+ P(\Phi_2, \Phi_3, \Phi_1)\widetilde{M}^{(-1, 1, 0)} + P(\Phi_2, \Phi_1, \Phi_3)\widetilde{M}^{(0, 1, -1)} \nonumber \\
&+ P(\Phi_3, \Phi_1, \Phi_2)\widetilde{M}^{(0, -1, 1)} + P(\Phi_3, \Phi_2, \Phi_1)\widetilde{M}^{(-1, 0, 1)}, 
\end{align} 
where $\mathcal{W} = S_3$ in our case.  A given dressed monopole can also be written in a number of different ways, depending on how the components of the GNO charges are Weyl-ordered: for example, $P(\Phi_1)\mathcal{M}^{(1, 0, 0)}$, $P(\Phi_2)\mathcal{M}^{(0, 1, 0)}$, and $P(\Phi_3)\mathcal{M}^{(0, 0, 1)}$ are all equivalent.  Starting with \eqref{straightforwardfirst}, we can construct $\mathcal{M}^{(1, 0, -1)}$ dressed by any polynomial that is symmetric in the first two arguments by writing
\begin{equation}
P(\Phi_1, \Phi_2, \Phi_3)\mathcal{M}^{(0, 0, -1)}\star \mathcal{M}^{(1, 0, 0)} = \frac{P(\Phi_1, \Phi_2, \Phi_3) + P(\Phi_2, \Phi_1, \Phi_3)}{2}\mathcal{M}^{(1, 0, -1)}.
\end{equation}
Similarly, for $P$ symmetric in the last two arguments, the dressing of $\mathcal{M}^{(1, 0, -1)}$ by $P$ is given by the leading term of
\begin{equation}
\mathcal{M}^{(0, 0, -1)}\star P(\Phi_1, \Phi_2, \Phi_3)\mathcal{M}^{(1, 0, 0)}
\end{equation}
(as $P$ is now acted on by the shift operators in $\mathcal{M}^{(0, 0, -1)}$).  Starting with \eqref{straightforwardsecond}, we find that
\begin{equation}
P(\Phi_1, \Phi_2, \Phi_3)\mathcal{M}^{(0, -1, -1)}\star (\mathcal{M}^{(1, 0, 0)})^2 = \frac{P(\Phi_1, \Phi_2, \Phi_3) + P(\Phi_1, \Phi_3, \Phi_2)}{2}\mathcal{M}^{(2, -1, -1)}.
\end{equation}
Unlike in the case of $\mathcal{M}^{(1, 0, -1)}$, this procedure for constructing dressings of $\mathcal{M}^{(2, -1, -1)}$ is completely general because $\mathcal{M}^{(2, -1, -1)}$, like $\mathcal{M}^{(0, -1, -1)}$, is sensitive only to the part of the dressing polynomial that is Weyl-symmetric in the last two arguments.  Similar statements hold for $\mathcal{M}^{(1, 1, -2)}$.  Note that in this theory, all non-bubbling monopoles have nontrivial stabilizers under the Weyl group.  In particular, dressing the minuscule monopoles in \eqref{straightforwardfirst} individually is not enough to extract the abelianized bubbling coefficient $Z_{(1, 0, -1)\to (0, 0, 0)}^\text{ab}$ in \eqref{10m1example}.  However, by dressing both $\mathcal{M}^{(0, 0, -1)}$ and $\mathcal{M}^{(1, 0, 0)}$ at the same time, one can in principle extract $Z_{(1, 0, -1)\to (0, 0, 0)}^\text{ab}$ itself (we will not need to do so).

Moving on to the chiral ring, we have listed 14 generators (\eqref{straightforwardfirst}--\eqref{straightforwardlast}, \eqref{ratherfirst}--\eqref{ratherlast}, and the Casimir invariants), while the Coulomb branch has complex dimension 4.  Hence we must find at least 10 relations.  If the moduli space is not a complete intersection (meaning that the relations could be redundant at generic points but may all be needed to describe the whole variety), then we will find strictly more than 10 relations.  In this case, we cannot read off the degrees of the relations and generators from the Hilbert series.\footnote{Indeed, from (8.38) of \cite{Hanany:2016ezz}, although we see that the difference between the degrees of the denominator and the numerator of the Hilbert series equals the dimension of the moduli space in this case, the degrees appearing in the denominator only include the dimensions of $\tr\Phi^2$, $\tr\Phi^3$, and the bare monopoles $(1, 0, -1)$, $(1, 1, -2)$, $(2, -1, -1)$; the dimensions of the dressed monopoles are missing.}  By contrast, the moduli space of $U$ with fundamental hypers is known to be a complete intersection \cite{Cremonesi:2013lqa}.  To proceed, one can write all possible relations according to dimension and solve for the coefficients.  Rather than presenting an exhaustive analysis, let us simply determine the lowest-dimension relation that relates operators of different GNO charges.  Clearly, this relation has $\Delta = 2N_f - 6$.  As a further simplification, we work with the commutative limit of the shift operators (in which the order of the multiplications in \eqref{straightforwardfirst}--\eqref{straightforwardlast} and \eqref{ratherfirst}--\eqref{ratherlast} is immaterial), as our interest is in the chiral ring and not its quantization.  Then we find that
\begin{align}
0 &= \mathcal{M}^{(2, -1, -1)} + \mathcal{M}^{(1, 1, -2)} \nonumber \\
&+ (g_1^{(1, 0, -1)})^2 - g_1^{(1, 0, -1)}[(\Phi_1 + \Phi_2)\mathcal{M}^{(1, 0, -1)}] + [(\Phi_1 + \Phi_2)\mathcal{M}^{(1, 0, -1)}]^2 \\
&- \frac{1}{2}\tr\Phi^2(\mathcal{M}^{(1, 0, -1)})^2 + P(\Phi_1, \Phi_2)\mathcal{M}^{(1, 0, -1)} \nonumber
\end{align}
in the chiral ring, where $P$ is a polynomial of degree $N_f - 2$, expressible in terms of the Casimirs, which we do not write explicitly (note that $P$ necessarily vanishes for $N_f = 0, 1$).

\section{Correlation Functions} \label{mirrorappendix}

\subsection{Mirror Symmetry Check for an $\cN = 8$ SCFT}

In this section, we give more details on the derivation of the mirror maps~\eqref{mirrormap1}--\eqref{mirrormapUN} in the main text.

When discussing an $\cN = 8$ SCFT in $\cN = 4$ language, it is useful to embed the $\mathcal{N} = 4$ superconformal algebra $\mathfrak{osp}(4|4)$ into the $\mathcal{N} = 8$ superconformal algebra $\mathfrak{osp}(8|4)$.  To this end, we have\footnote{The subscripts on the right are also called $H, C, F, F'$, respectively, in \cite{Agmon:2017lga}.}
\begin{equation}
\mathfrak{so}(8)_R\supset \mathfrak{su}(2)_L\oplus \mathfrak{su}(2)_R\oplus \mathfrak{su}(2)_1\oplus \mathfrak{su}(2)_2,
\end{equation}
where $\mathfrak{su}(2)_L\oplus \mathfrak{su}(2)_R$ is the $\cN = 4$ R-symmetry algebra and $ \mathfrak{su}(2)_1\oplus \mathfrak{su}(2)_2$ is a flavor symmetry from the $\cN = 4$ point of view \cite{Chester:2014mea}.  For 3D $\mathcal{N} = 8$ theories, the 1D topological theory has a global $\mathfrak{su}(2)_F$ symmetry \cite{Chester:2014mea}, which can be identified with $\mathfrak{su}(2)_1$ for the Higgs branch TQFT and $\mathfrak{su}(2)_2$ for the Coulomb branch TQFT\@.  To write correlation functions concisely, it is convenient to organize operators in the 1D theory into representations of this $\mathfrak{su}(2)_F$ symmetry and to contract their $\mathfrak{su}(2)_F$ indices with commuting $\mathfrak{su}(2)_F$ polarization vectors $z^i$ ($i = 1, 2$) transforming under $\mathfrak{su}(2)_F$ as a doublet.  For an operator $\cO_{i_1\cdots i_{2j}} (\vphi)$ in the spin-$j$ representation of $\mathfrak{su}(2)_F$, we define
 \es{ODef}{
  \cO(\vphi, z) \equiv \cO_{i_1\cdots i_{2j}}(\vphi) z^{i_1} \cdots z^{i_{2j}} \,.
 }
We thus label operators in the 1D theory by $\mathcal{O}^{(\Delta, j)}(\varphi, z)$ and subscripts $H, C$ depending on whether they belong to the Higgs or Coulomb branch TQFT, respectively.  The label $\Delta$ corresponds to the scaling dimension of the 3D operator from which $\cO^{(\Delta, j)}$ originates, and $j$ is the $\mathfrak{su}(2)_F$ spin.  We further normalize the two-point functions as
 \es{TwoPoint}{
\langle\mathcal{O}^{(\Delta, j)}(\varphi_1, z_1)\mathcal{O}^{(\Delta, j)}(\varphi_2, z_2)\rangle = \langle z_1, z_2\rangle^{2j}(\operatorname{sign}\varphi_{12})^{2\Delta} \,,
 }
with all other two-point functions vanishing, where we have defined the $\mathfrak{su}(2)_F$ invariant
\begin{equation}
\langle z_A, z_B\rangle\equiv \epsilon_{ij}z_A^i z_B^j\,, \qquad \epsilon^{12} = -\epsilon_{12} = 1,
\end{equation}
denoting the $\mathfrak{su}(2)_F$ singlet that can be formed from two polarizations $z_A$ and $z_B$.  With the normalization \eqref{TwoPoint}, three-point functions are fixed by the $\mathfrak{su}(2)_F$ symmetry as follows: for spins $j_1, j_2, j_3$ satisfying the triangle inequality,
\begin{align}
\langle\mathcal{O}^{(\Delta_1, j_1)}&(\varphi_1, z_1)\mathcal{O}^{(\Delta_2, j_2)}(\varphi_2, z_2)\mathcal{O}^{(\Delta_3, j_3)}(\varphi_3, z_3)\rangle = \lambda_{(\Delta_1, j_1), (\Delta_2, j_2), (\Delta_3, j_3)} \nonumber \\
&\times \langle z_1, z_2\rangle^{j_{123}}\langle z_2, z_3\rangle^{j_{231}}\langle z_3, z_1\rangle^{j_{312}}(\operatorname{sign}\varphi_{12})^{\Delta_{123}}(\operatorname{sign}\varphi_{23})^{\Delta_{231}}(\operatorname{sign}\varphi_{31})^{\Delta_{312}} \,,
\end{align}
where $j_{abc}\equiv j_a + j_b - j_c$ (the correlator vanishes otherwise).  The sign factors are fixed by conformal symmetry, while $\mathfrak{su}(2)_F$ symmetry requires that the polarizations appear as they do by power counting.

In the following, we specialize to the IR limit of $U(N_c)$ SQCD with $N_a = N_f = 1$.

\subsubsection{$U(2)$ with $N_a = N_f = 1$}

The 1D Higgs branch theory in the case $N_c = 2$ was partially analyzed in \cite{Dedushenko:2016jxl}; our discussion here is self-contained.  On the Higgs branch side, there is a single $j = 1/2$ operator
\begin{equation}
\widetilde{\mathcal{O}}_H^{(1/2, 1/2)}(\varphi, z) = z^1\tr Q(\varphi) + z^2\tr\tilde Q(\varphi)
\end{equation}
and two distinct operators with $j = 1$:\footnote{Since the natural index structure is $Q_i{}^j$ and $\tilde{Q}^i{}_j$, when we write $\tr Q\tilde{Q}$, we really mean $\tr Q\tilde{Q}^T$.}
\begin{align}
\widetilde{\mathcal{O}}_{H, 1}^{(1, 1)}(\varphi, z) &= (z^1)^2(\tr Q)^2(\varphi) + (z^2)^2(\tr\tilde Q)^2(\varphi) + 2z^1 z^2\tr Q\tr\tilde Q(\varphi), \nonumber \\
\widetilde{\mathcal{O}}_{H, 2}^{(1, 1)}(\varphi, z) &= (z^1)^2\tr Q^2(\varphi) + (z^2)^2\tr\tilde Q^2(\varphi) + 2z^1 z^2\tr Q\tilde Q(\varphi).
\end{align}
The tildes indicate that these operators are unnormalized and do not necessarily obey \eqref{TwoPoint}.  To find which linear combinations obey \eqref{TwoPoint}, we compute their two-point functions using \eqref{CorrFnHiggs}.  Performing Wick contractions in the theory at fixed $\sigma$ using \eqref{Wick}, we obtain
\begin{align}
\langle\tr Q(\varphi_1)\tr\tilde Q(\varphi_2)\rangle_\sigma &= -\frac{\operatorname{sign}\varphi_{12}}{4\pi r}, \nonumber \\
\langle(\tr Q)^2(\varphi_1)(\tr\tilde Q)^2(\varphi_2)\rangle_\sigma &= -2\langle\tr Q\tr\tilde Q(\varphi_1)\tr Q\tr\tilde Q(\varphi_2)\rangle_\sigma = \frac{1}{8\pi^2 r^2}, \nonumber \\
\langle\tr Q^2(\varphi_1)\tr\tilde Q^2(\varphi_2)\rangle_\sigma &= -2\langle\tr Q\tilde Q(\varphi_1)\tr Q\tilde Q(\varphi_2)\rangle_\sigma = \frac{1}{16\pi^2 r^2}\left[1 + \frac{1}{\cosh^2(\pi\sigma_{12})}\right], \\
\langle(\tr Q)^2(\varphi_1)\tr\tilde Q^2(\varphi_2)\rangle_\sigma &= \langle\tr Q^2(\varphi_1)(\tr\tilde Q)^2(\varphi_2)\rangle_\sigma = -2\langle\tr Q\tr\tilde Q(\varphi_1)\tr Q\tilde Q(\varphi_2)\rangle_\sigma \nonumber \\
&= \frac{1}{16\pi^2 r^2}. \nonumber
\end{align}
Substituting these expressions into \eqref{CorrFnHiggs},\footnote{Some useful integrals are as follows.  The partition function is
\begin{equation}
Z_{S^3} = \frac{1}{32}\int d\sigma_1\, d\sigma_2\, \frac{\sinh^2(\pi\sigma_{12})}{\cosh^2(\pi\sigma_{12})\cosh(\pi\sigma_1)\cosh(\pi\sigma_2)} = \frac{1}{16\pi}. \label{zs3}
\end{equation}
Denoting $Z_{S^3}$ with an extra insertion of $f(\sigma_1, \sigma_2)$ in the integrand by $Z_{S^3}[f(\sigma_1, \sigma_2)]$, we have
\begin{equation}
Z_{S^3}[\sigma_1\sigma_2] = -\frac{1}{384\pi}, \quad Z_{S^3}[\sigma_1^2 + \sigma_2^2] = \frac{1}{24\pi}, \quad Z_{S^3}\left[\frac{1}{\cosh^2(\pi\sigma_{12})}\right] = \frac{1}{96\pi}. \label{integrals}
\end{equation}
In particular, if $\langle\rangle_\sigma$ is a constant, then $\langle\rangle = \langle\rangle_\sigma$, while
\begin{equation}
\langle\tr Q^2(\varphi_1)\tr\tilde{Q}^2(\varphi_2)\rangle = -2\langle\tr Q\tilde{Q}(\varphi_1)\tr Q\tilde{Q}(\varphi_2)\rangle = \frac{7}{96\pi^2 r^2}. \label{twoptx2withy2}
\end{equation}
\label{FootnoteZ}} we obtain that the orthonormalized operators
\begin{align}
\mathcal{O}_H^{(1/2, 1/2)}(\varphi, z) &= \sqrt{4\pi r}\widetilde{\mathcal{O}}_{H, 1}^{(1, 1)}(\varphi, z), \vphantom{\frac{}{1}} \nonumber \\
\mathcal{O}_{H, \text{free}}^{(1, 1)}(\varphi, z) &= \sqrt{8\pi^2 r^2}\widetilde{\mathcal{O}}_{H, 1}^{(1, 1)}(\varphi, z), \label{OHNormalized} \\
\mathcal{O}_{H, \text{int}}^{(1, 1)}(\varphi, z) &= \sqrt{24\pi^2 r^2}\left[\widetilde{\mathcal{O}}_{H, 2}^{(1, 1)}(\varphi, z) - \frac{1}{2}\widetilde{\mathcal{O}}_{H, 1}^{(1, 1)}(\varphi, z)\right] \nonumber
\end{align}
obey \eqref{TwoPoint}.  The subscripts ``free'' and ``int'' indicate that this theory flows to the product of a free sector and an interacting sector.  Similarly, the nontrivial three-point functions with $(j_1, j_2, j_3) = (1/2, 1/2, 1)$ follow from
\begin{align}
\langle\tr Q(\varphi_1)\tr Q(\varphi_2)(\tr\tilde{Q})^2(\varphi_3)\rangle_\sigma &= \langle\tr\tilde{Q}(\varphi_1)\tr\tilde{Q}(\varphi_2)(\tr Q)^2(\varphi_3)\rangle_\sigma \nonumber \\
&= -2\langle\tr Q(\varphi_1)\tr\tilde{Q}(\varphi_2)\tr Q\tr\tilde{Q}(\varphi_3)\rangle_\sigma \nonumber \\
&= \frac{\operatorname{sign}(\varphi_{13}\varphi_{23})}{8\pi^2 r^2}, \\
\langle\tr Q(\varphi_1)\tr Q(\varphi_2)\tr\tilde{Q}^2(\varphi_3)\rangle_\sigma &= \langle\tr\tilde{Q}(\varphi_1)\tr\tilde{Q}(\varphi_2)\tr Q^2(\varphi_3)\rangle_\sigma \nonumber \\
&= -2\langle\tr Q(\varphi_1)\tr\tilde{Q}(\varphi_2)\tr Q\tilde{Q}(\varphi_3)\rangle_\sigma \nonumber \\
&= \frac{\operatorname{sign}(\varphi_{13}\varphi_{23})}{16\pi^2 r^2},
\end{align}
while those with $(j_1, j_2, j_3) = (1, 1, 1)$ each involve six distinct contractions: for example,
\begin{equation}
\langle\tr Q^2(\varphi_1)\tr\tilde{Q}^2(\varphi_2)\tr Q\tilde{Q}(\varphi_3)\rangle_\sigma = -\frac{\operatorname{sign}(\varphi_{12}\varphi_{23}\varphi_{31})}{64\pi^3 r^3}\left[1 + \frac{1}{\cosh^2(\pi\sigma_{12})}\right].
\end{equation}
In the end, these results can be packaged together in an $\mathfrak{su}(2)_F$-symmetric way as
\begin{align}
\langle\mathcal{O}_H^{(1/2, 1/2)}&(\varphi_1, z_1)\mathcal{O}_H^{(1/2, 1/2)}(\varphi_2, z_2)\mathcal{O}_{H, \text{free}}^{(1, 1)}(\varphi_3, z_3)\rangle \nonumber \\
&= \lambda_{(1/2, 1/2), (1/2, 1/2), (1, 1)}^{\text{free}}\langle z_2, z_3\rangle\langle z_3, z_1\rangle\operatorname{sign}(\varphi_{23}\varphi_{31}), \nonumber \\
\langle\mathcal{O}_{H, \text{free}}^{(1, 1)}&(\varphi_1, z_1)\mathcal{O}_{H, \text{free}}^{(1, 1)}(\varphi_2, z_2)\mathcal{O}_{H, \text{free}}^{(1, 1)}(\varphi_3, z_3)\rangle \nonumber \\
&= \lambda_{(1, 1), (1, 1), (1, 1)}^\text{free}\langle z_1, z_2\rangle\langle z_2, z_3\rangle\langle z_3, z_1\rangle\operatorname{sign}(\varphi_{12}\varphi_{23}\varphi_{31}), \nonumber \\
\langle\mathcal{O}_{H, \text{int}}^{(1, 1)}&(\varphi_1, z_1)\mathcal{O}_{H, \text{int}}^{(1, 1)}(\varphi_2, z_2)\mathcal{O}_{H, \text{int}}^{(1, 1)}(\varphi_3, z_3)\rangle \nonumber \\
&= \lambda_{(1, 1), (1, 1), (1, 1)}^\text{int}\langle z_1, z_2\rangle\langle z_2, z_3\rangle\langle z_3, z_1\rangle\operatorname{sign}(\varphi_{12}\varphi_{23}\varphi_{31}), \label{ThreePoint}
\end{align}
where we have defined the structure constants
\begin{equation}
\lambda_{(1/2, 1/2), (1/2, 1/2), (1, 1)}^{\text{free}} = \sqrt{2}, \quad \lambda_{(1, 1), (1, 1), (1, 1)}^\text{free} = (\sqrt{2})^3, \quad \lambda_{(1, 1), (1, 1), (1, 1)}^\text{int} = \sqrt{6}, \label{structureconstants}
\end{equation}
while
\begin{align}
&\langle\mathcal{O}_H^{(1/2, 1/2)}(\varphi_1, z_1)\mathcal{O}_H^{(1/2, 1/2)}(\varphi_2, z_2)\mathcal{O}_{H, \text{int}}^{(1, 1)}(\varphi_3, z_3)\rangle \nonumber \\
&\phantom{==} = \langle\mathcal{O}_{H, \text{free}}^{(1, 1)}(\varphi_1, z_1)\mathcal{O}_{H, \text{free}}^{(1, 1)}(\varphi_2, z_2)\mathcal{O}_{H, \text{int}}^{(1, 1)}(\varphi_3, z_3)\rangle \nonumber \\
&\phantom{==} = \langle\mathcal{O}_{H, \text{free}}^{(1, 1)}(\varphi_1, z_1)\mathcal{O}_{H, \text{int}}^{(1, 1)}(\varphi_2, z_2)\mathcal{O}_{H, \text{int}}^{(1, 1)}(\varphi_3, z_3)\rangle = 0
\end{align}
because the free and interacting sectors are decoupled.

On the Coulomb branch side, the monopoles $\cM^{(\pm 1, 0)}$ and $\cM^{\pm(1, 1)}$ do not bubble, so their shift operators $\mathcal{M}$ take the form of $M$ in \eqref{shiftAgain} averaged over the $\Z_2$ Weyl group.  The shift operators for bubbling monopoles can be constructed from these by taking star products (i.e., by composition), as described in Appendix \ref{sym2c2relationderivation}: see in particular \eqref{pm10squared} and \eqref{10withm10}.  We then find using \eqref{CBmatrixmodel} that, for instance,\footnote{The symmetry of the integral \eqref{zs3} under $\sigma_{1, 2}\leftrightarrow -\sigma_{1, 2}$ sometimes allows us to write these correlators in terms of $Z_{S^3}$ with simple insertions, such as
\begin{equation}
\langle\mathcal{M}^{(-1, 0)}(\varphi_1)\mathcal{M}^{(1, 0)}(\varphi_2)\rangle = \frac{Z_{S^3}[-\operatorname{sign}\varphi_{12}]}{rZ_{S^3}}, \quad \langle\mathcal{M}^{(-1, -1)}(\varphi_1)\mathcal{M}^{(1, 1)}(\varphi_2)\rangle = \frac{Z_{S^3}[1 - 4\sigma_1\sigma_2]}{4r^2 Z_{S^3}},
\end{equation}
which can be evaluated using \eqref{integrals}.}
\begin{align}
\textstyle \frac{1}{4\pi}\langle\mathcal{M}^{(-1, 0)}(\varphi_1)\mathcal{M}^{(1, 0)}(\varphi_2)\rangle &= \textstyle -\frac{\operatorname{sign}\varphi_{12}}{4\pi r}, \\
\textstyle \frac{1}{4\pi^2}\langle\mathcal{M}^{(-1, -1)}(\varphi_1)\mathcal{M}^{(1, 1)}(\varphi_2)\rangle &= \textstyle -2\left(-\frac{i}{4\pi}\right)^2\langle\tr\Phi(\varphi_1)\tr\Phi(\varphi_2)\rangle = \frac{7}{96\pi^2 r^2}, \\
\textstyle \left(\frac{1}{4\pi}\right)^2\langle\mathcal{M}^{(-2, 0)}(\varphi_1)\mathcal{M}^{(2, 0)}(\varphi_2)\rangle &= \textstyle -2\left(\frac{1}{4\pi}\right)^2\left\langle\left(\mathcal{M}^{(1, -1)} - \frac{1}{r}\right)(\varphi_1)\left(\mathcal{M}^{(1, -1)} - \frac{1}{r}\right)(\varphi_2)\right\rangle \nonumber \\
&= \textstyle \frac{1}{8\pi^2 r^2}, \\
\textstyle \left(\frac{1}{4\pi}\right)\left(\frac{1}{2\pi}\right)\langle\mathcal{M}^{(-2, 0)}(\varphi_1)\mathcal{M}^{(1, 1)}(\varphi_2)\rangle &= \textstyle \left(\frac{1}{2\pi}\right)\left(\frac{1}{4\pi}\right)\langle\mathcal{M}^{(-1, -1)}(\varphi_1)\mathcal{M}^{(2, 0)}(\varphi_2)\rangle \nonumber \\
&= \textstyle -2\left(\frac{1}{4\pi}\right)\left(-\frac{i}{4\pi}\right)\left\langle\left(\mathcal{M}^{(1, -1)} - \frac{1}{r}\right)(\varphi_1)\tr\Phi(\varphi_2)\right\rangle \nonumber \\
&= \textstyle \frac{1}{16\pi^2 r^2}.
\end{align}
It is straightforward to check that these correlation functions, as well as the various three-point functions, agree precisely with those of the 1D Higgs branch operators given in \eqref{mirrormap1}--\eqref{mirrormap5}, as extracted from the two-point functions of \eqref{OHNormalized} and the three-point functions \eqref{ThreePoint}.  This provides a derivation of \eqref{mirrormap1}--\eqref{mirrormap5}.

Here is an alternative method of deriving the mirror map.  Given the relations \eqref{pm10squared}--\eqref{complicatedrelation}, we may view the ``basic'' operators on the Coulomb branch side as $\mathcal{M}^{(\pm 1, 0)}$ and $\mathcal{M}^{\pm(1, 1)}$, for which we have already justified the mirror map in \eqref{mirrormap1} and \eqref{mirrormap3}.  The mirror map \eqref{mirrormap2} for $\mathcal{M}^{(\pm 2, 0)}$ then follows from $\tr Q\star \tr Q = (\tr Q)^2$ and \eqref{pm10squared}.  Next, using Wick contractions to define composite operators, we compute on the Higgs branch side that
\begin{equation}
\tr Q\star \tr\tilde Q + \frac{N_c}{8\pi r} = \tr\tilde Q\star \tr Q - \frac{N_c}{8\pi r} = \tr Q\tr\tilde Q
\label{compositetrxtry}
\end{equation}
for arbitrary $N_c$, which, in light of \eqref{10withm10}, is consistent with the identification \eqref{mirrormap4} for $\mathcal{M}^{(1, -1)}$ when $N_c = 2$.  Finally, on the Higgs branch side, we find that
\begin{equation}
\tr Q^2\star \tr\tilde Q^2 + \frac{1}{2\pi r}\tr Q\tilde Q = \tr\tilde Q^2\star \tr Q^2 - \frac{1}{2\pi r}\tr Q\tilde Q = \tr Q^2\tr\tilde Q^2 + \frac{1}{8\pi^2 r^2}.
\label{compositetrx2try2}
\end{equation}
Using \eqref{pm11withmp11} and \eqref{mirrormap3}, we deduce the mirror map \eqref{mirrormap5} for $\tr\Phi$ as well as
\begin{equation}
\frac{1}{8\pi^2}\left[\tr\Phi^2 - (\tr\Phi)^2 - \frac{1}{2r^2}\right]\leftrightarrow \tr Q^2\tr\tilde Q^2.
\label{trphi}
\end{equation}
Using \eqref{trphi} and the mirror map \eqref{mirrormap5}, we can further identify what $\tr\Phi^2$ corresponds to: on the Higgs branch side, we compute that
\begin{equation}
\tr Q\tilde Q\star \tr Q\tilde Q = (\tr Q\tilde Q)^2 - \frac{1}{16\pi^2 r^2},
\label{compositetrxysquared}
\end{equation}
so in light of $\tr\Phi\star \tr\Phi = (\tr\Phi)^2$ and \eqref{trphi}, we get that
\begin{equation}
\frac{1}{8\pi^2}\left(\tr\Phi^2 - \frac{3}{2r^2}\right)\leftrightarrow \tr Q^2\tr\tilde Q^2 - 2(\tr Q\tilde Q)^2.
\label{idtrphisquared}
\end{equation}
One can make further consistency checks of the identifications that we have derived by matching one-point functions of these composite operators.  As a consistency check of \eqref{mirrormap4}, we see from \eqref{compositetrxtry} that $\langle\tr Q\tr\tilde Q\rangle = 0$ for any $N_c$, so we expect that $\smash{\langle\mathcal{M}^{(1, -1)}\rangle} = \frac{1}{r}$, which is indeed the case.  As a consistency check of \eqref{mirrormap5}, we have that $\langle\tr Q\tilde Q\rangle = 0$, which is consistent with
\begin{equation}
\langle\tr\Phi\rangle = 0, \quad \langle\tr\Phi^2\rangle = \frac{Z_{S^3}[\sigma_1^2 + \sigma_2^2]}{r^2 Z_{S^3}} = \frac{2}{3r^2}, \quad \langle(\tr\Phi)^2\rangle = \frac{Z_{S^3}[(\sigma_1 + \sigma_2)^2]}{r^2 Z_{S^3}} = \frac{7}{12r^2}
\label{phistuff}
\end{equation}
(as follows from \eqref{integrals}), where the notation $Z_{S^3}[f(\sigma_1, \sigma_2)]$ is the same as in Footnote~\ref{FootnoteZ}.  As a consistency check of \eqref{idtrphisquared}, we may use \eqref{compositetrx2try2} and \eqref{compositetrxysquared} to rewrite \eqref{idtrphisquared} in terms of star products of elementary operators:
\begin{equation}
\frac{1}{8\pi^2}\left(\tr\Phi^2 + \frac{1}{2r^2}\right) = \tr Q^2\star \tr\tilde Q^2 - 2\tr Q\tilde Q\star \tr Q\tilde Q + \frac{1}{2\pi r}\tr Q\tilde Q.
\label{wanttocheck}
\end{equation}
Taking the expectation value of both sides of \eqref{wanttocheck} and using \eqref{twoptx2withy2} and $\langle\tr Q\tilde{Q}\rangle = 0$ results in $\langle\tr\Phi^2\rangle = 2/3r^2$, precisely as expected from \eqref{phistuff}.

\subsubsection{$U(N_c)$ with $N_a = N_f = 1$}

While we do not study the case $N_c > 2$ in detail, it is straightforward to match correlation functions in the free sector on the Higgs and Coulomb branch sides.  The free sector of $U(N_c)$ with $N_a = N_f = 1$ can be analyzed in the same way for all $N_c$ (compare to the analysis of $U(3)$ with $N_a = N_f = 1$ in \cite{Agmon:2017lga}).  First consider the Higgs branch side.  Letting tildes denote unnormalized operators, we set
\begin{equation}
\widetilde{\mathcal{O}}_{H, \text{free}}^{(1/2, 1/2)}(\varphi, z) = z^1\tr Q(\varphi) + z^2\tr\tilde Q(\varphi),
\end{equation}
so that all operators in the free sector of the 1D theory are simply powers of this operator:
\begin{equation}
\widetilde{\mathcal{O}}_{H, \text{free}}^{(j, j)}(\varphi, z) = [\widetilde{\mathcal{O}}_{H, \text{free}}^{(1/2, 1/2)}(\varphi, z)]^{2j}.
\end{equation}
The basic result is
\begin{equation}
\langle(\tr Q)^m(\varphi_1)(\tr\tilde Q)^m(\varphi_2)\rangle = \langle(\tr Q)^m(\varphi_1)(\tr\tilde Q)^m(\varphi_2)\rangle_\sigma = m!\left(-\frac{N_c\operatorname{sign}\varphi_{12}}{8\pi r}\right),
\end{equation}
by counting $m!$ equivalent contractions.  We compute the two-point functions
\begin{gather*}
\langle\widetilde{\mathcal{O}}_{H, \text{free}}^{(j, j)}(\varphi_1, z_1)\widetilde{\mathcal{O}}_{H, \text{free}}^{(j, j)}(\varphi_2, z_2)\rangle = (2j)!\left(\frac{N_c}{8\pi r}\right)^{2j}\langle z_1, z_2\rangle^{2j}(\operatorname{sign}\varphi_{12})^{2j}.
\end{gather*}
In terms of the normalized operators
\begin{equation}
\mathcal{O}_{H, \text{free}}^{(j, j)}(\varphi, z) = \frac{1}{\sqrt{(2j)!}}\left(\frac{8\pi r}{N_c}\right)^j\widetilde{\mathcal{O}}_{H, \text{free}}^{(j, j)}(\varphi, z),
\end{equation}
we then compute the three-point functions
\begin{align}
&\langle\mathcal{O}_{H, \text{free}}^{(j_1, j_1)}(\varphi_1, z_1)\mathcal{O}_{H, \text{free}}^{(j_2, j_2)}(\varphi_2, z_2)\mathcal{O}_{H, \text{free}}^{(j_3, j_3)}(\varphi_3, z_3)\rangle \\
&= \lambda_{(j_1, j_1), (j_2, j_2), (j_3, j_3)}^{\text{free}}\langle z_1, z_2\rangle^{j_{123}}\langle z_2, z_3\rangle^{j_{231}}\langle z_3, z_1\rangle^{j_{312}}(\operatorname{sign}\varphi_{12})^{j_{123}}(\operatorname{sign}\varphi_{23})^{j_{231}}(\operatorname{sign}\varphi_{31})^{j_{312}} \nonumber
\end{align}
for $j_1, j_2, j_3$ satisfying the triangle inequality, where
\begin{equation}
\lambda_{(j_1, j_1), (j_2, j_2), (j_3, j_3)}^{\text{free}} = \frac{j_{123}!j_{231}!j_{312}!}{\sqrt{(2j_1)!(2j_2)!(2j_3)!}}\binom{2j_1}{j_{123}}\binom{2j_2}{j_{231}}\binom{2j_3}{j_{312}}
\end{equation}
(compare to \eqref{structureconstants} for $N_c = 2$).  We claim that the corresponding operators on the Coulomb branch side are given by
\begin{equation}
\mathcal{O}_{C, \text{free}}^{(1/2, 1/2)}(\varphi, z) = \sqrt{\frac{2r}{N_c}}(z^1\mathcal{M}^{(-1, \vec{0})}(\varphi) + z^2\mathcal{M}^{(1, \vec{0})}(\varphi)).
\end{equation}
To see this, one can match two-point functions.  The shift operators are
\begin{align}
\mathcal{M}_N^{(-1, \vec{0})} &= \frac{1}{r^{1/2}}\sum_{I=1}^{N_c} \frac{\prod_{J\neq I} (\frac{1 + B_{IJ}}{2} - i\sigma_{IJ})}{\prod_{J\neq I} (-i\sigma_{IJ} + \frac{B_{IJ}}{2})}e^{\frac{i}{2}\partial_{\sigma_I} + \partial_{B_I}}, \\
\mathcal{M}_N^{(1, \vec{0})} &= -\frac{1}{r^{1/2}}\sum_{I=1}^{N_c} \frac{(\frac{1 - B_I}{2} + i\sigma_I)\prod_{J\neq I} (\frac{1 - B_{IJ}}{2} + i\sigma_{IJ})}{\prod_{J\neq I} (i\sigma_{IJ} - \frac{B_{IJ}}{2})}e^{-\frac{i}{2}\partial_{\sigma_I} - \partial_{B_I}},
\end{align}
from which we obtain
\[
\langle\mathcal{M}^{(\mp 1, \vec{0})}(\varphi_1)\mathcal{M}^{(\pm 1, \vec{0})}(\varphi_2)\rangle|_{\varphi_1 < \varphi_2} = \frac{Z_{S^3}[\mathcal{I}_\pm]}{Z_{S^3}}, \quad \mathcal{I}_\pm\equiv \pm\frac{1}{r}\sum_{I=1}^{N_c} \frac{(\frac{1}{2} + i\sigma_I)\prod_{J\neq I} (\frac{1}{2} + i\sigma_{IJ})^2}{\prod_{J\neq I} (i\sigma_{IJ})(1 + i\sigma_{IJ})}.
\]
Since the integrand of $Z_{S^3}$ without insertions is invariant under $\sigma_I\leftrightarrow -\sigma_I$, inserting $\mathcal{I}_\pm$ is equivalent to inserting
\begin{equation}
\frac{\mathcal{I}_\pm(\sigma_1, \ldots, \sigma_{N_c}) + \mathcal{I}_\pm(-\sigma_1, \ldots, -\sigma_{N_c})}{2} = \pm\frac{N_c}{2r} \,.
\label{trick}
\end{equation}
It follows that
\begin{equation}
\langle\mathcal{M}^{(-1, \vec{0})}(\varphi_1)\mathcal{M}^{(1, \vec{0})}(\varphi_2)\rangle = -\frac{N_c\operatorname{sign}\varphi_{12}}{2r} \,, 
\end{equation}
thus substantiating the stated map.

\subsection{An Abelian/Non-Abelian Mirror Symmetry Example}

Let us end with a simpler example where we can derive the correspondence between chiral ring generators in mirror dual pairs.  It is known that $SU(2)$ SQCD with three fundamental hypers is dual to $U(1)$ SQED with four charged hypers, because both theories are mirror dual to the $U(1)^4$ necklace quiver gauge theory \cite{Intriligator:1996ex}.  Their Coulomb branch is given by $\mathbb{C}^2/\mathbb{Z}_4$: it has three holomorphic generators $\mathcal{X}$, $\mathcal{Y}$, and $\mathcal{Z}$ subject to the chiral ring relation $\mathcal{X}\mathcal{Y} = \mathcal{Z}^4$, whose quantization is $\mathcal{X} \star \mathcal{Y} = (\mathcal{Z}^4)_\star + O(1/r)$.  The generators have dimensions $\Delta_{\mathcal{Z}} = 1$ and $\Delta_{\mathcal{X}} = \Delta_{\mathcal{Y}} = 2$.  Let us identify $\mathcal{X}$, $\mathcal{Y}$, and $\mathcal{Z}$ in the SQCD theory.

To compute correlation functions, we use that the vacuum wavefunction \eqref{vacuum} is
\begin{equation}
\Psi_0(\sigma, B) = \delta_{B, 0}\frac{[\frac{1}{2\pi}\Gamma(\frac{1}{2} - i\sigma)\Gamma(\frac{1}{2} + i\sigma)]^3}{\frac{1}{2\pi}\Gamma(1 - 2i\sigma)\Gamma(1 + 2i\sigma)} = \delta_{B, 0}\frac{\sinh(\pi\sigma)}{4\sigma\cosh^2(\pi\sigma)}
\end{equation}
and the gluing measure is
\begin{equation}
\mu(\sigma, B) = (-1)^{3|B|}(4\sigma^2 + B^2).
\end{equation}
Using $|\mathcal{W}| = 2$, this gives the $S^3$ partition function
\begin{equation}
Z = \frac{1}{2}\int d\sigma\, \mu(\sigma, 0)\Psi_0(\sigma, 0)^2 = \frac{1}{12\pi},
\end{equation}
in agreement with the $S^3$ partition function of the four-node quiver theory and SQED with four flavors (see, e.g., \cite{Dedushenko:2017avn}).

The Coulomb branch chiral ring operators are gauge-invariant products of $\Phi$ and GNO monopole operators with $b\in \mathbb{Z}$.  The smallest-dimension such operator is the GNO monopole $\mathcal{M}^{(1, -1)}$.  This operator has $\Delta = 1$, so it should correspond to $\mathcal{Z}$ in the four-node quiver theory.  Matching the normalization of the two-point function gives
\begin{equation}
\mathcal{Z} = \frac{1}{4\pi}\mathcal{M}^{(1, -1)}.
\end{equation}
There are three operators with $\Delta = 2$: $\mathcal{M}^{(1, -1)}\star \mathcal{M}^{(1, -1)}$, $\tr\Phi^2$, and the dressed monopole $\Phi\mathcal{M}^{(1, -1)}$.  Clearly, $\mathcal{M}^{(1, -1)}\star \mathcal{M}^{(1, -1)} = (4\pi)^2\mathcal{Z}\star \mathcal{Z}$, so we expect to obtain $\mathcal{X}$ and $\mathcal{Y}$ as linear combinations of $\tr\Phi^2$ and $\Phi\mathcal{M}^{(1, -1)}$.  We find that
\begin{align}
\mathcal{X} &= \frac{1}{64\pi^2}\left(\tr\Phi^2 - 4\mathcal{M}^{(1, -1)}\star \mathcal{M}^{(1, -1)} - \frac{1}{2r^2} + 4i\left(\Phi\mathcal{M}^{(1, -1)} - \frac{i}{2r}\mathcal{M}^{(1, -1)}\right)\right), \\
\mathcal{Y} &= \frac{1}{64\pi^2}\left(\tr\Phi^2 - 4\mathcal{M}^{(1, -1)}\star \mathcal{M}^{(1, -1)} - \frac{1}{2r^2} - 4i\left(\Phi\mathcal{M}^{(1, -1)} - \frac{i}{2r}\mathcal{M}^{(1, -1)}\right)\right)
\end{align}
obey the following relations:
\begin{equation}
[\mathcal{X}, \mathcal{Z}]_\star = \frac{1}{4\pi r}\mathcal{X}, \quad [\mathcal{Y}, \mathcal{Z}]_\star = -\frac{1}{4\pi r}\mathcal{Y}, \quad
\mathcal{X}\star \mathcal{Y} = \left(\mathcal{Z} + \frac{1}{8\pi r}\right)_\star^4.
\label{fournoderelations}
\end{equation}
These are precisely the relations obeyed in the four-node quiver theory.  In addition, one can check that $\langle\mathcal{X}\rangle = \langle\mathcal{Y}\rangle = \langle\mathcal{Z}\rangle = 0$, just as in the four-node quiver theory.  The last relation in \eqref{fournoderelations} shows that the Coulomb branch is indeed $\mathbb{C}^2/\mathbb{Z}_4$.

\bibliographystyle{utphys} 
\bibliography{NonAbel}

\providecommand{\href}[2]{#2}\begingroup\raggedright\begin{thebibliography}{10}

\bibitem{Borokhov:2002ib}
V.~Borokhov, A.~Kapustin, and X.-k. Wu, ``{Topological disorder operators in
  three-dimensional conformal field theory},''
  \href{http://dx.doi.org/10.1088/1126-6708/2002/11/049}{{\em JHEP} {\bfseries
  11} (2002) 049},
\href{http://arxiv.org/abs/hep-th/0206054}{{\ttfamily arXiv:hep-th/0206054
  [hep-th]}}.

\bibitem{Son:2015xqa}
D.~T. Son, ``{Is the Composite Fermion a Dirac Particle?},''
  \href{http://dx.doi.org/10.1103/PhysRevX.5.031027}{{\em Phys. Rev.}
  {\bfseries X5} no.~3, (2015) 031027},
\href{http://arxiv.org/abs/1502.03446}{{\ttfamily arXiv:1502.03446
  [cond-mat.mes-hall]}}.

\bibitem{Aharony:2015mjs}
O.~Aharony, ``{Baryons, monopoles and dualities in Chern-Simons-matter
  theories},'' \href{http://dx.doi.org/10.1007/JHEP02(2016)093}{{\em JHEP}
  {\bfseries 02} (2016) 093},
\href{http://arxiv.org/abs/1512.00161}{{\ttfamily arXiv:1512.00161 [hep-th]}}.

\bibitem{Karch:2016sxi}
A.~Karch and D.~Tong, ``{Particle-Vortex Duality from 3d Bosonization},''
  \href{http://dx.doi.org/10.1103/PhysRevX.6.031043}{{\em Phys. Rev.}
  {\bfseries X6} no.~3, (2016) 031043},
\href{http://arxiv.org/abs/1606.01893}{{\ttfamily arXiv:1606.01893 [hep-th]}}.

\bibitem{Murugan:2016zal}
J.~Murugan and H.~Nastase, ``{Particle-vortex duality in topological insulators
  and superconductors},'' \href{http://dx.doi.org/10.1007/JHEP05(2017)159}{{\em
  JHEP} {\bfseries 05} (2017) 159},
\href{http://arxiv.org/abs/1606.01912}{{\ttfamily arXiv:1606.01912 [hep-th]}}.

\bibitem{Seiberg:2016gmd}
N.~Seiberg, T.~Senthil, C.~Wang, and E.~Witten, ``{A Duality Web in 2+1
  Dimensions and Condensed Matter Physics},''
  \href{http://dx.doi.org/10.1016/j.aop.2016.08.007}{{\em Annals Phys.}
  {\bfseries 374} (2016) 395--433},
\href{http://arxiv.org/abs/1606.01989}{{\ttfamily arXiv:1606.01989 [hep-th]}}.

\bibitem{Hsin:2016blu}
P.-S. Hsin and N.~Seiberg, ``{Level/rank Duality and Chern-Simons-Matter
  Theories},'' \href{http://dx.doi.org/10.1007/JHEP09(2016)095}{{\em JHEP}
  {\bfseries 09} (2016) 095},
\href{http://arxiv.org/abs/1607.07457}{{\ttfamily arXiv:1607.07457 [hep-th]}}.

\bibitem{Radicevic:2016wqn}
D.~Radicevic, D.~Tong, and C.~Turner, ``{Non-Abelian 3d Bosonization and
  Quantum Hall States},'' \href{http://dx.doi.org/10.1007/JHEP12(2016)067}{{\em
  JHEP} {\bfseries 12} (2016) 067},
\href{http://arxiv.org/abs/1608.04732}{{\ttfamily arXiv:1608.04732 [hep-th]}}.

\bibitem{Kachru:2016rui}
S.~Kachru, M.~Mulligan, G.~Torroba, and H.~Wang, ``{Bosonization and Mirror
  Symmetry},'' \href{http://dx.doi.org/10.1103/PhysRevD.94.085009}{{\em Phys.
  Rev.} {\bfseries D94} no.~8, (2016) 085009},
\href{http://arxiv.org/abs/1608.05077}{{\ttfamily arXiv:1608.05077 [hep-th]}}.

\bibitem{Kachru:2016aon}
S.~Kachru, M.~Mulligan, G.~Torroba, and H.~Wang, ``{Nonsupersymmetric dualities
  from mirror symmetry},''
  \href{http://dx.doi.org/10.1103/PhysRevLett.118.011602}{{\em Phys. Rev.
  Lett.} {\bfseries 118} no.~1, (2017) 011602},
\href{http://arxiv.org/abs/1609.02149}{{\ttfamily arXiv:1609.02149 [hep-th]}}.

\bibitem{Karch:2016aux}
A.~Karch, B.~Robinson, and D.~Tong, ``{More Abelian Dualities in 2+1
  Dimensions},'' \href{http://dx.doi.org/10.1007/JHEP01(2017)017}{{\em JHEP}
  {\bfseries 01} (2017) 017},
\href{http://arxiv.org/abs/1609.04012}{{\ttfamily arXiv:1609.04012 [hep-th]}}.

\bibitem{Metlitski:2016dht}
M.~A. Metlitski, A.~Vishwanath, and C.~Xu, ``{Duality and bosonization of (2+1)
  -dimensional Majorana fermions},''
  \href{http://dx.doi.org/10.1103/PhysRevB.95.205137}{{\em Phys. Rev.}
  {\bfseries B95} no.~20, (2017) 205137},
\href{http://arxiv.org/abs/1611.05049}{{\ttfamily arXiv:1611.05049
  [cond-mat.str-el]}}.

\bibitem{Aharony:2016jvv}
O.~Aharony, F.~Benini, P.-S. Hsin, and N.~Seiberg, ``{Chern-Simons-matter
  dualities with $SO$ and $USp$ gauge groups},''
  \href{http://dx.doi.org/10.1007/JHEP02(2017)072}{{\em JHEP} {\bfseries 02}
  (2017) 072},
\href{http://arxiv.org/abs/1611.07874}{{\ttfamily arXiv:1611.07874
  [cond-mat.str-el]}}.

\bibitem{Benini:2017dus}
F.~Benini, P.-S. Hsin, and N.~Seiberg, ``{Comments on global symmetries,
  anomalies, and duality in (2 + 1)d},''
  \href{http://dx.doi.org/10.1007/JHEP04(2017)135}{{\em JHEP} {\bfseries 04}
  (2017) 135},
\href{http://arxiv.org/abs/1702.07035}{{\ttfamily arXiv:1702.07035
  [cond-mat.str-el]}}.

\bibitem{Komargodski:2017keh}
Z.~Komargodski and N.~Seiberg, ``{A symmetry breaking scenario for
  QCD$_{3}$},'' \href{http://dx.doi.org/10.1007/JHEP01(2018)109}{{\em JHEP}
  {\bfseries 01} (2018) 109},
\href{http://arxiv.org/abs/1706.08755}{{\ttfamily arXiv:1706.08755 [hep-th]}}.

\bibitem{Murthy:1989ps}
G.~Murthy and S.~Sachdev, ``{Action of Hedgehog Instantons in the Disordered
  Phase of the (2+1)-dimensional $\CP^N$ Model},''
\href{http://dx.doi.org/10.1016/0550-3213(90)90670-9}{{\em Nucl. Phys.}
  {\bfseries B344} (1990) 557--595}.

\bibitem{Metlitski:2008dw}
M.~A. Metlitski, M.~Hermele, T.~Senthil, and M.~P.~A. Fisher, ``{Monopoles in
  CP**(N-1) model via the state-operator correspondence},''
  \href{http://dx.doi.org/10.1103/PhysRevB.78.214418}{{\em Phys. Rev.}
  {\bfseries B78} (2008) 214418},
\href{http://arxiv.org/abs/0809.2816}{{\ttfamily arXiv:0809.2816
  [cond-mat.str-el]}}.

\bibitem{Pufu:2013vpa}
S.~S. Pufu, ``{Anomalous dimensions of monopole operators in three-dimensional
  quantum electrodynamics},''
  \href{http://dx.doi.org/10.1103/PhysRevD.89.065016}{{\em Phys. Rev.}
  {\bfseries D89} no.~6, (2014) 065016},
\href{http://arxiv.org/abs/1303.6125}{{\ttfamily arXiv:1303.6125 [hep-th]}}.

\bibitem{Dyer:2013fja}
E.~Dyer, M.~Mezei, and S.~S. Pufu, ``{Monopole Taxonomy in Three-Dimensional
  Conformal Field Theories},''
\href{http://arxiv.org/abs/1309.1160}{{\ttfamily arXiv:1309.1160 [hep-th]}}.

\bibitem{Dyer:2015zha}
E.~Dyer, M.~Mezei, S.~S. Pufu, and S.~Sachdev, ``{Scaling dimensions of
  monopole operators in the $ \mathbb{C}{\mathrm{\mathbb{P}}}^{N_b-1} $ theory
  in 2 $+$ 1 dimensions},'' \href{http://dx.doi.org/10.1007/JHEP03(2016)111,
  10.1007/JHEP06(2015)037}{{\em JHEP} {\bfseries 06} (2015) 037},
  \href{http://arxiv.org/abs/1504.00368}{{\ttfamily arXiv:1504.00368
  [hep-th]}}.
[Erratum: JHEP03,111(2016)].

\bibitem{Chester:2015wao}
S.~M. Chester, M.~Mezei, S.~S. Pufu, and I.~Yaakov, ``{Monopole operators from
  the $4-\epsilon$ expansion},''
  \href{http://dx.doi.org/10.1007/JHEP12(2016)015}{{\em JHEP} {\bfseries 12}
  (2016) 015},
\href{http://arxiv.org/abs/1511.07108}{{\ttfamily arXiv:1511.07108 [hep-th]}}.

\bibitem{2013PhRvL.111m7202B}
M.~S. {Block}, R.~G. {Melko}, and R.~K. {Kaul}, ``{Fate of CP$^{N-1}$ Fixed
  Points with q Monopoles},''
  \href{http://dx.doi.org/10.1103/PhysRevLett.111.137202}{{\em Physical Review
  Letters} {\bfseries 111} no.~13, (Sept., 2013) 137202},
  \href{http://arxiv.org/abs/1307.0519}{{\ttfamily arXiv:1307.0519
  [cond-mat.str-el]}}.

\bibitem{2015arXiv150205128K}
R.~K. Kaul and M.~S. Block, ``{Numerical studies of various N{\'{e}}el-{VBS}
  transitions in {SU}(N) anti-ferromagnets},''
  \href{http://dx.doi.org/10.1088/1742-6596/640/1/012041}{{\em Journal of
  Physics: Conference Series} {\bfseries 640} (Sep, 2015) 012041}.
  \url{https://doi.org/10.1088%2F1742-6596%2F640%2F1%2F012041}.

\bibitem{Benini:2009qs}
F.~Benini, C.~Closset, and S.~Cremonesi, ``{Chiral flavors and M2-branes at
  toric CY4 singularities},''
  \href{http://dx.doi.org/10.1007/JHEP02(2010)036}{{\em JHEP} {\bfseries 1002}
  (2010) 036},
\href{http://arxiv.org/abs/0911.4127}{{\ttfamily arXiv:0911.4127 [hep-th]}}.

\bibitem{Benini:2011cma}
F.~Benini, C.~Closset, and S.~Cremonesi, ``{Quantum moduli space of
  Chern-Simons quivers, wrapped D6-branes and AdS4/CFT3},''
  \href{http://dx.doi.org/10.1007/JHEP09(2011)005}{{\em JHEP} {\bfseries 09}
  (2011) 005},
\href{http://arxiv.org/abs/1105.2299}{{\ttfamily arXiv:1105.2299 [hep-th]}}.

\bibitem{Imamura:2011su}
Y.~Imamura and S.~Yokoyama, ``{Index for three dimensional superconformal field
  theories with general R-charge assignments},''
  \href{http://dx.doi.org/10.1007/JHEP04(2011)007}{{\em JHEP} {\bfseries 04}
  (2011) 007},
\href{http://arxiv.org/abs/1101.0557}{{\ttfamily arXiv:1101.0557 [hep-th]}}.

\bibitem{Kim:2009wb}
S.~Kim, ``{The Complete superconformal index for N=6 Chern-Simons theory},''
  \href{http://dx.doi.org/10.1016/j.nuclphysb.2012.07.015,
  10.1016/j.nuclphysb.2009.06.025}{{\em Nucl. Phys.} {\bfseries B821} (2009)
  241--284}, \href{http://arxiv.org/abs/0903.4172}{{\ttfamily arXiv:0903.4172
  [hep-th]}}.
[Erratum: Nucl. Phys.B864,884(2012)].

\bibitem{Aharony:2015pla}
O.~Aharony, P.~Narayan, and T.~Sharma, ``{On monopole operators in
  supersymmetric Chern-Simons-matter theories},''
  \href{http://dx.doi.org/10.1007/JHEP05(2015)117}{{\em JHEP} {\bfseries 05}
  (2015) 117},
\href{http://arxiv.org/abs/1502.00945}{{\ttfamily arXiv:1502.00945 [hep-th]}}.

\bibitem{Kapustin:2009kz}
A.~Kapustin, B.~Willett, and I.~Yaakov, ``{Exact Results for Wilson Loops in
  Superconformal Chern-Simons Theories with Matter},''
  \href{http://dx.doi.org/10.1007/JHEP03(2010)089}{{\em JHEP} {\bfseries 03}
  (2010) 089},
\href{http://arxiv.org/abs/0909.4559}{{\ttfamily arXiv:0909.4559 [hep-th]}}.

\bibitem{Gaiotto:2008ak}
D.~Gaiotto and E.~Witten, ``{S-Duality of Boundary Conditions In N=4 Super
  Yang-Mills Theory},''
  \href{http://dx.doi.org/10.4310/ATMP.2009.v13.n3.a5}{{\em Adv. Theor. Math.
  Phys.} {\bfseries 13} no.~3, (2009) 721--896},
\href{http://arxiv.org/abs/0807.3720}{{\ttfamily arXiv:0807.3720 [hep-th]}}.

\bibitem{Aharony:2008ug}
O.~Aharony, O.~Bergman, D.~L. Jafferis, and J.~Maldacena, ``{N=6 superconformal
  Chern-Simons-matter theories, M2-branes and their gravity duals},''
  \href{http://dx.doi.org/10.1088/1126-6708/2008/10/091}{{\em JHEP} {\bfseries
  10} (2008) 091},
\href{http://arxiv.org/abs/0806.1218}{{\ttfamily arXiv:0806.1218 [hep-th]}}.

\bibitem{Aharony:2008gk}
O.~Aharony, O.~Bergman, and D.~L. Jafferis, ``{Fractional M2-branes},''
  \href{http://dx.doi.org/10.1088/1126-6708/2008/11/043}{{\em JHEP} {\bfseries
  11} (2008) 043},
\href{http://arxiv.org/abs/0807.4924}{{\ttfamily arXiv:0807.4924 [hep-th]}}.

\bibitem{Imamura:2008dt}
Y.~Imamura and K.~Kimura, ``{N=4 Chern-Simons theories with auxiliary vector
  multiplets},'' \href{http://dx.doi.org/10.1088/1126-6708/2008/10/040}{{\em
  JHEP} {\bfseries 10} (2008) 040},
\href{http://arxiv.org/abs/0807.2144}{{\ttfamily arXiv:0807.2144 [hep-th]}}.

\bibitem{Hosomichi:2008jd}
K.~Hosomichi, K.-M. Lee, S.~Lee, S.~Lee, and J.~Park, ``{N=4 Superconformal
  Chern-Simons Theories with Hyper and Twisted Hyper Multiplets},''
  \href{http://dx.doi.org/10.1088/1126-6708/2008/07/091}{{\em JHEP} {\bfseries
  07} (2008) 091},
\href{http://arxiv.org/abs/0805.3662}{{\ttfamily arXiv:0805.3662 [hep-th]}}.

\bibitem{Chester:2014mea}
S.~M. Chester, J.~Lee, S.~S. Pufu, and R.~Yacoby, ``{Exact Correlators of BPS
  Operators from the 3d Superconformal Bootstrap},''
  \href{http://dx.doi.org/10.1007/JHEP03(2015)130}{{\em JHEP} {\bfseries 03}
  (2015) 130},
\href{http://arxiv.org/abs/1412.0334}{{\ttfamily arXiv:1412.0334 [hep-th]}}.

\bibitem{Beem:2016cbd}
C.~Beem, W.~Peelaers, and L.~Rastelli, ``{Deformation quantization and
  superconformal symmetry in three dimensions},''
  \href{http://dx.doi.org/10.1007/s00220-017-2845-6}{{\em Commun. Math. Phys.}
  {\bfseries 354} no.~1, (2017) 345--392},
\href{http://arxiv.org/abs/1601.05378}{{\ttfamily arXiv:1601.05378 [hep-th]}}.

\bibitem{Dedushenko:2016jxl}
M.~Dedushenko, S.~S. Pufu, and R.~Yacoby, ``{A one-dimensional theory for Higgs
  branch operators},'' \href{http://dx.doi.org/10.1007/JHEP03(2018)138}{{\em
  JHEP} {\bfseries 03} (2018) 138},
\href{http://arxiv.org/abs/1610.00740}{{\ttfamily arXiv:1610.00740 [hep-th]}}.

\bibitem{Dedushenko:2017avn}
M.~Dedushenko, Y.~Fan, S.~S. Pufu, and R.~Yacoby, ``{Coulomb Branch Operators
  and Mirror Symmetry in Three Dimensions},''
  \href{http://dx.doi.org/10.1007/JHEP04(2018)037}{{\em JHEP} {\bfseries 04}
  (2018) 037},
\href{http://arxiv.org/abs/1712.09384}{{\ttfamily arXiv:1712.09384 [hep-th]}}.

\bibitem{Nekrasov:2002qd}
N.~A. Nekrasov, ``{Seiberg-Witten prepotential from instanton counting},''
  \href{http://dx.doi.org/10.4310/ATMP.2003.v7.n5.a4}{{\em Adv. Theor. Math.
  Phys.} {\bfseries 7} no.~5, (2003) 831--864},
\href{http://arxiv.org/abs/hep-th/0206161}{{\ttfamily arXiv:hep-th/0206161
  [hep-th]}}.

\bibitem{Nekrasov:2009rc}
N.~A. Nekrasov and S.~L. Shatashvili,
  \href{http://dx.doi.org/10.1142/9789814304634_0015}{``{Quantization of
  Integrable Systems and Four Dimensional Gauge Theories},''} in {\em
  {Proceedings, 16th International Congress on Mathematical Physics (ICMP09):
  Prague, Czech Republic, August 3-8, 2009}}, pp.~265--289.
\newblock 2009.
\newblock \href{http://arxiv.org/abs/0908.4052}{{\ttfamily arXiv:0908.4052
  [hep-th]}}.
\newblock
\url{https://inspirehep.net/record/829640/files/arXiv:0908.4052.pdf}.
\newblock

\bibitem{Nekrasov:2010ka}
N.~Nekrasov and E.~Witten, ``{The Omega Deformation, Branes, Integrability, and
  Liouville Theory},'' \href{http://dx.doi.org/10.1007/JHEP09(2010)092}{{\em
  JHEP} {\bfseries 09} (2010) 092},
\href{http://arxiv.org/abs/1002.0888}{{\ttfamily arXiv:1002.0888 [hep-th]}}.

\bibitem{Yagi:2014toa}
J.~Yagi, ``{$\Omega$-deformation and quantization},''
  \href{http://dx.doi.org/10.1007/JHEP08(2014)112}{{\em JHEP} {\bfseries 08}
  (2014) 112},
\href{http://arxiv.org/abs/1405.6714}{{\ttfamily arXiv:1405.6714 [hep-th]}}.

\bibitem{Bullimore:2016nji}
M.~Bullimore, T.~Dimofte, D.~Gaiotto, and J.~Hilburn, ``{Boundaries, Mirror
  Symmetry, and Symplectic Duality in 3d $\mathcal{N}=4$ Gauge Theory},''
  \href{http://dx.doi.org/10.1007/JHEP10(2016)108}{{\em JHEP} {\bfseries 10}
  (2016) 108},
\href{http://arxiv.org/abs/1603.08382}{{\ttfamily arXiv:1603.08382 [hep-th]}}.

\bibitem{Bullimore:2016hdc}
M.~Bullimore, T.~Dimofte, D.~Gaiotto, J.~Hilburn, and H.-C. Kim, ``{Vortices
  and Vermas},'' \href{http://dx.doi.org/10.4310/ATMP.2018.v22.n4.a1}{{\em Adv.
  Theor. Math. Phys.} {\bfseries 22} (2018) 803--917},
\href{http://arxiv.org/abs/1609.04406}{{\ttfamily arXiv:1609.04406 [hep-th]}}.

\bibitem{Dedushenko:2018tgx}
M.~Dedushenko, ``{Gluing II: Boundary Localization and Gluing Formulas},''
\href{http://arxiv.org/abs/1807.04278}{{\ttfamily arXiv:1807.04278 [hep-th]}}.

\bibitem{Yaakov:2013fza}
I.~Yaakov, ``{Redeeming Bad Theories},''
  \href{http://dx.doi.org/10.1007/JHEP11(2013)189}{{\em JHEP} {\bfseries 11}
  (2013) 189},
\href{http://arxiv.org/abs/1303.2769}{{\ttfamily arXiv:1303.2769 [hep-th]}}.

\bibitem{Assel:2017jgo}
B.~Assel and S.~Cremonesi, ``{The Infrared Physics of Bad Theories},''
  \href{http://dx.doi.org/10.21468/SciPostPhys.3.3.024}{{\em SciPost Phys.}
  {\bfseries 3} no.~3, (2017) 024},
\href{http://arxiv.org/abs/1707.03403}{{\ttfamily arXiv:1707.03403 [hep-th]}}.

\bibitem{Dey:2017fqs}
A.~Dey and P.~Koroteev, ``{Good IR Duals of Bad Quiver Theories},''
  \href{http://dx.doi.org/10.1007/JHEP05(2018)114}{{\em JHEP} {\bfseries 05}
  (2018) 114},
\href{http://arxiv.org/abs/1712.06068}{{\ttfamily arXiv:1712.06068 [hep-th]}}.

\bibitem{Assel:2018exy}
B.~Assel and S.~Cremonesi, ``{The Infrared Fixed Points of 3d $\mathcal{N}=4$
  $USp(2N)$ SQCD Theories},''
  \href{http://dx.doi.org/10.21468/SciPostPhys.5.2.015}{{\em SciPost Phys.}
  {\bfseries 5} no.~2, (2018) 015},
\href{http://arxiv.org/abs/1802.04285}{{\ttfamily arXiv:1802.04285 [hep-th]}}.

\bibitem{Jafferis:2010un}
D.~L. Jafferis, ``{The Exact Superconformal R-Symmetry Extremizes Z},''
  \href{http://dx.doi.org/10.1007/JHEP05(2012)159}{{\em JHEP} {\bfseries 05}
  (2012) 159},
\href{http://arxiv.org/abs/1012.3210}{{\ttfamily arXiv:1012.3210 [hep-th]}}.

\bibitem{Hama:2010av}
N.~Hama, K.~Hosomichi, and S.~Lee, ``{Notes on SUSY Gauge Theories on
  Three-Sphere},'' \href{http://dx.doi.org/10.1007/JHEP03(2011)127}{{\em JHEP}
  {\bfseries 03} (2011) 127},
\href{http://arxiv.org/abs/1012.3512}{{\ttfamily arXiv:1012.3512 [hep-th]}}.

\bibitem{Kapustin:2006pk}
A.~Kapustin and E.~Witten, ``{Electric-Magnetic Duality And The Geometric
  Langlands Program},''
  \href{http://dx.doi.org/10.4310/CNTP.2007.v1.n1.a1}{{\em Commun. Num. Theor.
  Phys.} {\bfseries 1} (2007) 1--236},
\href{http://arxiv.org/abs/hep-th/0604151}{{\ttfamily arXiv:hep-th/0604151
  [hep-th]}}.

\bibitem{Gomis:2011pf}
J.~Gomis, T.~Okuda, and V.~Pestun, ``{Exact Results for 't Hooft Loops in Gauge
  Theories on $S^4$},'' \href{http://dx.doi.org/10.1007/JHEP05(2012)141}{{\em
  JHEP} {\bfseries 05} (2012) 141},
\href{http://arxiv.org/abs/1105.2568}{{\ttfamily arXiv:1105.2568 [hep-th]}}.

\bibitem{Ito:2011ea}
Y.~Ito, T.~Okuda, and M.~Taki, ``{Line operators on $S^1 \times R^3$ and
  quantization of the Hitchin moduli space},''
  \href{http://dx.doi.org/10.1007/JHEP03(2016)085,
  10.1007/JHEP04(2012)010}{{\em JHEP} {\bfseries 04} (2012) 010},
  \href{http://arxiv.org/abs/1111.4221}{{\ttfamily arXiv:1111.4221 [hep-th]}}.
[Erratum: JHEP03,085(2016)].

\bibitem{Gang:2012yr}
D.~Gang, E.~Koh, and K.~Lee, ``{Line Operator Index on $S^{1}\times S^{3}$},''
  \href{http://dx.doi.org/10.1007/JHEP05(2012)007}{{\em JHEP} {\bfseries 05}
  (2012) 007},
\href{http://arxiv.org/abs/1201.5539}{{\ttfamily arXiv:1201.5539 [hep-th]}}.

\bibitem{Brennan:2018yuj}
T.~D. Brennan, A.~Dey, and G.~W. Moore, ``{On 't Hooft defects, monopole
  bubbling and supersymmetric quantum mechanics},''
  \href{http://dx.doi.org/10.1007/JHEP09(2018)014}{{\em JHEP} {\bfseries 09}
  (2018) 014},
\href{http://arxiv.org/abs/1801.01986}{{\ttfamily arXiv:1801.01986 [hep-th]}}.

\bibitem{Brennan:2018moe}
T.~D. Brennan, ``{Monopole Bubbling via String Theory},''
  \href{http://dx.doi.org/10.1007/JHEP11(2018)126}{{\em JHEP} {\bfseries 11}
  (2018) 126},
\href{http://arxiv.org/abs/1806.00024}{{\ttfamily arXiv:1806.00024 [hep-th]}}.

\bibitem{Brennan:2018rcn}
D.~T. Brennan, A.~Dey, and G.~W. Moore, ``{'t Hooft Defects and Wall Crossing
  in SQM},''
\href{http://arxiv.org/abs/1810.07191}{{\ttfamily arXiv:1810.07191 [hep-th]}}.

\bibitem{Bullimore:2015lsa}
M.~Bullimore, T.~Dimofte, and D.~Gaiotto, ``{The Coulomb Branch of 3d
  ${\mathcal{N}= 4}$ Theories},''
  \href{http://dx.doi.org/10.1007/s00220-017-2903-0}{{\em Commun. Math. Phys.}
  {\bfseries 354} no.~2, (2017) 671--751},
\href{http://arxiv.org/abs/1503.04817}{{\ttfamily arXiv:1503.04817 [hep-th]}}.

\bibitem{Nakajima:2015txa}
H.~Nakajima, ``{Towards a mathematical definition of Coulomb branches of
  $3$-dimensional $\mathcal{N}=4$ gauge theories, I},''
  \href{http://dx.doi.org/10.4310/ATMP.2016.v20.n3.a4}{{\em Adv. Theor. Math.
  Phys.} {\bfseries 20} (2016) 595--669},
\href{http://arxiv.org/abs/1503.03676}{{\ttfamily arXiv:1503.03676 [math-ph]}}.

\bibitem{Braverman:2016wma}
A.~Braverman, M.~Finkelberg, and H.~Nakajima, ``{Towards a mathematical
  definition of Coulomb branches of $3$-dimensional $\mathcal{N} = 4$ gauge
  theories, II},'' \href{http://dx.doi.org/10.4310/ATMP.2018.v22.n5.a1}{{\em
  Adv. Theor. Math. Phys.} {\bfseries 22} (2018) 1071--1147},
\href{http://arxiv.org/abs/1601.03586}{{\ttfamily arXiv:1601.03586 [math.RT]}}.

\bibitem{Braverman:2016pwk}
A.~Braverman, M.~Finkelberg, and H.~Nakajima, ``{Coulomb branches of $3d$
  $\mathcal N=4$ quiver gauge theories and slices in the affine Grassmannian
  (with appendices by Alexander Braverman, Michael Finkelberg, Joel Kamnitzer,
  Ryosuke Kodera, Hiraku Nakajima, Ben Webster, and Alex Weekes)},''
\href{http://arxiv.org/abs/1604.03625}{{\ttfamily arXiv:1604.03625 [math.RT]}}.

\bibitem{Braverman:2017ofm}
A.~Braverman, M.~Finkelberg, and H.~Nakajima, ``{Ring objects in the
  equivariant derived Satake category arising from Coulomb branches},''
\href{http://arxiv.org/abs/1706.02112}{{\ttfamily arXiv:1706.02112 [math.RT]}}.

\bibitem{Borokhov:2002cg}
V.~Borokhov, A.~Kapustin, and X.-k. Wu, ``{Monopole operators and mirror
  symmetry in three-dimensions},''
  \href{http://dx.doi.org/10.1088/1126-6708/2002/12/044}{{\em JHEP} {\bfseries
  12} (2002) 044},
\href{http://arxiv.org/abs/hep-th/0207074}{{\ttfamily arXiv:hep-th/0207074
  [hep-th]}}.

\bibitem{Goddard:1976qe}
P.~Goddard, J.~Nuyts, and D.~I. Olive, ``{Gauge Theories and Magnetic
  Charge},''
\href{http://dx.doi.org/10.1016/0550-3213(77)90221-8}{{\em Nucl. Phys.}
  {\bfseries B125} (1977) 1--28}.

\bibitem{Dedushenko:2018aox}
M.~Dedushenko, ``{Gluing I: Integrals and Symmetries},''
\href{http://arxiv.org/abs/1807.04274}{{\ttfamily arXiv:1807.04274 [hep-th]}}.

\bibitem{DiPietro:2014bca}
L.~Di~Pietro and Z.~Komargodski, ``{Cardy formulae for SUSY theories in $d =$ 4
  and $d =$ 6},'' \href{http://dx.doi.org/10.1007/JHEP12(2014)031}{{\em JHEP}
  {\bfseries 12} (2014) 031},
\href{http://arxiv.org/abs/1407.6061}{{\ttfamily arXiv:1407.6061 [hep-th]}}.

\bibitem{Benini:2016qnm}
F.~Benini and B.~Le~Floch, ``{Supersymmetric localization in two dimensions},''
  in {\em Localization techniques in quantum field theories}.
\newblock 2016.
\newblock \href{http://arxiv.org/abs/1608.02955}{{\ttfamily arXiv:1608.02955
  [hep-th]}}.
\newblock
\url{http://inspirehep.net/record/1480382/files/arXiv:1608.02955.pdf}.
\newblock

\bibitem{invariants}
R.~Kane, {\em Reflection Groups and Invariant Theory}.
\newblock Springer US, 2011.

\bibitem{Dimofte:2018abu}
T.~Dimofte and N.~Garner, ``{Coulomb Branches of Star-Shaped Quivers},''
  \href{http://dx.doi.org/10.1007/JHEP02(2019)004}{{\em JHEP} {\bfseries 02}
  (2019) 004},
\href{http://arxiv.org/abs/1808.05226}{{\ttfamily arXiv:1808.05226 [hep-th]}}.

\bibitem{Lichnerowicz1979}
A.~Lichnerowicz, ``Existence and equivalence of twisted products on a
  symplectic manifold,'' \href{http://dx.doi.org/10.1007/BF00401931}{{\em
  Letters in Mathematical Physics} {\bfseries 3} no.~6, (Nov, 1979) 495--502}.
  \url{https://doi.org/10.1007/BF00401931}.

\bibitem{fedosov1994}
B.~V. Fedosov, ``A simple geometrical construction of deformation
  quantization,'' \href{http://dx.doi.org/10.4310/jdg/1214455536}{{\em J.
  Differential Geom.} {\bfseries 40} no.~2, (1994) 213--238}.
  \url{https://doi.org/10.4310/jdg/1214455536}.

\bibitem{Nest1995}
R.~Nest and B.~Tsygan, ``Algebraic index theorem,''
  \href{http://dx.doi.org/10.1007/BF02099427}{{\em Communications in
  Mathematical Physics} {\bfseries 172} no.~2, (Sep, 1995) 223--262}.
  \url{https://doi.org/10.1007/BF02099427}.

\bibitem{NEST1995151}
R.~Nest and B.~Tsygan, ``Algebraic index theorem for families,''
  \href{http://dx.doi.org/https://doi.org/10.1006/aima.1995.1037}{{\em Advances
  in Mathematics} {\bfseries 113} no.~2, (1995) 151 -- 205}.
  \url{http://www.sciencedirect.com/science/article/pii/S0001870885710377}.

\bibitem{SB_1993-1994__36__389_0}
A.~Weinstein, ``Deformation quantization,'' in {\em S\'eminaire Bourbaki :
  volume 1993/94, expos\'es 775-789}, no.~227 in Ast\'erisque, pp.~389--409.
\newblock Soci\'et\'e math\'ematique de France, 1995.
\newblock \url{http://http://www.numdam.org/item/SB_1993-1994__36__389_0}.
\newblock talk:789.

\bibitem{Kontsevich:1997vb}
M.~Kontsevich, ``{Deformation quantization of Poisson manifolds. 1.},''
  \href{http://dx.doi.org/10.1023/B:MATH.0000027508.00421.bf}{{\em Lett. Math.
  Phys.} {\bfseries 66} (2003) 157--216},
\href{http://arxiv.org/abs/q-alg/9709040}{{\ttfamily arXiv:q-alg/9709040
  [q-alg]}}.

\bibitem{formality}
C.~Esposito, {\em Formality theory}, vol.~2 of Springer Briefs in Mathematical
  Physics.
\newblock Springer, 2015.

\bibitem{Intriligator:1996ex}
K.~A. Intriligator and N.~Seiberg, ``{Mirror symmetry in three-dimensional
  gauge theories},'' \href{http://dx.doi.org/10.1016/0370-2693(96)01088-X}{{\em
  Phys. Lett.} {\bfseries B387} (1996) 513--519},
\href{http://arxiv.org/abs/hep-th/9607207}{{\ttfamily arXiv:hep-th/9607207
  [hep-th]}}.

\bibitem{deBoer:1996mp}
J.~de~Boer, K.~Hori, H.~Ooguri, and Y.~Oz, ``{Mirror symmetry in
  three-dimensional gauge theories, quivers and D-branes},''
  \href{http://dx.doi.org/10.1016/S0550-3213(97)00125-9}{{\em Nucl. Phys.}
  {\bfseries B493} (1997) 101--147},
\href{http://arxiv.org/abs/hep-th/9611063}{{\ttfamily arXiv:hep-th/9611063
  [hep-th]}}.

\bibitem{deBoer:1996ck}
J.~de~Boer, K.~Hori, H.~Ooguri, Y.~Oz, and Z.~Yin, ``{Mirror symmetry in
  three-dimensional theories, SL(2,Z) and D-brane moduli spaces},''
  \href{http://dx.doi.org/10.1016/S0550-3213(97)00115-6}{{\em Nucl. Phys.}
  {\bfseries B493} (1997) 148--176},
\href{http://arxiv.org/abs/hep-th/9612131}{{\ttfamily arXiv:hep-th/9612131
  [hep-th]}}.

\bibitem{Bashkirov:2010kz}
D.~Bashkirov and A.~Kapustin, ``{Supersymmetry enhancement by monopole
  operators},'' \href{http://dx.doi.org/10.1007/JHEP05(2011)015}{{\em JHEP}
  {\bfseries 05} (2011) 015},
\href{http://arxiv.org/abs/1007.4861}{{\ttfamily arXiv:1007.4861 [hep-th]}}.

\bibitem{Cremonesi:2013lqa}
S.~Cremonesi, A.~Hanany, and A.~Zaffaroni, ``{Monopole operators and Hilbert
  series of Coulomb branches of $3d$ $\mathcal{N} = 4$ gauge theories},''
  \href{http://dx.doi.org/10.1007/JHEP01(2014)005}{{\em JHEP} {\bfseries 01}
  (2014) 005},
\href{http://arxiv.org/abs/1309.2657}{{\ttfamily arXiv:1309.2657 [hep-th]}}.

\bibitem{Dey:2014tka}
A.~Dey, A.~Hanany, P.~Koroteev, and N.~Mekareeya, ``{Mirror Symmetry in Three
  Dimensions via Gauged Linear Quivers},''
  \href{http://dx.doi.org/10.1007/JHEP06(2014)059}{{\em JHEP} {\bfseries 06}
  (2014) 059},
\href{http://arxiv.org/abs/1402.0016}{{\ttfamily arXiv:1402.0016 [hep-th]}}.

\bibitem{Kapustin:1999ha}
A.~Kapustin and M.~J. Strassler, ``{On mirror symmetry in three-dimensional
  Abelian gauge theories},''
  \href{http://dx.doi.org/10.1088/1126-6708/1999/04/021}{{\em JHEP} {\bfseries
  04} (1999) 021},
\href{http://arxiv.org/abs/hep-th/9902033}{{\ttfamily arXiv:hep-th/9902033
  [hep-th]}}.

\bibitem{Beem:2018fng}
C.~Beem, D.~Ben-Zvi, M.~Bullimore, T.~Dimofte, and A.~Neitzke, ``{Secondary
  products in supersymmetric field theory},''
\href{http://arxiv.org/abs/1809.00009}{{\ttfamily arXiv:1809.00009 [hep-th]}}.

\bibitem{Festuccia:2018rew}
G.~Festuccia, J.~Qiu, J.~Winding, and M.~Zabzine, ``{Twisting with a Flip (the
  Art of Pestunization)},''
\href{http://arxiv.org/abs/1812.06473}{{\ttfamily arXiv:1812.06473 [hep-th]}}.

\bibitem{Balasubramanian:2018pbp}
A.~Balasubramanian and J.~Distler, ``{Masses, Sheets and Rigid SCFTs},''
\href{http://arxiv.org/abs/1810.10652}{{\ttfamily arXiv:1810.10652 [hep-th]}}.

\bibitem{Hanany:2016ezz}
A.~Hanany and M.~Sperling, ``{Coulomb branches for rank 2 gauge groups in 3d $
  \mathcal{N}=4 $ gauge theories},''
  \href{http://dx.doi.org/10.1007/JHEP08(2016)016}{{\em JHEP} {\bfseries 08}
  (2016) 016},
\href{http://arxiv.org/abs/1605.00010}{{\ttfamily arXiv:1605.00010 [hep-th]}}.

\bibitem{Hanany:2018xth}
A.~Hanany and D.~Miketa, ``{Nilpotent orbit Coulomb branches of types AD},''
  \href{http://dx.doi.org/10.1007/JHEP02(2019)113}{{\em JHEP} {\bfseries 02}
  (2019) 113},
\href{http://arxiv.org/abs/1807.11491}{{\ttfamily arXiv:1807.11491 [hep-th]}}.

\bibitem{Hanany:1996ie}
A.~Hanany and E.~Witten, ``{Type IIB superstrings, BPS monopoles, and
  three-dimensional gauge dynamics},''
  \href{http://dx.doi.org/10.1016/S0550-3213(97)00157-0,
  10.1016/S0550-3213(97)80030-2}{{\em Nucl. Phys.} {\bfseries B492} (1997)
  152--190},
\href{http://arxiv.org/abs/hep-th/9611230}{{\ttfamily arXiv:hep-th/9611230
  [hep-th]}}.

\bibitem{Assel:2017hck}
B.~Assel, ``{Ring Relations and Mirror Map from Branes},''
  \href{http://dx.doi.org/10.1007/JHEP03(2017)152}{{\em JHEP} {\bfseries 03}
  (2017) 152},
\href{http://arxiv.org/abs/1701.08766}{{\ttfamily arXiv:1701.08766 [hep-th]}}.

\bibitem{Agmon:2017lga}
N.~B. Agmon, S.~M. Chester, and S.~S. Pufu, ``{A new duality between $
  \mathcal{N} $ = 8 superconformal field theories in three dimensions},''
  \href{http://dx.doi.org/10.1007/JHEP06(2018)005}{{\em JHEP} {\bfseries 06}
  (2018) 005},
\href{http://arxiv.org/abs/1708.07861}{{\ttfamily arXiv:1708.07861 [hep-th]}}.

\end{thebibliography}\endgroup

\end{document}